\documentclass[conference, draftcls, onecolumn]{IEEEtran} 
\usepackage{cite}

\ifCLASSINFOpdf
\else
\fi
\usepackage{amsmath}
  \usepackage[caption=false,font=footnotesize]{subfig}
\usepackage{url}

\usepackage{amssymb}
\usepackage{amsthm}

\theoremstyle{plain}
\newtheorem{definition}{Definition}
\newtheorem{lemma}{Lemma}
\newtheorem{conjecture}{Conjecture}
\newtheorem{theorem}{Theorem}
\newtheorem{corollary}{Corollary}
\newtheorem{remark}{Remark}
\newtheorem{fact}{Fact}

\usepackage[pdftex]{graphicx}
\usepackage{subfig}
\usepackage{overpic}
\usepackage{pict2e}
\usepackage{fixltx2e}

\usepackage{color}
\definecolor{burgundy}{rgb}{0.545098,0,0}
\definecolor{navyblue}{rgb}{0.0, 0.0, 0.5}
\definecolor{leafgreen}{rgb}{0.290196, 0.470588, 0.0}
\definecolor{bluegreen}{rgb}{0, 0.470588, 0.415686}
\definecolor{zuhl}{rgb}{0.1875, 0.26171875, 0.46484375}
\definecolor{orange}{rgb}{1, 0.6470588235, 0}

\newcommand{\bvec}[1]{\mbox{\boldmath $#1$}}
\newcommand{\sbvec}[1]{\mbox{\scriptsize \boldmath $#1$}}
\newcommand{\sgn}{\operatorname{sgn}}

\allowdisplaybreaks[4]

\begin{document}
%
\title{Relations Between Conditional Shannon \\ Entropy and Expectation of $\ell_{\alpha}$-Norm}

\author{\IEEEauthorblockN{Yuta Sakai and Ken-ichi Iwata}
\IEEEauthorblockA{Department of Information Science, University of Fukui, \\
3-9-1 Bunkyo, Fukui, Fukui, 910-8507, Japan, \\
E-mail: \url{{ji140117, k-iwata}@u-fukui.ac.jp}}
}


%


\maketitle

\begin{abstract}
The paper examines relationships between the conditional Shannon entropy and the expectation of $\ell_{\alpha}$-norm for joint probability distributions.
More precisely, we investigate the tight bounds of the expectation of $\ell_{\alpha}$-norm with a fixed conditional Shannon entropy, and vice versa.
As applications of the results, we derive the tight bounds between the conditional Shannon entropy and several information measures which are determined by the expectation of $\ell_{\alpha}$-norm. 
Moreover, we apply these results to discrete memoryless channels under a uniform input distribution.
Then, we show the tight bounds of Gallager's $E_{0}$ functions with a fixed mutual information under a uniform input distribution.
\end{abstract}


%
\IEEEpeerreviewmaketitle

\section{Introduction}

Inequalities for information measures are widely used in many applications. 
In \cite{part1, part1_arxiv}, we investigated tight bounds between the Shannon entropy \cite{shannon} and the $\ell_{\alpha}$-norm, as shown in Theorems \ref{th:extremes} and \ref{th:extremes2} of Section \ref{subsect:prev}.
Using Theorems \ref{th:extremes} and \ref{th:extremes2}, we \cite{part1, part1_arxiv} showed tight bounds between the Shannon entropy and several information measures \cite{renyi, tsallis2, boekee}. 

In this study, we extend the previous work \cite{part1, part1_arxiv} from information measures of $n$-ary probability vector to \emph{conditional} information measures of joint probability distributions.
Accurately, we provide the tight bounds of the expectation of $\ell_{\alpha}$-norm with a fixed conditional Shannon entropy in Theorem \ref{th:cond_extremes}, and vice versa in Theorem \ref{th:cond_extremes2}.
Directly extending Theorem \ref{th:cond_extremes} to Corollary \ref{cor:cond_extremes}, we obtain the tight bounds of several conditional entropies, which are related to the expectation of $\ell_{\alpha}$-norm, with a fixed conditional Shannon entropy.
In Section \ref{subsect:DMC}, we consider applications of Corollary \ref{cor:cond_extremes} for discrete memoryless channels (DMCs) under a uniform input distribution.
On the other hand, Section \ref{subsect:alpha_half} provides the exact formula of the bounds of Theorems \ref{th:cond_extremes} and \ref{th:cond_extremes2} with $\alpha = \frac{1}{2}$. 


\section{Preliminaries}

\subsection{Probability distributions and its information measures}

Let $\mathcal{P}_{n}$ denotes the set of all $n$-ary probability vectors for an integer $n \ge 2$.
In particular, we define the $n$-ary equiprobable distribution
\begin{align}
\bvec{u}_{n}
\triangleq
(u_{1}, u_{2}, \dots, u_{n}) \in \mathcal{P}_{n}
\end{align}
as $u_{i} = \frac{1}{n}$ for $i \in \{ 1, 2, \dots, n \}$.
Moreover, we define the following two $n$-ary probability vectors:
(i) the $n$-ary probability vector
\begin{align}
\bvec{v}_{n}( p )
\triangleq
(v_{1}(p), v_{2}(p), \dots v_{n}(p)) \in \mathcal{P}_{n}
\end{align}
for $p \in [0, \frac{1}{n}]$ is defined as
\begin{align}
v_{i}( p )
=
\begin{cases}
1 - (n-1) p
& \mathrm{if} \ i = 1 , \\
p
& \mathrm{otherwise} ,
\end{cases}
\end{align}
and (ii) the $n$-ary probability vector%
\footnote{The definition of $\bvec{w}_{n}( \cdot )$ is similar to the definition of \cite[Eq. (26)]{verdu}.}
\begin{align}
\bvec{w}_{n}( p )
\triangleq
(w_{1}( p ), w_{2}( p ), \dots, w_{n}( p )) \in \mathcal{P}_{n}
\end{align}
for $p \in [\frac{1}{n}, 1]$ is defined as
\begin{align}
w_{i}( p )
=
\begin{cases}
p
& \mathrm{if} \ 1 \le i \le \lfloor p^{-1} \rfloor , \\
1 - \lfloor p^{-1} \rfloor p
& \mathrm{if} \ i = \lfloor p^{-1} \rfloor + 1 , \\
0
& \mathrm{otherwise} ,
\end{cases}
\end{align}
where $\lfloor \cdot \rfloor$ denotes the floor function.

For an $n$-ary random variable $X \sim \bvec{p} \in \mathcal{P}_{n}$, we define the Shannon entropy \cite{shannon} of $X \sim \bvec{p} \in \mathcal{P}_{n}$ as
\begin{align}
H( X )
=
H( \bvec{p} )
\triangleq
- \sum_{i=1}^{n} p_{i} \ln p_{i} ,
\end{align}
where $\ln$ denotes the natural logarithm and assume that
$0 \ln 0 = 0$.
Moreover, we define the $\ell_{\alpha}$-norm of $\bvec{p} \in \mathcal{P}_{n}$ as
\begin{align}
\| \bvec{p} \|_{\alpha}
\triangleq
\left( \sum_{i=1}^{n} p_{i}^{\alpha} \right)^{\frac{1}{\alpha}}
\end{align}
for $\alpha \in (0, \infty)$.
Note that $\lim_{\alpha \to \infty} \| \bvec{p} \|_{\alpha} = \| \bvec{p} \|_{\infty} \triangleq \max \{ p_{1}, p_{2}, \dots, p_{n} \}$ for $\bvec{p} \in \mathcal{P}_{n}$.


We next introduce the conditional entropies.
For a pair of random variables%
\footnote{The random variable $Y$ can be considered both as discrete and continuous.}
$(X, Y) \sim P_{X|Y} P_{Y}$ which $X$ follows an $n$-ary distribution, i.e., $P_{X|Y}( \cdot \mid y ) \in \mathcal{P}_{n}$ for all realization $y$ of $Y$, let the conditional R\'{e}nyi entropy \cite{arimoto} of order $\alpha \in (0, 1) \cup (1, \infty)$ be denoted by
\begin{align}
H_{\alpha}( X \mid Y )
\triangleq
\frac{ \alpha }{ 1 - \alpha } \ln \mathbb{E}[ \| P_{X|Y}( \cdot \mid Y ) \|_{\alpha} ]
\label{eq:cond_Renyi}
\end{align}
where $\mathbb{E}[ \cdot ]$ denotes the expectation of the random variable.
Besides, if $\alpha = 1$, then it is defined that
\begin{align}
H_{1}( X \mid Y ) = H(X \mid Y) \triangleq \mathbb{E}[ H( P_{X|Y}( \cdot \mid Y ) ) ]
\end{align}
is the conditional Shannon entropy \cite{shannon}.
In this study, we examine relationships between $H(X \mid Y)$ and $\mathbb{E}[ \| P_{X|Y}(\cdot \mid Y) \|_{\alpha} ]$ to evaluate relations between the conditional Shannon entropy and several information measures which are related to the expectations of $\ell_{\alpha}$-norm. 

\subsection{Bounds on Shannon entropy and $\ell_{\alpha}$-norm}
\label{subsect:prev}

In this subsection, we introduce the results of our previous work \cite{part1, part1_arxiv}. 
For simplicity, we define
$
H_{\sbvec{v}_{n}}( p )
\triangleq
H( \bvec{v}_{n}( p ) )
$
and
$
H_{\sbvec{w}_{n}}( p )
\triangleq
H( \bvec{w}_{n}( p ) )
$.
Moreover, we denote by $H_{\sbvec{v}_{n}}^{-1} : [0, \ln n] \to [0, \frac{1}{n}]$ the inverse function of $H_{\sbvec{v}_{n}}( p )$ for $p \in [0, \frac{1}{n}]$
and we also denote by $H_{\sbvec{w}_{n}}^{-1} : [0, \ln n] \to [\frac{1}{n}, 1]$ the inverse function of $H_{\sbvec{w}_{n}}( p )$ for $p \in [\frac{1}{n}, 1]$.
The following two theorems were derived in \cite{part1, part1_arxiv}.

\begin{theorem}
\label{th:extremes}
Let $\bar{\bvec{v}}_{n}( \bvec{p} ) \triangleq \bvec{v}_{n}( H_{\sbvec{v}_{n}}^{-1}( H( \bvec{p} ) ) )$ and $\bar{\bvec{w}}_{n}( \bvec{p} ) \triangleq \bvec{w}_{n}( H_{\sbvec{w}_{n}}^{-1}( H( \bvec{p} ) ) )$.
Then, we observe that
\begin{align}
\| \bar{\bvec{w}}_{n}( \bvec{p} ) \|_{\alpha} \le \| \bvec{p} \|_{\alpha} \le \| \bar{\bvec{v}}_{n}( \bvec{p} ) \|_{\alpha}
\label{ineq:extremes}
\end{align}
for any $n \ge 2$, any $\bvec{p} \in \mathcal{P}_{n}$, and any $\alpha \in (0, \infty)$.
\end{theorem}

\begin{theorem}
\label{th:extremes2}
Let $p \in [0, \frac{1}{n}]$ and $p^{\prime} \in [\frac{1}{n}, 1]$ be chosen to satisfy
\begin{align}
\| \bvec{v}_{n}( p ) \|_{\alpha}
=
\| \bvec{p} \|_{\alpha}
=
\| \bvec{w}_{n}( p^{\prime} ) \|_{\alpha}
\label{ineq:norm_v_to_w}
\end{align}
for fixed $n \ge 2$, $\bvec{p} \in \mathcal{P}_{n}$, and $\alpha \in (0, 1) \cup (1, \infty)$.
Then, we observe that
\begin{align}
0 < \alpha < 1
\ & \Longrightarrow \
H_{\sbvec{v}_{n}}( p ) \le H( \bvec{p} ) \le H_{\sbvec{w}_{n}}( p^{\prime} ) ,
\\
\alpha > 1
\ & \Longrightarrow \
H_{\sbvec{w}_{n}}( p^{\prime} ) \le H( \bvec{p} ) \le H_{\sbvec{v}_{n}}( p ) .
\end{align}
\end{theorem}

\begin{figure}[!t]
\centering
\subfloat[The case $\alpha = \frac{1}{2}$.]{
\begin{overpic}[width = 0.45\hsize, clip]{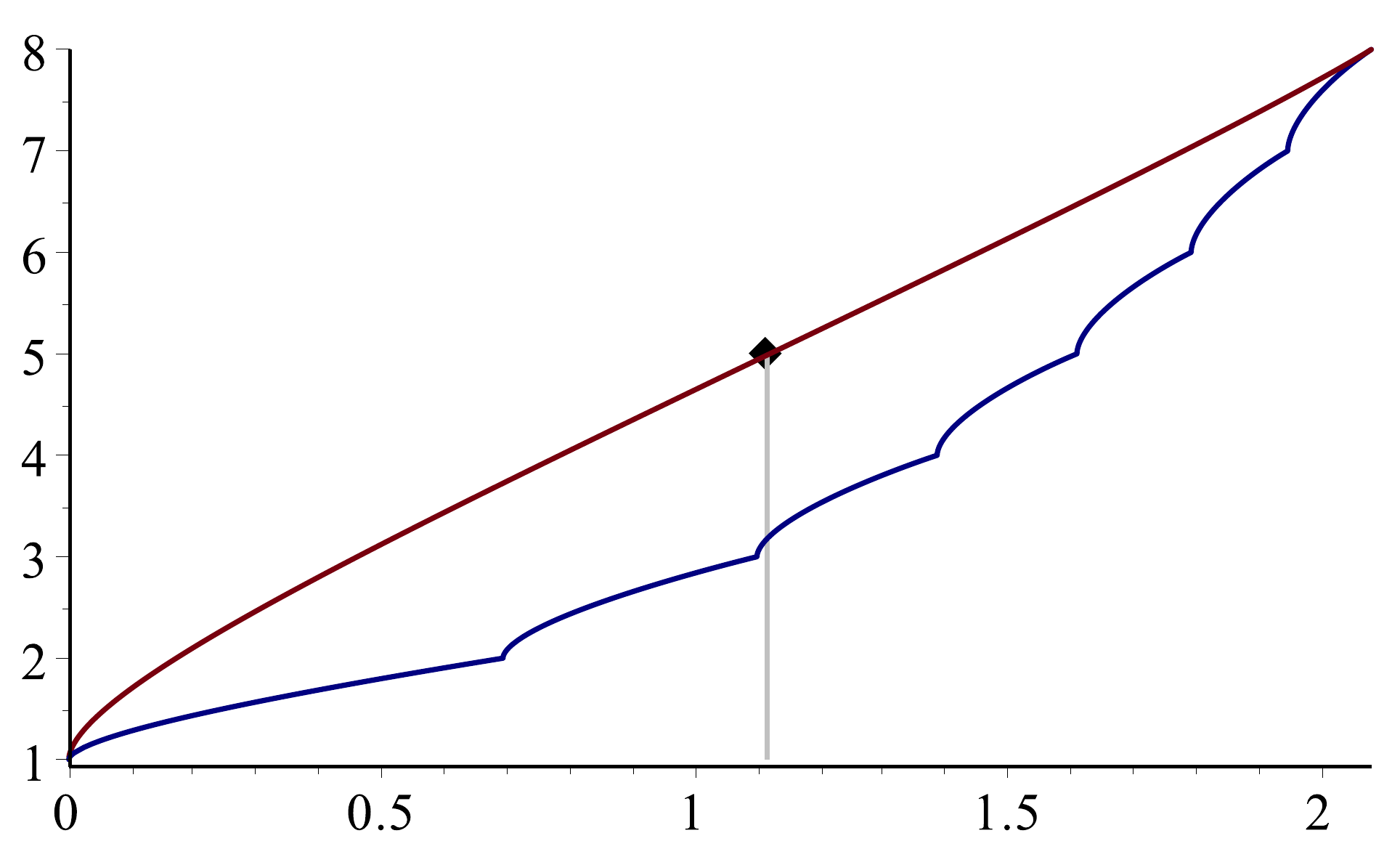}
\put(-5, 30){\rotatebox{90}{$\| \bvec{p} \|_{\alpha}$}}
\put(75, -2){$H( \bvec{p} )$}
\put(97, 1.5){\scriptsize [nats]}
\put(29, 32){\color{burgundy} $\bvec{v}_{n}( \cdot )$}
\put(70, 29){\color{navyblue} $\bvec{w}_{n}( \cdot )$}
\put(67, 12){\small $H_{\sbvec{v}_{n}}( p ) = \chi_{n}( \alpha )$}
\put(66, 13){\vector(-2, -1){11}}
\put(40, 46){inflection}
\put(46, 41){point}
\end{overpic}
}\hspace{0.05\hsize}
\subfloat[The case $\alpha = 2$.]{
\begin{overpic}[width = 0.45\hsize, clip]{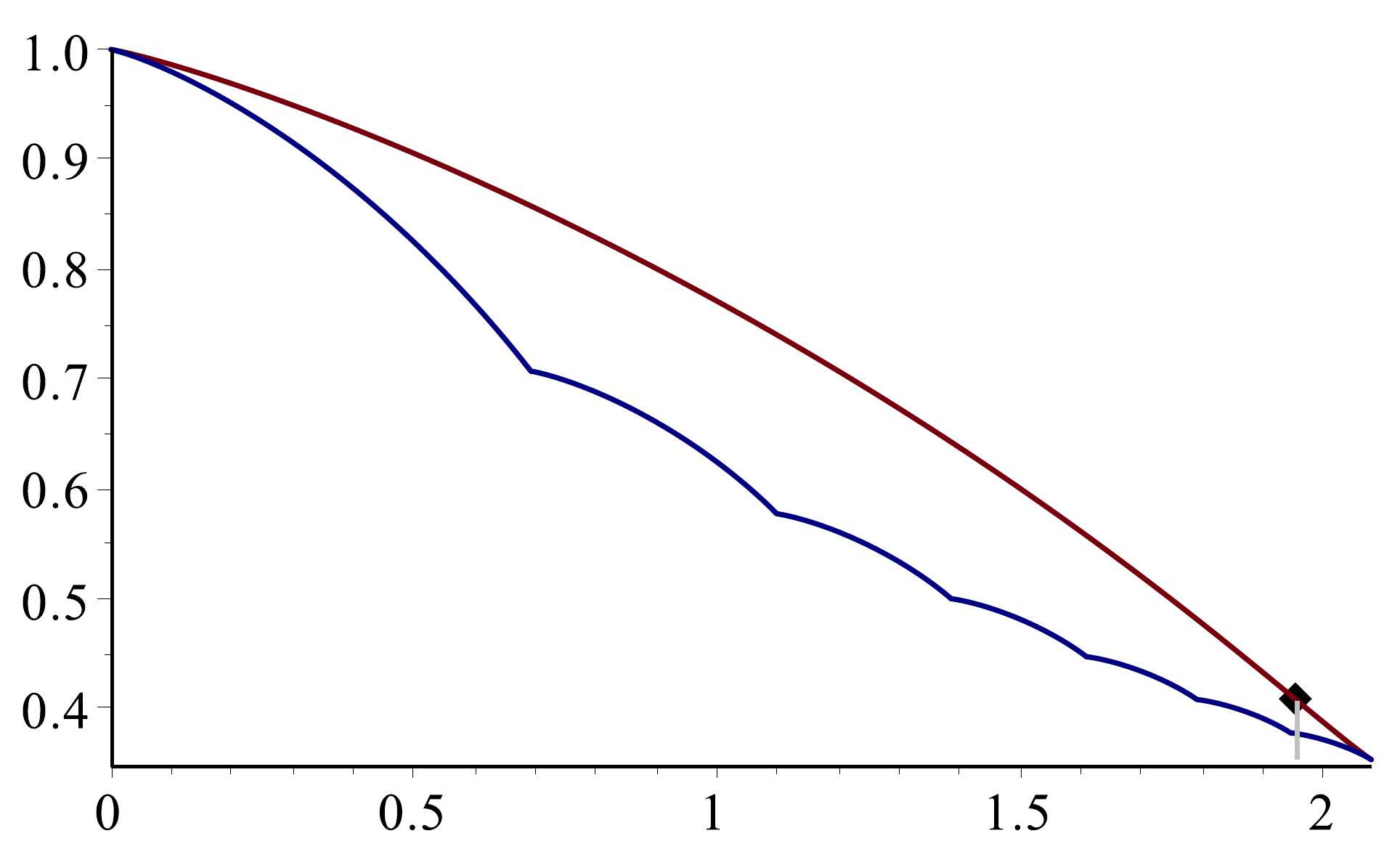}
\put(-5, 30){\rotatebox{90}{$\| \bvec{p} \|_{\alpha}$}}
\put(75, -2.5){$H( \bvec{p} )$}
\put(97, 1.5){\scriptsize [nats]}
\put(60, 41){\color{burgundy} $\bvec{v}_{n}( \cdot )$}
\put(40, 23){\color{navyblue} $\bvec{w}_{n}( \cdot )$}
\put(45, 11){\small $H_{\sbvec{v}_{n}}( p ) = \chi_{n}( \alpha )$}
\put(75, 12){\vector(4, -1){17}}
\put(88, 20){inflection}
\put(95, 15){point}
\end{overpic}
}
\caption{
Plot of the boundary of $\mathcal{R}_{n}( \alpha )$ with $n = 8$.
If $0 < \alpha < 1$, then the upper- and lower-boundaries correspond to distributions $\bvec{v}_{n}( \cdot )$ and $\bvec{w}_{n}( \cdot )$, respectively.
If $\alpha > 1$, then these correspondences are reversed.
The inflection point of the curve $p \mapsto (H_{\sbvec{v}_{n}}( p ), \| \bvec{v}_{n}( p ) \|_{\alpha})$ is at $H_{\sbvec{v}_{n}}( p ) = \chi_{n}( \alpha )$ (see Lemma \ref{lem:convex_v}).}
\label{fig:region_P6_half}
\end{figure}
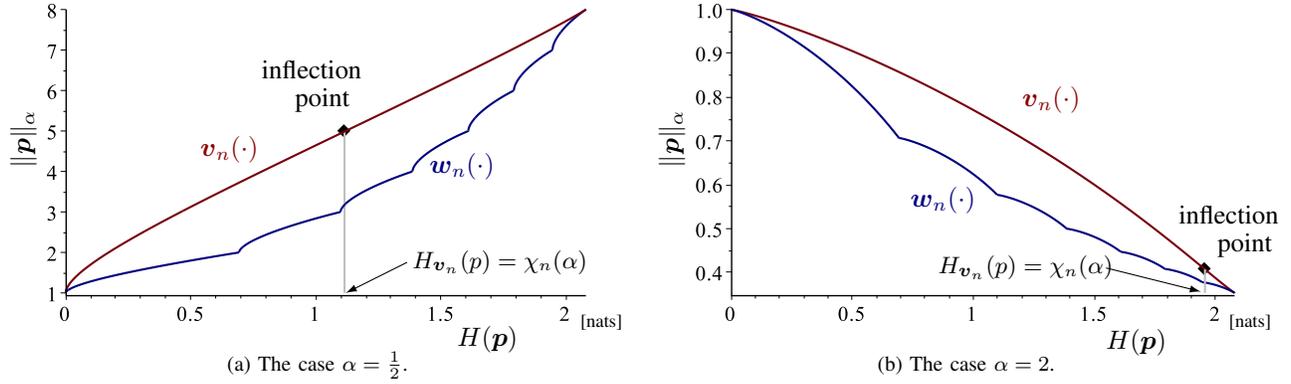

Theorems \ref{th:extremes} and \ref{th:extremes2} imply the exact feasible region of
\begin{align}
\mathcal{R}_{n}( \alpha )
\triangleq
\{ (H( \bvec{p} ), \| \bvec{p} \|_{\alpha}) \mid \bvec{p} \in \mathcal{P}_{n} \}
\label{def:region}
\end{align}
for $n \ge 2$ and $\alpha \in (0, 1) \cup (1, \infty)$.
We illustrate feasible regions of $\mathcal{R}_{n}( \alpha )$ in Fig. \ref{fig:region_P6_half}.
In this study, we extend $\mathcal{R}_{n}( \alpha )$ to the region between the \emph{conditional} Shannon entropy and the \emph{expectation} of $\ell_{\alpha}$-norm.

\section{Bounds on conditional Shannon entropy and expectation of $\ell_{\alpha}$-norm}

We define%
\footnote{Note that the alphabet $\mathcal{Y}$ of $\mathcal{R}_{n}^{\mathrm{cond}}( \alpha )$ must has elements more than one.}
\begin{align}
\mathcal{R}_{n}^{\mathrm{cond}}( \alpha )
& \triangleq
\left\{ \left. \left( \vphantom{\sum} H(X \mid Y), \mathbb{E}[ \| P_{X|Y}(\cdot \mid Y) \|_{\alpha} ] \right) \; \right| \, P_{XY} \in \mathcal{P}(\mathcal{X} \times \mathcal{Y} ) \ \mathrm{and} \ |\mathcal{X}| = n \right\} ,
\end{align}
where $\mathcal{P}( \cdot )$ denotes the set of all probability distributions on the alphabet and $| \cdot |$ denotes the cardinality of the finite set.
Using Theorems \ref{th:extremes} and \ref{th:extremes2}, Theorem \ref{th:convexhull} establishes the exact feasible regions of $\mathcal{R}_{n}^{\mathrm{cond}}( \alpha )$ as follows:

\begin{theorem}
\label{th:convexhull}
For any $n \ge 2$ and any $\alpha \in (0, \infty)$, we observe that
\begin{align}
\mathcal{R}_{n}^{\mathrm{cond}}( \alpha )
=
\mathrm{Conv} ( \mathcal{R}_{n}( \alpha ) ) ,
\end{align}
where $\mathrm{Conv}( \mathcal{R} )$ denotes the convex hull of the set $\mathcal{R}$. 
\end{theorem}

\begin{IEEEproof}[Proof of Theorem \ref{th:convexhull}]
We provide the proof of Theorem \ref{th:convexhull} in a similar manner to \cite[p. 517]{tebbe} or \cite[Theorem 1]{feder}.
First note that, for any $n \ge 2$ and any $\alpha \in (0, +\infty)$, the set $\mathcal{R}_{n}( \alpha )$ is a subset of $\mathbb{R}^{2}$ since $0 \le H( \bvec{p} ) \le \ln n$ and $\min\{ 1, n^{\frac{1}{\alpha}-1} \} \le \| \bvec{p} \|_{\alpha} \le \max\{ 1, n^{\frac{1}{\alpha}-1} \}$ for $\bvec{p} \in \mathcal{P}_{n}$.
Moreover, we see that arbitrary point of $\mathcal{R}_{n}^{\mathrm{cond}}( \alpha )$ is a convex combination of points of $\mathcal{R}_{n}( \alpha )$.
Therefore, it follows from \cite[Theorem 2.3]{rockafellar} that $\mathcal{R}_{n}^{\mathrm{cond}}( \alpha )$ is the convex hull of $\mathcal{R}_{n}( \alpha )$.
\end{IEEEproof}

Therefore, Theorem \ref{th:convexhull} obtains the exact feasible region of $\mathcal{R}_{n}^{\mathrm{cond}}( \alpha )$ from Theorems \ref{th:extremes} and \ref{th:extremes2}.
Moreover, we will investigate the exact boundary of $\mathcal{R}_{n}^{\mathrm{cond}}( \alpha )$ in the paper.
More precisely, we examine the tight bounds between the conditional Shannon entropy and the expectation of $\ell_{\alpha}$-norm, as with Theorems \ref{th:extremes} and \ref{th:extremes2}.

To accomplished the end, we now derive some lemmas.
The $\alpha$-logarithm function \cite{tsallis} is defined by
\begin{align}
\ln_{\alpha} x
\triangleq
\frac{ x^{1-\alpha} - 1 }{ 1 - \alpha }
\end{align}
for $\alpha \neq 1$ and $x > 0$;
besides, since $\lim_{\alpha \to 1} \ln_{\alpha} x = \ln x$ by L'H\^{o}pital's rule, it is defined that $\ln_{1} x \triangleq \ln x$.
For the $\alpha$-logarithm function, we can see the following useful lemma.

\begin{lemma}
\label{lem:IT_ineq}
For $\alpha < \beta$ and $x > 0$, we observe that
\begin{align}
\ln_{\alpha} x \ge \ln_{\beta} x
\end{align}
with equality if and only if $x = 1$.
\end{lemma}

\begin{IEEEproof}[Proof of Lemma \ref{lem:IT_ineq}]
We consider the monotonicity of $\ln_{\alpha} x$ with respect to $\alpha$.
Direct calculation yields
\begin{align}
\frac{ \partial \ln_{\alpha} x }{ \partial \alpha }
& =
\frac{ \partial }{ \partial \alpha } \left( \frac{ x^{1-\alpha} - 1 }{ 1 - \alpha } \right)
\\
& =
\frac{ \partial }{ \partial \alpha } \left( \frac{ x^{1-\alpha} }{ 1 - \alpha } \right) - \frac{ \partial }{ \partial \alpha } \left( \frac{ 1 }{ 1 - \alpha } \right)
\\
& =
\left( \frac{ \left( \frac{ \partial (x^{1-\alpha}) }{ \partial \alpha } \right) (1 - \alpha) - x^{1-\alpha} \left( \frac{ \partial (1-\alpha) }{ \partial \alpha } \right) }{ (1 - \alpha)^{2} } \right) - \left( - \frac{ \left( \frac{ \partial (1 - \alpha) }{ \partial \alpha } \right) }{ (1 - \alpha)^{2} } \right)
\\
& =
\left( \frac{ (\ln x) (-1) x^{1-\alpha} (1-\alpha) - x^{1-\alpha} (-1) }{ (1-\alpha)^{2} } \right) - \left( - \frac{ -1 }{ (1-\alpha)^{2} } \right)
\\
& =
\left( \frac{ - (\ln x) x^{1-\alpha} (1-\alpha) + x^{1-\alpha} }{ (1-\alpha)^{2} } \right) - \left( \frac{ 1 }{ (1-\alpha)^{2} } \right)
\\
& =
\frac{ - (\ln x) x^{1-\alpha} (1-\alpha) + x^{1-\alpha} - 1 }{ (1-\alpha)^{2} }
\\
& =
\frac{ - (\ln x) x^{1-\alpha} (1-\alpha) + (1 - \alpha) \ln_{\alpha} x }{ (1-\alpha)^{2} }
\\
& =
\frac{ \ln_{\alpha} x - x^{1-\alpha} \ln x }{ 1-\alpha }
\\
& =
\frac{ x^{\alpha} \ln_{\alpha} x - x \ln x }{ x^{\alpha} (1-\alpha) } .
\label{eq:diff1_lnq} 
\end{align}
Then, we can see that
\begin{align}
\sgn \! \left( \frac{ \partial \ln_{\alpha} x }{ \partial \alpha } \right)
& =
\sgn \! \left( \frac{ x^{\alpha} \ln_{\alpha} x - x \ln x }{ x^{\alpha} (1-\alpha) } \right)
\\
& =
\sgn \! \left( \frac{ 1 }{ x^{\alpha} (1-\alpha) } \right) \cdot \, \sgn \! \left( \vphantom{\sum} x^{\alpha} \ln_{\alpha} x - x \ln x \right) ,
\label{eq:sign_diff1_qlog}
\end{align}
where $\sgn : \mathbb{R} \to \{ -1, 0, 1 \}$ denote the sign function, i.e.,
\begin{align}
\sgn ( x )
\triangleq
\begin{cases}
1
& \mathrm{if} \ x > 0 , \\
0
& \mathrm{if} \ x = 0 , \\
-1
& \mathrm{if} \ x < 0 .
\end{cases}
\end{align}
Thus, to check the sign of $\frac{ \partial \ln_{\alpha} x }{ \partial \alpha }$, we now examine the functions $\frac{ 1 }{ x^{\alpha} (1-\alpha) }$ and $x^{\alpha} \ln_{\alpha} x - x \ln x$.
We readily see that
\begin{align}
\sgn \! \left( \frac{ 1 }{ x^{\alpha} (1-\alpha) } \right)
=
\begin{cases}
1
& \mathrm{if} \ \alpha < 1 , \\
-1
& \mathrm{if} \ \alpha > 1
\end{cases}
\label{eq:sign_1_over_x^a(1-a)}
\end{align}
for $x > 0$.
To see the sign of $x^{\alpha} \ln_{\alpha} x - x \ln x$, we calculate the following derivatives:
\begin{align}
\frac{ \partial }{ \partial \alpha } \left( \vphantom{\sum} x^{\alpha} \ln_{\alpha} x - x \ln x \right)
& =
\frac{ \partial }{ \partial \alpha } \left( \vphantom{\sum} x^{\alpha} \ln_{\alpha} x \right)
\\
& =
\frac{ \partial }{ \partial \alpha } \left( x^{\alpha} \left( \frac{ x^{1-\alpha} - 1 }{ 1 - \alpha } \right) \right)
\\
& =
\frac{ \partial }{ \partial \alpha } \left( \frac{ x - x^{\alpha} }{ 1 - \alpha } \right)
\\
& =
\frac{ \partial }{ \partial \alpha } \left( \frac{ x }{ 1 - \alpha } \right) - \frac{ \partial }{ \partial \alpha } \left( \frac{ x^{\alpha} }{ 1 - \alpha } \right)
\\
& =
\left( - \frac{ - x }{ (1 - \alpha)^{2} } \right) - \left( \frac{ \left( \frac{ \partial x^{\alpha} }{ \partial \alpha } \right) (1-\alpha) - x^{\alpha} \left( \frac{ \partial (1-\alpha) }{ \partial \alpha } \right) }{ (1 - \alpha)^{2} } \right)
\\
& =
\frac{ x }{ (1 - \alpha)^{2} } - \frac{ (\ln x) x^{\alpha} (1-\alpha) + x^{\alpha} }{ (1 - \alpha)^{2} }
\\
& =
\frac{ x - (\ln x) x^{\alpha} (1-\alpha) - x^{\alpha} }{ (1 - \alpha)^{2} }
\\
& =
\frac{ x^{\alpha} (x^{1-\alpha} - 1) - x^{\alpha} (1-\alpha) \ln x }{ (1 - \alpha)^{2} }
\\
& =
\frac{ x^{\alpha} (1 - \alpha) \ln_{\alpha} x - x^{\alpha} (1-\alpha) \ln x }{ (1 - \alpha)^{2} }
\\
& =
\frac{ x^{\alpha} \ln_{\alpha} x - x^{\alpha} \ln x }{ 1 - \alpha }
\\
& =
\frac{ x^{\alpha} (\ln_{\alpha} x - \ln x) }{ 1 - \alpha } ,
\label{eq:diff1_partial_tsallis}
\\
\frac{ \partial (\ln_{\alpha} x - \ln x) }{ \partial x }
& =
\frac{ \partial \ln_{\alpha} x }{ \partial x } - \frac{ \partial \ln x }{ \partial x }
\\
& =
\frac{ \partial }{ \partial x } \left( \frac{ x^{1-\alpha} - 1 }{ 1 - \alpha } \right) - \frac{ 1 }{ x }
\\
& =
\frac{ 1 }{ 1 - \alpha } \left( \frac{ \partial x^{1-\alpha} }{ \partial x } \right) - \frac{ 1 }{ x }
\\
& =
\frac{ (1 - \alpha) x^{-\alpha} }{ 1 - \alpha } - \frac{ 1 }{ x }
\\
& =
\frac{ 1 }{ x^{\alpha} } - \frac{ 1 }{ x }
\\
& =
x^{-1} \left( \vphantom{\sum} x^{1-\alpha} - 1 \right) .
\label{eq:diff_qlog-log}
\end{align}
By the monotonicity of the exponential function, it follows from \eqref{eq:diff_qlog-log} that, if $0 < x \le 1$, then
\begin{align}
\sgn \! \left( \frac{ \partial (\ln_{\alpha} x - \ln x) }{ \partial x } \right)
=
\begin{cases}
1
& \mathrm{if} \ x \neq 1 \ \mathrm{and} \ \alpha > 1 , \\
0
& \mathrm{if} \ x = 1 \ \mathrm{or} \ \alpha = 1 , \\
-1
& \mathrm{if} \ x \neq 1 \ \mathrm{and} \ \alpha < 1 ,
\end{cases}
\label{eq:sign_gap_qlog1}
\end{align}
and, if $x \ge 1$, then
\begin{align}
\sgn \! \left( \frac{ \partial (\ln_{\alpha} x - \ln x) }{ \partial x } \right)
=
\begin{cases}
1
& \mathrm{if} \ x \neq 1 \ \mathrm{and} \ \alpha < 1 , \\
0
& \mathrm{if} \ x = 1 \ \mathrm{or} \ \alpha = 1 , \\
-1
& \mathrm{if} \ x \neq 1 \ \mathrm{and} \ \alpha > 1 .
\end{cases}
\label{eq:sign_gap_qlog2}
\end{align}
It follows from \eqref{eq:sign_gap_qlog1} and \eqref{eq:sign_gap_qlog2} that the following monotonicity hold:
\begin{itemize}
\item
if $\alpha < 1$, then $\ln_{\alpha} x - \ln x$ is strictly decreasing for $x \in (0, 1]$ and strictly increasing for $x \ge 1$, and
\item
if $\alpha > 1$, then $\ln_{\alpha} x - \ln x$ is strictly increasing for $x \in (0, 1]$ and strictly decreasing for $x \ge 1$.
\end{itemize}
Hence, we have
\begin{align}
\sgn \! \left( \vphantom{\sum} \ln_{\alpha} x - \ln x \right)
=
\begin{cases}
1
& \mathrm{if} \ x \neq 1 \ \mathrm{and} \ \alpha < 1 , \\
0
& \mathrm{if} \ x = 1 \ \mathrm{or} \ \alpha = 1 , \\
-1
& \mathrm{if} \ x \neq 1 \ \mathrm{and} \ \alpha > 1
\end{cases}
\label{eq:gap_qlog_ln_sgn}
\end{align}
for $x > 0$ since $(\ln_{\alpha} x - \ln x) |_{x = 1} = 0$ for $\alpha \in (-\infty, +\infty)$.
Thus, we observe that
\begin{align}
\sgn \! \left( \frac{ \partial }{ \partial \alpha } \left( \vphantom{\sum} x^{\alpha} \ln_{\alpha} x - x \ln x \right) \right)
& \overset{\eqref{eq:diff1_partial_tsallis}}{=}
\sgn \! \left( \frac{ x^{\alpha} (\ln_{\alpha} x - \ln x) }{ 1 - \alpha } \right)
\\
& =
\sgn \! \left( \frac{ x^{\alpha} }{ 1 - \alpha } \right) \cdot \, \sgn \! \left( \vphantom{\sum} \ln_{\alpha} x - \ln x \right)
\\
& =
\begin{cases}
1
& \mathrm{if} \ x \neq 1 \ \mathrm{and} \ \alpha \neq 1 , \\
0
& \mathrm{if} \ x = 1 ,
\end{cases}
\label{eq:sign_diff1_partial_tsallis}
\end{align}
where the last equality follows from \eqref{eq:gap_qlog_ln_sgn} and
\begin{align}
\sgn \! \left( \frac{ x^{\alpha} }{ 1 - \alpha } \right)
=
\begin{cases}
1
& \mathrm{if} \ \alpha < 1 , \\
-1
& \mathrm{if} \ \alpha > 1 .
\end{cases}
\end{align}
Note that $\lim_{\alpha \to 1} \ln_{\alpha} x = \ln_{1} x = \ln x$.
It follows from \eqref{eq:sign_diff1_partial_tsallis} that $x^{\alpha} \ln_{\alpha} x - x \ln x$ with a fixed $x > 0$ is strictly increasing for $\alpha \in (-\infty, +\infty)$ unless $x = 1$.
Hence, we have
\begin{align}
\sgn \! \left( \vphantom{\sum} x^{\alpha} \ln_{\alpha} x - x \ln x \right)
=
\begin{cases}
1
& \mathrm{if} \ x \neq 1 \ \mathrm{and} \ \alpha > 1 , \\
0
& \mathrm{if} \ x = 1 \ \mathrm{or} \ \alpha = 1 , \\
-1
& \mathrm{if} \ x \neq 1 \ \mathrm{and} \ \alpha < 1
\end{cases}
\label{eq:sign_partial_tsallis}
\end{align}
for $x > 0$ since $(x^{\alpha} \ln_{\alpha} x - x \ln x) |_{\alpha = 1} = 0$ for $x > 0$.
Concluding the above calculus, we obtain
\begin{align}
\sgn \! \left( \frac{ \partial \ln_{\alpha} x }{ \partial \alpha } \right)
& \overset{\eqref{eq:sign_diff1_qlog}}{=}
\sgn \! \left( \frac{ 1 }{ x^{\alpha} (1-\alpha) } \right) \cdot \, \sgn \! \left( \vphantom{\sum} x^{\alpha} \ln_{\alpha} x - x \ln x \right)
\\
& =
\begin{cases}
0
& \mathrm{if} \ x = 1 , \\
-1
& \mathrm{if} \ x \neq 1 \ \mathrm{and} \ \alpha \neq 1 ,
\end{cases}
\end{align}
where the last equality follows from \eqref{eq:sign_1_over_x^a(1-a)} and \eqref{eq:sign_partial_tsallis}.
Therefore, we have that $\ln_{\alpha} x$ with a fixed $x > 0$ is strictly decreasing for $\alpha \in (-\infty, +\infty)$ unless $x = 1$, which implies Lemma \ref{lem:IT_ineq}.
\end{IEEEproof}

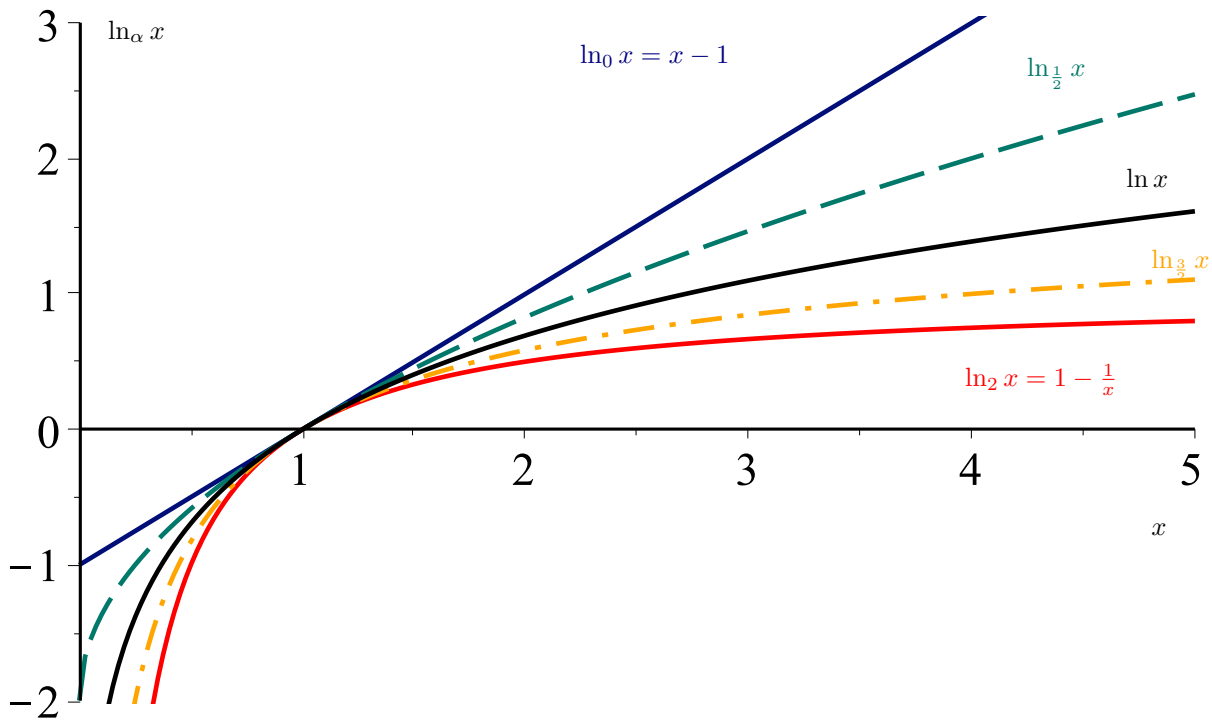
\begin{figure}[!t]
\centering
\begin{overpic}[width = 1\hsize, clip]{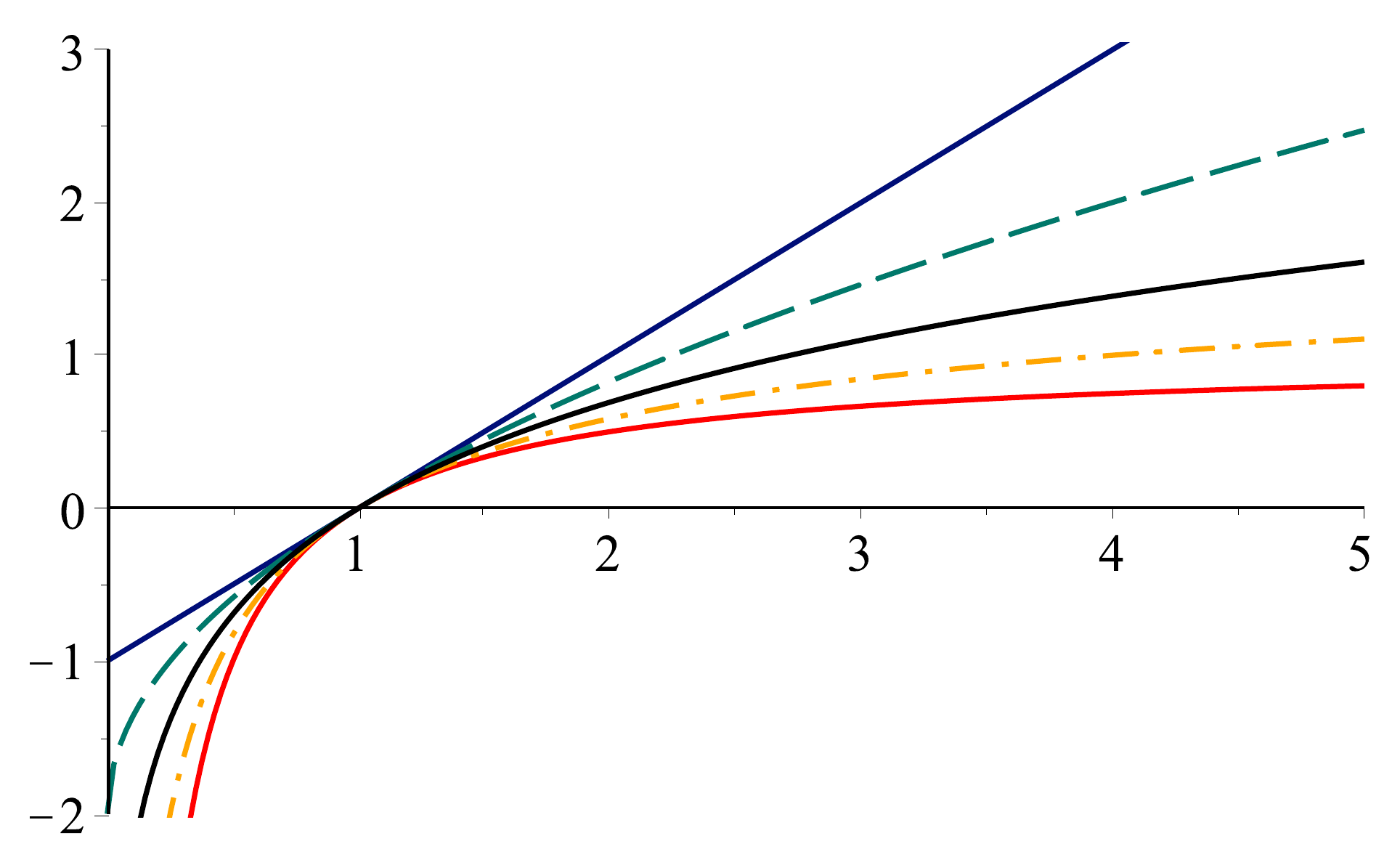}
\put(94, 17){$x$}
\put(10, 57){$\ln_{\alpha} x$}
\put(48, 55){\color{navyblue} $\ln_{0} x = x - 1$}
\put(84, 54){\color{bluegreen} $\ln_{\frac{1}{2}} x$}
\put(92, 45){$\ln x$}
\put(94, 38.5){\color{orange} $\ln_{\frac{3}{2}} x$}
\put(79, 29){\color{red} $\ln_{2} x = 1 - \frac{1}{x}$}
\end{overpic}
\caption{Plots of $\alpha$-logarithm functions with $\alpha \in \{ 0, \frac{1}{2}, 1, \frac{3}{2}, 2 \}$.}
\label{fig:qlog}
\end{figure}

Note that it is easy to see that
\begin{align}
\ln_{0} x
& =
x - 1 ,
\\
\ln_{1} x
& =
\ln x ,
\\
\ln_{2} x
& =
1 - \frac{1}{x}
\end{align}
for $x > 0$;
that is, Lemma \ref{lem:IT_ineq} implies that
\begin{align}
1 - \frac{1}{x} \le \ln x \le x - 1
\label{eq:ITineq}
\end{align}
for $x > 0$, which are famous inequalities in information theory.
We illustrate Lemma \ref{lem:IT_ineq} in Fig. \ref{fig:qlog}.
In this study, we use Lemma \ref{lem:IT_ineq} to prove Lemmas \ref{lem:convex_v} and \ref{lem:Lmin}.

The following lemma shows the convexities of $\| \bvec{v}_{n}( p ) \|_{\alpha}$ with respect to $H_{\sbvec{v}_{n}}( p )$ for $p \in [0, \frac{1}{n}]$.

\begin{lemma}
\label{lem:convex_v}
For any $\alpha \in (0, 1) \cup (1, \infty)$, $\| \bvec{v}_{2}( p ) \|_{\alpha}$ is strictly concave in $H( \bvec{v}_{2}( p ) ) \in [0, \ln 2]$.
Moreover, for any $n \ge 3$ and any $\alpha \in [\frac{1}{2}, 1) \cup (1, \infty)$, if $p \in [0, \frac{1}{n}]$, then there exists an inflection point $\chi_{n}( \alpha )$ such that
\begin{itemize}
\setlength{\itemindent}{-1em}
\item
$\| \bvec{v}_{n}( p ) \|_{\alpha}$ is strictly concave in $H_{\sbvec{v}_{n}}( p ) \in [0, \chi_{n}( \alpha )]$ and
\item
$\| \bvec{v}_{n}( p ) \|_{\alpha}$ is strictly convex in $H_{\sbvec{v}_{n}}( p ) \in [\chi_{n}( \alpha ), \ln n]$.
\end{itemize}
In addition, for $n \ge 3$ and $\alpha \in [\frac{1}{2}, 1) \cup (1, \infty)$, the value $\chi_{n}( \alpha )$ satisfies the following statements:
\begin{itemize}
\setlength{\itemindent}{-1em}
\item
$\chi_{n}( \alpha )$ is strictly increasing for $\alpha \in [\frac{1}{2}, 1) \cup (1, \infty)$,
\item
$\chi_{n}( \frac{1}{2} ) > \ln n - (1 - \frac{2}{n}) \ln (n-1)$,
\item
$\lim_{\alpha \to 1} \chi_{n}( \alpha ) = \ln 2 + \ln \sqrt{n-1}$, and
\item
$\lim_{\alpha \to \infty} \chi_{n}( \alpha ) = \ln n$.
\end{itemize}
\end{lemma}

We provide the proof of Lemma \ref{lem:convex_v} in Appendix \ref{app:convex_v} by following \cite[Appendix I]{fabregas}.
In Fig. \ref{fig:region_P6_half}, we illustrate the convexity and concavity of $\| \bvec{v}_{n}( p ) \|_{\alpha}$ in $H_{\sbvec{v}_{n}}( p ) \in [0, \ln n]$ with its inflection points.

\begin{remark}
We now consider what happens when $\alpha \to \infty$ in Lemma \ref{lem:convex_v}.
Assume that $n \ge 3$ and $p \in [0, \frac{1}{n}]$.
Note that $\frac{1}{n} \le \| \bvec{v}_{n}( p ) \|_{\infty} \le 1$.
We verify that%
\footnote{The right-hand side of \eqref{eq:infty_binary} is equivalent to the lower bound of Fano's inequality \cite{fano} since $P_{\mathrm{e}}( \sbvec{p} ) = 1 - \| \sbvec{p} \|_{\infty}$.}
\begin{align}
H_{\sbvec{v}_{n}}( p )
=
h_{2}( \| \bvec{v}_{n}( p ) \|_{\infty} ) + (1 - \| \bvec{v}_{n}( p ) \|_{\infty}) \ln (n-1) ,
\label{eq:infty_binary}
\end{align}
where $h_{2}( x ) \triangleq - x \ln x - (1-x) \ln (1-x)$ denotes the binary entropy function.
Since the right-hand side of \eqref{eq:infty_binary} is strictly decreasing for $\| \bvec{v}_{n}( p ) \|_{\infty} \in [\frac{1}{n}, 1]$ and strictly concave in $\| \bvec{v}_{n}( p ) \|_{\infty} \in [\frac{1}{n}, 1]$, we observe that $\| \bvec{v}_{n}( p ) \|_{\infty}$ is always strictly concave in $H_{\sbvec{v}_{n}}( p ) \in [0, \ln n]$.
Therefore, this concavity of $\| \bvec{v}_{n}( p ) \|_{\infty}$ is consistent with the limiting value $\lim_{\alpha \to \infty} \chi_{n}( \alpha ) = \ln n$ shown in Lemma \ref{lem:convex_v}.
\end{remark}

In addition, the following lemma shows the concavities of $\| \bvec{w}_{n}( p ) \|_{\alpha}$ with respect to $H_{\sbvec{w}_{n}}( p )$ for $p \in [\frac{1}{n}, 1]$.

\begin{lemma}
\label{lem:concave_w}
For fixed integers $n \ge m \ge 2$ and $\alpha \in (0, 1) \cup (1, \infty)$, if $p \in [\frac{1}{m}, \frac{1}{m-1}]$, then $\| \bvec{w}_{n}( p ) \|_{\alpha}$ is strictly concave in $H_{\sbvec{w}_{n}}( p ) \in [\ln (m-1), \ln m]$.
\end{lemma}

Lemma \ref{lem:concave_w} is proved in Appendix \ref{app:concave_w}.
Lemma \ref{lem:concave_w} implies that $\| \bvec{w}_{n}( p ) \|_{\alpha}$ is a piecewise concave function of $H_{\sbvec{w}_{n}}( p )$, composed of $n-1$ segments.
Note that Fig. \ref{fig:region_P6_half} is consistent with Lemma \ref{lem:concave_w}.

Employing Lemmas \ref{lem:convex_v} and \ref{lem:concave_w}, we can provide the tight bounds between the conditional Shannon entropy and the expectation of $\ell_{\alpha}$-norm, as with Theorems \ref{th:extremes} and \ref{th:extremes2}.
We now define the extremal functions $L_{\min}^{\alpha}(X \mid Y)$ and $L_{\max}^{\alpha}(X \mid Y)$ which take the boundary of $\mathcal{R}_{n}^{\mathrm{cond}}( \alpha )$ as follows:
(i) For a pair of random variable $(X, Y) \sim P_{X|Y} P_{Y}$, we define
\begin{align}
L_{\min}^{\alpha}(X \mid Y)
\triangleq
\lambda \| \bvec{u}_{m} \|_{\alpha} + (1 - \lambda) \| \bvec{u}_{m+1} \|_{\alpha} ,
\end{align}
where
\begin{align}
m
=
\left\lfloor \mathrm{e}^{H(X \mid Y)} \right\rfloor
\quad \mathrm{and} \quad
\lambda
=
\cfrac{ \ln (m+1) - H(X \mid Y) }{ \ln (m+1) - \ln m } .
\notag
\end{align}
Note that the quantity $L_{\min}^{\alpha}(X \mid Y)$ is determined by $\alpha \in (0, 1) \cup (1, \infty)$ and $H(X \mid Y) \in [0, \ln n]$.
In addition, if $|\mathcal{X}| = n$, then we define
\begin{align}
L_{\max}^{\alpha}(X \mid Y)
\triangleq
\begin{cases}
\| \hat{\bvec{v}}_{n}(X \mid Y) \|_{\alpha}
& \mathrm{if} \ H(X \mid Y) \le H_{\sbvec{v}_{n}}( p_{n}^{\ast}( \alpha ) ) , \\
T_{n, \alpha} (X \mid Y)
& \mathrm{if} \ H(X \mid Y) > H_{\sbvec{v}_{n}}( p_{n}^{\ast}( \alpha ) ) ,
\end{cases}
\label{def:Lmax}
\end{align}
where $\hat{\bvec{v}}_{n}(X \mid Y) \triangleq \bvec{v}_{n}( H_{\sbvec{v}_{n}}^{-1}(H(X \mid Y)) )$, the value $p_{n}^{\ast}( \alpha )$ denotes the root of the equation
\begin{align}
(\ln n - H_{\sbvec{v}_{n}}( p )) \left( \! \frac{ \partial \| \bvec{v}_{n}( p ) \|_{\alpha} }{ \partial H_{\sbvec{v}_{n}}( p ) } \! \right)
=
\| \bvec{u}_{n} \|_{\alpha} - \| \bvec{v}_{n}( p ) \|_{\alpha}
\label{eq:equation_p^ast}
\end{align}
with respect to $p \in (0, \frac{1}{n})$ for $n \ge 3$, and $T_{n, \alpha} (X \mid Y)$ is defined as
\begin{align}
T_{n, \alpha} (X \mid Y)
& \triangleq
\lambda \| \bvec{v}_{n}( p_{n}^{\ast}( \alpha ) ) \|_{\alpha} + (1-\lambda) \| \bvec{u}_{n} \|_{\alpha} ,
\\
\lambda
& =
\cfrac{ \ln n - H(X \mid Y) }{ \ln n - H_{\sbvec{v}_{n}}( p_{n}^{\ast}( \alpha ) ) } .
\end{align}
Note that $p_{2}^{\ast}( \alpha ) = \frac{1}{2}$ for $\alpha \in (0, 1) \cup (1, \infty)$ if $n = 2$.
In \eqref{eq:equation_p^ast}, the derivative $\frac{ \partial \| \bvec{v}_{n}( p ) \|_{\alpha} }{ \partial H_{\sbvec{v}_{n}}( p ) }$ is calculated as
\begin{align}
\frac{ \partial \| \bvec{v}_{n}( p ) \|_{\alpha} }{ \partial H_{\sbvec{v}_{n}}( p ) }
& =
\left( \vphantom{\sum} (n-1) \, p^{\alpha} + (1 - (n-1)p)^{\alpha} \right)^{\frac{1}{\alpha} - 1} \left( \frac{ p^{\alpha-1} - (1 - (n-1)p)^{\alpha-1} }{ \ln (1 - (n-1) p) - \ln p } \right) ,
\label{eq:first-order}
\end{align}
where the right-hand side of \eqref{eq:first-order} is derived in \cite[Eq. (59)]{part1_arxiv}.
We defer to present the special case of $p_{n}^{\ast}( \alpha )$ which can be solved easily in Section \ref{subsect:alpha_half}.
Note that $L_{\max}^{\alpha}(X \mid Y) = \| \hat{\bvec{v}}_{n}(X \mid Y) \|_{\alpha}$ when $n = 2$.
For $n \ge 3$, the quantity $L_{\max}^{\alpha}(X \mid Y)$ is determined by $\alpha \in [\frac{1}{2}, 1) \cup (1, \infty)$ and $H(X \mid Y) \in [0, \ln n]$;
the quantity $T_{n, \alpha} (X \mid Y)$ is linear in $H(X \mid Y) \in [H_{\sbvec{v}_{n}}( p_{n}^{\ast}( \alpha ) ), \ln n]$.

We give the properties of $L_{\min}^{\alpha}(X \mid Y)$ in the following lemma.

\begin{lemma}
\label{lem:Lmin}
$L_{\min}^{\alpha}(X \mid Y)$ is a piecewise linear function of $H(X \mid Y)$, composed of $n-1$ segments.
More precisely, we observe that
\begin{itemize}
\item[(i)]
if $\alpha \in (0, 1)$, then $L_{\min}^{\alpha}(X \mid Y)$ is strictly increasing for $H(X \mid Y) \in [0, \ln n]$,
\item[(ii)]
if $\alpha \in (1, \infty)$, then $L_{\min}^{\alpha}(X \mid Y)$ is strictly decreasing for $H(X \mid Y) \in [0, \ln n]$, and
\item[(iii)]
$L_{\min}^{\alpha}(X \mid Y)$ is convex in $H(X \mid Y) \in [0, \ln n]$.
\end{itemize}
\end{lemma}

\begin{IEEEproof}[Proof of Lemma \ref{lem:Lmin}]
Let $m = \lfloor \mathrm{e}^{H(X \mid Y)} \rfloor$, i.e., we choose an integer $m$ as $m \le \mathrm{e}^{H(X \mid Y)} < m+1$.
Note that $1 \le m < n$ since $X$ follows an $n$-ary distribution, i.e., $H(X \mid Y) \in [0, \ln n]$.
We readily see that $L_{\min}^{\alpha}(X \mid Y)$ is linear in $H(X \mid Y) \in [\ln m, \ln (m+1))$.
Moreover, we see that
\begin{align}
\lim_{H(X \mid Y) \to (\ln m)^{-}} L_{\min}^{\alpha}(X \mid Y)
& =
\lim_{H(X \mid Y) \to (\ln m)^{+}} L_{\min}^{\alpha}(X \mid Y)
\\
& =
\| \bvec{u}_{m} \|_{\alpha} ;
\end{align}
and therefore, we get that $L_{\min}^{\alpha}(X \mid Y)$ is a piecewise linear continuous function of $H(X \mid Y)$, composed of $n-1$ segments.

Next, we calculate the derivative of $L_{\min}^{\alpha}(X \mid Y)$ with respect to $H(X \mid Y)$ as follows:
\begin{align}
\frac{ \partial L_{\min}^{\alpha}(X \mid Y) }{ \partial H(X \mid Y) }
& =
\frac{ \partial \left( \vphantom{\sum} \lambda \| \bvec{u}_{m} \|_{\alpha} + (1-\lambda) \| \bvec{u}_{m+1} \|_{\alpha} \right) }{ \partial H(X \mid Y) }
\\
& =
\frac{ \| \bvec{u}_{m+1} \|_{\alpha} - \| \bvec{u}_{m} \|_{\alpha} }{ \ln (m+1) - \ln m }
\\
& \overset{\text{(a)}}{=}
\frac{ (m+1)^{\beta-1} - m^{\beta-1} }{ \ln (m+1) - \ln m }
\\
& =
\frac{ \left( \frac{m+1}{m} \right)^{\beta-1} - 1 }{ m^{1-\beta} \left( \ln \frac{ m+1 }{ m } \right) }
\\
& =
- \frac{ \left( \frac{m}{m+1} \right)^{1 - \beta} - 1 }{ m^{1-\beta} \left( \ln \frac{ m }{ m+1 } \right) }
\\
& =
- \frac{ (1 - \beta) \ln_{\beta} \frac{m}{m+1} }{ m^{1-\beta} \left( \ln \frac{m}{m+1} \right) }
\\
& =
\left( \frac{ \beta - 1 }{ m^{1-\beta} } \right) \frac{ \ln_{\beta} \frac{m}{m+1} }{ \ln \frac{m}{m+1} } ,
\label{eq:diff1_Lmin_H}
\end{align}
where (a) follows by the change of variable: $\beta = \frac{1}{\alpha}$.
Then, we can see that
\begin{align}
\sgn \! \left( \frac{ \partial L_{\min}^{\alpha}(X \mid Y) }{ \partial H(X \mid Y) } \right)
& =
\sgn \! \left( \frac{ \beta - 1 }{ m^{1-\beta} } \right) \! \cdot \sgn \! \left( \frac{ \ln_{\beta} \frac{m}{m+1} }{ \ln \frac{m}{m+1} } \right)
\\
& =
\sgn( \beta - 1 )
\\
& =
\begin{cases}
1
& \mathrm{if} \ \alpha \in (0, 1) , \\
-1
& \mathrm{if} \ \alpha \in (1, \infty) ,
\end{cases}
\end{align}
which implies the monotonicity of $L_{\min}^{\alpha}(X \mid Y)$ with respect to $H(X \mid Y)$.

Finally, we consider the monotonicity of the right-hand side of \eqref{eq:diff1_Lmin_H} with respect to $m \ge 1$.
Noting that
\begin{align}
\frac{ m }{ m+1 } < \frac{ m+1 }{ m+2 } ,
\end{align}
we can check that
\begin{align}
\left( \frac{ \beta - 1 }{ (m+1)^{1-\beta} } \right) \frac{ \ln_{\beta} \frac{m+1}{m+2} }{ \ln \frac{m+1}{m+2} } - \left( \frac{ \beta - 1 }{ m^{1-\beta} } \right) \frac{ \ln_{\beta} \frac{m}{m+1} }{ \ln \frac{m}{m+1} }
& \overset{\text{(a)}}{\ge}
\left( \frac{ \beta - 1 }{ (m+1)^{1-\beta} } \right) \frac{ \ln_{1} \frac{m+1}{m+2} }{ \ln \frac{m+1}{m+2} } - \left( \frac{ \beta - 1 }{ m^{1-\beta} } \right) \frac{ \ln_{\beta} \frac{m}{m+1} }{ \ln \frac{m}{m+1} }
\\
& =
\left( \frac{ \beta - 1 }{ (m+1)^{1-\beta} } \right) - \left( \frac{ \beta - 1 }{ m^{1-\beta} } \right) \frac{ \ln_{\beta} \frac{m}{m+1} }{ \ln \frac{m}{m+1} }
\\
& =
(\beta - 1) \left( \frac{ m^{1-\beta} \left( \ln \frac{m}{m+1} \right) - (m+1)^{1-\beta} \left( \ln_{\beta} \frac{m}{m+1} \right) }{ ( m (m+1) )^{1-\beta} \left( \ln \frac{m}{m+1} \right) } \right)
\\
& =
(\beta - 1) \left( \frac{ \ln \frac{m}{m+1} - \left( \frac{ m+1 }{ m } \right)^{1-\beta} \left( \ln_{\beta} \frac{m}{m+1} \right) }{ (m+1)^{1-\beta} \left( \ln \frac{m}{m+1} \right) } \right)
\\
& =
(\beta - 1) \left( \frac{ \ln \frac{m}{m+1} - \left( \frac{ m }{ m+1 } \right)^{\beta-1} \left( \ln_{\beta} \frac{m}{m+1} \right) }{ (m+1)^{1-\beta} \left( \ln \frac{m}{m+1} \right) } \right)
\\
& \overset{\text{(b)}}{=}
(\beta - 1) \left( \frac{ \ln \frac{m}{m+1} + \ln_{\beta} \frac{m+1}{m} }{ (m+1)^{1-\beta} \left( \ln \frac{m}{m+1} \right) } \right)
\\
& =
(\beta - 1) \left( \frac{ \ln_{\beta} \frac{m+1}{m} - \ln \frac{m+1}{m} }{ (m+1)^{1-\beta} \left( \ln \frac{m}{m+1} \right) } \right)
\\
& \overset{\text{(c)}}{\ge}
(\beta - 1) \left( \frac{ \ln_{1} \frac{m+1}{m} - \ln \frac{m+1}{m} }{ (m+1)^{1-\beta} \left( \ln \frac{m}{m+1} \right) } \right)
\\
& =
0 ,
\label{eq:sign_diff1_Lmin_H}
\end{align}
where
\begin{itemize}
\item
(a) and (c) follow by Lemma \ref{lem:IT_ineq} and
\item
(b) follows from the fact that $- \ln_{\alpha} \frac{1}{x} = x^{\alpha-1} \ln_{\alpha} x$.
\end{itemize}
Hence, for any fixed $\alpha \in (0, 1) \cup (1, \infty)$, the right-hand side of \eqref{eq:diff1_Lmin_H} is strictly increasing for $m \ge 1$.
Since a piecewise linear function is convex if its successive slopes are nondecreasing, the bounds \eqref{eq:sign_diff1_Lmin_H} implies that $L_{\min}^{\alpha}(X \mid Y)$ is convex in $H(X \mid Y) \in [0, \ln n]$.
\end{IEEEproof}

As with Lemma \ref{lem:Lmin}, we also give the properties of $L_{\max}^{\alpha}(X \mid Y)$ for $n \ge 3$ in the following lemma.

\begin{lemma}
\label{lem:Lmax}
For any $n \ge 3$ and any $\alpha \in [\frac{1}{2}, 1) \cup (1, \infty)$, the value $p_{n}^{\ast}( \alpha )$ is uniquely determined and $L_{\max}^{\alpha}(X \mid Y)$ is a piecewise continuous function of $H(X \mid Y)$.
More precisely, we observe that
\begin{itemize}
\item[(i)]
if $\alpha \in [\frac{1}{2}, 1)$, then $L_{\max}^{\alpha}(X \mid Y)$ is strictly increasing for $H(X \mid Y) \in [0, \ln n]$,
\item[(ii)]
if $\alpha \in (1, \infty)$, then $L_{\max}^{\alpha}(X \mid Y)$ is strictly decreasing for $H(X \mid Y) \in [0, \ln n]$, and
\item[(iii)]
$L_{\max}^{\alpha}(X \mid Y)$ is concave in $H(X \mid Y) \in [0, \ln n]$.
\end{itemize}
\end{lemma}

\begin{IEEEproof}[Proof of Lemma \ref{lem:Lmax}]
If $\alpha > 1$, then we see that
\begin{align}
\lim_{p \to 0^{+}} \left( \! \frac{ \partial \| \bvec{v}_{n}( p ) \|_{\alpha} }{ \partial H_{\sbvec{v}_{n}}( p ) } \! \right)
& \overset{\eqref{eq:first-order}}{=}
\lim_{p \to 0^{+}} \left( \vphantom{\sum} (n-1) \, p^{\alpha} + (1 - (n-1)p)^{\alpha} \right)^{\frac{1}{\alpha} - 1} \left( \frac{ p^{\alpha-1} - (1 - (n-1)p)^{\alpha-1} }{ \ln (1 - (n-1) p) - \ln p } \right)
\\
& =
0 .
\end{align}
On the other hand, if $\alpha < 1$, then we see that
\begin{align}
\lim_{p \to 0^{+}} \left( \! \frac{ \partial \| \bvec{v}_{n}( p ) \|_{\alpha} }{ \partial H_{\sbvec{v}_{n}}( p ) } \! \right)
& =
\lim_{p \to 0^{+}} \left( \vphantom{\sum} (n-1) \, p^{\alpha} + (1 - (n-1)p)^{\alpha} \right)^{\frac{1}{\alpha} - 1} \left( \frac{ p^{\alpha-1} - (1 - (n-1)p)^{\alpha-1} }{ \ln (1 - (n-1) p) - \ln p } \right)
\\
& \overset{\text{(a)}}{=}
\lim_{p \to 0^{+}} \left( \vphantom{\sum} (n-1) \, p^{\alpha} + (1 - (n-1)p)^{\alpha} \right)^{\frac{1}{\alpha} - 1} \left( \frac{ \frac{ \partial }{ \partial p } (p^{\alpha-1} - (1 - (n-1)p)^{\alpha-1}) }{ \frac{ \partial }{ \partial p } (\ln (1 - (n-1) p) - \ln p) } \right)
\\
& =
\lim_{p \to 0^{+}} \left( \vphantom{\sum} (n-1) \, p^{\alpha} + (1 - (n-1)p)^{\alpha} \right)^{\frac{1}{\alpha} - 1}
\notag \\
& \qquad \qquad \qquad \qquad \qquad \times
\left( \frac{ (\alpha-1) p^{\alpha-2} + (\alpha-1) (n-1) (1 - (n-1)p)^{\alpha-2} }{ \left( \frac{p}{1 - (n-1) p} \right) \left( \frac{ \partial }{ \partial p } \left( \frac{ 1 - (n-1) p}{ p } \right) \right) } \right)
\\
& =
\lim_{p \to 0^{+}} \left( \vphantom{\sum} (n-1) \, p^{\alpha} + (1 - (n-1)p)^{\alpha} \right)^{\frac{1}{\alpha} - 1} (\alpha-1)
\notag \\
& \qquad \qquad \qquad \qquad \qquad \times
\left( \frac{ p^{\alpha-2} + (n-1) (1 - (n-1)p)^{\alpha-2} }{ \left( \frac{p}{1 - (n-1) p} \right) \left( - \frac{ 1 }{ p^{2} } \right) } \right)
\\
& =
\lim_{p \to 0^{+}} \left( \vphantom{\sum} (n-1) \, p^{\alpha} + (1 - (n-1)p)^{\alpha} \right)^{\frac{1}{\alpha} - 1} (\alpha-1)
\notag \\
& \qquad \qquad \qquad \qquad \qquad \times
\left( \frac{ p^{\alpha-2} + (n-1) (1 - (n-1)p)^{\alpha-2} }{ - \frac{1}{p (1 - (n-1) p)} } \right)
\\
& =
\lim_{p \to 0^{+}} \left( \vphantom{\sum} (n-1) \, p^{\alpha} + (1 - (n-1)p)^{\alpha} \right)^{\frac{1}{\alpha} - 1} (\alpha-1)
\notag \\
& \qquad \qquad \qquad \qquad \qquad \times
\left( \vphantom{\sum} - p^{\alpha-1} (1 - (n-1) p) - (n-1) p (1 - (n-1)p)^{\alpha-1} \right)
\\
& =
+\infty ,
\end{align}
where (a) follows by L'H\^{o}pital's rule.
Note that the sign of the slope of the line through the two points $(H_{\sbvec{v}_{n}}( 0 ), \| \bvec{v}_{n}( 0 ) \|_{\alpha})$ and $(H_{\sbvec{v}_{n}}( \frac{1}{n} ), \| \bvec{v}_{n}( \frac{1}{n} ) \|_{\alpha})$ is
\begin{align}
\sgn \! \left( \frac{  \| \bvec{v}_{n}( \frac{1}{n} ) \|_{\alpha} -  \| \bvec{v}_{n}( 0 ) \|_{\alpha} }{ H_{\sbvec{v}_{n}}( \frac{1}{n} ) - H_{\sbvec{v}_{n}}( 0 ) } \right)
& =
\sgn \! \left( \frac{ n^{\frac{1}{\alpha}-1} - 1 }{ \ln n } \right)
\\
& =
\begin{cases}
1
& \mathrm{if} \ \alpha \in (0, 1) , \\
0
& \mathrm{if} \ \alpha = 1 , \\
-1
& \mathrm{if} \ \alpha \in (1, \infty) ;
\end{cases}
\end{align}
namely, we get
\begin{align}
\lim_{p \to 0^{+}} \left( \! \frac{ \partial \| \bvec{v}_{n}( p ) \|_{\alpha} }{ \partial H_{\sbvec{v}_{n}}( p ) } \! \right)
>
\frac{  \| \bvec{v}_{n}( \frac{1}{n} ) \|_{\alpha} -  \| \bvec{v}_{n}( 0 ) \|_{\alpha} }{ H_{\sbvec{v}_{n}}( \frac{1}{n} ) - H_{\sbvec{v}_{n}}( 0 ) }
\end{align}
for $\alpha \in (0, 1) \cup (1, \infty)$.
Moreover, we see from Lemma \ref{lem:convex_v} that, for a fixed $\alpha \in [\frac{1}{2}, 1) \cup (1, \infty)$, $\| \bvec{v}_{n}( p ) \|_{\alpha}$ is a strict concave/convex function of $H_{\sbvec{v}_{n}}( p ) \in [0, \ln n]$ which the inflection point is at $H_{\sbvec{v}_{n}}( p ) = \chi_{n}( \alpha )$.
Therefore, for any $\alpha \in [\frac{1}{2}, 1) \cup (1, \infty)$, there exists a unique $p_{n}^{\ast}( \alpha ) \in (0, \chi_{n}( \alpha ))$ such that the tangent line of the curve $p \mapsto (H_{\sbvec{v}_{n}}( p ), \| \bvec{v}_{n}( p ) \|_{\alpha})$ for $p \in [0, \frac{1}{n}]$ at $H_{\sbvec{v}_{n}}( p_{n}^{\ast}( \alpha ) )$ passes through the point $(H_{\sbvec{v}_{n}}( \frac{1}{n} ), \| \bvec{v}_{n}( \frac{1}{n} ) \|_{\alpha})$.
Note that this tangent line corresponds to $T_{n, \alpha} (X \mid Y)$.
It follows from \cite[Lemma 3]{part1, part1_arxiv} that
\begin{itemize}
\item
if $\alpha \in (0, 1)$, then $T_{n, \alpha} (X \mid Y)$ is strictly increasing for $H(X \mid Y) \in [H_{\sbvec{v}_{n}}(p_{n}^{\ast}( \alpha )), \ln n ]$ and
\item
if $\alpha \in (1, \infty)$, then $T_{n, \alpha} (X \mid Y)$ is strictly decreasing for $H(X \mid Y) \in [H_{\sbvec{v}_{n}}(p_{n}^{\ast}( \alpha )), \ln n ]$.
\end{itemize}
Hence, the monotonicity of $L_{\max}^{\alpha}(X \mid Y)$, i.e., (i) and (ii) of Lemma \ref{lem:Lmax}, holds.

Finally, since $p_{n}^{\ast}( \alpha ) \in (0, \chi_{n}(\alpha))$ and $L_{\max}^{\alpha}(X \mid Y)$ is linear in $H(X \mid Y) \in [H_{\sbvec{v}_{n}}(p_{n}^{\ast}( \alpha )), \ln n ]$, we obtain the concavity of $L_{\max}^{\alpha}(X \mid Y)$, i.e., (iii) of Lemma \ref{lem:Lmax}.
\end{IEEEproof}

Using $L_{\min}^{\alpha}(X \mid Y)$ and $L_{\max}^{\alpha}(X \mid Y)$, as with Theorem \ref{th:extremes}, we provide the tight bounds of the expectation of $\ell_{\alpha}$-norm with a fixed conditional Shannon entropy as follows:.

\begin{theorem}
\label{th:cond_extremes}
For $(X, Y) \sim P_{X|Y} P_{Y}$ and $\alpha \in [\frac{1}{2}, 1) \cup (1, \infty)$,
\begin{align}
L_{\min}^{\alpha}(X \mid Y)
\le
\mathbb{E}[ \| P_{X|Y}(\cdot \mid Y) \|_{\alpha} ]
\le
L_{\max}^{\alpha}(X \mid Y) .
\label{eq:cond_extremes}
\end{align}
In particular, the left-hand inequality also holds for $\alpha \in (0, \frac{1}{2})$.
Furthermore, if $|\mathcal{X}| = 2$, then the right-hand inequality also holds for $\alpha \in (0, \frac{1}{2})$.
\end{theorem}

\begin{IEEEproof}[Proof of Theorem \ref{th:cond_extremes}]
We first prove the lower bound.
Since $L_{\min}^{\alpha}(X \mid Y)$ is a piecewise linear function through the points $(H_{\sbvec{w}_{n}}( m^{-1} ), \| \bvec{w}_{n}( m^{-1} ) \|_{\alpha})$ for integers $m \in [1, n]$, it follows from Lemmas \ref{lem:concave_w} and \ref{lem:Lmin} that
\begin{align}
\| \bvec{w}_{n}( H_{\sbvec{w}_{n}}^{-1}( H(X \mid Y) ) ) \|_{\alpha}
\ge
L_{\min}^{\alpha}(X \mid Y) .
\label{eq:cond_extremes_Lmin}
\end{align}
Moreover, since $L_{\min}^{\alpha}(X \mid Y)$ is convex in $H(X \mid Y) \in [0, \ln n]$, it follows from Theorem \ref{th:convexhull} and Lemma \ref{lem:Lmin} that the right-hand side of \eqref{eq:cond_extremes_Lmin} correspond to the lower-boundary of the convex hull of $\mathcal{R}_{n}( \alpha )$.
Therefore, we have the lower bound.

We second prove the upper bound.
Since $T_{n, \alpha}(X \mid Y)$ is the tangent line of the curve $p \mapsto (H_{\sbvec{v}_{n}}( p ), \| \bvec{v}_{n}( p ) \|_{\alpha})$ for $p \in [0, \frac{1}{n}]$ such that passes through the point $(H_{\sbvec{v}_{n}}( \frac{1}{n} ), \| \bvec{v}_{n}( \frac{1}{n} ) \|_{\alpha})$, it follows from Lemma \ref{lem:convex_v} that
\begin{align}
\| \bvec{v}_{n}( H_{\sbvec{v}_{n}}^{-1}( H(X \mid Y) ) ) \|_{\alpha}
\le
L_{\max}^{\alpha}(X \mid Y) .
\label{eq:cond_extremes_Lmax}
\end{align}
As with \eqref{eq:cond_extremes_Lmin}, since $L_{\max}^{\alpha}(X \mid Y)$ is concave in $H(X \mid Y) \in [0, \ln n]$, it follows from Theorem \ref{th:convexhull} and Lemma \ref{lem:Lmax} that the right-hand side of \eqref{eq:cond_extremes_Lmax} correspond to the upper-boundary of the convex hull of $\mathcal{R}_{n}( \alpha )$.
Therefore, we have the upper bound.
That completes the proof of Theorem \ref{th:cond_extremes}.
\end{IEEEproof}

Theorem \ref{th:cond_extremes} shows the bounds of the expectation of $\ell_{\alpha}$-norm with a fixed conditional Shannon entropy.
Note that, if $|\mathcal{X}| = 2$, then Theorem \ref{th:cond_extremes} is reduced to \cite[Theorem 1]{fabregas}.
Thus, henceforth, we omit the case $|\mathcal{X}| = 2$ in the analyses of this study.
Moreover, since Lemma \ref{lem:convex_v} provides the inflection point $\chi_{n}( \alpha )$ for only $\alpha \in [\frac{1}{2}, 1) \cup (1, \infty)$, Theorem \ref{th:cond_extremes} does not establish the upper bound of the expectation of $\ell_{\alpha}$-norm for $\alpha \in (0, \frac{1}{2})$ and $n \ge 3$ with a fixed conditional Shannon entropy.

We now prove that the bounds of Theorem \ref{th:cond_extremes} is tight;
namely, we consider the two pairs of random variables $(X^{\prime}, Y^{\prime})$ and $(X^{\prime\prime}, Y^{\prime\prime})$ which attain each equality of \eqref{eq:cond_extremes}.

\begin{definition}
\label{def:RVs_prime1}
Let the pair of random variables $(X^{\prime}, Y^{\prime}) \in \mathcal{X} \times \mathcal{Y}$ be defined as follows:
Set $\mathcal{X} = \{ 1, 2, \dots, n \}$ and $\mathcal{Y} = \{ 0, 1 \}$.
We define the conditional distribution $P_{X^{\prime}|Y^{\prime}}$ as $P_{X^{\prime}|Y^{\prime}}(\cdot \mid y) = \bvec{w}_{n}( (m+y)^{-1} )$ for $y \in \mathcal{Y}$.
\end{definition}

\begin{definition}
\label{def:RVs_prime2}
Let the pair of random variables $(X^{\prime\prime}, Y^{\prime\prime}) \in \mathcal{X} \times \mathcal{Y}$ be defined as follows:
Set $\mathcal{X} = \{ 1, 2, \dots, n \}$ and $\mathcal{Y} = \{ 0, 1 \}$.
If $H(X^{\prime\prime} \mid Y^{\prime\prime}) \le H_{\sbvec{v}_{n}}( p_{n}^{\ast}( \alpha ) )$ for a given $\alpha \in [\frac{1}{2}, 1) \cup (1, \infty)$, then we set $P_{X^{\prime\prime}|Y^{\prime\prime}}(\cdot \mid y) = \bvec{v}_{n}( p )$ for all $y \in \mathcal{Y}$ and some $p \in [0, p_{n}^{\ast}( \alpha )]$.
On the other hand, if $H(X^{\prime\prime} \mid Y^{\prime\prime}) > H_{\sbvec{v}_{n}}( p_{n}^{\ast}( \alpha ) )$ for a given $\alpha \in [\frac{1}{2}, 1) \cup (1, \infty)$, then we set $P_{X^{\prime\prime}|Y^{\prime\prime}}(\cdot \mid 0) = \bvec{v}_{n}( p_{\ast}^{n}( \alpha ) )$ and $P_{X^{\prime\prime}|Y^{\prime\prime}}(\cdot \mid 1) = \bvec{u}_{n}$.
\end{definition}

In Definitions \ref{def:RVs_prime1} and \ref{def:RVs_prime2}, note that the marginal distribution $P_{Y^{\prime}}$ and $P_{Y^{\prime\prime}}$ can be chosen arbitrarily and appropriately.
For these pair of random variables $(X^{\prime}, Y^{\prime})$ and $(X^{\prime\prime}, Y^{\prime\prime})$ of Definitions \ref{def:RVs_prime1} and \ref{def:RVs_prime2}, respectively, the following fact holds.

\begin{fact}
\label{fact:RVs_prime}
For $(X^{\prime}, X^{\prime})$ and $(X^{\prime\prime}, Y^{\prime\prime})$, we observe that
$
\mathbb{E}[ \| P_{X^{\prime}|Y^{\prime}}(\cdot \mid Y^{\prime}) \|_{\alpha} ]
=
L_{\min}^{\alpha}(X^{\prime} \mid Y^{\prime})
$
and
$
\mathbb{E}[ \| P_{X^{\prime\prime}|Y^{\prime\prime}}(\cdot \mid Y^{\prime\prime}) \|_{\alpha} ]
=
L_{\max}^{\alpha}(X^{\prime\prime} \mid Y^{\prime\prime})
$.
\end{fact}

\begin{IEEEproof}[Proof of Fact \ref{fact:RVs_prime}]
For $(X^{\prime}, Y^{\prime})$ of Definition \ref{def:RVs_prime1}, it can be seen that
\begin{align}
H(X^{\prime} \mid Y^{\prime})
& =
\sum_{y \in \{ 0, 1 \}} \! P_{Y^{\prime}}( y ) H(X^{\prime} \mid Y^{\prime} = y)
\\
& =
\sum_{y \in \{ 0, 1 \}} \! P_{Y^{\prime}}( y ) H_{\sbvec{w}_{n}}( (m+y)^{-1} )
\\
& =
\sum_{y \in \{ 0, 1 \}} P_{Y^{\prime}}( y ) \ln (m + y) ,
\label{eq:example_Lmin_H} \\
\mathbb{E}[ \| P_{X^{\prime}|Y^{\prime}}(\cdot \mid Y^{\prime}) \|_{\alpha} ]
& =
\sum_{y \in \{ 0, 1 \}} \! P_{Y^{\prime}}( y ) \| \bvec{w}_{n}( (m+y)^{-1} ) \|_{\alpha}
\\
& =
\sum_{y \in \{ 0, 1 \}} P_{Y^{\prime}}( y ) \| \bvec{u}_{(m+y)} \|_{\alpha} .
\label{eq:example_Lmin_N}
\end{align}
Since $P_{Y^{\prime}}( 0 ) + P_{Y^{\prime}}( 1 ) = 1$, it follows from \eqref{eq:example_Lmin_H} and \eqref{eq:example_Lmin_N} that
\begin{align}
\mathbb{E}[ \| P_{X^{\prime}|Y^{\prime}}(\cdot \mid Y^{\prime}) \|_{\alpha} ]
=
L_{\min}^{\alpha}(X^{\prime} \mid Y^{\prime})
\end{align}
for $\alpha \in (0, 1) \cup (1, \infty)$.
Therefore, the lower bound of \eqref{eq:cond_extremes} is tight.

Moreover, we consider $(X^{\prime\prime}, Y^{\prime\prime})$ of Definition \ref{def:RVs_prime2}.
If $H(X^{\prime\prime} \mid Y^{\prime\prime}) \le H_{\sbvec{v}_{n}}( p_{n}^{\ast}( \alpha ) )$, then 
it immediately holds that
\begin{align}
\mathbb{E}[ \| P_{X^{\prime\prime}|Y^{\prime\prime}}(\cdot \mid Y^{\prime\prime}) \|_{\alpha} ]
=
L_{\min}^{\alpha}(X^{\prime\prime} \mid Y^{\prime\prime}) .
\end{align}
On the other hand, if $H(X^{\prime\prime} \mid Y^{\prime\prime}) > H_{\sbvec{v}_{n}}( p_{n}^{\ast}( \alpha ) )$, then it can be seen that
\begin{align}
H(X^{\prime\prime} \mid Y^{\prime\prime})
& =
\sum_{y \in \{ 0, 1 \}} \! P_{Y^{\prime\prime}}( y ) H(X^{\prime\prime} \mid Y^{\prime\prime} = y)
\\
& =
P_{Y^{\prime\prime}}( 0 ) H_{\sbvec{v}_{n}}( p_{n}^{\ast}( \alpha ) ) + P_{Y^{\prime\prime}}( 1 ) H( \bvec{u}_{n} )
\\
& =
P_{Y^{\prime\prime}}( 0 ) H_{\sbvec{v}_{n}}( p_{n}^{\ast}( \alpha ) ) + P_{Y^{\prime\prime}}( 1 ) \ln n ,
\label{eq:example_Lmax_H} \\
\!\!\! \mathbb{E}[ \| P_{X^{\prime\prime}|Y^{\prime\prime}}(\cdot \mid Y^{\prime\prime}) \|_{\alpha} ]
& =
P_{Y^{\prime\prime}}( 0 ) \| \bvec{v}_{n}( p_{n}^{\ast}( \alpha ) ) \|_{\alpha} + P_{Y^{\prime\prime}}( 1 ) \| \bvec{u}_{n} \|_{\alpha} .
\label{eq:example_Lmax_N}
\end{align}
Since $P_{Y^{\prime\prime}}( 0 ) + P_{Y^{\prime\prime}}( 1 ) = 1$, it follows from \eqref{eq:example_Lmax_H} and \eqref{eq:example_Lmax_N} that
\begin{align}
\mathbb{E}[ \| P_{X^{\prime\prime}|Y^{\prime\prime}}(\cdot \mid Y^{\prime\prime}) \|_{\alpha} ]
=
L_{\max}^{\alpha}(X^{\prime\prime} \mid Y^{\prime\prime})
\end{align}
for $\alpha \in [\frac{1}{2}, 1) \cup (1, \infty)$.
Therefore, the upper bound of \eqref{eq:cond_extremes} is also tight.
\end{IEEEproof}

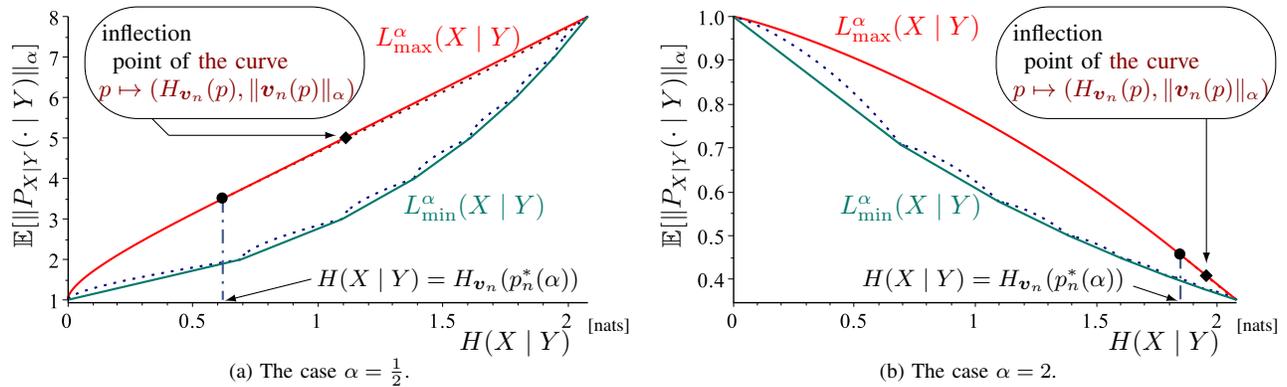
\begin{figure}[!t]
\centering
\subfloat[The case $\alpha = \frac{1}{2}$.]{
\label{subfig:norm_half}
\begin{overpic}[width = 0.45\hsize, clip]{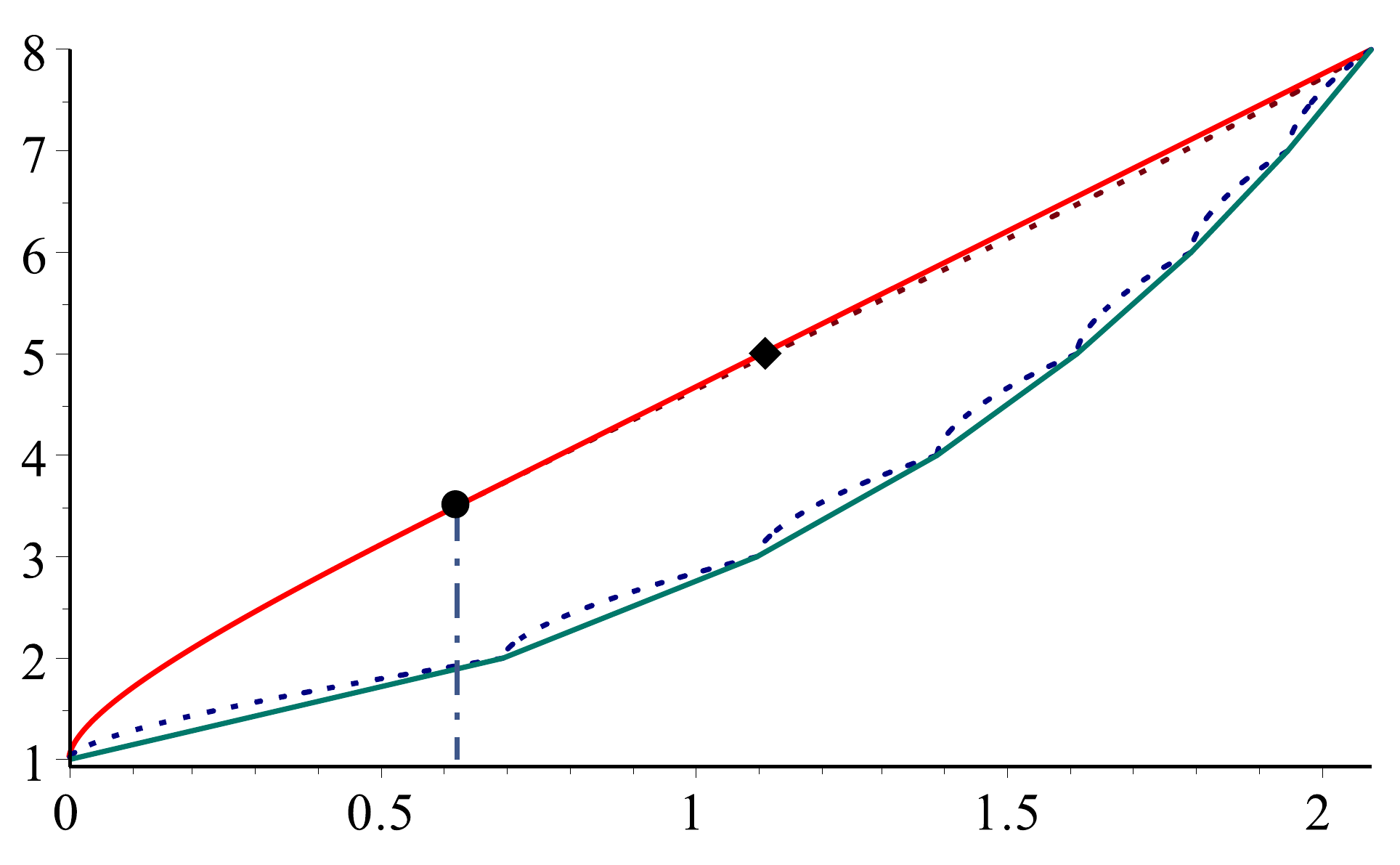}
\put(75, -1.5){$H(X \mid Y)$}
\put(98, 2){\scriptsize [nats]}
\put(-5, 17){\rotatebox{90}{$\mathbb{E}[ \| P_{X|Y}(\cdot \mid Y) \|_{\alpha} ]$}}
\put(60, 53){\color{red} $L_{\max}^{\alpha}(X \mid Y)$}
\put(65, 23){\color{bluegreen} $L_{\min}^{\alpha}(X \mid Y)$}
\put(49.5, 10){\small $H(X \mid Y) = H_{\sbvec{v}_{n}}( p_{n}^{\ast}( \alpha ) )$}
\put(48.5, 11.25){\vector(-4, -1){15.5}}
\put(11, 54){\small inflection}
\put(13, 49){\small point of {\color{burgundy} the curve}}
\put(10.5, 44){\small \color{burgundy} $p \mapsto (H_{\sbvec{v}_{n}}( p ), \| \bvec{v}_{n}( p ) \|_{\alpha})$}
\put(20, 40){\line(1, -1){3}}
\put(33, 50){\oval(50, 20)}
\put(23, 37){\vector(1, 0){30}}
\end{overpic}
}\hspace{0.05\hsize}
\subfloat[The case $\alpha = 2$.]{
\begin{overpic}[width = 0.45\hsize, clip]{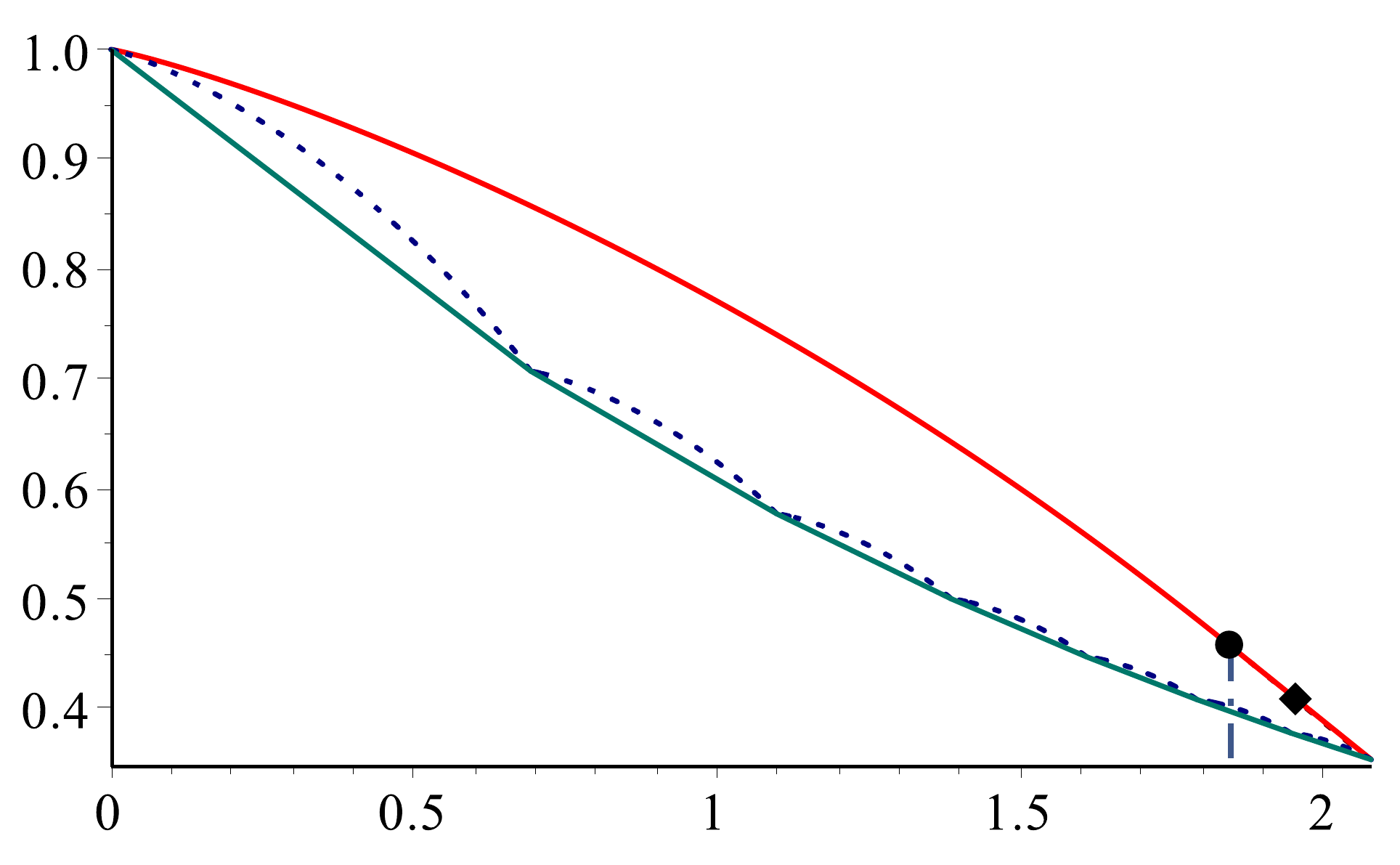}
\put(75, -1.5){$H(X \mid Y)$}
\put(98, 2){\scriptsize [nats]}
\put(-5, 17){\rotatebox{90}{$\mathbb{E}[ \| P_{X|Y}(\cdot \mid Y) \|_{\alpha} ]$}}
\put(26, 55){\color{red} $L_{\max}^{\alpha}(X \mid Y)$}
\put(27, 23){\color{bluegreen} $L_{\min}^{\alpha}(X \mid Y)$}
\put(31, 10){\small $H(X \mid Y) = H_{\sbvec{v}_{n}}( p_{n}^{\ast}( \alpha ) )$}
\put(79.5, 10){\vector(3, -1){8}}
\put(58, 54){\small inflection}
\put(60, 49){\small point of {\color{burgundy} the curve}}
\put(58, 44){\small \color{burgundy} $p \mapsto (H_{\sbvec{v}_{n}}( p ), \| \bvec{v}_{n}( p ) \|_{\alpha})$}
\put(80, 50){\oval(50, 20)}
\put(92.6, 40){\vector(0, -1){26}}
\end{overpic}
}
\caption{Plots of the boundaries of $\mathcal{R}_{n}^{\mathrm{cond}}( \alpha )$ with $n = 8$.
The upper- and lower-boundaries correspond to $L_{\max}^{\alpha}(X \mid Y)$ and $L_{\min}^{\alpha}(X \mid Y)$, respectively.
The left- and right-hand sides of $L_{\max}^{\alpha}(X \mid Y)$ from the circle points \textbullet \ correspond to $\| \hat{\bvec{v}}_{n}(X \mid Y) \|_{\alpha}$ and $T_{n, \alpha}(X \mid Y)$, respectively.
Note that $T_{n, \alpha}(X \mid Y)$ is the tangent line of the curve $p \mapsto (H_{\sbvec{v}_{n}}( p ), \| \bvec{v}_{n}( p ) \|_{\alpha})$ at $H_{\sbvec{v}_{n}}( p ) = H_{\sbvec{v}_{n}}( p_{n}^{\ast}( \alpha ) )$.
The dotted lines correspond to the boundaries of $\mathcal{R}_{n}( \alpha )$;
in particular, the dotted lines of (a) are same as Fig. \ref{fig:region_P6_half}.}
\label{fig:LminLmax}
\end{figure}

Fact \ref{fact:RVs_prime} implies that the bounds of Theorem \ref{th:cond_extremes} are tight.
Namely, Theorem \ref{th:cond_extremes} shows that the boundaries of $\mathcal{R}_{n}^{\mathrm{cond}}( \alpha )$ are correspond to $L_{\min}^{\alpha}(X \mid Y)$ and $L_{\max}^{\alpha}(X \mid Y)$ for $n \ge 3$ and $\alpha \in [\frac{1}{2}, 1) \cup (1, \infty)$.
We illustrate the boundaries of $\mathcal{R}_{n}^{\mathrm{cond}}( \alpha )$ are correspond to $L_{\min}^{\alpha}(X \mid Y)$ and $L_{\max}^{\alpha}(X \mid Y)$ in Fig. \ref{fig:LminLmax}.
From Theorem \ref{th:convexhull}, note that Fig. \ref{fig:LminLmax}-\subref{subfig:norm_half} is convex hulls of Fig. \ref{fig:region_P6_half}.
Furthermore, as with Theorem \ref{th:cond_extremes}, we also provide tight bounds of the conditional Shannon entropy with a fixed expectation of $\ell_{\alpha}$-norm in the following theorem.

\begin{theorem}
\label{th:cond_extremes2}
For a given $(X, Y)$ and a fixed $\alpha \in [\frac{1}{2}, 1) \cup (1, \infty)$, if $X, X^{\prime}, X^{\prime\prime} \in \mathcal{X}$ and
\begin{align}
\mathbb{E}[ \| P_{X|Y}(\cdot \mid Y) \|_{\alpha} ]
=
\mathbb{E}[ \| P_{X^{\prime}|Y^{\prime}}(\cdot \mid Y^{\prime}) \|_{\alpha} ]
=
\mathbb{E}[ \| P_{X^{\prime\prime}|Y^{\prime\prime}}(\cdot \mid Y^{\prime\prime}) \|_{\alpha} ] ,
\end{align}
then we observe that
\begin{align}
\frac{1}{2} \le \alpha < 1
\ & \Longrightarrow \
H(X^{\prime\prime} \mid Y^{\prime\prime})
\le
H(X \mid Y)
\le
H(X^{\prime} \mid Y^{\prime}) ,
\label{ineq:cond_H_less1} \\
\alpha > 1
\ & \Longrightarrow \
H(X^{\prime} \mid Y^{\prime})
\le
H(X \mid Y)
\le
H(X^{\prime\prime} \mid Y^{\prime\prime}) ,
\label{ineq:cond_H_greater1}
\end{align}
where the upper bound of \eqref{ineq:cond_H_less1} also holds for $\alpha \in (0, \frac{1}{2})$.
\end{theorem}

\begin{IEEEproof}[Proof of Theorem \ref{th:cond_extremes2}]
Using the monotonicity of Lemmas \ref{lem:Lmin} and \ref{lem:Lmax}, we can prove Theorem \ref{th:cond_extremes2} from Theorem \ref{th:cond_extremes}, as with the proof of \cite[Theorem \ref{th:extremes2}]{part1_arxiv}.
\end{IEEEproof}

Since there exists $(X^{\prime}, Y^{\prime})$ and $(X^{\prime\prime}, Y^{\prime\prime})$ such that the bounds \eqref{ineq:cond_H_less1} and \eqref{ineq:cond_H_greater1} hold with equalities, Theorem \ref{th:cond_extremes} also provides the tight bounds of the conditional Shannon entropy with a fixed expectation of $\ell_{\alpha}$-norm.

We now extend Theorem \ref{th:cond_extremes} in a similar manner with \cite[Theorem 2]{fabregas} and \cite[Corollary 1]{part1, part1_arxiv} as follows:

\begin{corollary}
\label{cor:cond_extremes}
Let $f( \cdot )$ be a strictly monotonic function. Then, for $\alpha \in [\frac{1}{2}, 1) \cup (1, \infty)$, we observe that
\begin{itemize}
\item
if $f( \cdot )$ is strictly increasing, then
\begin{align}
\hspace{-2em}
f( L_{\min}^{\alpha}(X|Y) )
\le
f( \mathbb{E}[ \| P_{X|Y}(\cdot|Y) \|_{\alpha} ] )
\le
f( \vphantom{\sum} L_{\max}^{\alpha}(X|Y) ) ,
\notag
\end{align}
\item
if $f( \cdot )$ is strictly decreasing, then
\begin{align}
\hspace{-2em}
f( L_{\max}^{\alpha}(X|Y) )
\le
f( \mathbb{E}[ \| P_{X|Y}(\cdot|Y) \|_{\alpha} ] )
\le
f( L_{\min}^{\alpha}(X|Y) ) .
\notag
\end{align}
\end{itemize}
The bounds with $f(L_{\min}^{\alpha}(X \mid Y))$ also hold for $\alpha \in (0, \frac{1}{2})$.
\end{corollary}

\begin{IEEEproof}[Proof of Corollary \ref{cor:cond_extremes}]
We can proof Corollary \ref{cor:cond_extremes} in the same manner with the proof of \cite[Corollary 1]{part1_arxiv}.
\end{IEEEproof}

\begin{figure}[!t]
\centering
\subfloat[The case $\alpha = \frac{1}{2}$.]{
\label{subfig:Renyi_half}
\begin{overpic}[width = 0.45\hsize, clip]{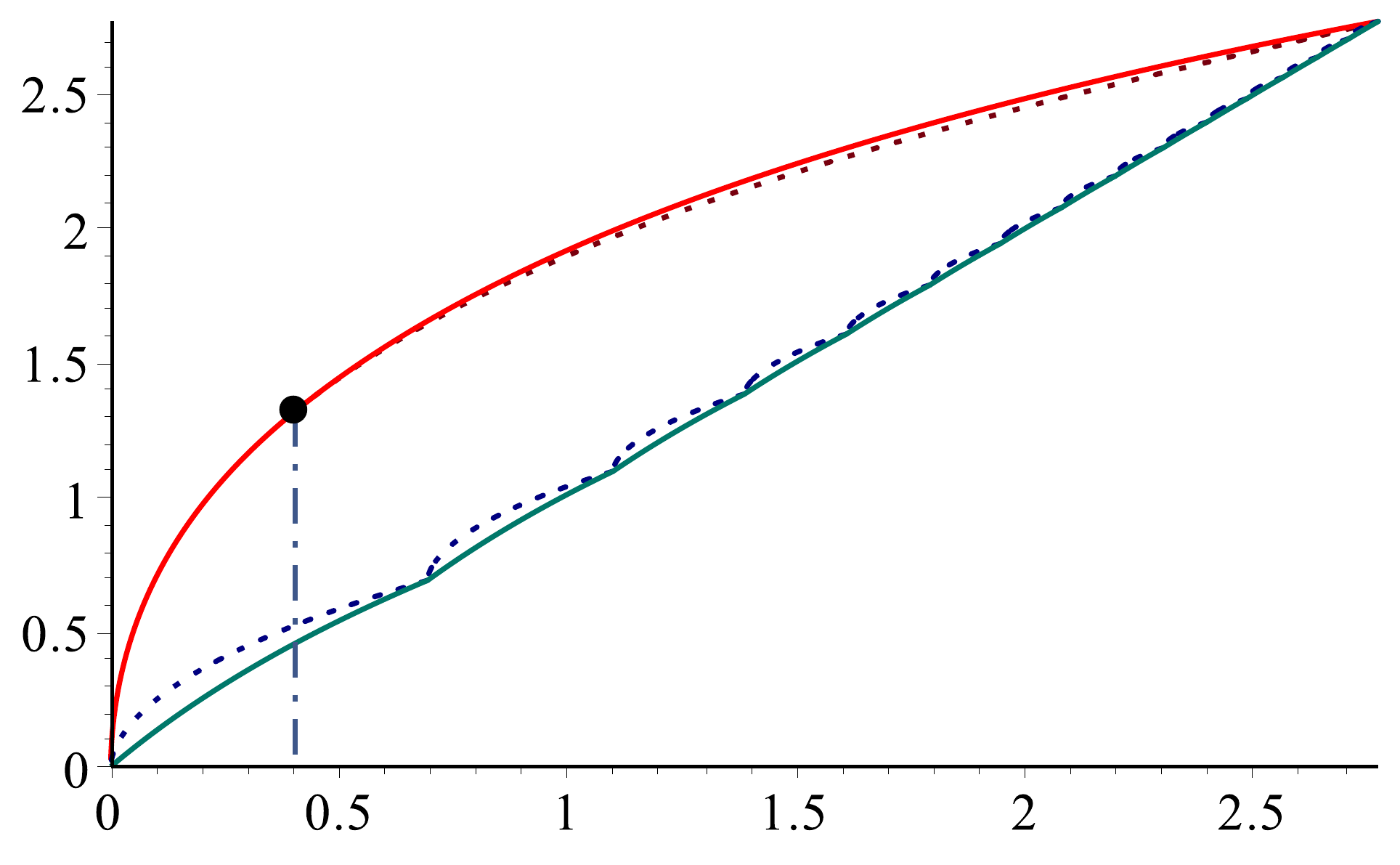}
\put(75, -1.5){$H(X \mid Y)$}
\put(98, 2){\scriptsize [nats]}
\put(-4, 23){\rotatebox{90}{$H_{\alpha}(X \mid Y)$}}
\put(-1, 60){\scriptsize [nats]}
\put(60, 32){\color{bluegreen} $H_{\alpha}(X^{\prime} \mid Y^{\prime})$}
\put(24, 52){\color{red} $H_{\alpha}(X^{\prime\prime} \mid Y^{\prime\prime})$}
\put(45, 15){\small $H(X \mid Y) = H_{\sbvec{v}_{n}}( p_{n}^{\ast}( \alpha ) )$}
\put(44, 15){\vector(-3, -1){22}}
\end{overpic}
}\hspace{0.05\hsize}
\subfloat[The case $\alpha = 2$.]{
\begin{overpic}[width = 0.45\hsize, clip]{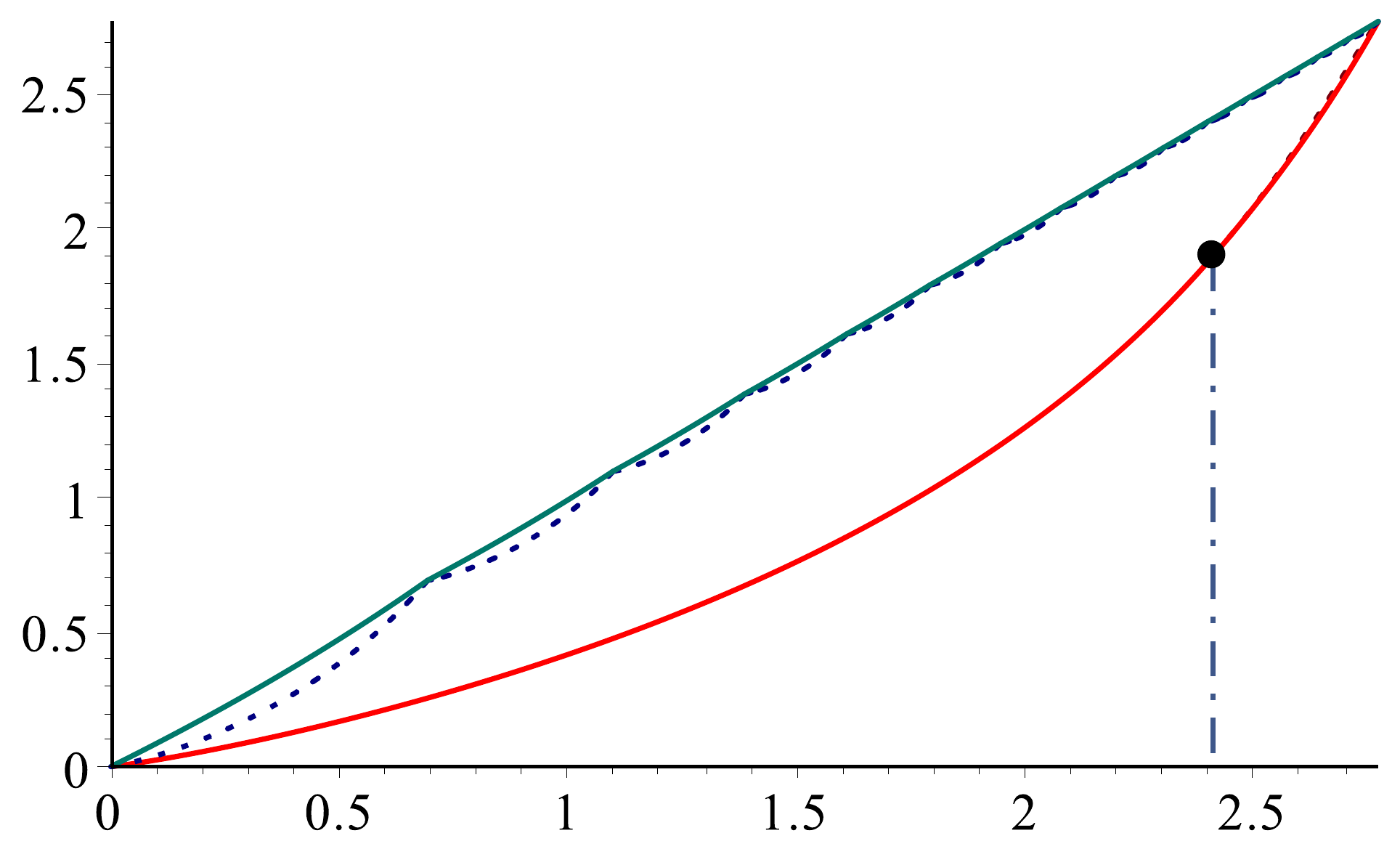}
\put(75, -1.5){$H(X \mid Y)$}
\put(98, 2){\scriptsize [nats]}
\put(-4, 23){\rotatebox{90}{$H_{\alpha}(X \mid Y)$}}
\put(-1, 60){\scriptsize [nats]}
\put(25, 34){\color{bluegreen} $H_{\alpha}(X^{\prime} \mid Y^{\prime})$}
\put(70, 24){\color{red} $H_{\alpha}(X^{\prime\prime} \mid Y^{\prime\prime})$}
\put(37, 11){\footnotesize $H(X \mid Y) = H_{\sbvec{v}_{n}}( p_{n}^{\ast}( \alpha ) )$}
\put(81, 11){\vector(3, -2){5.5}}
\end{overpic}
}
\caption{Plots of the boundaries of $\{ (H(X \mid Y), H_{\alpha}(X \mid Y)) \mid P_{XY} \in \mathcal{P}( \mathcal{X} \times \mathcal{Y} ), |\mathcal{X}| = n \}$ with $n = 16$.
The upper- and lower-boundaries correspond to $(X^{\prime\prime}, Y^{\prime\prime})$ of Definition \ref{def:RVs_prime2} and $(X^{\prime}, Y^{\prime})$ of Definition \ref{def:RVs_prime1}, respectively.
The dotted lines correspond to the boundary of $\{ (H( \bvec{p} ), H_{\alpha}( \bvec{p} )) \mid \bvec{p} \in \mathcal{P}_{n} \}$.}
\label{fig:LminLmax_Renyi}
\end{figure}

\begin{figure}[!t]
\centering
\subfloat[The case $\alpha = \frac{1}{2}$.]{
\begin{overpic}[width = 0.45\hsize, clip]{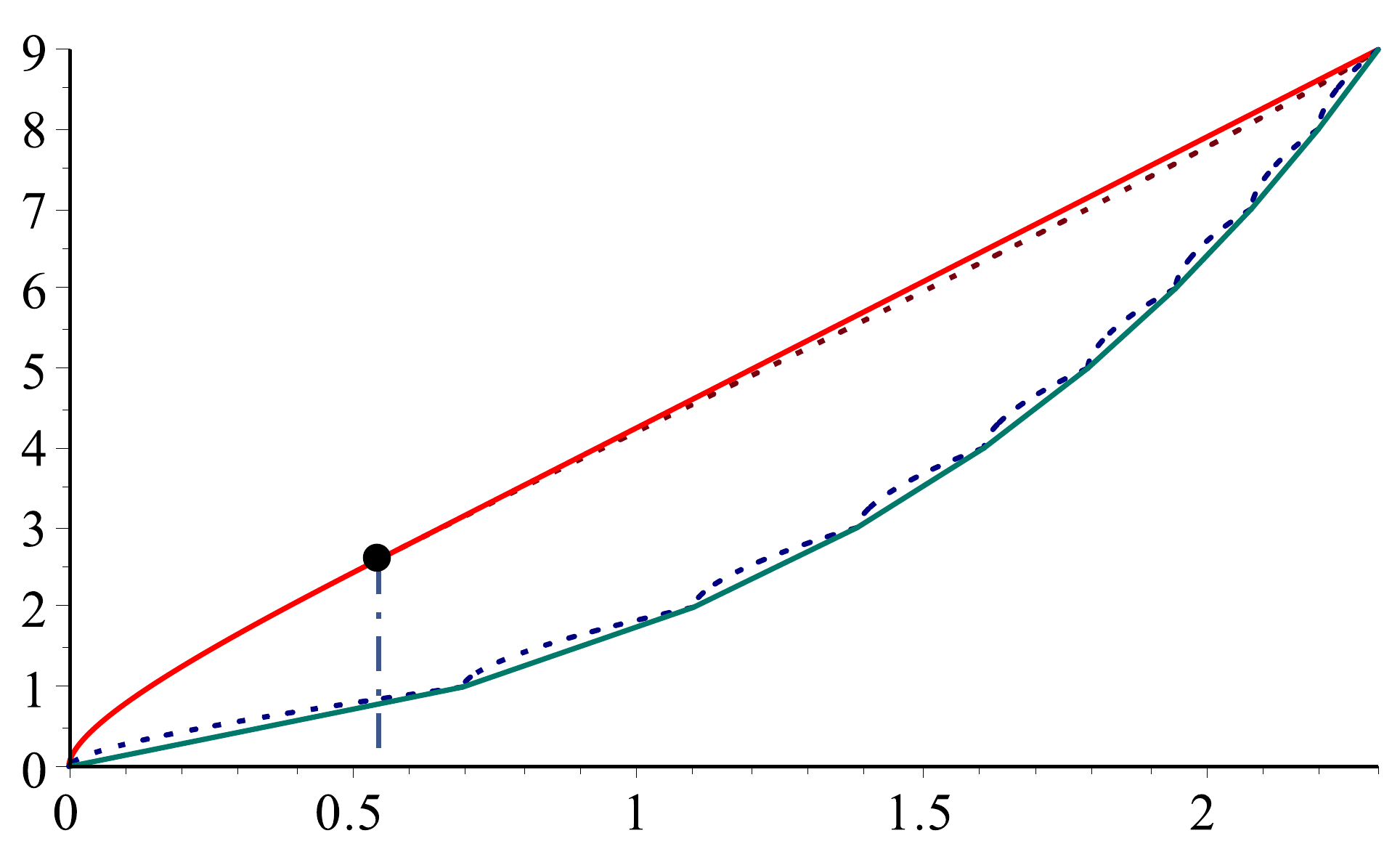}
\put(75, -1.5){$H(X \mid Y)$}
\put(98, 2){\scriptsize [nats]}
\put(-4, 23){\rotatebox{90}{$H_{R}(X \mid Y)$}}
\put(71, 25){\color{bluegreen} $H_{R}(X^{\prime} \mid Y^{\prime})$}
\put(29, 40){\color{red} $H_{R}(X^{\prime\prime} \mid Y^{\prime\prime})$}
\put(50, 14.75){\small $H(X \mid Y) = H_{\sbvec{v}_{n}}( p_{n}^{\ast}( \alpha ) )$}
\put(49, 14.75){\vector(-3, -1){22}}
\end{overpic}
}\hspace{0.05\hsize}
\subfloat[The case $\alpha = 2$.]{
\begin{overpic}[width = 0.45\hsize, clip]{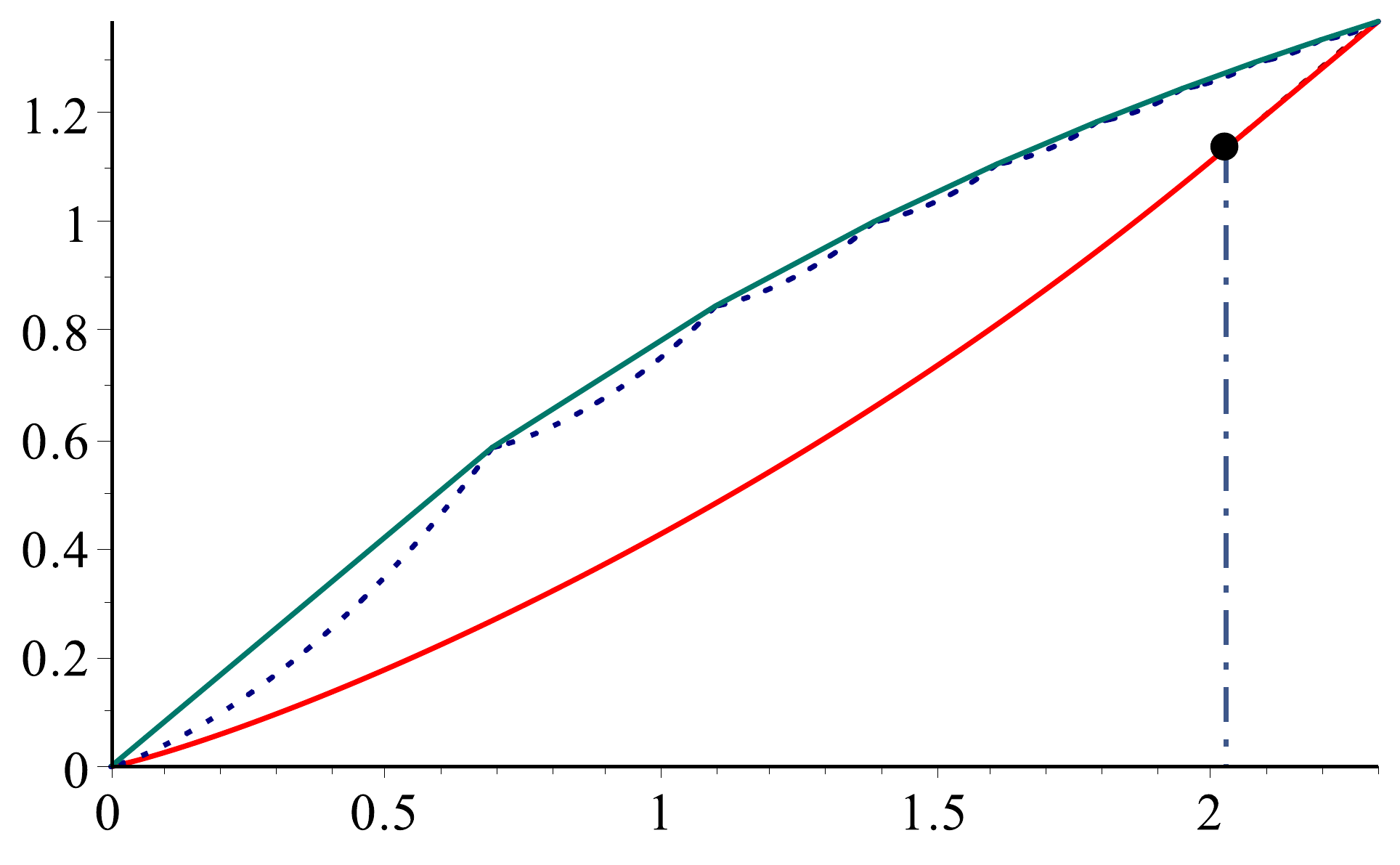}
\put(75, -1.5){$H(X \mid Y)$}
\put(98, 2){\scriptsize [nats]}
\put(-4, 23){\rotatebox{90}{$H_{R}(X \mid Y)$}}
\put(25, 42){\color{bluegreen} $H_{\alpha}(X^{\prime} \mid Y^{\prime})$}
\put(60, 26){\color{red} $H_{\alpha}(X^{\prime\prime} \mid Y^{\prime\prime})$}
\put(37.5, 11){\footnotesize $H(X \mid Y) = H_{\sbvec{v}_{n}}( p_{n}^{\ast}( \alpha ) )$}
\put(81.5, 11){\vector(3, -2){5.5}}
\end{overpic}
}
\caption{Plots of the boundaries of $\{ (H(X \mid Y), H_{R}(X \mid Y)) \mid P_{XY} \in \mathcal{P}( \mathcal{X} \times \mathcal{Y} ), |\mathcal{X}| = n \}$ with $n = 10$.
The upper- and lower-boundaries correspond to $(X^{\prime\prime}, Y^{\prime\prime})$ of Definition \ref{def:RVs_prime2} and $(X^{\prime}, Y^{\prime})$ of Definition \ref{def:RVs_prime1}, respectively.
The dotted lines correspond to the boundary of $\{ (H( \bvec{p} ), H_{R}( \bvec{p} )) \mid \bvec{p} \in \mathcal{P}_{n} \}$, where $H_{R}( \bvec{p} ) \triangleq \frac{ R }{ R - 1 } (1 - \| \bvec{p} \|_{R})$ denotes the $R$-norm information \cite{boekee} of $\bvec{p} \in \mathcal{P}_{n}$.}
\label{fig:LminLmax_R}
\end{figure}
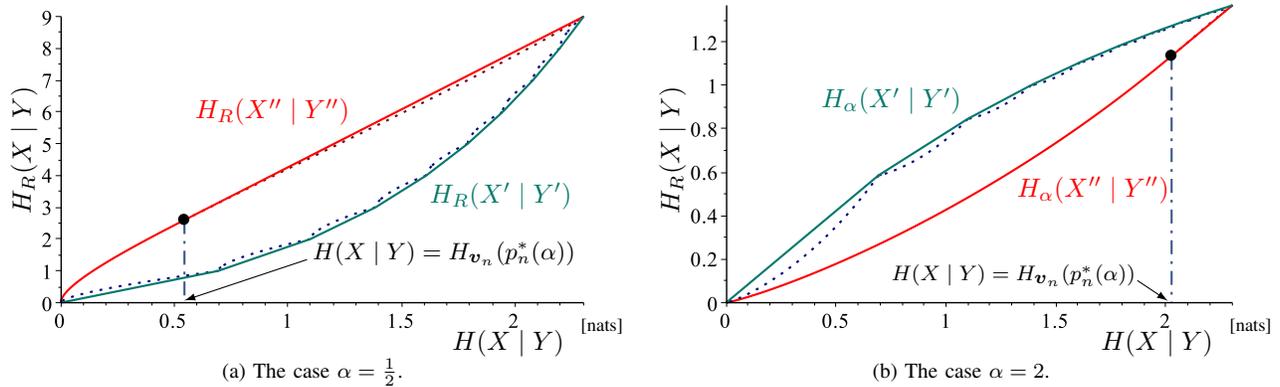

Therefore, we can obtain the tight bounds of several information measures, determined by the expectation of $\ell_{\alpha}$-norm, with a fixed conditional Shannon entropy.
As an instance, we introduce the application of Corollary \ref{cor:cond_extremes} to the conditional R\'{e}nyi entropy as follows:
Let $f_{\alpha}( x ) = \frac{\alpha}{1-\alpha} \ln x$.
Then, we readily see that
\begin{align}
H_{\alpha}(X \mid Y)
=
f_{\alpha}( \mathbb{E}[ \| P_{X|Y}(\cdot \mid Y) \|_{\alpha} ] ) .
\label{eq:renyi_f}
\end{align}
It can be easily seen that $f_{\alpha}( x )$ is strictly increasing for $x \ge 0$ when $\alpha \in (0, 1)$ and strictly decreasing for $x \ge 0$ when $\alpha \in (1, \infty)$.
Hence, assuming that $(X^{\prime}, Y^{\prime})$ and $(X^{\prime\prime}, Y^{\prime\prime})$ satisfy $X, X^{\prime}, X^{\prime\prime} \in \mathcal{X}$ and
\begin{align}
H(X \mid Y)
=
H(X^{\prime} \mid Y^{\prime})
=
H(X^{\prime\prime} \mid Y^{\prime\prime})
\end{align}
for a given $(X, Y)$, it follows from Corollary \ref{cor:cond_extremes} that
\begin{align}
\hspace{-0.5em}
\frac{1}{2} \le \alpha < 1 \
& \Longrightarrow \
H_{\alpha}(X^{\prime\prime}|Y^{\prime\prime})
\le
H_{\alpha}(X|Y)
\le
H_{\alpha}(X^{\prime}|Y^{\prime}) ,
\label{eq:cond_Renyi_bound1} \\
\hspace{-0.5em}
\alpha > 1 \
& \Longrightarrow \
H_{\alpha}(X^{\prime}|Y^{\prime})
\le
H_{\alpha}(X|Y)
\le
H_{\alpha}(X^{\prime\prime}|Y^{\prime\prime}) ,
\label{eq:cond_Renyi_bound2}
\end{align}
where note that the left-hand inequality of \eqref{eq:cond_Renyi_bound1} also holds for $\alpha \in (0, \frac{1}{2})$.
Moreover, we can provide the tight bounds of the conditional Shannon entropy with a fixed conditional R\'{e}nyi entropy by using Theorem \ref{th:cond_extremes2}.
We illustrate \eqref{eq:cond_Renyi_bound1} and \eqref{eq:cond_Renyi_bound2} in Fig. \ref{fig:LminLmax_Renyi}, as with Fig. \ref{fig:LminLmax}.
As another application, letting $f_{R}( x ) = \frac{R}{R-1} (1 - x)$, we can also provide tight bounds of the conditional $R$-norm information \cite[Eq. (39)]{boekee}, defined by
\begin{align}
H_{R}(X \mid Y)
\triangleq
f_{R}( \mathbb{E}[ \| P_{X|Y}(\cdot \mid Y) \|_{R} ] ) .
\label{eq:R_f}
\end{align}
We illustrate the exact feasible regions between the conditional Shannon entropy and the conditional $R$-norm information in Fig. \ref{fig:LminLmax_R}.

\subsection{Applications for discrete memoryless channels}
\label{subsect:DMC}

In this subsection, we consider applications of Corollary \ref{cor:cond_extremes} for discrete memoryless channels (DMCs).
We define DMCs as follows:
Let the discrete random variables $X \in \mathcal{X}$ and $Y \in \mathcal{Y}$ denote the input and output of a DMC, respectively, where $\mathcal{X}$ and $\mathcal{Y}$ denote the finite input and output alphabets, respectively.
Let
$P_{Y|X}(y \mid x)$ denotes the transition probability of a DMC $(X, Y)$ for $(x, y) \in \mathcal{X} \times \mathcal{Y}$.
For a DMC $(X, Y)$, the mutual information of order $\alpha \in (0, \infty)$ \cite{arimoto} is defined as
\begin{align}
I_{\alpha}(X; Y)
\triangleq
H_{\alpha}(X) - H_{\alpha}(X \mid Y) ,
\end{align}
where $I_{1}(X; Y) \triangleq I(X; Y)$ denotes the (ordinary) mutual information.
Since $H_{\alpha}( \bvec{u}_{n} ) = \ln n$ for $\alpha \in (0, \infty)$, if the input $X$ follows a uniform distribution, i.e., $X \sim \bvec{u}_{|\mathcal{X}|}$, then we observe that
\begin{align}
I_{\alpha}(X; Y)
=
\ln |\mathcal{X}| - H_{\alpha}(X \mid Y) .
\end{align}
Therefore, by using \eqref{eq:cond_Renyi_bound1} and \eqref{eq:cond_Renyi_bound2}, we can obtain the tight bounds of $I_{\alpha}(X; Y)$ for a DMC $(X, Y)$ which the input $X$ follows a uniform distribution.
We summarize the bounds of $I_{\alpha}(X; Y)$ which follows from \eqref{eq:cond_Renyi_bound1} and \eqref{eq:cond_Renyi_bound2} in the following corollary.

\begin{corollary}
\label{cor:mutual}
For a given DMC $(X, Y)$, if the channels $(X^{\prime}, Y^{\prime})$ and $(X^{\prime\prime}, Y^{\prime\prime})$ of Definitions \ref{def:RVs_prime1} and \ref{def:RVs_prime2}, respectively, satisfy $X, X^{\prime}, X^{\prime\prime} \sim \bvec{u}_{|\mathcal{X}|}$ and
\begin{align}
I(X; Y)
=
I(X^{\prime}, Y^{\prime})
=
I(X^{\prime\prime}, Y^{\prime\prime}) ,
\end{align}
then we observe that
\begin{align}
\hspace{-0.5em}
\frac{1}{2} \le \alpha < 1
& \Longrightarrow
I_{\alpha}(X^{\prime}, Y^{\prime})
\le
I_{\alpha}(X; Y)
\le
I_{\alpha}(X^{\prime\prime}, Y^{\prime\prime}) , \!
\label{eq:mutual1} \\
\hspace{-0.5em}
\alpha > 1
& \Longrightarrow
I_{\alpha}(X^{\prime\prime}, Y^{\prime\prime})
\le
I_{\alpha}(X; Y)
\le
I_{\alpha}(X^{\prime}, Y^{\prime}) , \!
\label{eq:mutual1}
\end{align}
where the lower bound of \eqref{eq:mutual1} also holds for $\alpha \in (0, \frac{1}{2})$.
\end{corollary}

\begin{figure}[!t]
\centering
\subfloat[The case $\alpha = \frac{1}{2}$.]{
\begin{overpic}[width = 0.45\hsize, clip]{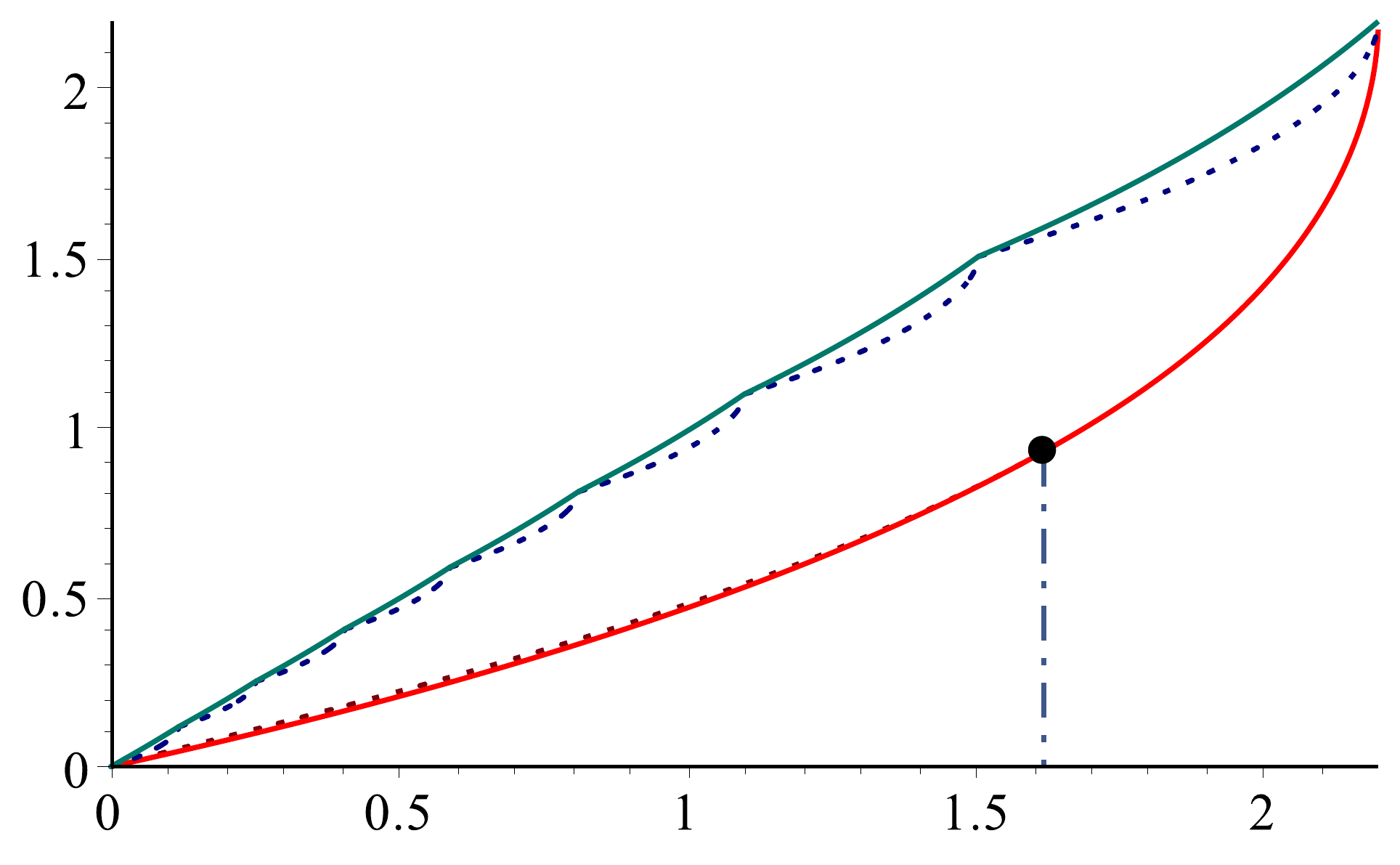}
\put(75, -1.5){$I(X ; Y)$}
\put(98, 2){\scriptsize [nats]}
\put(-4, 23){\rotatebox{90}{$I_{\alpha}(X; Y)$}}
\put(-1, 60){\scriptsize [nats]}
\put(30, 35){\color{bluegreen} $I_{\alpha}(X^{\prime} ; Y^{\prime})$}
\put(80, 27){\color{red} $I_{\alpha}(X^{\prime\prime} ; Y^{\prime\prime})$}
\put(50, 14.5){\small $I(X ; Y) = \ln n - H_{\sbvec{v}_{n}}( p_{n}^{\ast}( \alpha ) )$}
\put(53, 13){\vector(4, -1){21}}
\end{overpic}
}\hspace{0.05\hsize}
\subfloat[The case $\alpha = 2$.]{
\begin{overpic}[width = 0.45\hsize, clip]{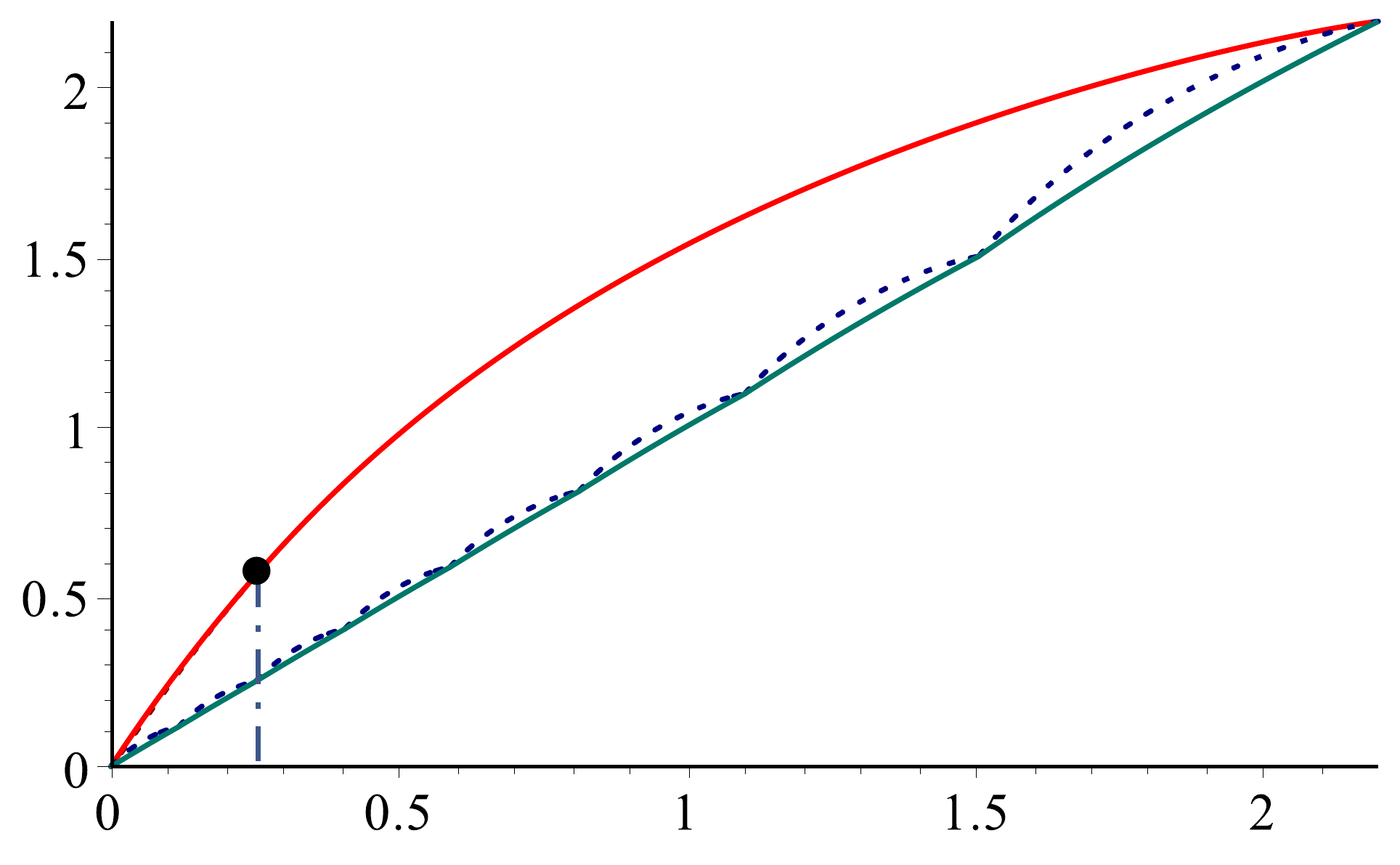}
\put(75, -1.5){$I(X ; Y)$}
\put(98, 2){\scriptsize [nats]}
\put(-4, 23){\rotatebox{90}{$I_{\alpha}(X ; Y)$}}
\put(-1, 60){\scriptsize [nats]}
\put(60, 32){\color{bluegreen} $I_{\alpha}(X^{\prime} ; Y^{\prime})$}
\put(25, 47){\color{red} $I_{\alpha}(X^{\prime\prime} ; Y^{\prime\prime})$}
\put(33, 12){\small $I(X; Y) = \ln n - H_{\sbvec{v}_{n}}( p_{n}^{\ast}( \alpha ) )$}
\put(32, 12){\vector(-3, -1){13.5}}
\end{overpic}
}
\caption{Plots of the boundaries of $\{ (I(X ; Y), I_{\alpha}(X ; Y)) \mid P_{XY} \in \mathcal{P}( \mathcal{X} \times \mathcal{Y} ), |\mathcal{X}| = n, P_{X}( \cdot ) = \bvec{u}_{n} \}$ with $n = 9$.
The upper- and lower-boundaries correspond to $(X^{\prime\prime}, Y^{\prime\prime})$ of Definition \ref{def:RVs_prime2} and $(X^{\prime}, Y^{\prime})$ of Definition \ref{def:RVs_prime1}, respectively.
The dotted lines correspond to the boundary of $\{ (H( \bvec{p} ), H_{\alpha}( \bvec{p} )) \mid \bvec{p} \in \mathcal{P}_{n} \}$.}
\label{fig:LminLmax_MI}
\end{figure}

Furthermore, we consider Gallager's $E_{0}$ function \cite{gallager} of a DMC $(X, Y)$, defined by
\begin{align}
E_{0}(\rho, X, Y)
& \triangleq
- \ln \sum_{y \in \mathcal{Y}} \left( \sum_{x \in \mathcal{X}} P_{X}( x ) P_{Y|X}(y \mid x)^{\frac{1}{1+\rho}} \right)^{1+\rho}
\notag
\end{align}
for $\rho \in (-1, \infty)$.
We can conclude an extremality of the $E_{0}$ function by using Definitions \ref{def:RVs_prime1} and \ref{def:RVs_prime2} in the following theorem.

\begin{theorem}
\label{th:E0_symmetric}
Assume that channels $(X^{\prime}, Y^{\prime})$ and $(X^{\prime\prime}, Y^{\prime\prime})$ satisfy $X, X^{\prime}, X^{\prime\prime} \sim \bvec{u}_{|\mathcal{X}|}$ and
\begin{align}
I(X; Y)
=
I(X^{\prime}; Y^{\prime})
=
I(X^{\prime\prime}; Y^{\prime\prime})
\end{align}
for a given DMC $(X, Y)$.
Then, we observe that
\begin{align}
E_{0}(\rho, X^{\prime\prime}, Y^{\prime\prime})
\le
E_{0}(\rho, X, Y)
\le
E_{0}(\rho, X^{\prime}, Y^{\prime})
\label{ineq:E0_symmetric}
\end{align}
for any $\rho \in (-1, 1]$.
In particular, the right-hand inequality of \eqref{ineq:E0_symmetric} also holds for $\rho \in (1, \infty)$.
\end{theorem}

\begin{figure}[!t]
\centering
\subfloat[The case $n = 5$ and $\rho = 1$ (cutoff rate).]{
\begin{overpic}[width = 0.45\hsize, clip]{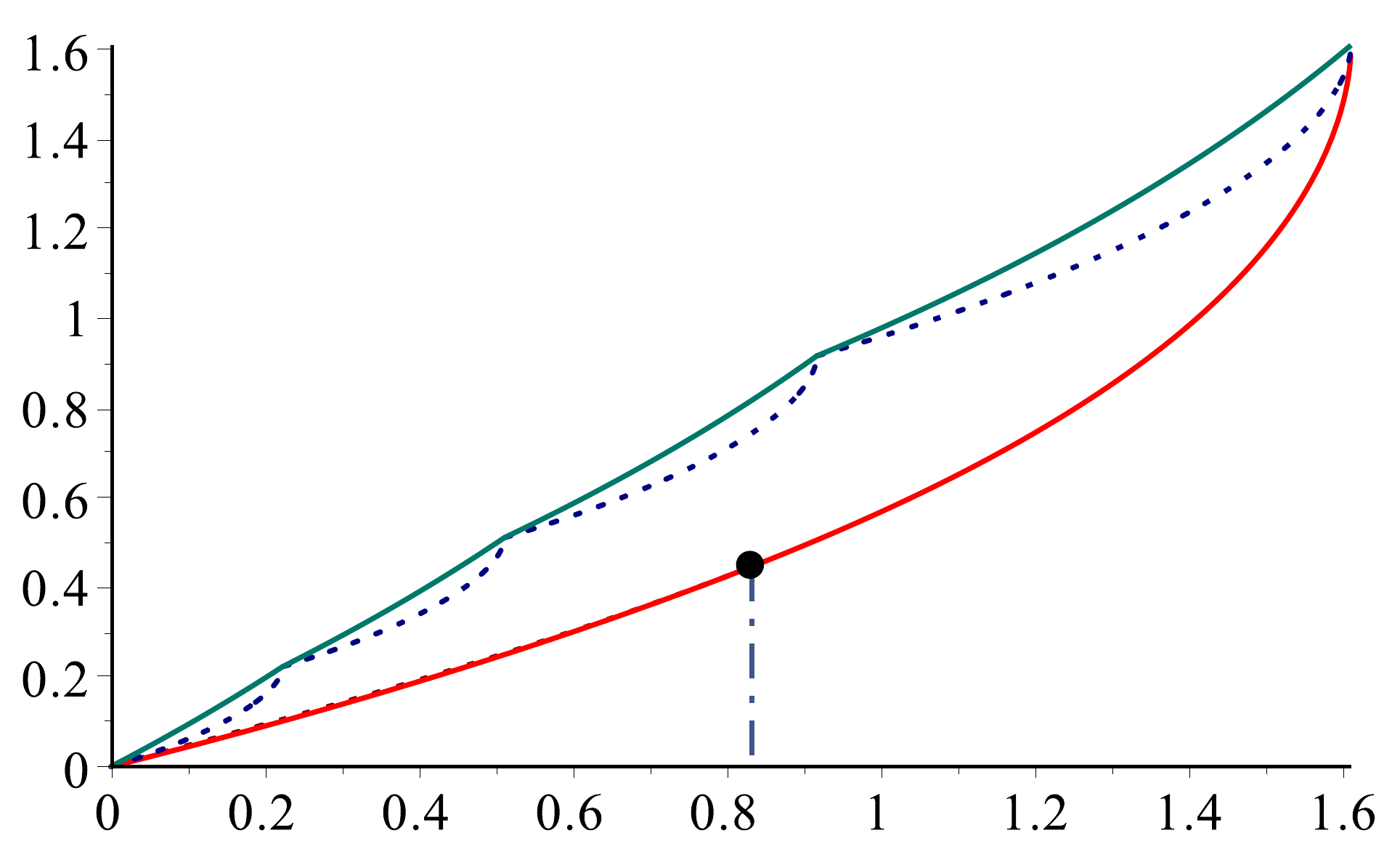}
\put(75, -1.5){$I(X ; Y)$}
\put(95, -1){\scriptsize [nats]}
\put(-4, 23){\rotatebox{90}{$E_{0}(\rho, X, Y)$}}
\put(-1, 60.5){\scriptsize [nats]}
\put(29, 37){\color{bluegreen} $E_{0}(\rho, X^{\prime}, Y^{\prime})$}
\put(71, 25){\color{red} $E_{0}(\rho, X^{\prime\prime}, Y^{\prime\prime})$}
\put(44, 13.5){\small $I(X ; Y) = \ln n - H_{\sbvec{v}_{n}}( p_{n}^{\ast}( \alpha ) )$}
\put(46.5, 12){\vector(3, -2){7.25}}
\end{overpic}
}\hspace{0.05\hsize}
\subfloat[The case $n = 5$ and $\rho = - \frac{1}{2}$.]{
\begin{overpic}[width = 0.45\hsize, clip]{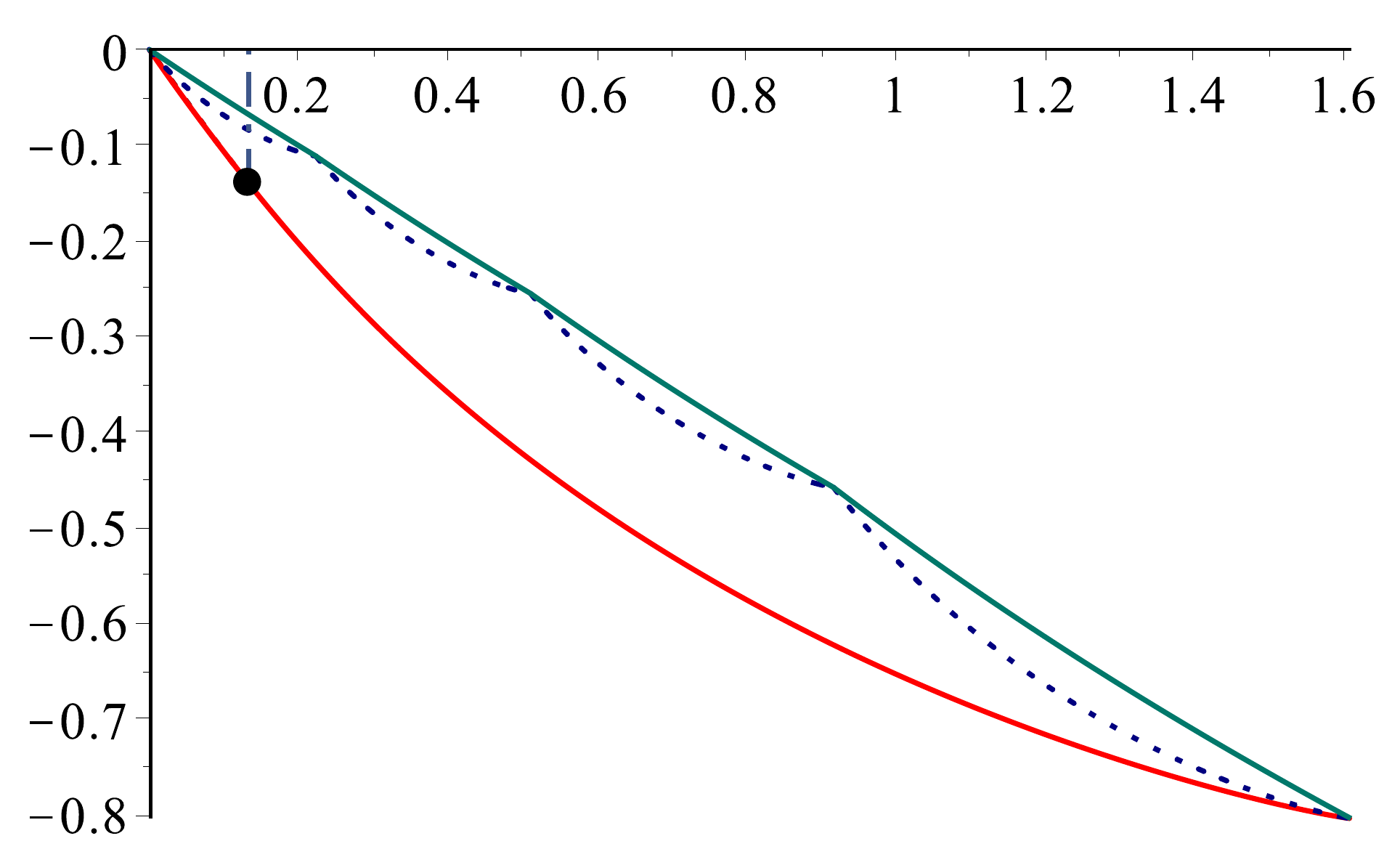}
\put(75, 48){$I(X ; Y)$}
\put(95, 50){\scriptsize [nats]}
\put(-4, 20){\rotatebox{90}{$E_{0}(\rho, X, Y)$}}
\put(-2, -1){\scriptsize [nats]}
\put(70, 23){\color{bluegreen} $E_{0}(\rho, X^{\prime}, Y^{\prime})$}
\put(20, 15){\color{red} $E_{0}(\rho, X^{\prime\prime}, Y^{\prime\prime})$}
\put(50, 40){\small $I(X; Y) = \ln n - H_{\sbvec{v}_{n}}( p_{n}^{\ast}( \alpha ) )$}
\put(49, 42){\vector(-2, 1){31}}
\end{overpic}
}\\
\subfloat[The case $n = 256$ and $\rho = 1$ (cutoff rate).]{
\begin{overpic}[width = 0.45\hsize, clip]{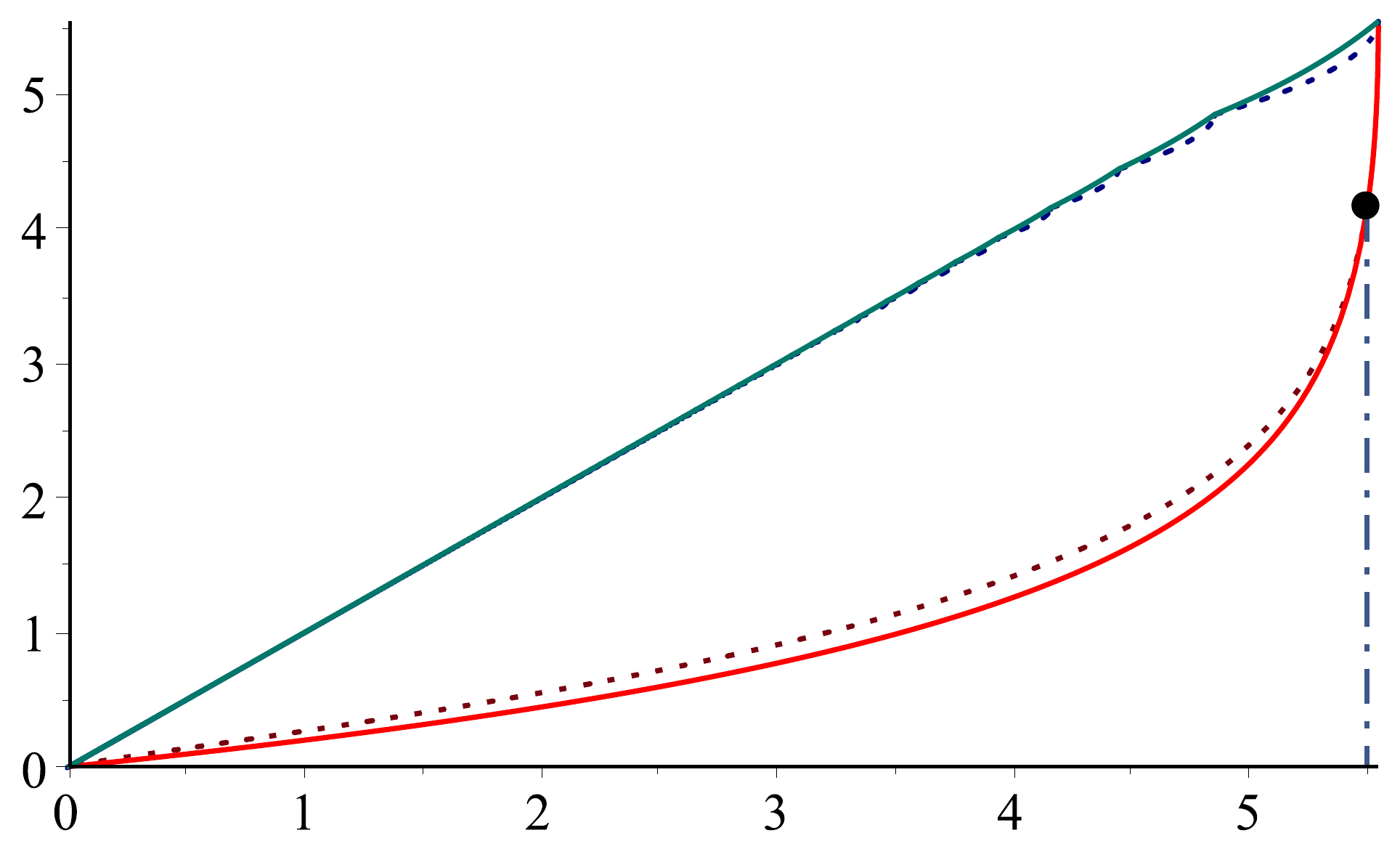}
\put(75, -1.5){$I(X ; Y)$}
\put(98, 2){\scriptsize [nats]}
\put(-4, 23){\rotatebox{90}{$E_{0}(\rho, X, Y)$}}
\put(-3, 59){\scriptsize [nats]}
\put(29, 38){\color{bluegreen} $E_{0}(\rho, X^{\prime}, Y^{\prime})$}
\put(88, 20){\rotatebox{60}{\color{red} $E_{0}(\rho, X^{\prime\prime}, Y^{\prime\prime})$}}
\put(35, 13.5){\small $I(X ; Y) = \ln n - H_{\sbvec{v}_{n}}( p_{n}^{\ast}( \alpha ) )$}
\put(91, 11.75){\vector(3, -2){6.5}}
\end{overpic}
}\hspace{0.05\hsize}
\subfloat[The case $n = 256$ and $\rho = - \frac{1}{2}$.]{
\begin{overpic}[width = 0.45\hsize, clip]{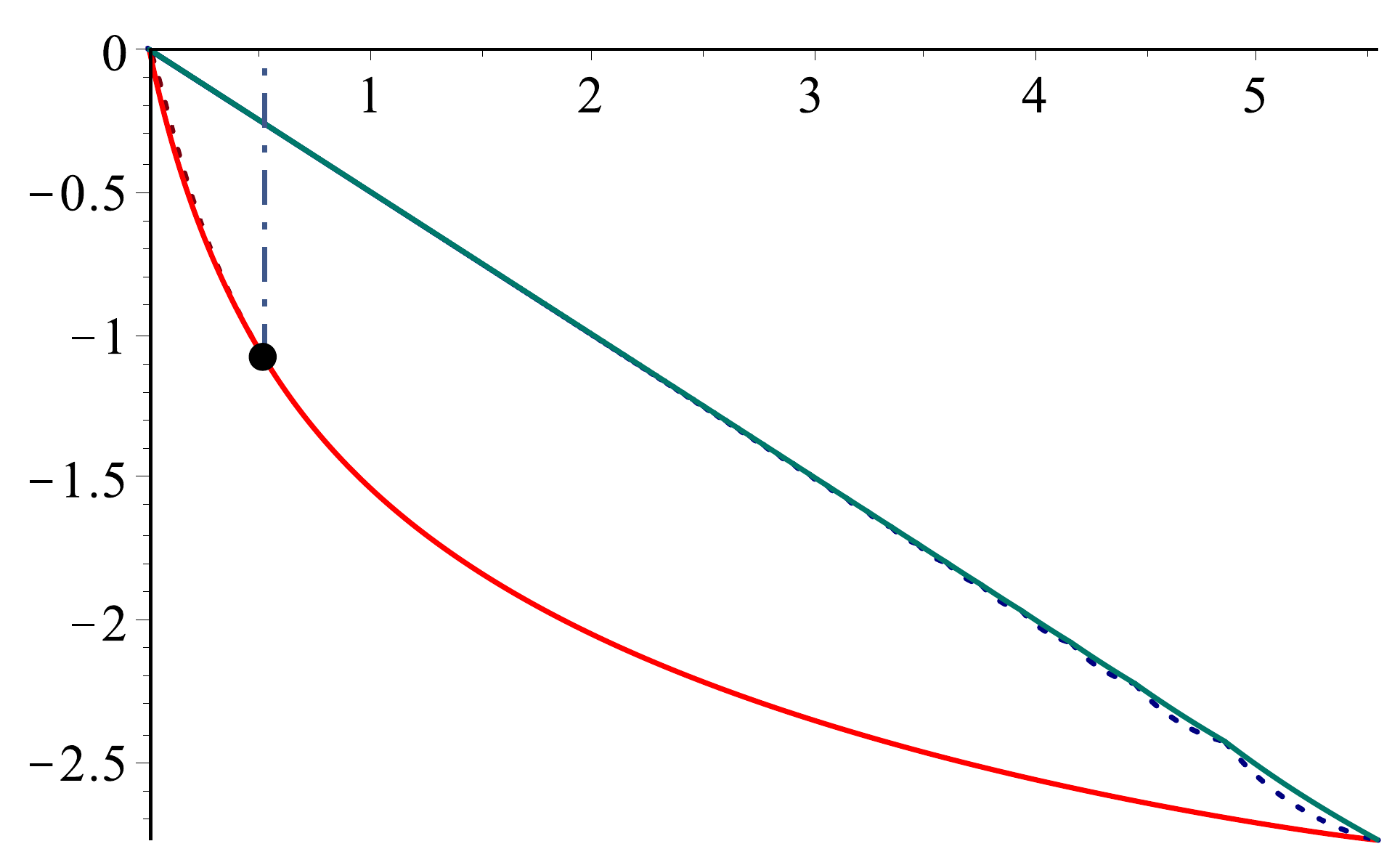}
\put(75, 48){$I(X ; Y)$}
\put(98, 54){\scriptsize [nats]}
\put(-4, 20){\rotatebox{90}{$E_{0}(\rho, X, Y)$}}
\put(1, 0){\scriptsize [nats]}
\put(70, 23){\color{bluegreen} $E_{0}(\rho, X^{\prime}, Y^{\prime})$}
\put(20, 8){\color{red} $E_{0}(\rho, X^{\prime\prime}, Y^{\prime\prime})$}
\put(51, 40){\small $I(X; Y) = \ln n - H_{\sbvec{v}_{n}}( p_{n}^{\ast}( \alpha ) )$}
\put(50, 42){\vector(-2, 1){31}}
\end{overpic}
}
\caption{Plots of the boundaries of $\{ (I(X ; Y), E_{0}(\rho, X, Y)) \mid P_{XY} \in \mathcal{P}( \mathcal{X} \times \mathcal{Y} ), |\mathcal{X}| = n, P_{X}( \cdot ) = \bvec{u}_{n} \}$.
The upper- and lower-boundaries correspond to $(X^{\prime\prime}, Y^{\prime\prime})$ of Definition \ref{def:RVs_prime2} and $(X^{\prime}, Y^{\prime})$ of Definition \ref{def:RVs_prime1}, respectively.
The dotted lines correspond to the boundary of $\{ (H( \bvec{p} ), H_{\alpha}( \bvec{p} )) \mid \bvec{p} \in \mathcal{P}_{n} \}$.}
\label{fig:LminLmax_MI}
\end{figure}
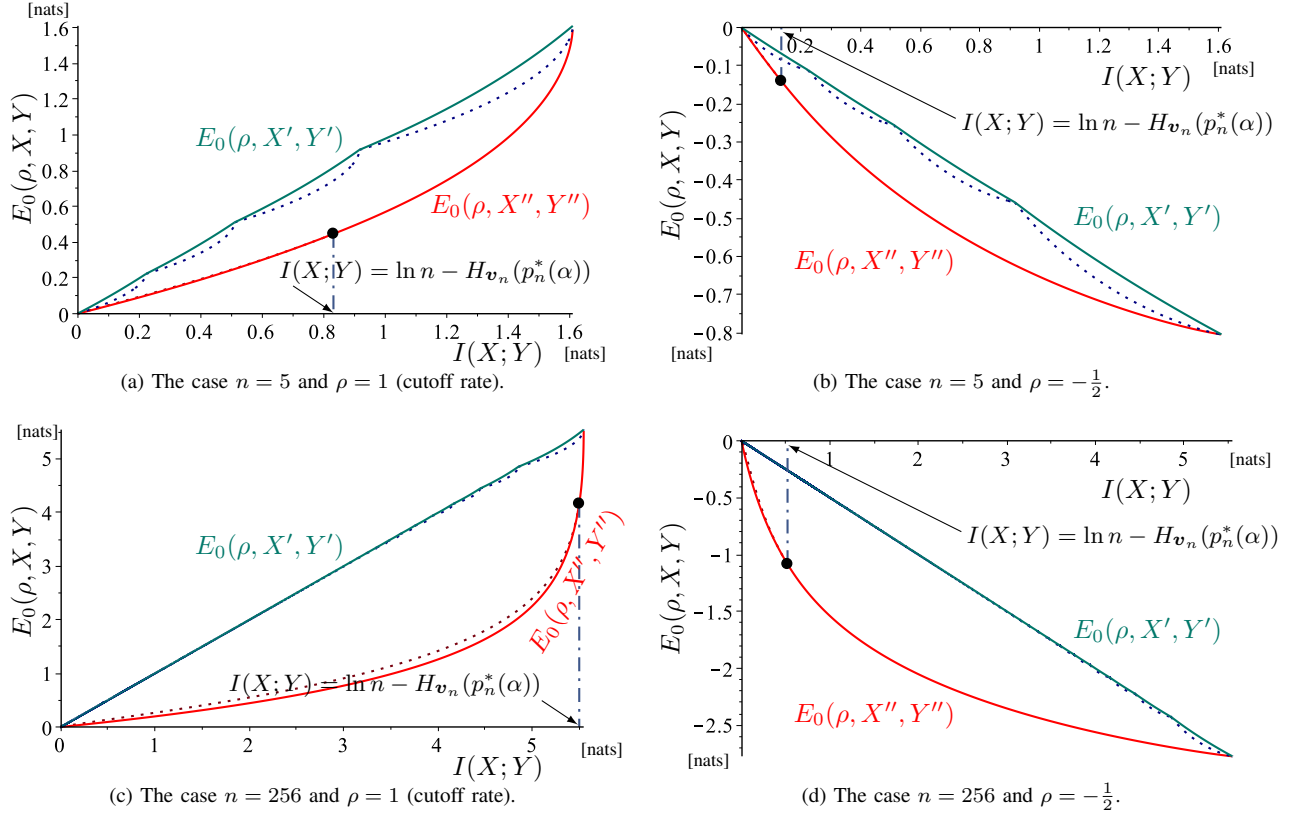

\begin{IEEEproof}[Proof of Theorem \ref{th:E0_symmetric}]
It can be seen from \cite[Eq. (6)]{alsan} that, if the input $X$ follows a uniform distribution, then
\begin{align}
E_{0}(\rho, X, Y)
& =
\rho \, I_{\frac{1}{1+\rho}}(X; Y) .
\label{E0_mutual}
\end{align}
Therefore, for $\alpha = \frac{1}{1+\rho}$, noting the relations
\begin{align}
-1 < \rho \le 0
& \iff
1 \le \alpha < \infty ,
\\
0 \le \rho \le 1
& \iff
\frac{1}{2} \le \alpha \le 1 ,
\\
1 < \rho < \infty
& \iff
0 < \alpha < \frac{1}{2} ,
\end{align}
we can obtain Theorem \ref{th:E0_symmetric} from Corollary \ref{cor:mutual}.
\end{IEEEproof}

Theorem \ref{th:E0_symmetric} is a generalization of \cite[Theorem 2]{fabregas} and \cite[Corollary 2]{alsan} from binary-input DMCs to non-binary input DMCs under a uniform input distribution.
In addition, Theorem \ref{th:E0_symmetric} contains \cite[Corollary 1]{isit2015}.

\subsection{A special case of $L_{\max}^{\alpha}(X \mid Y)$: $\alpha = \frac{1}{2}$}
\label{subsect:alpha_half}

In Theorem \ref{th:cond_extremes}, we saw that $L_{\max}^{\alpha}(X \mid Y)$ is the maximum of expectations of $\ell_{\alpha}$-norm with a fixed conditional Shannon entropy.
Note that $L_{\max}^{\alpha}(X \mid Y)$ is defined by a piecewise continuous function of $H(X \mid Y) \in [0, \ln n]$, composed of the two segments separated by $H(X \mid Y) = H_{\sbvec{v}_{n}}( p_{n}^{\ast}( \alpha ) )$.
We now show an example of the solution $p_{n}^{\ast}( \alpha )$ of the equation \eqref{eq:equation_p^ast} with respect to $p \in [0, \frac{1}{n}]$ by the simple form as follows:

\begin{fact}
\label{fact:alpha_half}
If $\alpha = \frac{1}{2}$, then $p_{n}^{\ast}( \frac{1}{2} ) = \frac{1}{n (n-1)}$ for any $n \ge 2$.
\end{fact}

\begin{IEEEproof}[Proof of Fact \ref{fact:alpha_half}]
Simple calculations yield
\begin{align}
H_{\sbvec{v}_{n}} \! \left( \frac{ 1 }{ n (n-1) } \right)
& =
\left. \left( \vphantom{\sum} - (1 - (n-1) p ) \ln (1 - (n-1) p) - (n-1) p \ln p \right) \right|_{p = \frac{1}{n (n-1)}}
\\
& =
- \left( 1 - (n-1) \left( \frac{1}{n (n-1)} \right) \right) \ln \left( 1 - (n-1) \left( \frac{1}{n (n-1)} \right) \right)
\notag \\
& \qquad \qquad \qquad \qquad \qquad \qquad \qquad 
- (n-1) \left( \frac{1}{n (n-1)} \right) \ln \left( \frac{1}{n (n-1)} \right)
\\
& =
\ln n - \left( 1 - \frac{2}{n} \right) \ln (n-1) ,
\\
\| \bvec{u}_{n} \|_{( \frac{1}{2} )}
& =
\left. \left( n^{\frac{1}{\alpha} - 1} \right) \right|_{\alpha = \frac{1}{2}}
\\
& =
n ,
\\
\left\| \bvec{v}_{n} \! \left( \frac{ 1 }{ n (n-1) } \right) \right\|_{( \frac{1}{2} )}
& =
\left. \left( \vphantom{\sum} (1 - (n-1) p)^{\alpha} + (n-1) p^{\alpha} \right)^{\frac{1}{\alpha}} \right|_{(p, \alpha) = (\frac{1}{n (n-1)}, \frac{1}{2})}
\\
& =
\left( \left( 1 - (n-1) \left( \frac{1}{n (n-1)} \right) \right)^{\frac{1}{2}} + (n-1) \left( \frac{1}{n (n-1)} \right)^{\frac{1}{2}} \right)^{2}
\\
& =
\left( 2 \left( \frac{n-1}{n} \right)^{\frac{1}{2}} \right)^{2}
\\
& =
4 \left( \frac{n-1}{n} \right) .
\end{align}
Substituting $(p, \alpha) = (\frac{1}{n (n-1)}, \frac{1}{2})$ into the left-hand side of \eqref{eq:equation_p^ast}, we have
\begin{align}
&
\left. (\ln n - H_{\sbvec{v}_{n}}( p )) \left( \frac{ \partial \| \bvec{v}_{n}( p ) \|_{\alpha} }{ H_{\sbvec{v}_{n}}( p ) } \right) \right|_{(p, \alpha) = (\frac{1}{n (n-1)}, \frac{1}{2})}
\notag \\
& \quad =
\left. (\ln n - H_{\sbvec{v}_{n}}( p )) \left( \vphantom{\sum} (n-1) \, p^{\alpha} + (1 - (n-1)p)^{\alpha} \right)^{\frac{1}{\alpha} - 1} \left( \frac{ p^{\alpha-1} - (1 - (n-1)p)^{\alpha-1} }{ \ln \frac{ 1 - (n-1) p}{ p } } \right) \right|_{(p, \alpha) = (\frac{1}{n (n-1)}, \frac{1}{2})}
\\
& \quad =
\left( \ln n - H_{\sbvec{v}_{n}} \!\! \left( \frac{1}{n (n-1)} \right) \right) \left( (n-1) \left( \frac{1}{n (n-1)} \right)^{\frac{1}{2}} + \left( 1 - (n-1) \left( \frac{1}{n (n-1)} \right) \right)^{\frac{1}{2}} \right)^{1}
\notag \\
& \qquad \qquad \qquad \qquad \qquad \qquad \qquad \qquad \qquad \qquad \qquad \times
\left( \frac{ \left( \frac{1}{n (n-1)} \right)^{-\frac{1}{2}} - \left( 1 - (n-1) \left( \frac{1}{n (n-1)} \right) \right)^{-\frac{1}{2}} }{ \ln ( (n-1)^{2} ) } \right)
\\
& \quad =
\left( \ln n - \left( \ln n - \left( 1 - \frac{2}{n} \right) \ln (n-1) \right) \right) \left( 2 \left( \frac{n-1}{n} \right)^{\frac{1}{2}} \right) \left( \frac{ ( n (n-1) )^{\frac{1}{2}} - \left( \frac{n}{n-1} \right)^{\frac{1}{2}} }{ 2 \ln (n-1) } \right)
\\
& \quad =
\left( \left( \frac{n - 2}{n} \right) \ln (n-1) \right) (n-1)^{\frac{1}{2}} \left( \frac{ (n-1)^{\frac{1}{2}} - \left( \frac{1}{n-1} \right)^{\frac{1}{2}} }{ \ln (n-1) } \right)
\\
& \quad =
\left( \frac{n-2}{n} \ln (n-1) \right) \left( \frac{ (n-1) - 1 }{ \ln (n-1) } \right)
\\
& \quad =
\frac{ (n-2)^{2} }{ n } .
\label{eq:LHS_equation_p^ast_1/2}
\end{align}
Similarly, substituting $(p, \alpha) = (\frac{1}{n (n-1)}, \frac{1}{2})$ into the right-hand side of \eqref{eq:equation_p^ast}, we have
\begin{align}
\left. \left( \vphantom{\sum} \| \bvec{u}_{n} \|_{\alpha} - \| \bvec{v}_{n}( p ) \|_{\alpha} \right) \right|_{(p, \alpha) = (\frac{1}{n (n-1)}, \frac{1}{2})}
& =
n - 4 \left( \frac{n-1}{n} \right)
\\
& =
\frac{n^{2} - 4 n + 4}{n}
\\
& =
\frac{(n-2)^{2}}{n} .
\label{eq:RHS_equation_p^ast_1/2}
\end{align}
Since \eqref{eq:LHS_equation_p^ast_1/2} and \eqref{eq:RHS_equation_p^ast_1/2} are the same, we have $p_{n}^{\ast}( \frac{1}{2} ) = \frac{1}{n (n-1)}$.
\end{IEEEproof}

Therefore, after some algebra, we can obtain
\begin{align}
& \hspace{-0.5em}
L_{\max}^{\frac{1}{2}}(X \mid Y)
 =
\begin{cases}
\| \hat{\bvec{v}}_{n}(X \mid Y) \|_{\frac{1}{2}}
& \mathrm{if} \ H(X \mid Y) \le H_{\sbvec{v}_{n}}( \frac{1}{n (n-1)} ) , \\
T_{n, \frac{1}{2}} (X \mid Y)
& \mathrm{if} \ H(X \mid Y) > H_{\sbvec{v}_{n}}( \frac{1}{n (n-1)} ) ,
\end{cases}
\label{eq:Lmax_alpha_half}
\end{align}
where
\begin{align}
H_{\sbvec{v}_{n}}( {\textstyle \frac{1}{n (n-1)}} )
& =
\ln n - \left( 1 - \frac{2}{n} \right) \ln (n-1) ,
\\
T_{n, \frac{1}{2}} (X \mid Y)
& =
n - \frac{ (n-2) (\ln n - H(X \mid Y)) }{ \ln (n-1) } .
\end{align}
Note that Figs. \ref{fig:LminLmax}-\subref{subfig:norm_half} and \ref{fig:LminLmax_Renyi}-\subref{subfig:Renyi_half} are plotted by using \eqref{eq:Lmax_alpha_half}.

Since $\alpha = \frac{1}{1+\rho}$ in the $E_{0}$ function, the order $\alpha = \frac{1}{2}$ implies $\rho = 1$.
Thus, we can obtain the tight lower bound of the \emph{cutoff rate} $E_{0}(1, X, Y)$ by using \eqref{eq:Lmax_alpha_half}.
More precisely, if $\rho = 1$, then the lower bound of \eqref{ineq:E0_symmetric} can be calculated as
\begin{align}
E_{0}(1, X^{\prime\prime}, Y^{\prime\prime})
=
\ln n - \ln L_{\max}^{\frac{1}{2}}(X^{\prime\prime} \mid Y^{\prime\prime})
\label{eq:cutoffrate}
\end{align}
for a fixed $H(X^{\prime\prime} \mid Y^{\prime\prime}) \in [0, \ln n]$.

\section{Conclusion}

In this study, we investigated extremal relations between the conditional Shannon entropy and the expectation of $\ell_{\alpha}$-norm for joint probability distributions in Theorems \ref{th:cond_extremes} and \ref{th:cond_extremes2}.
Extending Theorem \ref{th:cond_extremes} to Corollary \ref{cor:cond_extremes}, we obtained tight bounds of some conditional entropies \eqref{eq:renyi_f} and \eqref{eq:R_f} with a fixed conditional Shannon entropy.
In Section \ref{subsect:DMC}, we applied Corollary \ref{cor:cond_extremes} for DMCs under a uniform input distribution.
Then, we showed tight bounds of the $E_{0}$ function with a fixed mutual information.


\appendices

\section{Proof of Lemma \ref{lem:convex_v}}
\label{app:convex_v}

\begin{IEEEproof}[Proof of Lemma \ref{lem:convex_v}]
By the chain rule of the derivation, we have
\begin{align}
\frac{ \partial^{2} \| \bvec{v}_{n}( p ) \|_{\alpha} }{ \partial H_{\sbvec{v}_{n}}( p )^{2} }
& =
\left( \frac{ \partial^{2} \| \bvec{v}_{n}( p ) \|_{\alpha} }{ \partial p^{2} } \right) \cdot \left( \frac{ \partial p }{ \partial H_{\sbvec{v}_{n}}( p ) } \right)^{2} + \left( \frac{ \partial \| \bvec{v}_{n}( p ) \|_{\alpha} }{ \partial p } \right) \cdot \left( \frac{ \partial^{2} p }{ \partial H_{\sbvec{v}_{n}}( p )^{2} } \right) .
\label{eq:diff2}
\end{align}
We can see from the proofs of \cite[Lemmas 1 and 3]{part1_arxiv} that
\begin{align}
\frac{ \mathrm{d} H_{\sbvec{v}_{n}}( p ) }{ \mathrm{d} p }
& =
(n-1) \ln \frac{ 1 - (n-1) p }{ p } ,
\label{eq:diff1_Hv} \\
\frac{ \partial \| \bvec{v}_{n}( p ) \|_{\alpha} }{ \partial p }
& =
(n-1) \left( \vphantom{\sum} (n-1) \, p^{\alpha} + (1 - (n-1)p)^{\alpha} \right)^{\frac{1}{\alpha} - 1} \left( \vphantom{\sum} p^{\alpha-1} - (1 - (n-1)p)^{\alpha-1} \right) .
\label{eq:norm_diff1}
\end{align}
Direct calculation shows
\begin{align}
\frac{ \partial^{2} \| \bvec{v}_{n}( p ) \|_{\alpha} }{ \partial p^{2} }
& =
\frac{ \partial }{ \partial p } \left( \frac{ \partial \| \bvec{v}_{n}( p ) \|_{\alpha} }{ \partial p } \right)
\\
& \overset{\eqref{eq:norm_diff1}}{=}
\frac{ \partial }{ \partial p } \left( (n-1) \left( \vphantom{\sum} (n-1) \, p^{\alpha} + (1 - (n-1)p)^{\alpha} \right)^{\frac{1}{\alpha} - 1} \left( \vphantom{\sum} p^{\alpha-1} - (1 - (n-1)p)^{\alpha-1} \right) \right)
\\
& \overset{\text{(a)}}{=}
(n-1) \left[ \left( \frac{ \partial }{ \partial p } \left( \vphantom{\sum} (n-1) \, p^{\alpha} + (1 - (n-1)p)^{\alpha} \right)^{\frac{1}{\alpha} - 1} \right) \left( \vphantom{\sum} p^{\alpha-1} - (1 - (n-1)p)^{\alpha-1} \right) \right.
\notag \\
& \left. \qquad \qquad \quad +
\left( \vphantom{\sum} (n-1) \, p^{\alpha} + (1 - (n-1)p)^{\alpha} \right)^{\frac{1}{\alpha} - 1} \left( \frac{ \partial }{ \partial p } \left( \vphantom{\sum} p^{\alpha-1} - (1 - (n-1)p)^{\alpha-1} \right) \right) \right]
\\
& =
(n-1) \left[ (1 - \alpha) (n-1) \left( \vphantom{\sum} (n-1) \, p^{\alpha} + (1 - (n-1)p)^{\alpha} \right)^{\frac{1}{\alpha} - 2} \left( \vphantom{\sum} p^{\alpha-1} - (1 - (n-1)p)^{\alpha-1} \right)^{2} \right.
\notag \\
& \left. \qquad + \,
(\alpha - 1) \left( \vphantom{\sum} (n-1) \, p^{\alpha} + (1 - (n-1)p)^{\alpha} \right)^{\frac{1}{\alpha} - 1} \left( \vphantom{\sum} p^{\alpha-2} + (n-1) (1 - (n-1)p)^{\alpha-2} \right) \right]
\\
& =
(n-1) (\alpha - 1) \left( \vphantom{\sum} (n-1) \, p^{\alpha} + (1 - (n-1)p)^{\alpha} \right)^{\frac{1}{\alpha} - 2} \left[ - (n-1) \left( \vphantom{\sum} p^{\alpha-1} - (1 - (n-1)p)^{\alpha-1} \right)^{2} \right.
\notag \\
& \left. \qquad +
\left( \vphantom{\sum} (n-1) \, p^{\alpha} + (1 - (n-1)p)^{\alpha} \right) \left( \vphantom{\sum} p^{\alpha-2} + (n-1) (1 - (n-1)p)^{\alpha-2} \right) \vphantom{\left( \vphantom{\sum} p^{\alpha-1} - (1 - (n-1)p)^{\alpha-1} \right)^{2}} \right]
\label{eq:norm_diff2_halfway} \\
& \overset{\text{(b)}}{=}
(n-1) (\alpha - 1) \left( \vphantom{\sum} (n-1) \, p^{\alpha} + (1 - (n-1)p)^{\alpha} \right)^{\frac{1}{\alpha} - 2} \left( \vphantom{\sum} p (1 - (n-1)p) \right)^{\alpha - 2} ,
\label{eq:norm_diff2}
\end{align}
where 
\begin{itemize}
\item
(a) follows by the product rule and
\item
(b) follows from the fact that the bracket $[ \cdot ]$ of the right-hand side of \eqref{eq:norm_diff2_halfway} is
\begin{align}
&
\left[ \left( \vphantom{\sum} (n-1) \, p^{\alpha} + (1 - (n-1)p)^{\alpha} \right) \left( \vphantom{\sum} p^{\alpha-2} + (n-1) (1 - (n-1)p)^{\alpha-2} \right) \right.
\notag \\
& \left. \qquad \qquad \qquad \qquad \qquad \qquad \qquad \qquad \qquad \qquad \qquad \qquad \qquad -
(n-1) \left( \vphantom{\sum} p^{\alpha-1} - (1 - (n-1)p)^{\alpha-1} \right)^{2} \right]
\notag \\
& \quad =
\left( \vphantom{\sum} (n-1) p^{2(\alpha-1)} + (n-1)^{2} p^{\alpha} (1 - (n-1)p)^{\alpha-2} + p^{\alpha-2} (1 - (n-1)p)^{\alpha} + (n-1) (1 - (n-1)p)^{2(\alpha-1)} \right)
\notag \\
& \qquad \qquad \qquad \qquad \qquad \qquad \qquad
- (n-1) \left( \vphantom{\sum} p^{2(\alpha-1)} - 2 p^{\alpha-1} (1 - (n-1)p)^{\alpha-1} + (1 - (n-1)p)^{2(\alpha-1)} \right)
\\
& \quad =
(n-1)^{2} p^{\alpha} (1 - (n-1)p)^{\alpha-2} + p^{\alpha-2} (1 - (n-1)p)^{\alpha} + 2 (n-1) p^{\alpha-1} (1 - (n-1)p)^{\alpha-1}
\\
& \quad =
\left( \vphantom{\sum} p (1 - (n-1)p) \right)^{\alpha-2} \left( \vphantom{\sum} 2 (n-1) p (1 - (n-1)p) + (n-1)^{2} p^{2} + (1 - (n-1)p)^{2} \right)
\\
& \quad =
\left( \vphantom{\sum} p (1 - (n-1)p) \right)^{\alpha-2} \underbrace{ \left( \vphantom{\sum} 2 (n-1) p - 2 (n-1)^{2} p^{2} + (n-1)^{2} p^{2} + 1 - 2 (n-1)p + (n-1)^{2} p^{2} \right) }_{ = 1 }
\\
& \quad =
\left( \vphantom{\sum} p (1 - (n-1)p) \right)^{\alpha-2} .
\end{align}
\end{itemize}
Moreover, we see that
\begin{align}
\frac{ \mathrm{d}^{2} p }{ \mathrm{d} H_{\sbvec{v}_{n}}( p )^{2} }
& =
\left[ \frac{ \mathrm{d} }{ \mathrm{d} p } \left( \frac{ \mathrm{d} p }{ \mathrm{d} H_{\sbvec{v}_{n}}( p ) } \right) \right] \left( \frac{ \mathrm{d} p }{ \mathrm{d} H_{\sbvec{v}_{n}}( p ) } \right)
\\
& \overset{\text{(a)}}{=}
\left[ \frac{ \mathrm{d} }{ \mathrm{d} p } \left( \frac{ 1 }{ \frac{ \mathrm{d} h_{n}( \mbox{\boldmath \scriptsize $p$}^{\ast} ) }{ \mathrm{d} p } } \right) \right] \left( \frac{ 1 }{ \frac{ \mathrm{d} h_{n}( \mbox{\boldmath \scriptsize $p$}^{\ast} ) }{ \mathrm{d} p } } \right)
\\
& \overset{\eqref{eq:diff1_Hv}}{=}
\left[ \frac{ \mathrm{d} }{ \mathrm{d} p } \left( \frac{ 1 }{ (n-1) \ln \frac{ 1 - (n-1) p}{ p } } \right) \right] \left( \frac{ 1 }{ (n-1) \ln \frac{ 1 - (n-1) p}{ p } } \right)
\\
& \overset{\text{(b)}}{=}
\left[ - \frac{ 1 }{ (n-1) \left( \ln \frac{ 1 - (n-1) p}{ p } \right)^{2} } \left( \frac{ \mathrm{d} \left( \ln \frac{ 1 - (n-1) p}{ p } \right) }{ \mathrm{d} p } \right) \right] \left( \frac{ 1 }{ (n-1) \ln \frac{ 1 - (n-1) p}{ p } } \right)
\\
& =
- \left( \frac{ 1 }{ (n-1)^{2} \left( \ln \frac{ 1 - (n-1) p}{ p } \right)^{3} } \right) \left( \frac{ \mathrm{d} \left( \ln \frac{ 1 - (n-1) p}{ p } \right) }{ \mathrm{d} p } \right)
\\
& \overset{\text{(c)}}{=}
- \left( \frac{ 1 }{ (n-1)^{2} \left( \ln \frac{ 1 - (n-1) p}{ p } \right)^{3} } \right) \left( \frac{ 1 }{ \frac{ 1 - (n-1) p }{ p } } \left( - \frac{1}{p^{2}} \right) \right)
\\
& =
\left( \frac{ 1 }{ (n-1)^{2} \left( \ln \frac{ 1 - (n-1) p}{ p } \right)^{3} } \right) \left( \frac{ 1 }{ p (1 - (n-1)p) } \right)
\\
& =
\frac{ 1 }{ p (1 - (n-1)p) (n-1)^{2} \left( \ln \frac{ 1 - (n-1) p}{ p } \right)^{3} } ,
\label{eq:hn_diff2}
\end{align}
where
\begin{itemize}
\item
(a) follows by the inverse function theorem,
\item
(b) follows from the fact that $\frac{ \mathrm{d} }{ \mathrm{d} x } \left( \frac{1}{f(x)} \right) = - \frac{ 1 }{ (f(x))^{2} } \left( \frac{ \mathrm{d} f(x) }{ \mathrm{d} x } \right)$, and
\item
(c) follows from the fact that $\frac{ \mathrm{d} \ln f(x) }{ \mathrm{d} x } = \frac{ 1 }{ f(x) } \left( \frac{ \mathrm{d} f(x) }{ \mathrm{d} x } \right)$.
\end{itemize}
Substituting \eqref{eq:norm_diff1}, \eqref{eq:diff1_Hv}, \eqref{eq:norm_diff2}, and \eqref{eq:hn_diff2} into \eqref{eq:diff2}, we obtain
\begin{align}
&
\frac{ \partial^{2} \| \bvec{v}_{n}( p ) \|_{\alpha} }{ \partial H_{\sbvec{v}_{n}}( p )^{2} }
\notag \\
& =
\underbrace{ (n-1) (\alpha - 1) \left( \vphantom{\sum} (n-1) \, p^{\alpha} + (1 - (n-1)p)^{\alpha} \right)^{\frac{1}{\alpha} - 2} \left( \vphantom{\sum} p (1 - (n-1)p) \right)^{\alpha - 2} }_{= \, \text{\eqref{eq:norm_diff2}}} \times \left( \vphantom{\frac{ 1 }{ (n-1) \ln \left( \frac{ 1 - (n-1) p }{ p } \right) }} \right. \underbrace{ \frac{ 1 }{ (n-1) \ln  \frac{ 1 - (n-1) p }{ p } } }_{= \, \frac{ 1 }{ \text{\eqref{eq:diff1_Hv}} }} \left. \vphantom{\frac{ 1 }{ (n-1) \ln \left( \frac{ 1 - (n-1) p }{ p } \right) }} \right)^{2}
\notag \\
& \qquad \qquad \qquad +
\underbrace{ (n-1) \left( \vphantom{\sum} (n-1) \, p^{\alpha} + (1 - (n-1)p)^{\alpha} \right)^{\frac{1}{\alpha} - 1} \left( \vphantom{\sum} p^{\alpha-1} - (1 - (n-1)p)^{\alpha-1} \right) }_{ = \, \text{\eqref{eq:norm_diff1}} }
\notag \\
& \qquad \qquad \qquad \qquad \qquad \qquad \qquad \qquad \qquad \qquad \qquad \times
\underbrace{ \left( \frac{ 1 }{ p (1 - (n-1)p) (n-1)^{2} \left( \ln \frac{ 1 - (n-1) p}{ p } \right)^{3} } \right) }_{ = \, \text{\eqref{eq:hn_diff2}} }
\\
& =
\frac{ (\alpha - 1) \left( \vphantom{\frac{ 1 - (n-1) p}{ p }} (n-1) \, p^{\alpha} + (1 - (n-1)p)^{\alpha} \right)^{\frac{1}{\alpha} - 2} \left( \vphantom{\frac{ 1 - (n-1) p}{ p }} p (1 - (n-1)p) \right)^{\alpha - 2} }{ (n-1) \left( \ln \left( \frac{ 1 - (n-1) p }{ p } \right) \right)^{2} }
\notag \\
& \qquad \qquad \qquad \qquad \qquad \qquad \qquad +
\frac{ \left( \vphantom{\frac{ 1 - (n-1) p}{ p }} (n-1) \, p^{\alpha} + (1 - (n-1)p)^{\alpha} \right)^{\frac{1}{\alpha} - 1} \left( \vphantom{\frac{ 1 - (n-1) p}{ p }} p^{\alpha-1} - (1 - (n-1)p)^{\alpha-1} \right) }{ p (1 - (n-1)p) (n-1) \left( \ln \frac{ 1 - (n-1) p}{ p } \right)^{3} }
\\
& =
\frac{ \left( \vphantom{\frac{ 1 - (n-1) p}{ p }} (n-1) \, p^{\alpha} + (1 - (n-1)p)^{\alpha} \right)^{\frac{1}{\alpha} - 2} \left( \vphantom{\frac{ 1 - (n-1) p}{ p }} p (1 - (n-1)p) \right)^{\alpha - 2} }{ (n-1) \left( \ln \frac{ 1 - (n-1) p}{ p } \right)^{2} }
\notag \\
& \qquad \qquad \qquad \qquad \qquad \times
\left[ (\alpha - 1) + \frac{ \left( \vphantom{\frac{ 1 - (n-1) p}{ p }} (n-1) \, p^{\alpha} + (1 - (n-1)p)^{\alpha} \right) \left( \vphantom{\frac{ 1 - (n-1) p}{ p }} p^{\alpha-1} - (1 - (n-1)p)^{\alpha-1} \right) }{ \left( \vphantom{\frac{ 1 - (n-1) p}{ p }} p (1 - (n-1)p) \right)^{\alpha - 1} \ln \frac{ 1 - (n-1) p}{ p } } \right]
\\
& =
\frac{ \left( \vphantom{\frac{ 1 - (n-1) p}{ p }} (n-1) \, p^{\alpha} + (1 - (n-1)p)^{\alpha} \right)^{\frac{1}{\alpha} - 2} \left( \vphantom{\frac{ 1 - (n-1) p}{ p }} p (1 - (n-1)p) \right)^{\alpha - 2} }{ (n-1) \left( \ln \frac{ 1 - (n-1) p}{ p } \right)^{2} }
\notag \\
& \quad \times
\left[ (\alpha - 1) + \frac{ (n-1) p^{2 \alpha - 1} - (n-1) p^{\alpha} (1 - (n-1) p)^{\alpha-1} + p^{\alpha-1} (1 - (n-1) p)^{\alpha} - (1 - (n-1) p)^{2\alpha-1} }{ \left( \vphantom{\frac{ 1 - (n-1) p}{ p }} p (1 - (n-1)p) \right)^{\alpha - 1} \ln \frac{ 1 - (n-1) p}{ p } } \right]
\\
& =
\frac{ \left( \vphantom{\frac{ 1 - (n-1) p}{ p }} (n-1) \, p^{\alpha} + (1 - (n-1)p)^{\alpha} \right)^{\frac{1}{\alpha} - 2} \left( \vphantom{\frac{ 1 - (n-1) p}{ p }} p (1 - (n-1)p) \right)^{\alpha - 2} }{ (n-1) \left( \ln \frac{ 1 - (n-1) p}{ p } \right)^{2} }
\notag \\
& \qquad \times
\left[ (\alpha - 1) + \frac{ (n-1) p^{\alpha} (1 - (n-1)p)^{1 - \alpha} - (n-1) p + (1 - (n-1) p) - p^{1 - \alpha} (1 - (n-1) p)^{\alpha} }{ \ln \frac{ 1 - (n-1) p}{ p } } \right]
\\
& =
\frac{ \left( \vphantom{\frac{ 1 - (n-1) p}{ p }} (n-1) \, p^{\alpha} + (1 - (n-1)p)^{\alpha} \right)^{\frac{1}{\alpha} - 2} \left( \vphantom{\frac{ 1 - (n-1) p}{ p }} p (1 - (n-1)p) \right)^{\alpha - 2} }{ (n-1) \left( \ln \frac{ 1 - (n-1) p}{ p } \right)^{2} }
\notag \\
& \qquad \qquad \qquad \quad \times
\left[ (\alpha - 1) + \frac{ 1 - 2 (n-1) p + (n-1) p^{\alpha} (1 - (n-1)p)^{1 - \alpha} - p^{1 - \alpha} (1 - (n-1) p)^{\alpha} }{ \ln \frac{ 1 - (n-1) p}{ p } } \right]
\\
& =
\left( \frac{ \left( \vphantom{\frac{ 1 - (n-1) p}{ p }} (n-1) \, p^{\alpha} + (1 - (n-1)p)^{\alpha} \right)^{\frac{1}{\alpha} - 2} \left( \vphantom{\frac{ 1 - (n-1) p}{ p }} p (1 - (n-1)p) \right)^{\alpha - 2} }{ (n-1) \left( \ln \frac{ 1 - (n-1) p}{ p } \right)^{2} } \right) \, g(n, p, \alpha) ,
\label{eq:diff2:N_H_v}
\end{align}
where the function $g(n, p, \alpha)$ is defined by
\begin{align}
g(n, p, \alpha)
\triangleq
(\alpha - 1) + \frac{ 1 - 2 (n-1) p + (n-1) p^{\alpha} (1 - (n-1)p)^{1 - \alpha} - p^{1 - \alpha} (1 - (n-1) p)^{\alpha} }{ \ln \frac{ 1 - (n-1) p}{ p } } .
\label{eq:g_p}
\end{align}
Since
\begin{align}
\sgn \! \left( \frac{ \left( \vphantom{\frac{ 1 - (n-1) p}{ p }} (n-1) \, p^{\alpha} + (1 - (n-1)p)^{\alpha} \right)^{\frac{1}{\alpha} - 2} \left( \vphantom{\frac{ 1 - (n-1) p}{ p }} p (1 - (n-1)p) \right)^{\alpha - 2} }{ (n-1) \left( \ln \frac{ 1 - (n-1) p}{ p } \right)^{2} } \right)
=
1
\end{align}
for $p \in (0, \frac{1}{n})$ and $\alpha \in (-\infty, 0) \cup (0, +\infty)$, we get from \eqref{eq:diff2:N_H_v} that
\begin{align}
\sgn \! \left( \frac{ \partial^{2} \| \bvec{v}_{n}( p ) \|_{\alpha} }{ \partial H_{\sbvec{v}_{n}}( p )^{2} } \right)
=
\sgn \! \left( \vphantom{\sum} g(n, p, \alpha) \right)
\label{eq:sgn_diff2_N_Hv}
\end{align}
for $p \in (0, \frac{1}{n})$ and $\alpha \in (-\infty, 0) \cup (0, +\infty)$.
After some algebra, we can rewrite $g(n, p, \alpha)$ as follows:
\begin{align}
g(n, p, \alpha)
& =
(\alpha - 1) + \frac{ 1 - 2 (n-1) p + (n-1) p^{\alpha} (1 - (n-1)p)^{1 - \alpha} - p^{1 - \alpha} (1 - (n-1) p)^{\alpha} }{ \ln \frac{ 1 - (n-1) p}{ p } }
\\
& =
(\alpha - 1) + \frac{ 1 - p \left( (n-1) \left(2 - \left( \frac{1 - (n-1)p}{p} \right)^{1 - \alpha} \right) + \left( \frac{1 - (n-1)p}{p} \right)^{\alpha} \right) }{ \ln \frac{ 1 - (n-1) p}{ p } }
\\
& \overset{\text{(a)}}{=}
(\alpha - 1) + \frac{ 1 - p \left( (n-1) \left(2 - z^{1 - \alpha} \right) + z^{\alpha} \right) }{ \ln z }
\\
& \overset{\text{(b)}}{=}
(\alpha - 1) + \frac{ 1 - \frac{1}{(n-1) + z} \left( (n-1) \left(2 - z^{1 - \alpha} \right) + z^{\alpha}
\right) }{ \ln z }
\\
& =
(\alpha - 1) + \frac{ ((n-1) + z) - \left( (n-1) \left(2 - z^{1 - \alpha} \right) + z^{\alpha}
\right) }{ ((n-1) + z) \ln z }
\\
& =
(\alpha - 1) + \frac{ z - (n-1) + (n-1) z^{1 - \alpha} - z^{\alpha} }{ ((n-1) + z) \ln z }
\\
& =
(\alpha - 1) + \frac{ z (1 - z^{\alpha-1}) - (n-1) (1 - z^{1 - \alpha}) }{ ((n-1) + z) \ln z }
\\
& =
(\alpha - 1) + \frac{ z^{\alpha} (z^{1-\alpha} - 1) + (n-1) (z^{1 - \alpha}-1) }{ ((n-1) + z) \ln z }
\\
& =
(\alpha - 1) + \frac{ ((n-1) + z^{\alpha}) (z^{1-\alpha} - 1) }{ ((n-1) + z) \ln z }
\end{align}
for $p \in (0, \frac{1}{n})$ and $\alpha \in (- \infty, 0) \cup (0, + \infty)$, where
\begin{itemize}
\item
(a) follows from the change of variable: $z = z(n, p) \triangleq \frac{ 1 - (n-1) p }{ p }$, and
\item
(b) follows from the fact that $z = \frac{ 1 - (n-1) p }{ p } \iff p = \frac{ 1 }{ (n-1) + z }$.
\end{itemize}
We define
\begin{align}
g(n, z, \alpha)
\triangleq
(\alpha - 1) + \frac{ ((n-1) + z^{\alpha}) (z^{1-\alpha} - 1) }{ ((n-1) + z) \ln z } ,
\label{eq:g_z}
\end{align}
where note that
\begin{itemize}
\item
$g(n, p, \alpha)$ denotes the right-hand side of \eqref{eq:g_p} and
\item
$g(n, z, \alpha)$ denotes the right-hand side of \eqref{eq:g_z}.
\end{itemize}
We now verify the relation between $p \in (0, \frac{1}{n})$ and $z(n, p)$.
It is easy to see that
\begin{align}
\frac{ \partial z(n, p) }{ \partial p }
& =
\frac{ \partial }{ \partial p } \left( \frac{ 1 - (n-1) p }{ p } \right)
\\
& \overset{\text{(a)}}{=}
\left( \frac{ \partial }{ \partial p } (1 - (n-1) p) \right) \frac{ 1 }{ p } + (1 - (n-1) p) \left( \frac{ \partial }{ \partial p } \left( \frac{1}{p} \right) \right)
\\
& =
( - (n-1) ) \frac{ 1 }{ p } + (1 - (n-1) p) \left(  - \frac{1}{p^{2}} \right)
\\
& =
\frac{  - (n-1) p - (1 - (n-1) p) }{ p^{2} }
\\
& =
- \frac{1}{p^{2}}
\\
& <
0
\label{eq:diff_z}
\end{align}
for $p \in (0, \frac{1}{n-1}]$, where (a) follows by the product rule;
namely, it follows from \eqref{eq:diff_z} that $z = z(n, p)$ is strictly decreasing for $p \in (0, \frac{1}{n-1}]$.
Moreover, we can see that
\begin{align}
\lim_{p \to 0^{+}} z(n, p)
& =
\lim_{p \to 0^{+}} \left( \frac{ 1 - (n-1) p }{ p } \right)
\\
& =
\lim_{p \to 0^{+}} \left( \frac{ 1 }{ p } - (n-1) \right)
\\
& =
+ \infty ,
\\
z(n, {\textstyle \frac{1}{n}})
& =
\left. \frac{ 1 - (n-1) p }{ p } \right|_{p = \frac{1}{n}}
\\
& =
\frac{ 1 - (n-1) \frac{1}{n} }{ \frac{1}{n} }
\\
& =
n - (n-1)
\\
& =
1 ,
\\
z(n, {\textstyle \frac{1}{n-1}})
& =
\left. \frac{ 1 - (n-1) p }{ p } \right|_{p = \frac{1}{n-1}}
\\
& =
\frac{ 1 - (n-1) \frac{1}{n-1} }{ \frac{1}{n-1} }
\\
& =
\frac{ 1 - 1 }{ \frac{1}{n-1} }
\\
& =
0 ,
\end{align}
which imply that
\begin{align}
0 < p \le \frac{1}{n}
& \iff
1 \le z < +\infty ,
\\
\frac{1}{n} \le p \le \frac{1}{n-1}
& \iff
0 \le z \le 1 .
\label{eq:range_z_w}
\end{align}
Therefore, it is enough to check the sign of $g(n, z, \alpha)$ for $z \in (1, +\infty)$ and $\alpha \in (0, 1) \cup (1, +\infty)$ rather than the sign of $g(n, p, \alpha)$ for $p \in (0, \frac{1}{n})$.
To analyze $g(n, z, \alpha)$, we calculate the partial derivatives of $g(n, z, \alpha)$ with respect to $\alpha$ as follows:
\begin{align}
\frac{ \partial g(n, z, \alpha) }{ \partial \alpha }
& \overset{\eqref{eq:g_z}}{=}
\frac{ \partial g(n, z, \alpha) }{ \partial \alpha } \left( (\alpha - 1) + \frac{ (z^{\alpha} + (n-1)) (z^{1-\alpha} - 1) }{ ((n-1) + z) \ln z } \right)
\\
& =
1 + \frac{ 1 }{ ((n-1) + z) \ln z } \left( \frac{ \partial }{ \partial \alpha } (z^{\alpha} + (n-1)) (z^{1-\alpha} - 1) \right)
\\
& =
1 + \frac{ 1 }{ ((n-1) + z) \ln z } \left( \frac{ \partial }{ \partial \alpha } ( z - z^{\alpha} + (n-1) z^{1 - \alpha} - (n-1) ) \right)
\\
& =
1 + \frac{ 1 }{ ((n-1) + z) \ln z } \left( (n-1) \left( \frac{ \partial z^{1 - \alpha} }{ \partial \alpha } \right) - \left( \frac{ \partial z^{\alpha} }{ \partial \alpha } \right) \right)
\\
& \overset{\text{(a)}}{=}
1 + \frac{ 1 }{ ((n-1) + z)  \ln z } \left( (n-1) (\ln z) (-1) z^{1-\alpha} - (\ln z) (1) z^{\alpha} \right)
\\
& =
1 - \frac{ (n-1) z^{1-\alpha} + z^{\alpha} }{ (n-1) + z } ,
\label{eq:diff1_g} \\
\frac{ \partial^{2} g(n, z, \alpha) }{ \partial \alpha^{2} }
& \overset{\eqref{eq:diff1_g}}{=}
\frac{ \partial }{ \partial \alpha } \left( 1 - \frac{ (n-1) z^{1-\alpha} + z^{\alpha} }{ (n-1) + z } \right)
\\
& =
- \frac{ \frac{ \partial }{ \partial \alpha } ((n-1) z^{1-\alpha} + z^{\alpha}) }{ (n-1) + z }
\\
& \overset{\text{(b)}}{=}
- \frac{ (n-1) (\ln z) (-1) z^{1-\alpha} + (\ln z) (1) z^{\alpha} }{ (n-1) + z }
\\
& =
\left( (n-1) z^{1-\alpha} - z^{\alpha} \right) \frac{ \ln z }{ (n-1) + z } ,
\label{eq:diff2_g}
\\
\frac{ \partial^{3} g(n, z, \alpha) }{ \partial \alpha^{3} }
& \overset{\eqref{eq:diff2_g}}{=}
\frac{ \partial }{ \partial \alpha } \left( \left( (n-1) z^{1-\alpha} - z^{\alpha} \right) \frac{ \ln z }{ (n-1) + z } \right)
\\
& \overset{\text{(c)}}{=}
( (n-1) (\ln z) (-1) z^{1-\alpha} - (\ln z) (1) z^{\alpha} ) \frac{ \ln z }{ (n-1) + z }
\\
& =
\underbrace{ - \left( (n-1) z^{1-\alpha} + z^{\alpha} \right) }_{< 0} \underbrace{ \frac{ (\ln z)^{2} }{ (n-1) + z } }_{\ge 0}
\\
& =
\begin{cases}
< 0
& \mathrm{if} \ z \in (0, 1) \cup (1, +\infty) , \\
= 0
& \mathrm{if} \ z = 1 ,
\end{cases}
\label{eq:diff3_g}
\end{align}
where (a), (b), and (c) follow from the fact that $\frac{ \mathrm{d} (a^{f(x)}) }{ \mathrm{d} x } = (\ln a) \left( \frac{ \mathrm{d} f(x) }{ \mathrm{d} x } \right) a^{f(x)}$ for $a > 0$.
It follows from \eqref{eq:diff3_g} that, if $z > 1$, then $\frac{ \partial^{2} g(n, z, \alpha) }{ \partial \alpha^{2} }$ is strictly decreasing for $\alpha \in (-\infty, +\infty)$.
We derive the solution of the equation $\frac{ \partial^{2} g(n, z, \alpha) }{ \partial \alpha^{2} } = 0$ with respect to $\alpha$ as follows:
\begin{align}
&&
\frac{ \partial^{2} g(n, z, \alpha) }{ \partial \alpha^{2} }
& =
0
\\
& \overset{\eqref{eq:diff2_g}}{\iff} &
\left( (n-1) z^{1-\alpha} - z^{\alpha} \right) \frac{ \ln z }{ (n-1) + z }
& =
0
\\
& \iff &
(n-1) z^{1-\alpha} - z^{\alpha}
& =
0
\\
& \iff &
(n-1) z^{1-\alpha}
& =
z^{\alpha}
\\
& \iff &
(n-1)
& =
z^{2\alpha-1}
\label{root:diff2_g}
\\
& \iff &
\ln (n-1)
& =
\ln z^{2\alpha-1} 
\\
& \iff &
\ln (n-1)
& =
( 2 \alpha - 1 ) \ln z
\\
& \iff &
2 \alpha - 1
& =
\frac{ \ln (n-1) }{ \ln z }
\\
& \iff &
\alpha
& =
\frac{1}{2} \left( 1 + \frac{ \ln (n-1) }{ \ln z } \right) .
\end{align}
Thus, we can denote by
\begin{align}
\alpha_{2}(n, z)
& \triangleq
\frac{ 1 }{ 2 } \left( 1 + \frac{ \ln (n-1) }{ \ln z } \right)
\label{def:a2}
\end{align}
the solution of $\frac{ \partial^{2} g(n, z, \alpha) }{ \partial \alpha^{2} } = 0$ with respect to $\alpha$ for $z \in (0, 1) \cup (1, +\infty)$.
Similarly, the solution of $\frac{ \partial^{2} g(n, z, \alpha) }{ \partial \alpha^{2} } = 0$ with respect to $z$ can also be derived as follows:
\begin{align}
&&
\frac{ \partial^{2} g(n, z, \alpha) }{ \partial \alpha^{2} }
& =
0
\\
& \overset{\eqref{root:diff2_g}}{\iff} &
(n-1)
& =
z^{2\alpha-1}
\\
& \iff &
z
& =
(n-1)^{\frac{1}{2\alpha-1}} .
\end{align}
Thus, we can also denote by
\begin{align}
z_{2}(n, \alpha)
& \triangleq
(n-1)^{\frac{1}{2 \alpha-1}}
\label{def:z2}
\end{align}
the root of $\frac{ \partial^{2} g(n, z, \alpha) }{ \partial \alpha^{2} } = 0$ with respect to $z$ for $\alpha \in (-\infty, \frac{1}{2}) \cup (\frac{1}{2}, +\infty)$.
Since $z_{2}(n, \cdot)$ is the inverse function of $\alpha_{2}(n, z)$ for $z \in (0, 1) \cup (1, +\infty)$, note that
\begin{align}
\alpha_{2}(n, z) = \alpha
\iff
z_{2}(n, \alpha) = z
\label{eq:inverse_a2z2}
\end{align}
for $z \in (0, 1) \cup (1, +\infty)$ and $\alpha \in (-\infty, \frac{1}{2}) \cup (\frac{1}{2}, +\infty)$.
Further, note that
\begin{align}
\alpha_{2}(2, z)
& \overset{\eqref{def:a2}}{=}
\left. \frac{1}{2} \left( 1 + \frac{ \ln (n-1) }{ \ln z } \right) \right|_{n = 2}
\\
& =
\frac{1}{2} \left( 1 + \frac{ \ln 1 }{ \ln z } \right)
\\
& =
\frac{1}{2} ,
\label{eq:alpha2_n2} \\
z_{2}(2, \alpha)
& \overset{\eqref{def:z2}}{=}
\left. (n-1)^{\frac{1}{2 \alpha - 1}} \right|_{n = 2}
\\
& =
1^{\frac{1}{2\alpha-1}}
\\
& =
1
\end{align}
for $z \in (0, 1) \cup (1, +\infty)$ and $\alpha \in (-\infty, \frac{1}{2}) \cup (\frac{1}{2}, +\infty)$.
If $n \ge 3$, we can see the following limiting values:
\begin{align}
\lim_{z \to 0^{+}} \alpha_{2}(n, z)
& =
\frac{1}{2} ,
\label{eq:alpha2_0} \\
\lim_{z \to 1^{-}} \alpha_{2}(n, z)
& =
- \infty,
\label{eq:alpha2_1_minus} \\
\lim_{z \to 1^{+}} \alpha_{2}(n, z)
& =
+ \infty,
\label{eq:alpha2_1_plus} \\
\lim_{z \to +\infty} \alpha_{2}(n, z)
& =
\frac{1}{2} ,
\label{eq:alpha2_infty} \\
\lim_{\alpha \to -\infty} z_{2}(n, \alpha)
& =
1 ,
\\
\lim_{\alpha \to (\frac{1}{2})^{-}} z_{2}(n, \alpha)
& =
0 ,
\\
\lim_{\alpha \to (\frac{1}{2})^{+}} z_{2}(n, \alpha)
& =
+\infty ,
\\
\lim_{\alpha \to +\infty} z_{2}(n, \alpha)
& =
1 .
\end{align}
Calculating the derivative of $\alpha_{2}(n, z)$ with respect to $z$ as
\begin{align}
\frac{ \partial \alpha_{2}(n, z) }{ \partial z }
& =
\frac{ \partial }{ \partial z } \left( \frac{ 1 }{ 2 } \left( 1 + \frac{ \ln (n-1) }{ \ln z } \right) \right)
\\
& =
\frac{ \ln (n-1) }{ 2 } \left( \frac{ \partial }{ \partial z } \left( \frac{ 1 }{ \ln z } \right) \right)
\\
& =
\frac{ \ln (n-1) }{ 2 } \left( - \frac{ 1 }{ (\ln z)^{2} } \left( \frac{ \partial \ln z }{ \partial z } \right) \right)
\\
& =
\frac{ \ln (n-1) }{ 2 } \left( - \frac{ 1 }{ (\ln z)^{2} } \frac{ 1 }{ z } \right)
\\
& =
- \frac{ \ln (n-1) }{ 2 z (\ln z)^{2} }
\\
& =
\begin{cases}
< 0
& \mathrm{if} \ z \in (0, 1) \cup (1, \infty) \ \mathrm{and} \ n \ge 3 , \\
= 0
& \mathrm{if} \ z \in (0, 1) \cup (1, \infty) \ \mathrm{and} \ n = 2 , \\
\end{cases}
\label{eq:diff_alpha1}
\end{align}
we see that, if $n \ge 3$, then
\begin{itemize}
\item
$\alpha_{2}(n, z)$ is strictly decreasing for $z \in (0, 1)$ and
\item
$\alpha_{2}(n, z)$ is strictly decreasing for $z \in (1, +\infty)$.
\end{itemize}
Moreover, the inverse function theorem shows that, if $n \ge 3$, then
\begin{itemize}
\item
$z_{2}(n, \alpha)$ is strictly decreasing for $\alpha \in (-\infty, \frac{1}{2})$ and
\item
$z_{2}(n, \alpha)$ is strictly decreasing for $\alpha \in (\frac{1}{2}, +\infty)$.
\end{itemize}
Since
\begin{itemize}
\item
if $z \in (0, 1) \cup (1, +\infty)$, then $\frac{ \partial^{2} g(n, z, \alpha) }{ \partial \alpha^{2} }$ is strictly decreasing for $\alpha \in (-\infty, +\infty)$ (see Eq. \eqref{eq:diff3_g}) and
\item
$\left. \frac{ \partial^{2} g(n, z, \alpha) }{ \partial \alpha^{2} } \right|_{\alpha = \alpha_{2}(n, z)} = 0$ for $z \in (0, 1) \cup (1, +\infty)$,
\end{itemize}
we can see that
\begin{itemize}
\item
if $z \in (0, 1) \cup (1, +\infty)$, then
\begin{align}
\sgn \! \left( \frac{ \partial^{2} g(n, z, \alpha) }{ \partial \alpha^{2} } \right)
=
\begin{cases}
1
& \mathrm{if} \ \alpha < \alpha_{2}(n, z) , \\
0
& \mathrm{if} \ \alpha = \alpha_{2}(n, z) , \\
-1
& \mathrm{if} \ \alpha > \alpha_{2}(n, z) .
\end{cases}
\label{eq:sign_diff2_g}
\end{align}
\end{itemize}
Hence, we have the following monotonicity:
\begin{itemize}
\item
for a fixed $z \in (0, 1) \cup (1, +\infty)$, the stationary point (global maximum) of $\frac{ \partial g(n, z, \alpha) }{ \partial \alpha }$ is at $\alpha = \alpha_{2}(n, z)$ and
\item
if $z \in (0, 1) \cup (1, +\infty)$, then
\begin{itemize}
\item
$\frac{ \partial g(n, z, \alpha) }{ \partial \alpha }$ is strictly increasing for $\alpha \in (- \infty, \alpha_{2}(n, z)]$ and
\item
$\frac{ \partial g(n, z, \alpha) }{ \partial \alpha }$ is strictly decreasing for $\alpha \in [\alpha_{2}(n, z), + \infty)$.
\end{itemize}
\end{itemize}

We now verify the sign of $\frac{ \partial g(n, z, \alpha) }{ \partial \alpha }$; namely, the monotonicity of $g(n, z, \alpha)$ with respect to $\alpha$ are examined.
Substituting $\alpha = 1$ into $\frac{ \partial g(n, z, \alpha) }{ \partial \alpha }$, we see that
\begin{align}
\left. \frac{ \partial g(n, z, \alpha) }{ \partial \alpha } \right|_{\alpha = 1}
& \overset{\eqref{eq:diff1_g}}{=}
\left. \left( 1 - \frac{ (n-1) z^{1-\alpha} + z^{\alpha} }{ (n-1) + z } \right) \right|_{\alpha = 1}
\\
& =
1 - \frac{ (n-1) + z }{ (n-1) + z }
\\
& =
1 - 1
\\
& =
0 .
\label{eq:diff1_g_a1}
\end{align}
Similarly, substituting $z = 1$ into $\frac{ \partial g(n, z, \alpha) }{ \partial \alpha }$, we see that
\begin{align}
\left. \frac{ \partial g(n, z, \alpha) }{ \partial \alpha } \right|_{z = 1}
& \overset{\eqref{eq:diff1_g}}{=}
\left. \left( 1 - \frac{ (n-1) z^{1-\alpha} + z^{\alpha} }{ (n-1) + z } \right) \right|_{z = 1}
\\
& =
1 - \frac{ (n-1) + 1 }{ (n-1) + 1 }
\\
& =
1 - 1
\\
& =
0 .
\label{eq:diff1_g_z1}
\end{align}
Moreover, we derive another solution of $\frac{ \partial g(n, z, \alpha) }{ \partial \alpha } = 0$ with respect to $\alpha$ as follows:
\begin{align}
&&
\frac{ \partial g(n, z, \alpha) }{ \partial \alpha }
& =
0
\\
& \overset{\eqref{eq:diff1_g}}{\iff} &
1 - \frac{ (n-1) z^{1-\alpha} + z^{\alpha} }{ (n-1) + z }
& =
0
\\
& \iff &
\frac{ (n-1) z^{1-\alpha} + z^{\alpha} }{ (n-1) + z }
& =
1
\\
& \iff &
(n-1) z^{1-\alpha} + z^{\alpha}
& =
(n-1) + z
\\
& \iff &
(n-1) z^{1-\alpha} - (n-1)
& =
z - z^{\alpha}
\\
& \iff &
(n-1) (z^{1-\alpha} - 1)
& =
z^{\alpha} (z^{1-\alpha} - 1) ;
\\
& \iff &
(n-1)
& =
z^{\alpha}
\label{eq:root_diff1_g} \\
& \iff &
\ln (n-1)
& =
\ln z^{\alpha}
\\
& \iff &
\ln (n-1)
& =
\alpha \ln z
\\
& \iff &
\alpha
& =
\frac{ \ln (n-1) }{ \ln z } .
\label{eq:diff1_g_root}
\end{align}
Thus, a solution of $\frac{ \partial g(n, z, \alpha) }{ \partial \alpha } = 0$ with respect to $\alpha$ can be denoted by
\begin{align}
\alpha_{1}(n, z)
\triangleq
\frac{ \ln (n-1) }{ \ln z }
\label{def:a1}
\end{align}
for $n \ge 2$ and $z \in (0, 1) \cup (1, +\infty)$.
We can also derive another solution of $\frac{ \partial g(n, z, \alpha) }{ \partial \alpha } = 0$ with respect to $z$ as follows:
\begin{align}
&&
\frac{ \partial g(n, z, \alpha) }{ \partial \alpha }
& =
0
\\
& \overset{\eqref{eq:root_diff1_g}}{\iff} &
(n-1)
& =
z^{\alpha}
\\
& \iff &
z
& =
(n-1)^{\frac{1}{\alpha}} .
\label{eq:unless_alpha=0} 
\end{align}
Thus, we can also denote by
\begin{align}
z_{1}(n, \alpha)
& \triangleq
(n-1)^{\frac{1}{\alpha}}
\label{def:z1}
\end{align}
a solution of $\frac{ \partial g(n, z, \alpha) }{ \partial \alpha } = 0$ with respect to $z$ for $\alpha \in (-\infty, 0) \cup (0, +\infty)$.
As with \eqref{eq:inverse_a2z2}, since $z_{1}(n, \cdot)$ is the inverse function of $\alpha_{1}(n, z)$ for $z \in (0, 1) \cup (1, +\infty)$, note that
\begin{align}
\alpha_{1}(n, z) = \alpha
\iff
z_{1}(n, \alpha) = z
\end{align}
for $z \in (0, 1) \cup (1, +\infty)$ and $\alpha \in (-\infty, 0) \cup (0, +\infty)$.
Further, note that
\begin{align}
\alpha_{1}(2, z)
& \overset{\eqref{def:a1}}{=}
\left. \frac{ \ln (n-1) }{ \ln z } \right|_{n = 2}
\\
& =
\frac{ \ln 1 }{ \ln z }
\\
& =
0 ,
\label{eq:alpha1_n2} \\
z_{1}(2, \alpha)
& \overset{\eqref{def:z1}}{=}
\left. (n-1)^{\frac{1}{\alpha}} \right|_{n = 2}
\\
& =
1^{\frac{1}{\alpha}}
\\
& =
1
\end{align}
for $z \in (0, 1) \cup (1, +\infty)$ and $\alpha \in (-\infty, 0) \cup (0, +\infty)$.
If $n \ge 3$, we can see the following limiting values:
\begin{align}
\lim_{z \to 0^{+}} \alpha_{1}(n, z)
& =
0 ,
\\
\lim_{z \to 1^{-}} \alpha_{1}(n, z)
& =
- \infty ,
\\
\lim_{z \to 1^{+}} \alpha_{1}(n, z)
& =
+ \infty ,
\label{eq:alpha1_lim_z=0+}
\\
\lim_{z \to +\infty} \alpha_{1}(n, z)
& =
0 ,
\\
\lim_{\alpha \to -\infty} z_{1}(n, \alpha)
& =
1 ,
\\
\lim_{\alpha \to 0^{-}} z_{1}(n, \alpha)
& =
0 ,
\\
\lim_{\alpha \to 0^{+}} z_{1}(n, \alpha)
& =
+\infty ,
\\
\lim_{\alpha \to +\infty} z_{1}(n, \alpha)
& =
1 .
\end{align}
Calculating the derivative of $\alpha_{1}(n, z)$ with respect to $z$ as
\begin{align}
\frac{ \partial \alpha_{1}(n, z) }{ \partial z }
& =
\frac{ \partial }{ \partial z } \left( \frac{ \ln (n-1) }{ \ln z } \right)
\\
& =
- \frac{ \ln (n-1) }{ (\ln z)^{2} } \left( \frac{ \mathrm{d} \ln z }{ \mathrm{d} z } \right)
\\
& =
- \frac{ \ln (n-1) }{ z (\ln z)^{2} }
\\
& =
\begin{cases}
< 0
& \mathrm{if} \ z \in (0, 1) \cup (1, +\infty) \ \mathrm{and} \ n \ge 3 , \\
= 0
& \mathrm{if} \ z \in (0, 1) \cup (1, +\infty) \ \mathrm{and} \ n = 2 , \\
\end{cases}
\label{eq:diff1_alpha1}
\end{align}
we can see that, if $n \ge 3$, then
\begin{itemize}
\item
$\alpha_{1}(n, z)$ is strictly decreasing for $z \in (0, 1)$ and
\item
$\alpha_{1}(n, z)$ is strictly decreasing for $z \in (1, +\infty)$.
\end{itemize}
Moreover, the inverse function theorem shows that, if $n \ge 3$, then
\begin{itemize}
\item
$z_{1}(n, \alpha)$ is strictly decreasing for $\alpha \in (-\infty, 0)$ and
\item
$z_{1}(n, \alpha)$ is strictly decreasing for $\alpha \in (0, +\infty)$.
\end{itemize}

We now check the magnitude relation between $\alpha_{1}(n, z)$ and $\alpha_{2}(n, z)$.
When $n = 2$, it follows from \eqref{eq:alpha2_n2} and \eqref{eq:alpha1_n2} that
\begin{align}
\underbrace{ \alpha_{1}(2, z) }_{ = 0 } < \underbrace{ \alpha_{2}(2, z) }_{ = \frac{1}{2} } < 1
\end{align}
for $z \in (0, 1) \cup (1, +\infty)$.
Hence, if $n = 2$, we readily see from \eqref{eq:diff1_g_z1} that
\begin{align}
\sgn \! \left( \frac{ \partial g(2, z, \alpha) }{ \partial \alpha } \right)
=
\begin{cases}
1
& \mathrm{if} \ \alpha \in (0, 1) , \\
0
& \mathrm{if} \ \alpha \in \{ 0, 1 \} , \\
-1
& \mathrm{if} \ \alpha \in (-\infty, 0) \cup (1, +\infty)
\end{cases}
\label{eq:diff1_g_n2}
\end{align}
for $z \in (0, 1) \cup (1, +\infty)$;
and therefore, for any $z \in (0, 1) \cup (1, +\infty)$, we have that
\begin{itemize}
\item
the stationary point (local minimum) $g(2, z, \alpha)$ is at $\alpha = \alpha_{1}(2, z) = 0$,
\item
the inflection point of $g(2, z, \alpha)$ is at $\alpha = \alpha_{2}(2, z) = \frac{1}{2}$,
\item
the stationary point (local maximum) $g(2, z, \alpha)$ is at $\alpha = 1$,
\item
$g(2, z, \alpha)$ is strictly decreasing for $\alpha \in (-\infty, 0]$,
\item
$g(2, z, \alpha)$ is strictly increasing for $\alpha \in [0, 1]$, and
\item
$g(2, z, \alpha)$ is strictly decreasing for $\alpha \in [1, +\infty)$.
\end{itemize}
We now consider the magunitude relation between $\alpha_{1}(n, z)$ and $\alpha_{2}(n, z)$ for $n \ge 3$.
Direct calculation yields
\begin{align}
\alpha_{2}(n, z) - \alpha_{1}(n, z)
& =
\frac{1}{2} \left( 1 + \frac{ \ln (n-1) }{ \ln z } \right) - \frac{ \ln (n-1) }{ \ln z }
\\
& =
\frac{1}{2} \left( 1 - \frac{ \ln (n-1) }{ \ln z } \right) ,
\\
\frac{ \partial \left( \vphantom{\frac{1}{x}} \alpha_{2}(n, z) - \alpha_{1}(n, z) \right) }{ \partial z }
& =
\frac{ \partial }{ \partial z } \left( \frac{1}{2} \left( 1 - \frac{ \ln (n-1) }{ \ln z } \right) \right)
\\
& =
- \frac{1}{2} \left( \frac{ \partial }{ \partial z } \left( \frac{ \ln (n-1) }{ \ln z } \right) \right)
\\
& =
- \frac{1}{2} \left( - \frac{ \ln (n-1) }{ (\ln z)^{2} } \right) \left( \frac{ \mathrm{d} \ln z }{ \mathrm{d} z } \right)
\\
& =
\frac{ \ln (n-1) }{ 2 z (\ln z)^{2} }
\\
& =
\begin{cases}
> 0
& \mathrm{if} \ z \in (0, 1) \cup (1, +\infty) \ \mathrm{and} \ n \ge 3 , \\
= 0
& \mathrm{if} \ z \in (0, 1) \cup (1, +\infty) \ \mathrm{and} \ n = 2 ,
\end{cases}
\\
\lim_{z \to 0^{+}} \left( \vphantom{\frac{1}{x}} \alpha_{2}(n, z) - \alpha_{1}(n, z) \right)
& =
\frac{1}{2} ,
\\
\lim_{z \to 1^{-}} \left( \vphantom{\frac{1}{x}} \alpha_{2}(n, z) - \alpha_{1}(n, z) \right)
& =
+ \infty ,
\\
\lim_{z \to 1^{+}} \left( \vphantom{\frac{1}{x}} \alpha_{2}(n, z) - \alpha_{1}(n, z) \right)
& =
- \infty ,
\\
\left. \left( \vphantom{\frac{1}{x}} \alpha_{2}(n, z) - \alpha_{1}(n, z) \right) \right|_{z = n-1}
& =
0 ,
\\
\lim_{z \to +\infty} \left( \vphantom{\frac{1}{x}} \alpha_{2}(n, z) - \alpha_{1}(n, z) \right)
& =
\frac{1}{2} .
\end{align}
From the above equations, the following magnitude relations hold:
\begin{align}
\left\{
\begin{array}{ll}
\alpha_{1}(n, z) < 0 < \alpha_{2}(n, z) < \frac{1}{2} & \mathrm{if} \ z \in (0, \frac{1}{n-1}) , \\
\alpha_{1}(n, z) = -1 \ \mathrm{and} \ \alpha_{2}(n, z) = 0 & \mathrm{if} \ z = \frac{1}{n-1} , \\
\alpha_{1}(n, z) < \alpha_{2}(n, z) < 0 < 1 & \mathrm{if} \ z = (\frac{1}{n-1}, 1) , \\
0 < 1 < \alpha_{2}(n, z) < \alpha_{1}(n, z) & \mathrm{if} \ z \in (1, n-1) , \\
\alpha_{1}(n, z) = \alpha_{2}(n, z) = 1 & \mathrm{if} \ z = n-1 , \\
0 < \alpha_{1}(n, z) < \alpha_{2}(n, z) < 1 & \mathrm{if} \ z \in (n-1, +\infty)
\end{array}
\right.
\label{eq:range_a1a2}
\end{align}
for $n \ge 3$.
Moreover, since $z_{1}(n, \alpha)$ and $z_{2}(n, \alpha)$ are the inverse functions of $\alpha_{1}(n, z)$ and $\alpha_{2}(n, z)$, respectively, with respect to $z \in (0, 1) \cup (1, +\infty)$, we also have the following magnitude relations:
\begin{align}
\left\{
\begin{array}{ll}
0 < z_{1}(n, \alpha) < z_{2}(n, \alpha) < 1
& \mathrm{if} \ \alpha \in (-\infty, 0) ,
\\
z_{1}(n, \alpha) \ \mathrm{is} \ \mathrm{undefined} \ \mathrm{and} \ z_{2}(n, \alpha) = \frac{1}{n-1}
& \mathrm{if} \ \alpha = 0 ,
\\
0 < z_{2}(n, \alpha) < \frac{1}{n-1} \ \mathrm{and} \ (n-1)^{2} < z_{1}(n, \alpha)
& \mathrm{if} \ \alpha \in (0, \frac{1}{2}) ,
\\
z_{1}(n, \alpha) = (n-1)^{2} \ \mathrm{and} \ z_{2}(n, \alpha) \ \mathrm{is} \ \mathrm{undefined}
& \mathrm{if} \ \alpha = \frac{1}{2} ,
\\
n-1 < z_{1}(n, \alpha) < z_{2}(n, \alpha)
& \mathrm{if} \ \alpha \in (\frac{1}{2}, 1) ,
\\
z_{1}(n, \alpha) = z_{1}(n, \alpha) = n-1
& \mathrm{if} \ \alpha = 1 ,
\\
1 < z_{2}(n, \alpha) < z_{1}(n, \alpha) < n-1
& \mathrm{if} \ \alpha \in (1, +\infty) .
\end{array}
\right.
\label{eq:range_z1z2_orig}
\end{align}

Noting the above magnitude relations, we can see from \eqref{eq:sign_diff2_g} that
\begin{itemize}

\item
if $z \in (0, 1) \cup (n-1, +\infty)$, then
\begin{align}
\sgn \! \left( \frac{ \partial g(n, z, \alpha) }{ \partial \alpha } \right)
=
\begin{cases}
1
& \mathrm{if} \ \alpha \in (\alpha_{1}(n, z), 1) , \\
0
& \mathrm{if} \ \alpha \in \{ \alpha_{1}(n, z), 1 \} , \\
-1
& \mathrm{if} \ \alpha \in (-\infty, \alpha_{1}(n, z)) \cup (1, +\infty) ,
\end{cases}
\label{eq:diff1_g_alpha_part1}
\end{align}

\item
if $z \in (1, n-1)$, then
\begin{align}
\sgn \! \left( \frac{ \partial g(n, z, \alpha) }{ \partial \alpha } \right)
=
\begin{cases}
1
& \mathrm{if} \ \alpha \in (1, \alpha_{1}(n, z)) , \\
0
& \mathrm{if} \ \alpha \in \{ 1, \alpha_{1}(n, z) \} , \\
-1
& \mathrm{if} \ \alpha \in (-\infty, 1) \cup (\alpha_{1}(n, z), +\infty) ,
\end{cases}
\label{eq:diff1_g_alpha_part2}
\end{align}
and

\item
if $z = n-1$, then
\begin{align}
\sgn \! \left( \frac{ \partial g(n, z, \alpha) }{ \partial \alpha } \right)
=
\begin{cases}
0
& \mathrm{if} \ \alpha = 1 , \\
-1
& \mathrm{if} \ \alpha \in (-\infty, 1) \cup (1, +\infty) .
\end{cases}
\label{eq:diff1_g_alpha_part3}
\end{align}
\end{itemize}
Therefore, if $n \ge 3$, we have the following monotonicity:
\begin{itemize}

\item
if $z \in (0, 1) \cup (n-1, +\infty)$, then
\begin{itemize}
\item
$\alpha_{1}(n, z) < \alpha_{2}(n, z) < 1$,
\item
the stationary point (local minimum) of $g(n, z, \alpha)$ is at $\alpha = \alpha_{1}(n, z)$,
\item
the inflection point of $g(n, z, \alpha)$ is at $\alpha = \alpha_{2}(n, z)$,
\item
the stationary point (local maximum) of $g(n, z, \alpha)$ is at $\alpha = 1$,
\item
$g(n, z, \alpha)$ is strictly decreasing for $\alpha \in (-\infty, \alpha_{1}(n, z)]$,
\item
$g(n, z, \alpha)$ is strictly increasing for $\alpha \in [\alpha_{1}(n, z), 1]$, and
\item
$g(n, z, \alpha)$ is strictly decreasing for $\alpha \in [1, +\infty)$,
\end{itemize}

\item
if $z \in (1, n-1)$, then
\begin{itemize}
\item
$1 < \alpha_{2}(n, z) < \alpha_{1}(n, z)$,
\item
the stationary point (local minimum) of $g(n, z, \alpha)$ is at $\alpha = 1$,
\item
the inflection point of $g(n, z, \alpha)$ is at $\alpha = \alpha_{2}(n, z)$,
\item
the stationary point (local maximum) of $g(n, z, \alpha)$ is at $\alpha = \alpha_{1}(n, z)$,
\item
$g(n, z, \alpha)$ is strictly decreasing for $\alpha \in (-\infty, 1]$,
\item
$g(n, z, \alpha)$ is strictly increasing for $\alpha \in [1, \alpha_{1}(n, z)]$, and
\item
$g(n, z, \alpha)$ is strictly decreasing for $\alpha \in [\alpha_{1}(n, z), +\infty)$,
\end{itemize}

\item
if $z = n-1$, then
\begin{itemize}
\item
$\alpha_{1}(n, z) = \alpha_{2}(n, z) = 1$,
\item
the stationary point (saddle point) of $g(n, z, \alpha)$ is at $\alpha = 1$,
\item
the inflection point of $g(n, z, \alpha)$ is at $\alpha = 1$, and
\item
$g(n, z, \alpha)$ is strictly decreasing for $\alpha \in (-\infty, +\infty)$.
\end{itemize}
\end{itemize}

We now calculate the limiting values of $g(n, z, \alpha)$ with respect to $\alpha$.
It can be seen that
\begin{align}
\lim_{\alpha \to -\infty} g(n, z, \alpha)
& =
\lim_{\alpha \to -\infty} \left( (\alpha-1) + \frac{ ((n-1) + z^{\alpha}) (z^{1-\alpha} - 1) }{ ((n-1) + z) \ln z } \right)
\\
& =
\lim_{\alpha \to -\infty} \left( (\alpha-1) + \frac{ (n-1) z^{1-\alpha} - (n-1) + z - z^{\alpha} }{ ((n-1) + z) \ln z } \right)
\\
& =
+\infty
\label{eq:g_lim_-}
\end{align}
for $z \in (0, 1) \cup (1, +\infty)$, where note that
\begin{align}
\lim_{\alpha \to -\infty} z^{1-\alpha}
& =
\begin{cases}
+\infty
& \mathrm{if} \ z \in (1, +\infty) , \\
0
& \mathrm{if} \ z \in (0, 1) ,
\end{cases}
\\
\lim_{\alpha \to -\infty} z^{\alpha}
& =
\begin{cases}
+\infty
& \mathrm{if} \ z \in (0, 1) , \\
0
& \mathrm{if} \ z \in (1, +\infty) .
\end{cases}
\end{align}
Similarly, we have
\begin{align}
\lim_{\alpha \to +\infty} g(n, z, \alpha)
& =
\lim_{\alpha \to +\infty} \left( (\alpha-1) + \frac{ ((n-1) + z^{\alpha}) (z^{1-\alpha} - 1) }{ ((n-1) + z) \ln z } \right)
\\
& = 
-\infty
\label{eq:g_lim_+}
\end{align}
for $z \in (0, 1) \cup (1, +\infty)$.
On the other hand, it easy see that
\begin{align}
g(n, z, 1)
& =
\left. \left( (\alpha-1) + \frac{ ((n-1) + z^{\alpha}) (z^{1-\alpha} - 1) }{ ((n-1) + z) \ln z } \right) \right|_{\alpha = 1}
\\
& =
(1-1) + \frac{ ((n-1) + z^{1}) (z^{1-1} - 1) }{ ((n-1) + z) \ln z }
\\
& =
\frac{ ((n-1) + z) (z^{0} - 1) }{ ((n-1) + z) \ln z }
\\
& =
\frac{ 1 - 1 }{ \ln z }
\\
& =
0
\label{eq:g_alpha_is_0}
\end{align}
for $n \ge 2$ and $z \in (0, 1) \cup (1, +\infty)$.

Using the above results, we now consider the sign of $g(n, z, \alpha)$ with $n = 2$.
Since
\begin{itemize}

\item
the following monotonicity of $g(2, z, \alpha)$ hold (see Eq. \eqref{eq:diff1_g_n2}):
\begin{itemize}
\item
$g(2, z, \alpha)$ is strictly decreasing for $\alpha \in (-\infty, 0]$,
\item
$g(2, z, \alpha)$ is strictly increasing for $\alpha \in [0, 1]$, and
\item
$g(2, z, \alpha)$ is strictly decreasing for $\alpha \in [1, +\infty)$,
\end{itemize}

\item
for $z \in (0, 1) \cup (1, +\infty)$, $g(2, z, 1) = 0$ (see Eq. \eqref{eq:g_alpha_is_0}), and

\item
for $z \in (0, 1) \cup (1, +\infty)$, $g(2, z, \alpha) \to -\infty$ as $\alpha \to +\infty$ (see Eq. \eqref{eq:g_lim_+}),

\end{itemize}
we observe that
\begin{align}
\sgn \! \left( \vphantom{\sum} g(2, z, \alpha) \right)
=
\begin{cases}
0
& \mathrm{if} \ \alpha = 1 , \\
-1
& \mathrm{if} \ \alpha \neq 1
\end{cases}
\end{align}
for $\alpha \in (0, 1) \cup (1, +\infty)$ and $z \in (0, 1) \cup (1, +\infty)$.
Note that this results for $n=2$ is same as \cite[Appendix I]{fabregas}.

We further consider the sign of $g(n, z, \alpha)$ for $n \ge 3$ by using the above analyses.
We first show the following lemma.

\begin{lemma}[The case of $\alpha = 0$]
\label{lem:g_a0}
For any $n \ge 3$, there exists $\kappa_{p}( n ) \in (\mathrm{e}^{-n}, \frac{1}{n(n-1)})$ such that
\begin{align}
\sgn \! \left( \vphantom{\sum} g(n, p, 0) \right)
=
\begin{cases}
1
& \mathrm{if} \ p \in (\kappa_{p}( n ), \frac{1}{n}) , \\
0
& \mathrm{if} \ p = \kappa_{p}( n ) , \\
-1
& \mathrm{if} \ p \in (0, \kappa_{p}( n )) \cup (\frac{1}{n}, \frac{1}{n-1}) .
\end{cases}
\label{eq:g_alpha0_p}
\end{align}
On the other hand, if $n = 2$, then $g(2, p, 0) < 0$ for $p \in (0, \frac{1}{2}) \cup (\frac{1}{2}, 1)$.
\end{lemma}

The proof of Lemma \ref{lem:g_a0} is given in Appendix \ref{app:g_a0}.
Since $z = \frac{1 - (n-1) p}{ p }$, we see that
\begin{align}
p = \mathrm{e}^{- n}
& \iff
z = \mathrm{e}^{n} - (n-1) ,
\\
p = \frac{1}{n (n-1)}
& \iff
z = (n-1)^{2} .
\end{align}
Thus, Eq. \eqref{eq:g_alpha0_p} of Lemma \ref{lem:g_a0} can be rewritten as follows:
for any $n \ge 3$, there exists $\kappa_{z}(n) \in (\mathrm{e}^{n} - (n-1), (n-1)^{2})$ such that
\begin{align}
\sgn \! \left( \vphantom{\sum} g(n, z, 0) \right)
=
\begin{cases}
1
& \mathrm{if} \ z \in (1, \kappa_{z}(n)) , \\
0
& \mathrm{if} \ z = \kappa_{z}(n) , \\
-1
& \mathrm{if} \ z \in (0, 1) \cup (\kappa_{z}(n), +\infty) .
\end{cases}
\label{eq:g_kappa_z}
\end{align}

By the intermediate value theorem and the monotonicity of $g(n, z, \alpha)$ with respect to $\alpha$ (see Eqs. \eqref{eq:diff1_g_alpha_part1}, \eqref{eq:diff1_g_alpha_part2}, and \eqref{eq:diff1_g_alpha_part3}), for a fixed $n \ge 3$, the sign of $g(n, z, \alpha)$ is evaluated as follows:
\begin{itemize}

\item
for any $z \in (1, n-1)$, there exists $\xi_{n}( z ) \in ( \alpha_{1}( n, z ), +\infty )$ such that
\begin{align}
\sgn \! \left( \vphantom{\sum} g(n, z, \alpha) \right)
=
\begin{cases}
1
& \mathrm{if} \ \alpha \in (-\infty, 1) \cup (1, \xi_{n}( z )) , \\
0
& \mathrm{if} \ \alpha \in \{ 1, \xi_{n}( z ) \} , \\
-1
& \mathrm{if} \ \alpha \in (\xi_{n}( z ), +\infty) ,
\end{cases}
\label{eq:sign_gz_part1}
\end{align}

\item
if $z = n - 1$, then
\begin{align}
\sgn \! \left( \vphantom{\sum} g(n, z, \alpha) \right)
=
\begin{cases}
1
& \mathrm{if} \ \alpha \in (-\infty, 1) , \\
0
& \mathrm{if} \ \alpha = 1 , \\
-1
& \mathrm{if} \ \alpha \in (1, +\infty) ,
\end{cases}
\label{eq:sign_gz_part2}
\end{align}

\item
for any $z \in (n-1, \kappa_{z}(n))$, there exists $\xi_{n}( z ) \in ( 0, \alpha_{1}( n, z ) )$ such that
\begin{align}
\sgn \! \left( \vphantom{\sum} g(n, z, \alpha) \right)
=
\begin{cases}
1
& \mathrm{if} \ \alpha \in (-\infty, \xi_{n}( z )) , \\
0
& \mathrm{if} \ \alpha \in \{ \xi_{n}( z ), 1 \} , \\
-1
& \mathrm{if} \ \alpha \in (\xi_{n}( z ), 1) \cup (1, +\infty) ,
\end{cases}
\label{eq:sign_gz_part3}
\end{align}

\item
if $z = \kappa_{z}(n)$, then
\begin{align}
\sgn \! \left( \vphantom{\sum} g(n, z, \alpha) \right)
=
\begin{cases}
1
& \mathrm{if} \ \alpha \in (-\infty, 0) , \\
0
& \mathrm{if} \ \alpha = \{ 0, 1 \} , \\
-1
& \mathrm{if} \ \alpha \in (0, 1) \cup (1, +\infty) ,
\end{cases}
\label{eq:sign_gz_part4}
\end{align}
and

\item
for any $z \in (\kappa_{z}(n), +\infty)$, there exists $\xi_{n}( z ) \in ( -\infty, 0 )$ such that
\begin{align}
\sgn \! \left( \vphantom{\sum} g(n, z, \alpha) \right)
=
\begin{cases}
1
& \mathrm{if} \ \alpha \in (-\infty, \xi_{n}( z )) , \\
0
& \mathrm{if} \ \alpha \in \{ \xi_{n}( z ), 1 \} , \\
-1
& \mathrm{if} \ \alpha \in (\xi_{n}( z ), 1) \cup (1, +\infty) .
\end{cases}
\label{eq:sign_gz_part5}
\end{align}

\end{itemize}

Note that, to prove the above statements, we use the following equalities:
\begin{align}
\lim_{\alpha \to -\infty} g(n, z, \alpha)
& \overset{\eqref{eq:g_lim_-}}{=}
+\infty ,
\\
\sgn \! \left( \vphantom{\sum} g(n, z, 0) \right)
& \overset{\eqref{eq:g_kappa_z}}{=}
\begin{cases}
1
& \mathrm{if} \ z \in (1, \kappa_{z}(n)) , \\
0
& \mathrm{if} \ z = \kappa_{z}(n) , \\
-1
& \mathrm{if} \ z \in (0, 1) \cup (\kappa_{z}(n), +\infty) ,
\end{cases}
\\
g(n, z, 1)
& \overset{\eqref{eq:g_alpha_is_0}}{=}
0 ,
\\
\lim_{\alpha \to +\infty} g(n, z, \alpha)
& \overset{\eqref{eq:g_lim_+}}{=}
-\infty .
\end{align}

Henceforth, for a fixed $n \ge 3$, we will prove that the value $\xi_{n}( z )$, used in \eqref{eq:sign_gz_part1} and \eqref{eq:sign_gz_part3}, is strictly decreasing for $z \in (1, (n-1)^{2}]$.
Note that, for any $\epsilon > 0$, there exists $\delta( \epsilon ) > 0$ such that $\xi_{n}( z ) > \epsilon$ for all $1 < z < 1 + \delta( \epsilon )$ since $\alpha_{1}(n, z) \to +\infty$ as $z \to 1^{+}$ (see Eq. \eqref{eq:alpha1_lim_z=0+}) and $\xi_{n}( z ) \in (\alpha_{1}(n, z), +\infty)$ when $z \in (1, n-1)$ (see Eq. \eqref{eq:sign_gz_part1}).
To put it simply, we see that
\begin{align}
\lim_{z \to 1^{+}} \xi_{n}( z )
=
+\infty .
\label{eq:xi_infty}
\end{align}
To show the monotonicity of $\xi_{n}( z )$ with respect to $z \in (1, (n-1)^{2}]$, we now provide the following three lemmas:

\begin{lemma}
\label{lem:dzda}

For any fixed $n \ge 3$, the following statements hold:
\begin{itemize}

\item
for any $\alpha \in (-\infty, 0)$, there exists $\gamma(n, \alpha) \in (z_{1}(n, 2 \alpha), z_{2}(n, \alpha))$ such that
\begin{align}
\sgn \! \left( \frac{ \partial^{2} g(n, z, \alpha) }{ \partial z \, \partial \alpha } \right)
=
\begin{cases}
1
& \mathrm{if} \ z \in (0, \gamma(n, \alpha)) , \\
0
& \mathrm{if} \ z = \gamma(n, \alpha) , \\
-1
& \mathrm{if} \ z \in (\gamma(n, \alpha), +\infty) ,
\end{cases}
\label{eq:g_dzda_1}
\end{align}
where note that $0< z_{1}(n, \alpha) < z_{1}(n, 2\alpha) < z_{2}(n, \alpha) < 1$ for $\alpha \in (-\infty, 0)$,

\item
if $\alpha = 0$, then $\frac{ \partial^{2} g(n, z, \alpha) }{ \partial z \, \partial \alpha } < 0$ for any $z \in (0, +\infty)$,

\item
for any $\alpha \in (0, \frac{1}{2})$, there exists $\gamma(n, \alpha) \in (n-1, z_{1}(n, 2 \alpha))$ such that
\begin{align}
\sgn \! \left( \frac{ \partial^{2} g(n, z, \alpha) }{ \partial z \, \partial \alpha } \right)
=
\begin{cases}
1
& \mathrm{if} \ z \in (\gamma(n, \alpha), +\infty) , \\
0
& \mathrm{if} \ z = \gamma(n, \alpha) , \\
-1
& \mathrm{if} \ z \in (0, \gamma(n, \alpha)) ,
\end{cases}
\label{eq:g_dzda_2}
\end{align}
where note that $n-1 < z_{1}(n, 2 \alpha) < z_{1}(n, \alpha)$ for $\alpha \in (0, \frac{1}{2})$,

\item
if $\alpha = \frac{1}{2}$, then
\begin{align}
\sgn \! \left( \frac{ \partial^{2} g(n, z, \alpha) }{ \partial z \, \partial \alpha } \right)
=
\begin{cases}
1
& \mathrm{if} \ z \in (n-1, +\infty) , \\
0
& \mathrm{if} \ z = n-1 , \\
-1
& \mathrm{if} \ z \in (0, n-1) ,
\end{cases}
\label{eq:g_dzda_3}
\end{align}

\item
for any $\alpha \in (\frac{1}{2}, 1)$, there exists $\gamma(n, \alpha) \in (z_{1}(n, 2 \alpha), n-1)$ such that
\begin{align}
\sgn \! \left( \frac{ \partial^{2} g(n, z, \alpha) }{ \partial z \, \partial \alpha } \right)
=
\begin{cases}
1
& \mathrm{if} \ z \in (\gamma(n, \alpha), +\infty) , \\
0
& \mathrm{if} \ z = \gamma(n, \alpha) , \\
-1
& \mathrm{if} \ z \in (0, \gamma(n, \alpha)) ,
\end{cases}
\label{eq:g_dzda_4}
\end{align}
where note that $\sqrt{n-1} < z_{1}(n, 2 \alpha) < n-1 < z_{1}(n, \alpha) < z_{2}(n, \alpha)$ for $\alpha \in (0, \frac{1}{2})$,

\item
if $\alpha = 1$, then $\frac{ \partial^{2} g(n, z, \alpha) }{ \partial z \, \partial \alpha } = 0$ for any $z \in (0, +\infty)$, and

\item
for any $\alpha \in (1, +\infty)$, there exists $\gamma(n, \alpha) \in (z_{1}(n, 2\alpha), z_{2}(n, \alpha))$ such that
\begin{align}
\sgn \! \left( \frac{ \partial^{2} g(n, z, \alpha) }{ \partial z \, \partial \alpha } \right)
=
\begin{cases}
1
& \mathrm{if} \ z \in (0, \gamma(n, \alpha)) , \\
0
& \mathrm{if} \ z = \gamma(n, \alpha) , \\
-1
& \mathrm{if} \ z \in (\gamma(n, \alpha), +\infty) ,
\end{cases}
\label{eq:g_dzda_5}
\end{align}
where note that $1 < z_{1}(n, 2 \alpha) < z_{2}(n, \alpha) < z_{1}(n, \alpha) < n-1$ for $\alpha \in (1, +\infty)$.

\end{itemize}
\end{lemma}

Lemma \ref{lem:dzda} is proved in Appendix \ref{app:dzda}.
In Lemma \ref{lem:dzda}, note that $\frac{ \partial^{2} g(n, z, \alpha) }{ \partial z \, \partial \alpha } = \frac{ \partial^{2} g(n, z, \alpha) }{ \partial \alpha \, \partial z }$ by Young's theorem.
We can verify it by calculating derivatives.

\begin{lemma}
\label{lem:diff_g_z}
For any $n \ge 3$ and any $z \in (1, +\infty)$, we observe that
\begin{align}
\left. \frac{ \partial g(n, z, \alpha) }{ \partial z } \right|_{\alpha = 1}
& =
0 ,
\\
\sgn \! \left( \left. \frac{ \partial g(n, z, \alpha) }{ \partial z } \right|_{\alpha = \alpha_{1}(n, z)} \right)
& =
\begin{cases}
0
& \mathrm{if} \ z = n-1 , \\
-1
& \mathrm{if} \ z \neq n-1 .
\end{cases}
\label{eq:diff_g_z}
\end{align}
\end{lemma}

Lemma \ref{lem:diff_g_z} is proved in Appendix \ref{app:diff_g_z}.

\begin{lemma}
\label{lem:ln(n-1)/2ln(z)}
For any $n \ge 3$ and any $z \in [n-1, (n-1)^{2}]$,
\begin{align}
\sgn \! \left( \vphantom{\sum} g(n, z, {\textstyle \frac{1}{2} \alpha_{1}(n, z)}) \right)
& =
1 .
\end{align}
\end{lemma}

Lemma \ref{lem:ln(n-1)/2ln(z)} is proved in Appendix \ref{app:ln(n-1)/2ln(z)}.

We divide the proof of the monotonicity of $\xi_{n}( z )$ for $z \in (1, (n-1)^{2}]$ into the case of $z \in (1, n-1]$ and the case of $z \in [n-1, (n-1)^{2}]$.
Firstly, we show that $\xi_{n}( z )$ is strictly decreasing for $z \in (1, n-1]$.
To prove this, we now provide that
\begin{align}
\sgn \! \left( \frac{ \partial^{2} g(n, z, \alpha) }{ \partial z \, \partial \alpha } \right)
=
-1
\label{eq:dzda_part5_>=alpha1}
\end{align}
for $z \in (1, n-1)$ and $\alpha \ge \alpha_{1}(n, z)$, which implies that $\frac{ \partial g(n, z, \alpha) }{ \partial z }$ with a fixed $z \in (1, n-1)$ is strictly decreasing for $\alpha \ge \alpha_{1}(n, z)$.
We can verify \eqref{eq:dzda_part5_>=alpha1} as follows:
Since $1 < \gamma( n, \alpha ) < z_{2}( n, \alpha ) < z_{1}( n, \alpha )$ for $\alpha \in (1, +\infty)$, it follows from \eqref{eq:g_dzda_5} that
\begin{align}
\sgn \! \left( \left. \frac{ \partial^{2} g(n, z, \alpha) }{ \partial z \, \partial \alpha } \right|_{z = z_{1}( n, \alpha)} \right)
=
-1
\label{eq:dzda_part5_>=alpha1_derive1}
\end{align}
for $\alpha \in (1, +\infty)$.
Then, since $z_{1}( n, \alpha )$ is strictly decreasing for $\alpha > 0$ (by the inverse function theorem and Eq. \eqref{eq:diff1_alpha1}), we see that $1 < \gamma( n, \alpha ) < z_{1}( n, \alpha ) \le z_{1}( n, \beta )$ for $1 < \beta \le \alpha < +\infty$;
and thus, we observe from \eqref{eq:dzda_part5_>=alpha1_derive1} that
\begin{align}
\sgn \! \left( \left. \frac{ \partial^{2} g(n, z, \alpha) }{ \partial z \, \partial \alpha } \right|_{z = z_{1}( n, \beta)} \right)
=
-1
\label{eq:dzda_part5_>=alpha1_derive2}
\end{align}
for $1 < \beta \le \alpha < +\infty$.
Moreover, since $\alpha_{1}( n, \cdot )$ is the inverse function of $z_{1}( n, \beta )$ for $\beta \in (-\infty, 0) \cup (0, +\infty)$, i.e.,
\begin{align}
z = z_{1}(n, \beta)
& \iff
\beta = \alpha_{1}(n, z) ,
\\
1 < \beta < +\infty
& \overset{\eqref{eq:range_z1z2_orig}}{\iff}
1 < z_{1}( n, \beta ) < n-1 ,
\end{align}
we obtain \eqref{eq:dzda_part5_>=alpha1} from \eqref{eq:dzda_part5_>=alpha1_derive2}.
On the other hand, from Lemma \ref{lem:diff_g_z}, we observe that
\begin{align}
\sgn \! \left( \left. \frac{ \partial g(n, z, \alpha) }{ \partial z } \right|_{\alpha = \alpha_{1}(n, z)} \right)
=
-1
\end{align}
for a fixed $z \in (1, n-1)$;
and therefore, it follows from \eqref{eq:dzda_part5_>=alpha1} that
\begin{align}
\sgn \! \left( \frac{ \partial g(n, z, \alpha) }{ \partial z } \right)
=
-1
\label{eq:diff1_z_1ton-1}
\end{align}
for $z \in (1, n-1)$ and $\alpha \ge \alpha_{1}( n, z )$, which implies that $g(n, z, \alpha)$ with a fixed $\alpha \ge \alpha_{1}( n, z )$ is strictly decreasing for $z \in (1, n-1)$.
Then, since
\begin{itemize}
\item
$\xi_{n}( z ) > \alpha_{1}( n, z )$ and $g( n, z, \xi_{n}( z ) ) = 0$ for $z \in (1, n-1)$ (see Eq. \eqref{eq:sign_gz_part1}), and
\item
$g( n, z, \alpha )$ with a fixed $\alpha \ge \alpha_{1}( n, z )$ is strictly decreasing for $z \in (1, n-1)$ (see Eq. \eqref{eq:diff1_z_1ton-1}),
\end{itemize}
we observe that
\begin{align}
g( n, z^{\prime}, \xi_{n}( z ) ) < 0
\label{eq:g_zprime_1ton-1}
\end{align}
for $1 < z < z^{\prime} < n-1$.
Moreover, since
\begin{itemize}
\item
if $1 < z^{\prime} < n-1$, then $g( n, z^{\prime}, \alpha )$ is strictly decreasing for $\alpha \ge \alpha_{1}( n, z^{\prime} )$ (see Eq. \eqref{eq:diff1_g_alpha_part2}),
\item
$g( n, z^{\prime}, \alpha_{1}( n, z^{\prime} ) ) > 0$ for $1 < z^{\prime} < n-1$ (see Eq. \eqref{eq:sign_gz_part1}), and
\item
$g( n, z^{\prime}, \xi_{n}( z ) ) < 0$ for $1 < z < z^{\prime} < n-1$ (see Eq. \eqref{eq:g_zprime_1ton-1}),
\end{itemize}
it follows by the intermediate value theorem that, for any $1 < z < z^{\prime} < n-1$, there exists $\xi_{n}( z^{\prime} ) \in (\alpha_{1}(n, z^{\prime}), \xi_{n}( z ))$ such that
\begin{align}
g( n, z^{\prime}, \xi_{n}( z^{\prime} ) )
=
0 ,
\end{align}
which implies that, if $1 < z < z^{\prime} < n-1$, then $\xi_{n}( z^{\prime} ) < \xi_{n}( z )$.
Note that
\begin{align}
\lim_{z \to 1^{+}} \xi_{n}( z )
& \overset{\eqref{eq:xi_infty}}{=}
+\infty ,
\label{eq:xi_infty2} \\
\xi_{n}( n-1 )
& \overset{\eqref{eq:sign_gz_part2}}{=}
1 .
\label{eq:xi_n-1_1}
\end{align}
Therefore, we obtain that $\xi_{n}( z ) \in [1, +\infty)$ is strictly decreasing for $z \in (1, n-1]$.

Secondly, we show that $\xi_{n}( z )$ is strictly decreasing for $z \in [n-1, (n-1)^{2}]$.
From \eqref{eq:range_a1a2}, note that
\begin{align}
0 & < \frac{1}{2} \alpha_{1}( n, z ) < \frac{1}{2} ,
\label{eq:magnitudes_1/2alpha1_1} \\
\frac{1}{2} \alpha_{1}( n, z ) & < \alpha_{1}( n, z ) < 1
\label{eq:magnitudes_1/2alpha1_2} 
\end{align}
for $z > n-1$.
Then, since
\begin{itemize}
\item
if $z > n-1$, then $g( n, z, \alpha )$ is strictly decreasing for $\alpha \in (-\infty, \alpha_{1}( n, z^{\prime} )]$ (see Eq. \eqref{eq:diff1_g_alpha_part1}),
\item
$g(n, z, \frac{1}{2} \alpha_{1}(n, z)) > 0$ for $z \in [n-1, (n-1)^{2}]$ (see Lemma \ref{lem:ln(n-1)/2ln(z)}), and
\item
$g(n, z, \alpha_{1}(n, z)) < 0$ for $z \in (n-1, (n-1)^{2}]$ (see Eq. \eqref{eq:sign_gz_part3}),
\end{itemize}
we can refine a part of the bounds of $\xi_{n}( z )$ used in \eqref{eq:sign_gz_part3} as follows:
for any $z \in (n-1, (n-1)^{2}]$, there exists $\xi_{n}( z ) \in (\frac{1}{2} \alpha_{1}(n, z), \alpha_{1}(n, z))$ such that
\begin{align}
\sgn \! \left( \vphantom{\sum} g(n, z, \alpha) \right)
=
\begin{cases}
1
& \mathrm{if} \ \alpha \in (-\infty, \xi_{n}( z )) , \\
0
& \mathrm{if} \ \alpha \in \{ \xi_{n}( z ), 1 \} , \\
-1
& \mathrm{if} \ \alpha \in (\xi_{n}( z ), 1) \cup (1, +\infty) .
\end{cases}
\label{eq:sign_gz_part3_2}
\end{align}
To prove the monotonicity of $\xi_{n}( z )$ for $z \in [n-1, (n-1)^{2}]$, we now provide that
\begin{align}
\sgn \! \left( \frac{ \partial g(n, z, \alpha) }{ \partial z } \right)
=
-1
\label{eq:diff1_gz_1/2alpha_to_1}
\end{align}
for $z > n-1$ and $\alpha \in [\frac{1}{2} \alpha_{1}( n, z ), 1)$.
It follows from \eqref{eq:g_dzda_3} and \eqref{eq:g_dzda_4} of Lemma \ref{lem:dzda} that
\begin{align}
\sgn \! \left( \frac{ \partial^{2} g(n, z, \alpha) }{ \partial z \, \partial \alpha } \right)
=
1
\label{eq:dzda_part5_>=1/2to1}
\end{align}
for $z > n-1$ and $\alpha \in [\frac{1}{2}, 1)$.
Moreover, we now prove that
\begin{align}
\sgn \! \left( \frac{ \partial^{2} g(n, z, \alpha) }{ \partial z \, \partial \alpha } \right)
=
1
\label{eq:dzda_part5_>=2alpha1}
\end{align}
for $z > n-1$ and $\alpha \in [\frac{1}{2} \alpha_{1}(n, z), \frac{1}{2}]$, as with the proof of \eqref{eq:dzda_part5_>=alpha1}.
Since $n-1 < \gamma( n, \alpha ) < z_{1}(n, 2 \alpha) < z_{1}(n, \alpha)$ for $\alpha \in (0, \frac{1}{2})$, it follows from \eqref{eq:g_dzda_2} that
\begin{align}
\sgn \! \left( \left. \frac{ \partial^{2} g(n, z, \alpha) }{ \partial z \, \partial \alpha } \right|_{z = z_{1}(n, 2\alpha)} \right)
=
1
\label{eq:dzda_part5_>=2alpha1_derive1}
\end{align}
for $\alpha \in (0, \frac{1}{2})$.
Then, since $z_{1}( n, \alpha )$ is strictly decreasing for $\alpha > 0$ (by the inverse function theorem and Eq. \eqref{eq:diff1_alpha1}), we see that $1 < \gamma( n, \alpha ) < z_{1}( n, 2\alpha ) \le z_{1}( n, 2\beta )$ for $0 < \beta \le \alpha < \frac{1}{2}$;
and thus, we observe from \eqref{eq:dzda_part5_>=2alpha1_derive1} that
\begin{align}
\sgn \! \left( \left. \frac{ \partial^{2} g(n, z, \alpha) }{ \partial z \, \partial \alpha } \right|_{z = z_{1}(n, 2\beta)} \right)
=
1
\label{eq:dzda_part5_>=2alpha1_derive2}
\end{align}
for $0 < \beta \le \alpha < \frac{1}{2}$.
Moreover, since $\alpha_{1}( n, \cdot )$ is the inverse function of $z_{1}( n, \beta )$ for $\beta \in (-\infty, 0) \cup (0, +\infty)$, i.e.,
\begin{align}
z = z_{1}(n, 2 \beta)
& \iff
\beta = \frac{1}{2} \alpha_{1}(n, z) ,
\\
0 < \beta < \frac{1}{2}
& \overset{\eqref{eq:range_z1z2_orig}}{\iff}
n-1 < z_{1}( n, 2 \beta ) < +\infty ,
\end{align}
we obtain \eqref{eq:dzda_part5_>=2alpha1} from \eqref{eq:dzda_part5_>=2alpha1_derive2}.
Combining \eqref{eq:dzda_part5_>=1/2to1} and \eqref{eq:dzda_part5_>=2alpha1}, we have
\begin{align}
\sgn \! \left( \frac{ \partial^{2} g(n, z, \alpha) }{ \partial z \, \partial \alpha } \right)
=
1
\end{align}
for $z > n-1$ and $\alpha \in [\frac{1}{2} \alpha_{1}(n, z), 1)$, which implies that $\frac{ \partial g(n, z, \alpha) }{ \partial z }$ with a fixed $z > n-1$ is strictly increasing for $\alpha \in [\frac{1}{2} \alpha_{1}(n, z), 1]$.
Then, since $\left. \frac{ \partial g(n, z, \alpha) }{ \partial z } \right|_{\alpha = 1} = 0$ (see Lemma \ref{lem:diff_g_z}), we obtain \eqref{eq:diff1_gz_1/2alpha_to_1}.
It follows from \eqref{eq:diff1_gz_1/2alpha_to_1} that $g(n, z, \alpha)$ with a fixed $\alpha \in [\frac{1}{2} \alpha_{1}(n, z), 1)$ is strictly decreasing for $z > n-1$.
Using this monotonicity, we can prove the monotonicity of $\xi_{n}( z )$ for $z \in [n-1, (n-1)^{2}]$ as follows:
Since
\begin{itemize}
\item
$\frac{1}{2} \alpha_{1}( n, z ) < \xi_{n}( z ) < \alpha_{1}( n, z ) < 1$ for $z \in (n-1, (n-1)^{2}]$ (see Eqs. \eqref{eq:magnitudes_1/2alpha1_2} and \eqref{eq:sign_gz_part3_2}),
\item
$g( n, z, \xi_{n}( z ) ) = 0$ for $z \in (n-1, (n-1)^{2}]$ (see Eq. \eqref{eq:sign_gz_part3_2}), and
\item
$g(n, z, \alpha)$ with a fixed $\alpha \in [\frac{1}{2} \alpha_{1}(n, z), 1)$ is strictly decreasing for $z > n-1$ (see Eq. \eqref{eq:diff1_gz_1/2alpha_to_1}),
\end{itemize}
we observe that
\begin{align}
g( n, z^{\prime}, \xi_{n}( z ) ) < 0
\label{eq:g_zprime_>n-1}
\end{align}
for $n-1 < z < z^{\prime}$.
Moreover, since
\begin{itemize}
\item
if $z^{\prime} > n-1$, then $g( n, z^{\prime}, \alpha )$ is strictly decreasing for $\alpha \in (-\infty, \alpha_{1}( n, z^{\prime} ))$ (see Eq. \eqref{eq:diff1_g_alpha_part1}),
\item
$g( n, z^{\prime}, \frac{1}{2} \alpha_{1}( n, z^{\prime} ) ) > 0$ for $z^{\prime} \in [n-1, (n-1)^{2}]$ (see Lemma \ref{lem:ln(n-1)/2ln(z)}), and
\item
$g( n, z^{\prime}, \xi_{n}( z ) ) < 0$ for $n-1 < z < z^{\prime}$ (see Eq. \eqref{eq:g_zprime_>n-1}),
\end{itemize}
it follows by the intermediate value theorem that, for any $n-1 < z < z^{\prime} \le (n-1)^{2}$, there exists $\xi_{n}( z^{\prime} ) \in (\frac{1}{2} \alpha_{1}(n, z^{\prime}), \xi_{n}( z ))$ such that
\begin{align}
g( n, z^{\prime}, \xi_{n}( z^{\prime} ) )
=
0 ,
\end{align}
which implies that, if $n-1 < z < z^{\prime} \le (n-1)^{2}$, then $\xi_{n}( z^{\prime} ) < \xi_{n}( z )$.
Note that
\begin{align}
\xi_{n}( n-1 )
& \overset{\eqref{eq:sign_gz_part2}}{=}
1 ,
\label{eq:xi_n-1_2} \\
\frac{1}{4}
<
\xi_{n}( (n-1)^{2} )
& <
\frac{1}{2} ,
\label{ineq:magnitudes_xi(n-1)^2}
\end{align}
where \eqref{ineq:magnitudes_xi(n-1)^2} follows by $\frac{1}{2} \alpha_{1}(n, (n-1)^{2}) < \xi_{n}( (n-1)^{2} ) < \alpha_{1}(n, (n-1)^{2})$ (see Eq. \eqref{eq:sign_gz_part3_2}).
Therefore, we obtain that $\xi_{n}( z )$ is strictly decreasing for $z \in [n-1, (n-1)^{2}]$.
Note that
\begin{align}
\xi_{n}( z ) < \frac{1}{2}
\label{ineq:magnitudes_xi>(n-1)^2}
\end{align}
for $z > (n-1)^{2}$ since $\xi_{n}( z ) < \alpha_{1}(n, z)$ and $\alpha_{1}(n, z)$ is strictly decreasing for $z > 1$.

Summarizing the above monotonicity of $\xi_{n}( z )$, we get that there exists the inverse function $\xi_{n}^{-1}( \cdot )$ of $\xi_{n}( z )$ for $z \in (1, (n-1)^{2}]$.
By the inverse function theorem, note that $\xi_{n}^{-1}( \alpha )$ is strictly decreasing for $\alpha \in [\frac{1}{2}, 1) \cup (1, +\infty)$.
Since
\begin{itemize}
\item
$\xi_{n}( z )$ is strictly decreasing for $z \in (1, n-1]$,
\item
$\lim_{z \to 1^{+}} \xi_{n}( z ) = +\infty$ (see Eq. \eqref{eq:xi_infty2}), and
\item
$\xi_{n}( n-1 ) = 1$ (see Eq. \eqref{eq:xi_n-1_1}),
\end{itemize}
we can rewrite \eqref{eq:sign_gz_part1} by using the inverse function $\xi_{n}^{-1}( \cdot )$ as follows:
for any $\alpha \in (1, +\infty)$, there exists $\xi_{n}^{-1}( \alpha ) \in (1, n-1)$ such that
\begin{align}
\sgn \! \left( \vphantom{\sum} g(n, z, \alpha) \right)
=
\begin{cases}
1
& \mathrm{if} \ z \in (1, \xi_{n}^{-1}( \alpha )) , \\
0
& \mathrm{if} \ z = \xi_{n}^{-1}( \alpha ) , \\
-1
& \mathrm{if} \ z \in (\xi_{n}^{-1}( \alpha ), +\infty) . 
\end{cases}
\label{eq:sign_gz_part1_rewrite}
\end{align}
Moreover, since
\begin{itemize}
\item
$\xi_{n}( z )$ is strictly decreasing for $z \in [n-1, (n-1)^{2}]$,
\item
$\xi_{n}( n-1 ) = 1$ (see Eq. \eqref{eq:xi_n-1_2}), and
\item
$\xi_{n}( z ) < \frac{1}{2}$ for $z \ge (n-1)^{2}$ (see Eq. \eqref{ineq:magnitudes_xi>(n-1)^2}),
\end{itemize}
we can also rewrite a part of \eqref{eq:sign_gz_part3} by using the inverse function $\xi_{n}^{-1}( \cdot )$ as follows:
for any $\alpha \in [\frac{1}{2}, 1)$ and any $n \ge 3$, there exists $\xi_{n}^{-1}( \alpha ) \in (n-1, (n-1)^{2})$ such that
\begin{align}
\sgn \! \left( \vphantom{\sum} g(n, z, \alpha) \right)
=
\begin{cases}
1
& \mathrm{if} \ z \in (1, \xi_{n}^{-1}( \alpha )) , \\
0
& \mathrm{if} \ z = \xi_{n}^{-1}( \alpha ) , \\
-1
& \mathrm{if} \ z \in (\xi_{n}^{-1}( \alpha ), +\infty) . 
\end{cases}
\label{eq:sign_gz_part3_rewrite}
\end{align}
Note that
\begin{align}
g(n, z, 1)
\overset{\eqref{eq:g_alpha_is_0}}{=}
0 .
\end{align}
for $z > 1$.
Combining \eqref{eq:sign_gz_part1_rewrite} and \eqref{eq:sign_gz_part3_rewrite}, we obtain that, for any $\alpha \in [\frac{1}{2}, 1) \cup (1, +\infty)$ and any $n \ge 3$, there exists $\xi_{n}^{-1}( \alpha ) \in (1, (n-1)^{2})$ such that
\begin{align}
\sgn \! \left( \vphantom{\sum} g(n, z, \alpha) \right)
=
\begin{cases}
1
& \mathrm{if} \ z \in (1, \xi_{n}^{-1}( \alpha )) , \\
0
& \mathrm{if} \ z = \xi_{n}^{-1}( \alpha ) , \\
-1
& \mathrm{if} \ z \in (\xi_{n}^{-1}( \alpha ), +\infty) . 
\end{cases}
\label{eq:sign_gz_rewrite}
\end{align}
Furthermore, since $p = \frac{1}{z + (n-1)}$ and $p$ is strictly decreasing for $z > 1$, we can rewrite \eqref{eq:sign_gz_rewrite} by using $g(n, z, \alpha)$ rather than $g(n, z, \alpha)$ as follows:
Let $\pi_{n}( \alpha ) = \frac{ 1 }{ \xi_{n}^{-1}( \alpha ) + (n-1) }$.
Then, for any $\alpha \in [\frac{1}{2}, 1) \cup (1, +\infty)$ and any $n \ge 3$, there exists $\pi_{n}( \alpha ) \in (\frac{1}{n (n-1)}, \frac{1}{n})$ such that
\begin{align}
\sgn \! \left( \vphantom{\sum} g(n, p, \alpha) \right)
=
\begin{cases}
1
& \mathrm{if} \ p \in (\pi_{n}( \alpha ), \frac{1}{n}) , \\
0
& \mathrm{if} \ p = \pi_{n}( \alpha) , \\
-1
& \mathrm{if} \ p \in (0, \pi_{n}( \alpha )) , 
\end{cases}
\label{eq:sign_gp_rewrite}
\end{align}
where note that $g(n, p, \alpha)$ denotes the right-hand side of \eqref{eq:g_p}.
Since $\xi_{n}^{-1}( \alpha )$ is strictly decreasing for $\alpha \in [\frac{1}{2}, 1) \cup (1, +\infty)$, note that $\pi_{n}( \alpha )$ is strictly increasing for $\alpha \in [\frac{1}{2}, 1) \cup (1, +\infty)$ and
\begin{align}
\lim_{\alpha \to +\infty} \pi_{n}( \alpha )
& =
\lim_{\alpha \to +\infty} \frac{1}{ \xi_{n}^{-1}( \alpha ) + (n-1) }
\\
& =
\frac{1}{ 1 + (n-1) }
\\
& =
\frac{1}{n} , \\
\lim_{\alpha \to 1} \pi_{n}( \alpha )
& =
\lim_{\alpha \to 1} \frac{1}{ \xi_{n}^{-1}( \alpha ) + (n-1) }
\\
& =
\frac{1}{ (n-1) + (n-1) }
\\
& =
\frac{1}{2 (n-1)} ,
\\
\pi_{n}( {\textstyle \frac{1}{2}} )
& =
\frac{1}{ \xi_{n}^{-1}( \frac{1}{2} ) + (n-1) }
\\
& >
\frac{1}{ (n-1)^{2} + (n-1) }
\\
& =
\frac{1}{n(n-1)} .
\end{align}
Therefore, we obtain that, for any $\alpha \in [\frac{1}{2}, 1) \cup (1, +\infty)$ and any $n \ge 3$, there exists $\pi_{n}( \alpha ) \in (\frac{1}{n (n-1)}, \frac{1}{n})$ such that
\begin{align}
\sgn \! \left( \frac{ \partial^{2} \| \bvec{v}_{n}( p ) \|_{\alpha} }{ \partial H_{\sbvec{v}_{n}}( p )^{2} } \right)
& \overset{\eqref{eq:sgn_diff2_N_Hv}}{=}
\sgn \! \left( \vphantom{\sum} g(n, p, \alpha) \right)
\\
& \overset{\eqref{eq:sign_gp_rewrite}}{=}
\begin{cases}
1
& \mathrm{if} \ p \in (\pi_{n}( \alpha ), \frac{1}{n}) , \\
0
& \mathrm{if} \ p = \pi_{n}( \alpha) , \\
-1
& \mathrm{if} \ p \in (0, \pi_{n}( \alpha )) .
\end{cases}
\label{eq:sgn_diff2_N_Hv_prime}
\end{align}
Since $H_{\sbvec{v}_{n}}( p )$ is strictly increasing for $p \in [0, \frac{1}{n}]$ (see \cite[Lemma 1]{part1, part1_arxiv}), we can further rewrite \eqref{eq:sgn_diff2_N_Hv_prime} by using $H_{\sbvec{v}_{n}}( p )$ rather than $p$ as follows:
Let $\chi_{n}( \alpha ) = H_{\sbvec{v}_{n}}( \pi_{n}( \alpha ) )$.
Then, for any $\alpha \in [\frac{1}{2}, 1) \cup (1, +\infty)$ and any $n \ge 3$, there exists $\chi_{n}( \alpha ) \in ( \ln n - (1 - \frac{2}{n}) \ln (n-1), \ln n )$ such that
\begin{align}
\sgn \! \left( \frac{ \partial^{2} \| \bvec{v}_{n}( p ) \|_{\alpha} }{ \partial H_{\sbvec{v}_{n}}( p )^{2} } \right)
& =
\begin{cases}
1
& \mathrm{if} \ H_{\sbvec{v}_{n}}( p ) \in (\chi_{n}( \alpha ), \ln n) , \\
0
& \mathrm{if} \ H_{\sbvec{v}_{n}}( p ) = \chi_{n}( \alpha) , \\
-1
& \mathrm{if} \ H_{\sbvec{v}_{n}}( p ) \in (0, \chi_{n}( \alpha )) .
\end{cases}
\end{align}
Since $\pi_{n}( \alpha )$ is strictly increasing for $\alpha \in [\frac{1}{2}, 1) \cup (1, +\infty)$, note that $\chi_{n}( \alpha )$ is also strictly increasing for $\alpha \in [\frac{1}{2}, 1) \cup (1, +\infty)$ and
\begin{align}
\lim_{\alpha \to +\infty} \chi_{n}( \alpha )
& =
\lim_{\alpha \to +\infty} H_{\sbvec{v}_{n}}( \pi_{n}( \alpha ) )
\\
& =
H_{\sbvec{v}_{n}}( {\textstyle \frac{1}{n}} )
\\
& =
\ln n ,
\\
\lim_{\alpha \to 1} \chi_{n}( \alpha )
& =
\lim_{\alpha \to 1} H_{\sbvec{v}_{n}}( \pi_{n}( \alpha ) )
\\
& =
H_{\sbvec{v}_{n}}( {\textstyle \frac{1}{2 (n-1)}} )
\\
& =
\left. \left( \vphantom{\sum} - (n-1) p \ln p - (1 - (n-1) p) \ln (1 - (n-1) p) \right) \right|_{p = \frac{1}{2 (n-1)}}
\\
& =
- (n-1) \left( \frac{1}{2(n-1)} \right) \ln \left( \frac{1}{2(n-1)} \right)
\notag \\
& \qquad \qquad \qquad \qquad \qquad \qquad
- \left(1 - (n-1) \left( \frac{1}{2(n-1)} \right)\right) \ln \left(1 - (n-1) \left( \frac{1}{2(n-1)} \right)\right)
\\
& =
- \frac{1}{2} \ln \left( \frac{1}{2(n-1)} \right) - \left( 1 - \frac{1}{2} \right) \ln \left( 1 - \frac{1}{2} \right)
\\
& =
- \frac{1}{2} \ln \frac{1}{2} + \frac{1}{2} \ln \frac{1}{n-1} - \frac{1}{2} \ln \frac{1}{2}
\\
& =
\ln 2 + \ln \sqrt{n-1} ,
\\
\chi_{n}( {\textstyle \frac{1}{2}} )
& =
H_{\sbvec{v}_{n}}( \pi_{n}( {\textstyle \frac{1}{2}} ) )
\\
& >
H_{\sbvec{v}_{n}}( {\textstyle \frac{1}{n (n-1)}} )
\\
& =
\left. \left( \vphantom{\sum} - (n-1) p \ln p - (1 - (n-1) p) \ln (1 - (n-1) p) \right) \right|_{p = \frac{1}{n (n-1)}}
\\
& =
- (n-1) \left( \frac{1}{n(n-1)} \right) \ln \left( \frac{1}{n(n-1)} \right)
\notag \\
& \qquad \qquad \qquad \qquad \qquad \qquad
- \left(1 - (n-1) \left( \frac{1}{n(n-1)} \right)\right) \ln \left(1 - (n-1) \left( \frac{1}{n(n-1)} \right)\right)
\\
& =
- \frac{1}{n} \ln \left( \frac{1}{n(n-1)} \right) - \left( 1 - \frac{1}{n} \right) \ln \left( 1 - \frac{1}{n} \right)
\\
& =
- \frac{1}{n} \ln \frac{1}{n} - \frac{1}{n} \ln \frac{1}{n-1} - \ln \left( 1 - \frac{1}{n} \right) + \frac{1}{n} \ln \left( 1 - \frac{1}{n} \right)
\\
& =
\frac{1}{n} \ln n + \frac{1}{n} \ln (n-1) - \ln \frac{n-1}{n} + \frac{1}{n} \ln \frac{n-1}{n}
\\
& =
\frac{1}{n} \ln n + \frac{1}{n} \ln (n-1) - \ln (n-1) + \ln n + \frac{1}{n} \ln (n-1) - \frac{1}{n} \ln n
\\
& =
\frac{2}{n} \ln (n-1) + \ln \frac{ n }{ n-1 }
\\
& =
\ln n - \left( 1 - \frac{2}{n} \right) \ln (n-1) .
\end{align}
That completes the proof of Lemma \ref{lem:convex_v}.
\end{IEEEproof}

\section{Proof of Lemma \ref{lem:concave_w}}
\label{app:concave_w}

\begin{IEEEproof}[Proof of Lemma \ref{lem:concave_w}]
Since $\bvec{w}_{n}( p ) = \bvec{v}_{n}( p )_{\downarrow}$ for $p \in [\frac{1}{n}, \frac{1}{n-1}]$ we can obtain immediately from \eqref{eq:sgn_diff2_N_Hv} that
\begin{align}
\sgn \! \left( \frac{ \partial^{2} \| \bvec{v}_{n}( p ) \|_{\alpha} }{ \partial H_{\sbvec{v}_{n}}( p )^{2} } \right)
=
\sgn \! \left( \vphantom{\sum} g(n, z, \alpha) \right)
\label{eq:sgn_diff2_N_Hw}
\end{align}
for $n \ge 2$, $z \in (0, 1)$, and $\alpha \in (0, 1) \cup (1, +\infty)$, where $z = z(n, \alpha) \triangleq \frac{1 - (n-1) p}{p}$ and the function $g(n, z, \alpha)$ is defined in \eqref{eq:g_z}.
Therefore, to analyze this, we can use the results of Appendix \ref{app:convex_v}.
Note that
\begin{align}
\frac{1}{n} \le p \le \frac{1}{n-1}
\overset{\eqref{eq:range_z_w}}{\iff}
0 \le z \le 1 .
\end{align}
Since
\begin{itemize}
\item
$g(n, z, 0) < 0$ for $n \ge 2$ and $z \in (0, 1)$ (see Lemma \ref{lem:g_a0} and Eq. \eqref{eq:g_kappa_z}),
\item
$g(n, z, 1) = 0$ for $n \ge 2$ and $z \in (0, 1)$ (see Eq. \eqref{eq:g_alpha_is_0}),
\item
$\lim_{\alpha \to +\infty} g(n, z, \alpha) = -\infty$ (see Eq. \eqref{eq:g_lim_+}), and
\item
the following monotonicity hold (see Eq. \eqref{eq:diff1_g_alpha_part1}):
\begin{itemize}
\item
$\alpha_{1}(n, z) < 0$ for $z \in (0, 1)$ (see Eq. \eqref{eq:range_a1a2}),
\item
$g(n, z, \alpha)$ is strictly increasing for $\alpha \in [0, 1]$, and
\item
$g(n, z, \alpha)$ is strictly decreasing for $\alpha \in [1, +\infty)$,
\end{itemize}
\end{itemize}
we can see that
\begin{align}
\sgn \! \left( \vphantom{\sum} g(n, z, \alpha) \right)
=
\begin{cases}
0
& \mathrm{if} \ \alpha = 1 , \\
-1
& \mathrm{if} \ \alpha \in [0, 1) \cup (1, +\infty)
\end{cases}
\label{eq:sign_g_w}
\end{align}
for $n \ge 2$ and $z \in (0, 1)$;
and therefore, we observe that
\begin{align}
\sgn \! \left( \frac{ \partial^{2} \| \bvec{v}_{n}( p ) \|_{\alpha} }{ \partial H_{\sbvec{v}_{n}}( p )^{2} } \right)
& \overset{\eqref{eq:sgn_diff2_N_Hw}}{=}
\sgn \! \left( \vphantom{\sum} g(n, z, \alpha) \right)
\\
& \overset{\eqref{eq:sign_g_w}}{=}
-1
\end{align}
for $n \ge 2$, $z \in (0, 1)$, and $\alpha \in (0, 1) \cup (1, +\infty)$, which implies that $\| \bvec{w}_{n}( p ) \|_{\alpha}$ is strictly concave in $H_{\sbvec{w}_{n}}( p ) \in [0, \ln n]$.

Finally, since
\begin{align}
H_{\sbvec{w}_{n}}( p )
& =
H_{\sbvec{w}_{n}}( p ) ,
\\
\| \bvec{w}_{n}( p ) \|_{\alpha}
& =
\| \bvec{w}_{m}( p ) \|_{\alpha}
\end{align}
for $n \ge m \ge 2$ and $p \in [\frac{1}{m}, \frac{1}{m-1}]$, we have Lemma \ref{lem:concave_w}.
\end{IEEEproof}

\section{Proof of Lemma \ref{lem:g_a0}}
\label{app:g_a0}

\begin{IEEEproof}[Proof of Lemma \ref{lem:g_a0}]
Direct calculation shows
\begin{align}
g(n, p, 0)
& \overset{\eqref{eq:g_p}}{=}
\left.
\left( (\alpha - 1) + \frac{ 1 - 2 (n-1) p + (n-1) p^{\alpha} (1 - (n-1)p)^{1 - \alpha} - p^{1 - \alpha} (1 - (n-1) p)^{\alpha} }{ \ln \left( \frac{ 1 - (n-1) p}{ p } \right) } \right) \right|_{\alpha = 0}
\\
& =
- 1 + \frac{ 1 - 2 (n-1) p + (n-1) (1 - (n-1) p) - p }{ \ln \left( \frac{ 1 - (n-1) p}{ p } \right) }
\\
& =
- 1 + \frac{ 1 - 2 n p + 2 p + n - 1 - n^{2} p + 2 n p - p - p }{ \ln \left( \frac{ 1 - (n-1) p}{ p } \right) }
\\
& =
- 1 + \frac{ n - n^{2} p }{ \ln \left( \frac{ 1 - (n-1) p }{ p } \right) }
\\
& =
- 1 - \frac{ n ( n p - 1 ) }{ \ln \left( \frac{ 1 - (n-1) p }{ p } \right) }
\\
& =
- \frac{ n ( n p - 1 ) + \ln \left( \frac{ 1 - (n-1) p }{ p } \right) }{ \ln \left( \frac{ 1 - (n-1) p }{ p } \right) }
\\
& =
- \frac{ s(n, p) }{ \ln \left( \frac{ 1 - (n-1) p }{ p } \right) } ,
\end{align}
where $s(n, p) \triangleq n ( n p - 1 ) + \ln \left( \frac{ 1 - (n-1) p }{ p } \right)$.
We now analyze $s(n, p)$ to prove Lemma \ref{lem:g_a0}.
We readily see that
\begin{align}
s(n, {\textstyle \frac{1}{n}})
& =
\left. \left( n ( n p - 1 ) + \ln \left( \frac{ 1 - (n-1) p }{ p } \right) \right) \right|_{p = \frac{1}{n}}
\\
& =
\left( n (1 - 1) + \ln 1 \right)
\\
& =
0 .
\label{eq:s_overn}
\end{align}
Substituting $p = \mathrm{e}^{-n}$ into $s(n, p)$, we have
\begin{align}
s(n, \mathrm{e}^{- n})
& =
\left. \left( n (n p - 1) + \ln \left( \frac{ 1 - (n-1) p }{ p } \right) \right) \right|_{p = \mathrm{e}^{-n}}
\\
& =
n (n \, \mathrm{e}^{-n} - 1) + \ln \left( \frac{ 1 - (n-1) \, \mathrm{e}^{-n} }{ \mathrm{e}^{-n} } \right)
\\
& =
n^{2} \, \mathrm{e}^{-n} - n + \ln (1 - (n-1) \, \mathrm{e}^{-n}) - \ln \mathrm{e}^{-n}
\\
& =
n^{2} \, \mathrm{e}^{-n} - n + \ln (1 - (n-1) \, \mathrm{e}^{-n}) - (-n)
\\
& =
n^{2} \, \mathrm{e}^{-n} + \ln (1 - (n-1) \, \mathrm{e}^{-n}) .
\end{align}
Then, we can see that
\begin{align}
\lim_{n \to +\infty} s(n, \mathrm{e}^{-n})
& =
\lim_{n \to +\infty} \left( \vphantom{\sum} n^{2} \, \mathrm{e}^{-n} + \ln (1 - (n-1) \, \mathrm{e}^{-n}) \right)
\\
& =
0 .
\end{align}
Moreover, the derivative of $s(n, \mathrm{e}^{- n})$ with respect to $n$ can be calculated as follows:
\begin{align}
\frac{ \mathrm{d} s(n, \mathrm{e}^{- n}) }{ \mathrm{d} n }
& =
\frac{ \mathrm{d} }{ \mathrm{d} n } \left( \vphantom{\sum} n^{2} \, \mathrm{e}^{-n} + \ln (1 - (n-1) \, \mathrm{e}^{-n}) \right)
\\
& =
\left[ \frac{ \mathrm{d} }{ \mathrm{d} n } \left( \vphantom{\sum} n^{2} \, \mathrm{e}^{-n} \right) \right] + \left[ \frac{ \mathrm{d} }{ \mathrm{d} n } \left( \vphantom{\sum} \ln (1 - (n-1) \, \mathrm{e}^{-n}) \right) \right]
\\
& =
\left[ \left( \frac{ \mathrm{d} }{ \mathrm{d} n } (n^{2}) \right) \mathrm{e}^{-n} + n^{2} \left( \frac{ \mathrm{d} }{ \mathrm{d} n } (\mathrm{e}^{-n}) \right) \right] + \left[ \frac{ \mathrm{d} }{ \mathrm{d} n } \left( \vphantom{\sum} \ln (1 - (n-1) \, \mathrm{e}^{-n}) \right) \right]
\\
& =
\left[ 2 n \, \mathrm{e}^{-n} - n^{2} \, \mathrm{e}^{-n} \right] + \left[ \frac{ \mathrm{d} }{ \mathrm{d} n } \left( \vphantom{\sum} \ln (1 - (n-1) \, \mathrm{e}^{-n}) \right) \right]
\\
& =
n \, \mathrm{e}^{-n} \, (2 - n) + \left[ \frac{ \mathrm{d} }{ \mathrm{d} n } \left( \vphantom{\sum} \ln (1 - (n-1) \, \mathrm{e}^{-n}) \right) \right]
\\
& =
n \, \mathrm{e}^{-n} \, (2 - n) + \left[ \frac{ \frac{ \mathrm{d} }{ \mathrm{d} n } \left( \vphantom{\sum} 1 - (n-1) \, \mathrm{e}^{-n} \right) }{ (1 - (n-1) \, \mathrm{e}^{-n}) } \right]
\\
& =
n \, \mathrm{e}^{-n} \, (2 - n) + \left[ \frac{ - \left( \frac{ \mathrm{d} }{ \mathrm{d} n } (n-1) \right) \mathrm{e}^{-n} - (n-1) \left( \frac{ \mathrm{d} }{ \mathrm{d} n } (\mathrm{e}^{-n}) \right) }{ 1 - (n-1) \, \mathrm{e}^{-n} } \right]
\\
& =
n \, \mathrm{e}^{-n} \, (2 - n) + \left[ \frac{ - \, \mathrm{e}^{-n} + (n-1) \, \mathrm{e}^{-n} }{ 1 - (n-1) \, \mathrm{e}^{-n} } \right]
\\
& =
n \, \mathrm{e}^{-n} \, (2 - n) + \left[ \frac{ \mathrm{e}^{-n} \, (- 1 + (n-1)) }{ 1 - (n-1) \, \mathrm{e}^{-n} } \right]
\\
& =
n \, \mathrm{e}^{-n} \, (2 - n) + \left[ \frac{ \mathrm{e}^{-n} \, (n-2) }{ 1 - (n-1) \, \mathrm{e}^{-n} } \right]
\\
& =
\frac{ n \, \mathrm{e}^{-n} \, (2 - n) (1 - (n-1) \, \mathrm{e}^{-n}) + \mathrm{e}^{-n} \, (n-2) }{ 1 - (n-1) \, \mathrm{e}^{-n} }
\\
& =
\mathrm{e}^{-n} \, (n-2) \, \frac{ n (1 - (n-1) \, \mathrm{e}^{-n}) + 1 }{ 1 - (n-1) \, \mathrm{e}^{-n} }
\\
& =
\mathrm{e}^{-n} \, (n-2) \, \frac{ n - n^{2} \, \mathrm{e}^{-n} + n \, \mathrm{e}^{-n} + 1 }{ 1 - (n-1) \, \mathrm{e}^{-n} }
\\
& =
\mathrm{e}^{-n} \, (n-2) \, \frac{ (n-1) ( n \, \mathrm{e}^{-n} - 1 ) }{ 1 - (n-1) \, \mathrm{e}^{-n} }
\\
& =
\mathrm{e}^{-n} \, (n-2) (n-1) \left( \frac{ n \, \mathrm{e}^{-n} - 1 }{ 1 - (n-1) \, \mathrm{e}^{-n} } \right)
\\
& =
\underbrace{ \mathrm{e}^{-n} \, (n-2) (n-1) }_{ > 0 \ \mathrm{for} \ n > 2 } \underbrace{ \left( - \frac{ 1 - n \, \mathrm{e}^{-n} }{ 1 - (n-1) \, \mathrm{e}^{-n} } \right) }_{ < 0 }
\\
& <
0 .
\end{align}
Since $s(n, \mathrm{e}^{-n})$ is strictly decreasing for $n \ge 2$ and $\lim_{n \to +\infty} s(n, \mathrm{e}^{-n}) = 0$, we obtain
\begin{align}
s(n, \mathrm{e}^{-n})
>
0
\label{eq:s_exp}
\end{align}
for $n \ge 2$.
We next calculate the derivative of $s(n, p)$ with respect to $z$ as follows:
\begin{align}
\frac{ \partial s(n, p) }{ \partial p }
& =
\frac{ \partial }{ \partial p } \left( n ( n p - 1 ) + \ln \left( \frac{ 1 - (n-1) p }{ p } \right) \right)
\\
& =
n^{2} + \frac{ \partial }{ \partial p } \ln \left( \frac{ 1 - (n-1) p }{ p } \right)
\\
& \overset{\text{(a)}}{=}
n^{2} + \frac{ 1 }{ \frac{ 1 - (n-1) p }{ p } } \left( - \frac{ 1 }{ p^{2} } \right)
\\
& =
n^{2} - \frac{ 1 }{ p (1 - (n-1) p) }
\\
& =
\frac{ n^{2} p (1 - (n-1) p) - 1 }{ p (1 - (n-1) p) }
\\
& =
\frac{ n^{2} p - n^{3} p^{2} + n^{2} p^{2} - 1 }{ p (1 - (n-1) p) }
\\
& =
\frac{ n^{2} \left( (1-n) p^{2} + p - \left( \frac{1}{n^{2}} \right) \right) }{ p (1 - (n-1) p) }
\\
& =
\frac{ n^{2} (1-n) \left( p^{2} - \left( \frac{1}{n-1} \right) p + \left( \frac{1}{n^{2} (n-1)} \right) \right) }{ p (1 - (n-1) p) }
\\
& =
\underbrace{ - \frac{ n^{2} (n-1) }{ p (1 - (n-1) p) } }_{ < 0 } \left( p - \frac{1}{n(n-1)} \right) \left( p - \frac{1}{n} \right)
\\
& \overset{\text{(b)}}{=}
\begin{cases}
< 0
& \mathrm{if} \ p \in (0, \frac{1}{n(n-1)}) \cup (\frac{1}{n}, \frac{1}{n-1}) , \\
= 0
& \mathrm{if} \ p \in \{ \frac{1}{n(n-1)}, \frac{1}{n} \} , \\
> 0
& \mathrm{if} \ p \in (\frac{1}{n(n-1)}, \frac{1}{n}) ,
\end{cases}
\label{eq:diff_s_sign}
\end{align}
where
\begin{itemize}
\item
(a) follows from the fact that $\frac{ \mathrm{d} \ln f(x) }{ \mathrm{d} x } = \frac{ 1 }{ f(x) } \left( \frac{ \mathrm{d} f(x) }{ \mathrm{d} x } \right)$ and
\item
(b) follows from the fact that
\begin{align}
\sgn \! \left( \left( p - \frac{1}{n(n-1)} \right) \left( p - \frac{1}{n} \right) \right)
=
\begin{cases}
1
& \mathrm{if} \ p \in (0, \frac{1}{n(n-1)}) \cup (\frac{1}{n}, \frac{1}{n-1}) , \\
0
& \mathrm{if} \ p \in \{ \frac{1}{n(n-1)}, \frac{1}{n} \} , \\
-1
& \mathrm{if} \ p \in (\frac{1}{n(n-1)}, \frac{1}{n}) .
\end{cases}
\end{align}
\end{itemize}
It follows from
\begin{itemize}
\item
the intermediate value theorem,
\item
the monotonicity of $s(n, p)$ with respect to $p$ (see Eq. \eqref{eq:diff_s_sign}):
\begin{itemize}
\item
$s(n, p)$ is strictly decreasing for $p \in (0, \frac{1}{n (n-1)}]$,
\item
$s(n, p)$ is strictly increasing for $p \in [\frac{1}{n (n-1)}, \frac{1}{n}]$,
\item
$s(n, p)$ is strictly decreasing for $p \in [\frac{1}{n}, \frac{1}{n-1})$,
\end{itemize}
\item
$s(n, \mathrm{e}^{-n}) > 0$ for $n \ge 2$ (see Eq. \eqref{eq:s_exp}),
\item
$s(n, \frac{1}{n}) = 0$ for $n \ge 2$ (see Eq. \eqref{eq:s_overn}), and
\item
$0 < \mathrm{e}^{-n} < \frac{1}{n (n-1)} < \frac{1}{n}$ for $n \ge 3$
\end{itemize}
that the following statements hold:
\begin{itemize}
\item
if $n = 2$, then
\begin{align}
\sgn \! \left( \vphantom{\sum} s(2, p) \right)
=
\begin{cases}
1
& \mathrm{if} \ p \in (0, \frac{1}{2}) , \\
0
& \mathrm{if} \ p = \frac{1}{2} , \\
-1
& \mathrm{if} \ p \in (\frac{1}{2}, 1) ,
\end{cases}
\end{align}
\item
if $n \ge 3$, there exists $\kappa_{p}( n ) \in (\mathrm{e}^{-n}, \frac{1}{n(n-1)})$ such that
\begin{align}
\sgn \! \left( \vphantom{\sum} s(n, p) \right)
=
\begin{cases}
1
& \mathrm{if} \ p \in (0, \kappa_{p}( n )) , \\
0
& \mathrm{if} \ p \in \{ \kappa_{p}( n ), \frac{1}{n} \} , \\
-1
& \mathrm{if} \ p \in (\kappa_{p}( n ), \frac{1}{n}) \cup (\frac{1}{n}, \frac{1}{n-1}) .
\end{cases}
\end{align}
\end{itemize}
Therefore, since
\begin{align}
\sgn \! \left( \vphantom{\sum} g(n, p, 0) \right)
& =
\sgn \! \left( \vphantom{\sum} s(n, p) \right) \cdot \, \sgn \! \left( - \frac{ 1 }{ \ln \left( \frac{ 1 - (n-1) p }{ p } \right) } \right) ,
\\
\sgn \! \left( - \frac{ 1 }{ \ln \left( \frac{ 1 - (n-1) p }{ p } \right) } \right)
& =
\begin{cases}
1
& \mathrm{if} \ p \in (\frac{1}{n}, \frac{1}{n-1}) , \\
-1
& \mathrm{if} \ p \in (0, \frac{1}{n}) ,
\end{cases}
\end{align}
we have that
\begin{itemize}
\item
if $n = 2$, then
\begin{align}
\sgn \! \left( \vphantom{\sum} g(2, p, 0) \right)
=
-1
\end{align}
for $p \in (0, \frac{1}{2}) \cup (\frac{1}{2}, 1)$,
\item
for any $n \ge 3$, there exists $\kappa_{p}( n ) \in (\mathrm{e}^{-n}, \frac{1}{n(n-1)})$ such that
\begin{align}
\sgn \! \left( \vphantom{\sum} g(n, p, 0) \right)
=
\begin{cases}
1
& \mathrm{if} \ p \in (\kappa_{p}( n ), \frac{1}{n}) , \\
0
& \mathrm{if} \ p = \kappa_{p}( n ) , \\
-1
& \mathrm{if} \ p \in (0, \kappa_{p}( n )) \cup (\frac{1}{n}, \frac{1}{n-1}) .
\end{cases}
\end{align}
\end{itemize}
Note that $\lim_{p \to \frac{1}{n}} g(n, p, 0) = 0$.
That concludes the proof of the lemma.
\end{IEEEproof}

\section{Proof of Lemma \ref{lem:dzda}}
\label{app:dzda}

\begin{IEEEproof}[Proof of Lemma \ref{lem:dzda}]
In the proof, assume that $n \ge 3$.
Direct calculation shows
\begin{align}
\frac{ \partial^{2} g(n, z, \alpha) }{ \partial z \, \partial \alpha }
& =
\frac{ \partial }{ \partial z } \left( \frac{ \partial g(n, z, \alpha) }{ \partial \alpha } \right)
\\
& =
\frac{ \partial }{ \partial z } \left( 1 - \frac{ (n-1) z^{1-\alpha} + z^{\alpha} }{ (n-1) + z } \right)
\\
& =
- \frac{ \partial }{ \partial z } \left( \frac{ (n-1) z^{1-\alpha} + z^{\alpha} }{ (n-1) + z } \right)
\\
& \overset{\text{(a)}}{=}
- \frac{ \left[ \frac{ \partial ( (n-1) z^{1-\alpha} + z^{\alpha} ) }{ \partial z } \right] ((n-1) + z) - ( (n-1) z^{1-\alpha} + z^{\alpha} ) \left[ \frac{ \partial ((n-1) + z) }{ \partial z } \right] }{ ((n-1) + z)^{2} }
\\
& =
- \frac{ [ (n-1) (1-\alpha) z^{-\alpha} + \alpha z^{\alpha-1} ] ((n-1) + z) - ( (n-1) z^{1-\alpha} + z^{\alpha} ) [ 1 ] }{ ((n-1) + z)^{2} }
\\
& =
- \frac{ (n-1)^{2} (1-\alpha) z^{-\alpha} + (n-1) (1-\alpha) z^{1-\alpha} + (n-1) \alpha z^{\alpha-1} + \alpha z^{\alpha} - (n-1) z^{1-\alpha} - z^{\alpha} }{ ((n-1) + z)^{2} }
\\
& =
- \frac{ (\alpha-1) z^{\alpha} + \alpha (n-1) z^{\alpha-1} - \alpha (n-1) z^{1-\alpha} - (\alpha-1) (n-1)^{2} z^{-\alpha} }{ ((n-1) + z)^{2} }
\\
& =
u(n, z, \alpha) \underbrace{ \left( - \frac{ 1 }{ ((n-1) + z)^{2} } \right) }_{ < 0 } ,
\label{eq:diff_g_za}
\end{align}
where $u(n, z, \alpha) \triangleq (\alpha-1) z^{\alpha} + \alpha (n-1) z^{\alpha-1} - \alpha (n-1) z^{1-\alpha} - (\alpha-1) (n-1)^{2} z^{-\alpha}$ and (a) follows from the fact that $\frac{ \mathrm{d} }{ \mathrm{d} x } \left( \frac{ f(x) }{ g(x) } \right) = \frac{ \left[ \frac{ \mathrm{d} f(x) }{ \mathrm{d} x } \right] g(x) - f(x) \left[ \frac{ \mathrm{d} g(x) }{ \mathrm{d} x } \right] }{ (g(x))^{2} }$.
It follows from \eqref{eq:diff_g_za} that it is enough to check the sign of $u(n, z, \alpha)$.
Simple calculation yields
\begin{align}
u(n, z, 0)
& =
\left. \left( \vphantom{\sum} (\alpha-1) z^{\alpha} + \alpha (n-1) z^{\alpha-1} - \alpha (n-1) z^{1-\alpha} - (\alpha-1) (n-1)^{2} z^{-\alpha} \right) \right|_{\alpha = 0}
\\
& =
(-1) z^{0} + 0 (n-1) z^{-1} - 0 (n-1) z^{1} - (-1) (n-1)^{2} z^{0}
\\
& =
-1 + (n-1)^{2}
\\
& >
0 ,
\\
u(n, z, 1)
& =
\left. \left( \vphantom{\sum} (\alpha-1) z^{\alpha} + \alpha (n-1) z^{\alpha-1} - \alpha (n-1) z^{1-\alpha} - (\alpha-1) (n-1)^{2} z^{-\alpha} \right) \right|_{\alpha = 1}
\\
& =
0 z^{1} + 1 (n-1) z^{0} - 1 (n-1) z^{0} - 0 (n-1)^{2} z^{-1}
\\
& =
(n-1) - (n-1)
\\
& =
0
\end{align}
for $z \in (0, +\infty)$.
Therefore, we readily see that
\begin{itemize}
\item
if $\alpha = 0$, then $\frac{ \partial^{2} g(n, z, \alpha) }{ \partial z \, \partial \alpha } < 0$ for $z \in (0, +\infty)$ and
\item
if $\alpha = 1$, then $\frac{ \partial^{2} g(n, z, \alpha) }{ \partial z \, \partial \alpha } = 0$ for $z \in (0, +\infty)$.
\end{itemize}
Thus, we will verify the sign of $u(n, z, \alpha)$ for $\alpha \in (-\infty, 0) \cup (0, 1) \cup (1, +\infty)$.
We first check the sign of the derivative of $u(n, z, \alpha)$ with respect to $z$ as
\begin{align}
\frac{ \partial u(n, z, \alpha) }{ \partial z }
& =
\frac{ \partial }{ \partial z } \left( \vphantom{\sum} (\alpha-1) z^{\alpha} + \alpha (n-1) z^{\alpha-1} - \alpha (n-1) z^{1-\alpha} - (\alpha-1) (n-1)^{2} z^{-\alpha} \right)
\\
& =
\alpha (\alpha-1) z^{\alpha-1} + \alpha (\alpha-1) (n-1) z^{\alpha-2}
\notag \\
& \qquad \qquad \qquad \qquad \qquad \qquad \qquad
+ \alpha (\alpha-1) (n-1) z^{-\alpha} + \alpha (\alpha-1) (n-1)^{2} z^{-1-\alpha}
\\
& =
\alpha (\alpha-1) \left( \vphantom{\sum} z^{\alpha-1} + (n-1) z^{\alpha-2} + (n-1) z^{-\alpha} + (n-1)^{2} z^{-1-\alpha} \right)
\\
& =
\alpha (\alpha-1) \underbrace{ \left( \frac{ (n-1) + z }{ z^{2} } \right) \left( \vphantom{\sum} z^{\alpha} + (n-1) z^{1-\alpha} \right) }_{ > 0 }
\\
& =
\begin{cases}
< 0
& \mathrm{if} \ \alpha \in (0, 1) , \\
= 0
& \mathrm{if} \ \alpha \in \{ 0, 1 \} , \\
> 0
& \mathrm{if} \ \alpha \in (-\infty, 0) \cup (1, +\infty) .
\end{cases}
\label{eq:diff_u}
\end{align}
Hence, we have that
\begin{itemize}
\item
if $\alpha \in (0, 1)$, then $u(n, z, \alpha)$ is strictly decreasing for $z \in (0, +\infty)$,
\item
if $\alpha \in (-\infty, 0) \cup (1, +\infty)$, then $u(n, z, \alpha)$ is strictly increasing for $z \in (0, +\infty)$, and
\item
if $\alpha \in \{ 0, 1 \}$, then $u(n, z, \alpha)$ is invariant for $z \in (0, +\infty)$.
\end{itemize}
We second check the some signs of $u(n, z, \alpha)$ as follows:
we first get
\begin{align}
u(n, n-1, \alpha)
& =
\left. \left( \vphantom{\sum} (\alpha-1) z^{\alpha} + \alpha (n-1) z^{\alpha-1} - \alpha (n-1) z^{1-\alpha} - (\alpha-1) (n-1)^{2} z^{-\alpha} \right) \right|_{z = n-1}
\\
& =
(\alpha-1) (n-1)^{\alpha} + \alpha (n-1) (n-1)^{\alpha-1} - \alpha (n-1) (n-1)^{1-\alpha}
\notag \\
& \qquad \qquad \qquad \qquad \qquad \qquad \qquad \qquad \qquad \qquad \qquad \qquad
- (\alpha-1) (n-1)^{2} (n-1)^{-\alpha}
\\
& =
(\alpha-1) (n-1)^{\alpha} + \alpha (n-1)^{\alpha} - \alpha (n-1)^{2-\alpha} - (\alpha-1) (n-1)^{2-\alpha}
\\
& =
(n-1)^{\alpha} \left( \vphantom{\sum} (\alpha-1) + \alpha \right) - (n-1)^{2-\alpha} \left( \vphantom{\sum} \alpha + (\alpha-1) \right)
\\
& =
(n-1)^{\alpha} (2 \alpha - 1) - (n-1)^{2-\alpha} (2 \alpha - 1)
\\
& =
(2 \alpha - 1) \left( \vphantom{\sum} (n-1)^{\alpha} - (n-1)^{2-\alpha} \right)
\\
& =
(2 \alpha - 1) \underbrace{ (n-1)^{2-\alpha} }_{ > 0 } \left( \vphantom{\sum} (n-1)^{2(\alpha-1)} - 1 \right)
\\
& \overset{\text{(a)}}{=}
\begin{cases}
< 0
& \mathrm{if} \ \alpha \in (\frac{1}{2}, 1) , \\
= 0
& \mathrm{if} \ \alpha \in \{ \frac{1}{2}, 1 \} , \\
> 0
& \mathrm{if} \ \alpha \in (-\infty, \frac{1}{2}) \cup (1, +\infty) ,
\end{cases}
\label{eq:u_n-1}
\end{align}
where (a) follows from the facts that
\begin{align}
\sgn \! \left( \vphantom{\sum} 2 \alpha - 1 \right)
& =
\begin{cases}
1
& \mathrm{if} \ \alpha \in (\frac{1}{2}, +\infty) , \\
0
& \mathrm{if} \ \alpha = \frac{1}{2} , \\
-1
& \mathrm{if} \ \alpha \in (-\infty, \frac{1}{2}) ,
\end{cases}
\\
\sgn \! \left( \vphantom{\sum} (n-1)^{2(\alpha - 1)} - 1 \right)
& =
\begin{cases}
1
& \mathrm{if} \ \alpha \in (1, +\infty) , \\
0
& \mathrm{if} \ \alpha = 1 , \\
-1
& \mathrm{if} \ \alpha \in (-\infty, 1) ,
\end{cases}
\end{align}
we second get
\begin{align}
u(n, z_{1}(n, \alpha) , \alpha)
& =
u(n, (n-1)^{\frac{1}{\alpha}}, \alpha)
\\
& =
\left. \left( \vphantom{\sum} (\alpha-1) z^{\alpha} + \alpha (n-1) z^{\alpha-1} - \alpha (n-1) z^{1-\alpha} - (\alpha-1) (n-1)^{2} z^{-\alpha} \right) \right|_{z = (n-1)^{\frac{1}{\alpha}}}
\\
& =
(\alpha - 1) \left( (n-1)^{\frac{1}{\alpha}} \right)^{\alpha} + \alpha (n-1) \left( (n-1)^{\frac{1}{\alpha}} \right)^{\alpha-1}
\notag \\
& \qquad \qquad \qquad \qquad \qquad
- \alpha (n-1) \left( (n-1)^{\frac{1}{\alpha}} \right)^{1-\alpha} - (\alpha-1) (n-1)^{2} \left( (n-1)^{\frac{1}{\alpha}} \right)^{-\alpha}
\\
& =
(\alpha-1) (n-1) + \alpha (n-1) (n-1)^{1 - \frac{1}{\alpha}}
\notag \\
& \qquad \qquad \qquad \qquad \qquad \qquad \qquad
- \alpha (n-1) (n-1)^{\frac{1}{\alpha} - 1} - (\alpha-1) (n-1)^{2} (n-1)^{-1}
\\
& =
(\alpha-1) (n-1) + \alpha (n-1)^{2 - \frac{1}{\alpha}} - \alpha (n-1)^{\frac{1}{\alpha}} - (\alpha-1) (n-1)
\\
& =
\alpha (n-1)^{2 - \frac{1}{\alpha}} - \alpha (n-1)^{\frac{1}{\alpha}}
\\
& =
\alpha (n-1)^{\frac{1}{\alpha}} \left( \vphantom{\sum} (n-1)^{2 - \frac{2}{\alpha}} - 1 \right)
\\
& =
\alpha \underbrace{ (n-1)^{\frac{1}{\alpha}} }_{ > 0 } \left( \vphantom{\sum} (n-1)^{\frac{2 (\alpha - 1)}{\alpha}} - 1 \right)
\\
& \overset{\text{(b)}}{=}
\begin{cases}
< 0
& \mathrm{if} \ \alpha \in (-\infty, 0) \cup (0, 1) , \\
= 0
& \mathrm{if} \ \alpha = 1 , \\
> 0
& \mathrm{if} \ \alpha \in (1, +\infty) ,
\end{cases}
\label{eq:u_z1}
\end{align}
where (b) follows from the facts that
\begin{align}
\sgn \! \left( \vphantom{\sum} \alpha \right)
& =
\begin{cases}
1
& \mathrm{if} \ \alpha \in (0, +\infty) , \\
0
& \mathrm{if} \ \alpha = 0 , \\
-1
& \mathrm{if} \ \alpha \in (-\infty, 0) ,
\end{cases}
\\
\sgn \! \left( \vphantom{\sum} (n-1)^{\frac{ 2 (\alpha-1) }{ \alpha }} - 1 \right)
& =
\begin{cases}
1
& \mathrm{if} \ \alpha \in (-\infty, 0) \cup (1, +\infty) , \\
0
& \mathrm{if} \ \alpha = 1 , \\
-1
& \mathrm{if} \ \alpha \in (0, 1) ,
\end{cases}
\end{align}
and we third get
\begin{align}
u(n, z_{2}(n, \alpha) , \alpha)
& =
u(n, (n-1)^{\frac{1}{2\alpha-1}}, \alpha)
\\
& =
\left. \left( \vphantom{\sum} (\alpha-1) z^{\alpha} + \alpha (n-1) z^{\alpha-1} - \alpha (n-1) z^{1-\alpha} - (\alpha-1) (n-1)^{2} z^{-\alpha} \right) \right|_{z = (n-1)^{\frac{1}{2\alpha-1}}}
\\
& =
(\alpha - 1) \left( (n-1)^{\frac{1}{2\alpha-1}} \right)^{\alpha} + \alpha (n-1) \left( (n-1)^{\frac{1}{2\alpha-1}} \right)^{\alpha-1}
\notag \\
& \qquad \qquad \qquad \qquad
- \alpha (n-1) \left( (n-1)^{\frac{1}{2\alpha-1}} \right)^{1-\alpha} - (\alpha-1) (n-1)^{2} \left( (n-1)^{\frac{1}{2\alpha-1}} \right)^{-\alpha}
\\
& =
(\alpha-1) (n-1)^{\frac{\alpha}{2\alpha-1}} + \alpha (n-1) (n-1)^{\frac{\alpha-1}{2\alpha-1}}
\notag \\
& \qquad \qquad \qquad \qquad \qquad \qquad
- \alpha (n-1) (n-1)^{\frac{1-\alpha}{2\alpha-1}} - (\alpha-1) (n-1)^{2} (n-1)^{-\frac{\alpha}{2\alpha-1}}
\\
& =
(\alpha-1) (n-1)^{\frac{\alpha}{2\alpha-1}} + \alpha (n-1)^{\frac{(2\alpha-1) + (\alpha-1)}{2\alpha-1}}
\notag \\
& \qquad \qquad \qquad \qquad \qquad \qquad \qquad
- \alpha (n-1)^{\frac{(2\alpha-1) + (1-\alpha)}{2\alpha-1}} - (\alpha-1) (n-1)^{\frac{2 (2\alpha-1) - \alpha}{2\alpha-1}}
\\
& =
(\alpha-1) (n-1)^{\frac{\alpha}{2\alpha-1}} + \alpha (n-1)^{\frac{3\alpha-2}{2\alpha-1}} - \alpha (n-1)^{\frac{\alpha}{2\alpha-1}} - (\alpha-1) (n-1)^{\frac{3\alpha-2}{2\alpha-1}}
\\
& =
(n-1)^{\frac{\alpha}{2\alpha-1}} \left( \vphantom{\sum} (\alpha-1) - \alpha \right) + (n-1)^{\frac{3\alpha-2}{2\alpha-1}} \left( \vphantom{\sum} \alpha - (\alpha - 1) \right)
\\
& =
- (n-1)^{\frac{\alpha}{2\alpha-1}} + (n-1)^{\frac{3\alpha-2}{2\alpha-1}}
\\
& =
(n-1)^{\frac{\alpha}{2\alpha-1}} \left( (n-1)^{\frac{3\alpha-2}{2\alpha-1} - \frac{\alpha}{2\alpha-1}} - 1 \right)
\\
& =
\underbrace{ (n-1)^{\frac{\alpha}{2\alpha-1}} }_{ > 0 } \left( (n-1)^{\frac{2(\alpha-1)}{2\alpha-1}} - 1 \right)
\\
& \overset{\text{(c)}}{=}
\begin{cases}
< 0
& \mathrm{if} \ \alpha \in (\frac{1}{2}, 1) , \\
= 0
& \mathrm{if} \ \alpha = 1 , \\
> 0
& \mathrm{if} \ \alpha \in (-\infty, \frac{1}{2}) \cup (1, +\infty) ,
\end{cases}
\label{eq:u_z2}
\end{align}
where (c) follows from the fact that
\begin{align}
\sgn \! \left( \vphantom{\sum} (n-1)^{\frac{2 (\alpha-1)}{2 \alpha - 1}} - 1 \right)
=
\begin{cases}
1
& \mathrm{if} \ \alpha \in (-\infty, \frac{1}{2}) \cup (1, +\infty) , \\
0
& \mathrm{if} \ \alpha = 1 , \\
-1
& \mathrm{if} \ \alpha \in (\frac{1}{2}, 1) .
\end{cases}
\end{align}
Since $u(n, z, \frac{1}{2})$ is strictly decreasing for $z \in (0, +\infty)$ (see Eq. \eqref{eq:diff_u}) and $u(n, n-1, \frac{1}{2}) = 0$ (see Eq. \eqref{eq:u_n-1}), we readily see that
\begin{align}
\sgn \! \left( \vphantom{\sum} u(n, z, {\textstyle \frac{1}{2}}) \right)
=
\begin{cases}
1
& \mathrm{if} \ z \in (0, n-1) , \\
0
& \mathrm{if} \ z = n-1 , \\
-1
& \mathrm{if} \ z \in (n-1, +\infty) ;
\end{cases}
\end{align}
and therefore, we have
\begin{align}
\sgn \! \left( \vphantom{\sum} \left. \frac{ \partial^{2} g(n, z, \alpha) }{ \partial z \, \partial \alpha } \right|_{\alpha = \frac{1}{2}} \right)
=
\begin{cases}
1
& \mathrm{if} \ z \in (n-1, +\infty) , \\
0
& \mathrm{if} \ z = n-1 , \\
-1
& \mathrm{if} \ z \in (0, n-1)
\end{cases}
\end{align}
from \eqref{eq:diff_g_za}.
Moreover, we check the sign of $u(n, z_{1}(n, 2 \alpha), \alpha)$ as follows:
Simple calculation yields
\begin{align}
u(n, z_{1}(n, 2 \alpha) , \alpha)
& =
u(n, (n-1)^{\frac{1}{2 \alpha}}, \alpha)
\\
& =
\left. \left( \vphantom{\sum} (\alpha-1) z^{\alpha} + \alpha (n-1) z^{\alpha-1} - \alpha (n-1) z^{1-\alpha} - (\alpha-1) (n-1)^{2} z^{-\alpha} \right) \right|_{z = (n-1)^{\frac{1}{2 \alpha}}}
\\
& =
(\alpha - 1) \left( (n-1)^{\frac{1}{2 \alpha}} \right)^{\alpha} + \alpha (n-1) \left( (n-1)^{\frac{1}{2 \alpha}} \right)^{\alpha-1}
\notag \\
& \qquad \qquad \qquad \qquad \quad
- \alpha (n-1) \left( (n-1)^{\frac{1}{2 \alpha}} \right)^{1-\alpha} - (\alpha-1) (n-1)^{2} \left( (n-1)^{\frac{1}{2 \alpha}} \right)^{-\alpha}
\\
& =
(\alpha - 1) (n-1)^{\frac{1}{2}} + \alpha (n-1) (n-1)^{\frac{1}{2} (1 - \frac{1}{\alpha})}
\notag \\
& \qquad \qquad \qquad \qquad \quad
- \alpha (n-1) (n-1)^{\frac{1}{2}(\frac{1}{\alpha} - 1)} - (\alpha-1) (n-1)^{2} (n-1)^{-\frac{1}{2}}
\\
& =
(\alpha - 1) (n-1)^{\frac{1}{2}} + \alpha (n-1)^{\frac{3}{2} - \frac{1}{2 \alpha}} - \alpha (n-1)^{\frac{1}{2 \alpha} + \frac{1}{2}} - (\alpha-1) (n-1)^{\frac{3}{2}}
\\
& =
(\alpha - 1) \sqrt{n-1} \, (1 - (n-1)) + \alpha (n-1)^{\frac{3}{2} - \frac{1}{2 \alpha}} - \alpha (n-1)^{\frac{1}{2 \alpha} + \frac{1}{2}}
\\
& =
(\alpha - 1) \sqrt{n-1} \, (2 - n) + \alpha \sqrt{n-1} \, \left( \vphantom{\sum} (n-1)^{1 - \frac{1}{2 \alpha}} - (n-1)^{\frac{1}{2 \alpha}} \right)
\\
& =
\sqrt{n-1} \, \left( \vphantom{\sum} (\alpha - 1) (2 - n) + \alpha (n-1)^{1 - \frac{1}{2 \alpha}} - \alpha (n-1)^{\frac{1}{2 \alpha}} \right)
\\
& =
\sqrt{n-1} \, \left( (\alpha - 1) (2 - n) + \alpha (n-1)^{\frac{1}{2\alpha}} \left( \vphantom{\sum} (n-1)^{1 - \frac{1}{\alpha}} - 1 \right) \right)
\\
& =
\sqrt{n-1} \, \left( (\alpha - 1) (2 - n) + \alpha \left( 1 - \frac{1}{\alpha} \right) (n-1)^{\frac{1}{2\alpha}} \ln_{\frac{1}{\alpha}} (n-1) \right)
\\
& =
(\alpha-1) \sqrt{n-1} \, \left( (2 - n) + (n-1)^{\frac{1}{2\alpha}} \ln_{\frac{1}{\alpha}} (n-1) \right)
\\
& \overset{\text{(a)}}{=}
(\alpha-1) \sqrt{n-1} \, \left( (2 - n) + (n-1)^{\frac{\beta}{2}} \ln_{\beta} (n-1) \right)
\\
& =
(\alpha-1) \sqrt{n-1} \, \cdot u_{1}( n, \beta ) ,
\label{eq:u_z1(n,2alpha)}
\end{align}
where (a) follows by the change of variable as $\beta = \beta( \alpha ) \triangleq \frac{1}{\alpha}$, and
\begin{align}
u_{1}( n, \beta )
\triangleq
(2 - n) + (n-1)^{\frac{\beta}{2}} \ln_{\beta} (n-1) .
\end{align}
We now calculate the following derivatives:
\begin{align}
\frac{ \partial u_{1}(n, \beta) }{ \partial \beta }
& =
\frac{ \partial }{ \partial \beta } \left( \vphantom{\sum} (2 - n) + (n-1)^{\frac{\beta}{2}} \ln_{\beta} (n-1) \right)
\\
& =
\frac{ \partial }{ \partial \beta } \left( \vphantom{\sum} (n-1)^{\frac{\beta}{2}} \ln_{\beta} (n-1) \right)
\\
& =
\left( \frac{ \partial }{ \partial \beta } ((n-1)^{\frac{\beta}{2}}) \right) \ln_{\beta} (n-1) + (n-1)^{\frac{\beta}{2}} \left( \frac{ \partial }{ \partial \beta } (\ln_{\beta} (n-1)) \right)
\\
& \overset{\eqref{eq:diff1_lnq}}{=}
\left( \frac{1}{2} (n-1)^{\frac{\beta}{2}} \ln (n-1) \right) \ln_{\beta} (n-1) + (n-1)^{\frac{\beta}{2}} \left( \frac{ (n-1)^{\beta} \ln_{\beta} (n-1) - (n-1) \ln (n-1) }{ (n-1)^{\beta} (1-\beta) } \right)
\\
& =
(n-1)^{\frac{\beta}{2}} \left( \frac{ (\ln (n-1)) (\ln_{\beta} (n-1)) }{2} + \frac{ \ln_{\beta} (n-1) - (n-1)^{1-\beta} \ln (n-1) }{ (1-\beta) } \right)
\\
& =
(n-1)^{\frac{\beta}{2}} \left( \frac{ ((n-1)^{1-\beta} - 1) \ln (n-1) }{2 (1-\beta)} + \frac{ \ln_{\beta} (n-1) - (n-1)^{1-\beta} \ln (n-1) }{ (1-\beta) } \right)
\\
& =
(n-1)^{\frac{\beta}{2}} \left( \frac{ ((n-1)^{1-\beta} - 1) \ln (n-1) }{2 (1-\beta)} + \frac{ ((n-1)^{1-\beta} - 1) - (1-\beta) (n-1)^{1-\beta} \ln (n-1) }{ (1-\beta)^{2} } \right)
\\
& =
(n-1)^{\frac{\beta}{2}} \left( \frac{ (1-\beta) ((n-1)^{1-\beta} - 1) \ln (n-1) }{2 (1-\beta)^{2}}
\right. \notag \\
& \left. \qquad \qquad \qquad \qquad \qquad \qquad \qquad
+ \frac{ 2 ((n-1)^{1-\beta} - 1) - 2 (1-\beta) (n-1)^{1-\beta} \ln (n-1) }{ 2 (1-\beta)^{2} } \right)
\\
& =
\frac{ (n-1)^{\frac{\beta}{2}} }{ 2 (1-\beta)^{2} } \left( \left[ \vphantom{\sum} (1-\beta) ((n-1)^{1-\beta} - 1) - 2 (1-\beta) (n-1)^{1-\beta} \right] \ln (n-1) + 2 ((n-1)^{1-\beta} - 1) \right)
\\
& =
\frac{ (n-1)^{\frac{\beta}{2}} }{ 2 (1-\beta)^{2} } \left( \vphantom{\sum} - (1-\beta) (n-1)^{1-\beta} \ln (n-1) - (1-\beta) \ln (n-1) + 2 ((n-1)^{1-\beta} - 1) \right)
\\
& =
\frac{ (n-1)^{\frac{\beta}{2}} }{ 2 (1-\beta)^{2} } \left( \vphantom{\sum} - (n-1)^{1-\beta} \ln (n-1)^{1-\beta} - \ln (n-1)^{1-\beta} + 2 ((n-1)^{1-\beta} - 1) \right)
\\
& \overset{\text{(a)}}{=}
\frac{ (n-1)^{\frac{\beta}{2}} }{ 2 (1-\beta)^{2} } \left( \vphantom{\sum} - k \ln k - \ln k + 2 (k - 1) \right) ,
\label{eq:diff1_u_beta}
\end{align}
where (a) follows by the change of variable: $k = k(n, \beta) \triangleq (n-1)^{1-\beta}$.
Note that $k = (n-1)^{1-\beta} > 0$ for $n \ge 2$ and $\beta \in (-\infty, +\infty)$.
Then, we readily see that
\begin{align}
\left. \left( \vphantom{\sum} - k \ln k - \ln k + 2 (k - 1) \right) \right|_{k = 1}
& =
0,
\\
\frac{ \mathrm{d} }{ \mathrm{d} k } \left( \vphantom{\sum} - k \ln k - \ln k + 2 (k - 1) \right)
& =
- \left( \frac{ \mathrm{d} }{ \mathrm{d} k } (k \ln k) \right) - \left( \frac{ \mathrm{d} }{ \mathrm{d} k } (\ln k) \right) + 2 \left( \frac{ \mathrm{d} }{ \mathrm{d} k } (k - 1) \right)
\\
& =
- \left( \vphantom{\sum} \ln k + 1 \right) - \left( \frac{1}{k} \right) + 2
\\
& =
\left( 1 - \frac{1}{k} \right) - \ln k
\\
& \overset{\text{(a)}}{\le}
\ln k - \ln k
\\
& =
0 ,
\end{align}
where note that (a) holds with equality if and only if $k = 1$.
Hence, we obtain
\begin{align}
\sgn \! \left( \vphantom{\sum} - k \ln k - \ln k + 2 (k - 1) \right)
=
\begin{cases}
1
& \mathrm{if} \ 0 < k < 1 , \\
0
& \mathrm{if} \ k = 1 , \\
-1
& \mathrm{if} \ k > 1 ;
\end{cases}
\label{eq:sgn_k}
\end{align}
and hence, we can rewrite \eqref{eq:sgn_k} as
\begin{align}
\sgn \left( \vphantom{\sum} - (n-1)^{1-\beta} \ln (n-1)^{1-\beta} - \ln (n-1)^{1-\beta} + 2 ((n-1)^{1-\beta} - 1) \right)
=
\begin{cases}
1
& \mathrm{if} \ \beta > 1 , \\
0
& \mathrm{if} \ \beta = 1 , \\
-1
& \mathrm{if} \ \beta < 1 ;
\end{cases}
\label{eq:sgn_k_prot}
\end{align}
and therefore, we obtain
\begin{align}
\sgn \! \left( \frac{ \partial u_{1}(n, \beta) }{ \partial \beta } \right)
& \overset{\eqref{eq:diff1_u_beta}}{=}
\underbrace{ \sgn \! \left( \frac{ (n-1)^{\frac{\beta}{2}} }{ 2 (1-\beta)^{2} } \right) }_{ = 1 } \cdot \, \sgn \left( \vphantom{\sum} - (n-1)^{1-\beta} \ln (n-1)^{1-\beta} - \ln (n-1)^{1-\beta} + 2 ((n-1)^{1-\beta} - 1) \right)
\\
& \overset{\eqref{eq:sgn_k_prot}}{=}
\begin{cases}
1
& \mathrm{if} \ \beta > 1 , \\
0
& \mathrm{if} \ \beta = 1 , \\
-1
& \mathrm{if} \ \beta < 1 ,
\end{cases}
\label{eq:sgn_diff1_u1_beta}
\end{align}
which implies that
\begin{itemize}
\item
$u_{1}(n, \beta)$ is strictly decreasing for $\beta \in (-\infty, 1]$ and
\item
$u_{1}(n, \beta)$ is strictly increasing for $\beta \in [1, +\infty)$.
\end{itemize}
Moreover, since
\begin{align}
u_{1}( n, 0 )
& =
\left. \left( \vphantom{\sum} (2-n) + (n-1)^{\frac{\beta}{2}} \ln_{\beta} (n-1) \right) \right|_{\beta=0}
\\
& =
-(n-2) + (n-1)^{0} \ln_{0} (n-1)
\\
& =
-(n-2) + (n-2)
\\
& =
0 ,
\\
u_{1}( n, 2 )
& =
\left. \left( \vphantom{\sum} (2-n) + (n-1)^{\frac{\beta}{2}} \ln_{\beta} (n-1) \right) \right|_{\beta=2}
\\
& =
- (n-2) + (n-1) \ln_{2} (n-1)
\\
& =
- (n-2) + (n-1) \left( 1 - \frac{1}{n-1} \right)
\\
& =
- (n-2) + (n-2)
\\
& =
0 ,
\end{align}
it follows from \eqref{eq:sgn_diff1_u1_beta} that
\begin{align}
\sgn \! \left( \vphantom{\sum} u_{1}( n, \beta ) \right)
=
\begin{cases}
1
& \mathrm{if} \ \beta \in (-\infty, 0) \cup (2, +\infty) , \\
0
& \mathrm{if} \ \beta \in \{ 0, 2 \} , \\
-1
& \mathrm{if} \ \beta \in (0, 2) .
\end{cases}
\label{eq:sgn_u1_beta}
\end{align}
Furthermore, since $\beta = \frac{1}{\alpha}$, we can rewrite \eqref{eq:sgn_u1_beta} as
\begin{align}
\sgn \! \left( \vphantom{\sum} u_{1}( n, {\textstyle \frac{1}{\alpha}} ) \right)
=
\begin{cases}
1
& \mathrm{if} \ \alpha \in (-\infty, 0) \cup (0, \frac{1}{2}) , \\
0
& \mathrm{if} \ \alpha = \frac{1}{2} , \\
-1
& \mathrm{if} \ \alpha \in (\frac{1}{2}, +\infty) ;
\end{cases}
\label{eq:sgn_u1_1_over_alpha}
\end{align}
and therefore, we obtain
\begin{align}
\sgn \! \left( \vphantom{\sum} u(n, z_{1}(n, 2 \alpha) , \alpha) \right)
& \overset{\eqref{eq:u_z1(n,2alpha)}}{=}
\sgn (\alpha-1) \cdot \, \underbrace{ \sgn \! \left( \vphantom{\sum} \sqrt{n-1} \right) }_{=1} \cdot \, \sgn \! \left( \vphantom{\sum} u_{1}( n, \beta ) \right)
\\
& \overset{\eqref{eq:sgn_u1_1_over_alpha}}{=}
\begin{cases}
1
& \mathrm{if} \ \alpha \in (\frac{1}{2}, 1) , \\
0
& \mathrm{if} \ \alpha \in \{ \frac{1}{2}, 1 \} , \\
-1
& \mathrm{if} \ \alpha \in (-\infty, 0) \cup (0, \frac{1}{2}) \cup (1, +\infty) .
\end{cases}
\label{eq:sgn_u_z1(n,2alpha)}
\end{align}

So far, we already provided the signs of $\frac{ \partial^{2} g(n, z, \alpha) }{ \partial z \, \partial \alpha }$ for $\alpha \in \{ 0, \frac{1}{2}, 1 \}$ in the proof.
We next show the signs of $u(n, z, \alpha)$ for $\alpha \in (-\infty, +\infty) \setminus \{ 0, \frac{1}{2}, 1 \}$.
Then, note that it follows from \eqref{eq:range_z1z2_orig} that
\begin{align}
\left\{
\begin{array}{ll}
0 < z_{1}(n, \alpha) < z_{1}(n, 2 \alpha) < z_{2}(n, \alpha) < 1
& \mathrm{if} \ \alpha \in (-\infty, 0) ,
\\
0 < z_{2}(n, \alpha) < \frac{1}{2} \ \mathrm{and} \ n-1 < z_{1}(n, 2 \alpha) < z_{1}(n, \alpha)
& \mathrm{if} \ \alpha \in (0, \frac{1}{2}) ,
\\
\sqrt{n-1} < z_{1}(n, 2 \alpha) < n-1 < z_{1}(n, \alpha) < z_{2}(n, \alpha)
& \mathrm{if} \ \alpha \in (\frac{1}{2}, 1) ,
\\
1 < z_{1}(n, 2 \alpha) < z_{2}(n, \alpha) < z_{1}(n, \alpha) < n-1
& \mathrm{if} \ \alpha \in (1, +\infty) .
\end{array}
\right.
\label{eq:range_z1z2}
\end{align}
By the intermediate value theorem, we can prove the signs of $u(n, z, \alpha)$ for $\alpha \in (-\infty, +\infty) \setminus \{ 0, \frac{1}{2}, 1 \}$ as follows:

\subsubsection*{The case of $\alpha \in (-\infty, 0)$}

Since
\begin{itemize}
\item
for a fixed $\alpha \in (-\infty, 0)$, $u(n, z, \alpha)$ is strictly increasing for $z \in (0, +\infty)$ (see Eq. \eqref{eq:diff_u}),
\item
if $\alpha \in (-\infty, 0)$, then $u(n, z_{1}(n, 2 \alpha), \alpha) < 0$ (see Eq. \eqref{eq:sgn_u_z1(n,2alpha)}),
\item
if $\alpha \in (-\infty, 0)$, then $u(n, z_{2}(n, \alpha), \alpha) > 0$ (see Eq. \eqref{eq:u_z2}), and
\item
if $\alpha \in (-\infty, 0)$, then $0 < z_{1}(n, \alpha) < z_{1}(n, 2\alpha) < z_{2}(n, \alpha) < 1$ (see Eq. \eqref{eq:range_z1z2}),
\end{itemize}
for any $\alpha \in (-\infty, 0)$, there exists $\gamma(n, \alpha) \in (z_{1}(n, 2\alpha), z_{2}(n, \alpha))$ such that
\begin{align}
\sgn \! \left( \vphantom{\sum} u(n, z, \alpha) \right)
=
\begin{cases}
1
& \mathrm{if} \ z \in (\gamma(n, \alpha), +\infty) , \\
0
& \mathrm{if} \ z = \gamma(n, \alpha) , \\
-1
& \mathrm{if} \ z \in (0, \gamma(n, \alpha)) .
\end{cases}
\label{eq:sign_u1}
\end{align}

\subsubsection*{The case of $\alpha \in (0, \frac{1}{2})$}

Since
\begin{itemize}
\item
for a fixed $\alpha \in (0, \frac{1}{2})$, $u(n, z, \alpha)$ is strictly decreasing for $z \in (0, +\infty)$ (see Eq. \eqref{eq:diff_u}),
\item
if $\alpha \in (0, \frac{1}{2})$, then $u(n, n-1, \alpha) > 0$ (see Eq. \eqref{eq:u_n-1}),
\item
if $\alpha \in (0, \frac{1}{2})$, then $u(n, z_{1}(n, 2 \alpha), \alpha) < 0$ (see Eq. \eqref{eq:sgn_u_z1(n,2alpha)}), and
\item
if $\alpha \in (0, \frac{1}{2})$, then $n-1 < z_{1}(n, 2 \alpha) < z_{1}(n, \alpha)$ (see Eq. \eqref{eq:range_z1z2}),
\end{itemize}
for any $\alpha \in (0, \frac{1}{2})$, there exists $\gamma(n, \alpha) \in (n-1, z_{1}(n, 2 \alpha))$ such that
\begin{align}
\sgn \! \left( \vphantom{\sum} u(n, z, \alpha) \right)
=
\begin{cases}
1
& \mathrm{if} \ z \in (0, \gamma(n, \alpha)) , \\
0
& \mathrm{if} \ z = \gamma(n, \alpha) , \\
-1
& \mathrm{if} \ z \in (\gamma(n, \alpha), +\infty) .
\end{cases}
\label{eq:sign_u2}
\end{align}

\subsubsection*{The case of $\alpha \in (\frac{1}{2}, 1)$}

Since
\begin{itemize}
\item
for a fixed $\alpha \in (\frac{1}{2}, 1)$, $u(n, z, \alpha)$ is strictly decreasing for $z \in (0, +\infty)$ (see Eq. \eqref{eq:diff_u}) ,
\item
if $\alpha \in (\frac{1}{2}, 1)$, then $u(n, z_{1}(n, 2 \alpha), \alpha) > 0$ (see Eq. \eqref{eq:sgn_u_z1(n,2alpha)}),
\item
if $\alpha \in (\frac{1}{2}, 1)$, then $u(n, n-1, \alpha) < 0$ (see Eq. \eqref{eq:u_n-1}), and
\item
$\sqrt{n-1} < z_{1}(n, 2 \alpha) < n-1$ (see Eq. \eqref{eq:range_z1z2}),
\end{itemize}
for any $\alpha \in (\frac{1}{2}, 1)$, there exists $\gamma(n, \alpha) \in (z_{1}(n, 2 \alpha), n-1)$ such that
\begin{align}
\sgn \! \left( \vphantom{\sum} u(n, z, \alpha) \right)
=
\begin{cases}
1
& \mathrm{if} \ z \in (0, \gamma(n, \alpha)) , \\
0
& \mathrm{if} \ z = \gamma(n, \alpha) , \\
-1
& \mathrm{if} \ z \in (\gamma(n, \alpha), +\infty) .
\end{cases}
\label{eq:sign_u3}
\end{align}

\subsubsection*{The case of $\alpha \in (1, +\infty)$}

Since
\begin{itemize}
\item
for a fixed $\alpha \in (1, +\infty)$, $u(n, z, \alpha)$ is strictly increasing for $z \in (0, +\infty)$ (see Eq. \eqref{eq:diff_u}) ,
\item
if $\alpha \in (1, +\infty)$, then $u(n, z_{1}(n, 2\alpha), \alpha) < 0$ (see Eq. \eqref{eq:sgn_u_z1(n,2alpha)}),
\item
if $\alpha \in (1, +\infty)$, then $u(n, z_{2}(n, \alpha), \alpha) > 0$ (see Eq. \eqref{eq:u_z2}), and
\item
if $\alpha \in (1, +\infty)$, then $1 < z_{1}(n, 2 \alpha) < z_{2}(n, \alpha) < z_{1}(n, \alpha)$ (see Eq. \eqref{eq:range_z1z2}),
\end{itemize}
for any $\alpha \in (1, +\infty)$, there exists $\gamma(n, \alpha) \in (z_{1}(n, 2\alpha), z_{2}(n, \alpha))$ such that
\begin{align}
\sgn \! \left( \vphantom{\sum} u(n, z, \alpha) \right)
=
\begin{cases}
1
& \mathrm{if} \ z \in (\gamma(n, \alpha), +\infty) , \\
0
& \mathrm{if} \ z = \gamma(n, \alpha) , \\
-1
& \mathrm{if} \ z \in (0, \gamma(n, \alpha)) .
\end{cases}
\label{eq:sign_u4}
\end{align}

Since
\begin{align}
\sgn \! \left( \vphantom{\sum} \frac{ \partial^{2} g(n, z, \alpha) }{ \partial z \, \partial \alpha } \right)
& \overset{\eqref{eq:diff_g_za}}{=}
- \sgn \! \left( \vphantom{\sum} u(n, z, \alpha) \right) ,
\end{align}
\begin{itemize}
\item
Eq. \eqref{eq:g_dzda_1} follows from \eqref{eq:sign_u1},
\item
Eq. \eqref{eq:g_dzda_2} follows from \eqref{eq:sign_u2},
\item
Eq. \eqref{eq:g_dzda_4} follows from \eqref{eq:sign_u3}, and
\item
Eq. \eqref{eq:g_dzda_5} follows from \eqref{eq:sign_u4},
\end{itemize}
Therefore, we obtain Lemma \ref{lem:dzda}.
\end{IEEEproof}

\section{Proof of Lemma \ref{lem:diff_g_z}}
\label{app:diff_g_z}

\begin{IEEEproof}[Proof of Lemma \ref{lem:diff_g_z}]
In the proof, assume that $n \ge 3$.
Direct calculation shows
\begin{align}
\frac{ \partial g(n, z, \alpha) }{ \partial z }
& \overset{\eqref{eq:g_z}}{=}
\frac{ \partial }{ \partial z } \left( (\alpha - 1) + \frac{ ((n-1) + z^{\alpha}) (z^{1-\alpha} - 1) }{ ((n-1) + z) \ln z } \right)
\\
& =
\frac{ \partial }{ \partial z } \left( \frac{ ((n-1) + z^{\alpha}) (z^{1-\alpha} - 1) }{ ((n-1) + z) \ln z } \right)
\\
& \overset{\text{(a)}}{=}
\frac{ 1 }{ ((n-1) + z)^{2} (\ln z)^{2} } \left( \left[ \frac{ \partial ((n-1) + z^{\alpha}) (z^{1-\alpha} - 1) }{ \partial z } \right] ((n-1) + z) \ln z
\right. \notag \\
& \left. \qquad \qquad \qquad \qquad \qquad \qquad \qquad \qquad
- \; ((n-1) + z^{\alpha}) (z^{1-\alpha} - 1) \left[ \frac{ \partial ((n-1) + z) \ln z }{ \partial z } \right] \right)
\\
& \overset{\text{(b)}}{=}
\frac{ 1 }{ ((n-1) + z)^{2} (\ln z)^{2} } \left( \left[ \vphantom{\sum} (n-1) (1-\alpha) z^{-\alpha} + 1 - \alpha z^{\alpha-1} \right] ((n-1) + z) \ln z
\right. \notag \\
& \left. \qquad \qquad \qquad \qquad \qquad \qquad \qquad \quad
- \; ((n-1) + z^{\alpha}) (z^{1-\alpha} - 1) \left[ \vphantom{\sum} (n-1) z^{-1} + (\ln z + 1) \right] \right)
\\
& =
\frac{ 1 }{ ((n-1) + z)^{2} (\ln z)^{2} } \left( \left[ \vphantom{\sum} (n-1) (1-\alpha) z^{-\alpha} + 1 - \alpha z^{\alpha-1} \right] ((n-1) \ln z + z \ln z)
\right. \notag \\
& \left. \qquad \qquad \qquad \qquad \qquad \qquad
- \; ((n-1) z^{1-\alpha} - (n-1) + z - z^{\alpha}) \left[ \vphantom{\sum} (n-1) z^{-1} + \ln z + 1 \right] \right)
\\
& =
\frac{ 1 }{ ((n-1) + z)^{2} (\ln z)^{2} } 
\notag \\
& \qquad \times
\left( \left[ \vphantom{\sum} (n-1)^{2} (1-\alpha) z^{-\alpha} \ln z + (n-1) (1-\alpha) z^{1-\alpha} \ln z
\right. \right. \notag \\
& \left. \qquad \qquad \qquad \qquad \qquad \vphantom{\sum}
+ (n-1) \ln z + z \ln z - (n-1) \alpha z^{\alpha-1} \ln z - \alpha z^{\alpha} \ln z \right]
\notag \\
& \qquad \qquad \quad
- \left[ \vphantom{\sum} (n-1)^{2} z^{-\alpha} + (n-1) z^{1-\alpha} \ln z + (n-1) z^{1-\alpha} - (n-1)^{2} z^{-1} - (n-1) \ln z
\right. \notag \\
& \left. \left. \qquad \qquad \qquad \qquad \qquad \vphantom{\sum}
- (n-1) + (n-1) + z \ln z + z - (n-1) z^{\alpha-1} - z^{\alpha} \ln z - z^{\alpha} \right] \right)
\\
& =
\frac{ 1 }{ ((n-1) + z)^{2} (\ln z)^{2} } 
\notag \\
& \qquad \times
\left( \left[ \vphantom{\sum} (n-1)^{2} (1-\alpha) z^{-\alpha} \ln z + (n-1) (1-\alpha) z^{1-\alpha} \ln z
\right. \right. \notag \\
& \left. \qquad \qquad \qquad \qquad \qquad \qquad \qquad \vphantom{\sum}
+ (n-1) \ln z - (n-1) \alpha z^{\alpha-1} \ln z - \alpha z^{\alpha} \ln z \right]
\notag \\
& \qquad \qquad \quad
- \left[ \vphantom{\sum} (n-1)^{2} z^{-\alpha} + (n-1) z^{1-\alpha} \ln z + (n-1) z^{1-\alpha} - (n-1)^{2} z^{-1} - (n-1) \ln z
\right. \notag \\
& \left. \left. \qquad \qquad \qquad \qquad \qquad \qquad \quad \qquad \qquad \qquad \qquad \quad \vphantom{\sum}
+ z - (n-1) z^{\alpha-1} - z^{\alpha} \ln z - z^{\alpha} \right] \right)
\\
& =
\frac{ 1 }{ ((n-1) + z)^{2} (\ln z)^{2} } 
\notag \\
& \qquad \times
\left( \left[ \vphantom{\sum} (n-1)^{2} (1-\alpha) z^{-\alpha} + (n-1) (1-\alpha) z^{1-\alpha} + (n-1) - (n-1) \alpha z^{\alpha-1} - \alpha z^{\alpha} \right] \ln z
\right. \notag \\
& \qquad \qquad \quad
- \left[ \vphantom{\sum} (n-1)^{2} z^{-\alpha} + (n-1) z^{1-\alpha} \ln z + (n-1) z^{1-\alpha} - (n-1)^{2} z^{-1} - (n-1) \ln z
\right. \notag \\
& \left. \left. \qquad \qquad \qquad \qquad \qquad \qquad \quad \qquad \qquad \qquad \qquad \quad \vphantom{\sum}
+ z - (n-1) z^{\alpha-1} - z^{\alpha} \ln z - z^{\alpha} \right] \right)
\\
& =
\frac{ 1 }{ ((n-1) + z)^{2} (\ln z)^{2} } 
\notag \\
& \qquad \times
\left( \left[ \vphantom{\sum} (n-1)^{2} (1-\alpha) z^{-\alpha} + (n-1) (1-\alpha) z^{1-\alpha} + (n-1) - (n-1) \alpha z^{\alpha-1} - \alpha z^{\alpha} \right] \ln z
\right. \notag \\
& \qquad \qquad \quad
- \left[ \vphantom{\sum} (n-1)^{2} z^{-\alpha} + z - (n-1) z^{\alpha-1} - z^{\alpha} + (n-1) z^{1-\alpha} - (n-1)^{2} z^{-1} \right]
\notag \\
& \left. \qquad \qquad \qquad \qquad \qquad \qquad \qquad \qquad \qquad \qquad \qquad 
- \left[ \vphantom{\sum} (n-1) z^{1-\alpha} - (n-1) - z^{\alpha} \right] \ln z \right)
\\
& =
\frac{ 1 }{ ((n-1) + z)^{2} (\ln z)^{2} } 
\notag \\
& \qquad \times
\left( \left[ \vphantom{\sum} (n-1)^{2} (1-\alpha) z^{-\alpha} + (n-1) (1-\alpha) z^{1-\alpha} + (n-1) - (n-1) \alpha z^{\alpha-1} - \alpha z^{\alpha}
\right. \right. \notag \\
& \left. \qquad \qquad \qquad \qquad \qquad \qquad \qquad \qquad \qquad \qquad \qquad \quad \vphantom{\sum}
- (n-1) z^{1-\alpha} + (n-1) + z^{\alpha} \right] \ln z
\notag \\
& \left. \qquad \qquad \qquad \quad
- \left[ \vphantom{\sum} (n-1)^{2} z^{-\alpha} + z - (n-1) z^{\alpha-1} - z^{\alpha} + (n-1) z^{1-\alpha} - (n-1)^{2} z^{-1} \right] \right)
\\
& =
\frac{ 1 }{ ((n-1) + z)^{2} (\ln z)^{2} } 
\notag \\
& \qquad \times
\left( \left[ \vphantom{\sum} (n-1)^{2} (1-\alpha) z^{-\alpha} + (n-1) z^{1-\alpha} ((1-\alpha) - 1)
\right. \right. \notag \\
& \left. \vphantom{\sum} \qquad \qquad \qquad \qquad \qquad \qquad \qquad \qquad
+ 2 (n-1) - (n-1) \alpha z^{\alpha-1} + z^{\alpha} (-\alpha + 1) \right] \ln z
\notag \\
& \left. \qquad \qquad \qquad \quad
- \left[ \vphantom{\sum} (n-1)^{2} z^{-\alpha} + z - (n-1) z^{\alpha-1} - z^{\alpha} + (n-1) z^{1-\alpha} - (n-1)^{2} z^{-1} \right] \right)
\\
& =
\frac{ 1 }{ ((n-1) + z)^{2} (\ln z)^{2} } 
\notag \\
& \qquad \times
\left( \left[ \vphantom{\sum} (n-1)^{2} (1-\alpha) z^{-\alpha} - (n-1) \alpha z^{1-\alpha} + 2 (n-1) - (n-1) \alpha z^{\alpha-1} + (1 - \alpha) z^{\alpha} \right] \ln z
\right. \notag \\
& \left. \qquad \qquad \qquad \quad
- \left[ \vphantom{\sum} (n-1)^{2} z^{-\alpha} + z - (n-1) z^{\alpha-1} - z^{\alpha} + (n-1) z^{1-\alpha} - (n-1)^{2} z^{-1} \right] \right)
\\
& =
\frac{ 1 }{ ((n-1) + z)^{2} (\ln z)^{2} } 
\notag \\
& \qquad \times
\left( \left[ \vphantom{\sum} (n-1)^{2} (1-\alpha) z^{-\alpha} - (n-1) \alpha (z^{1-\alpha} + z^{\alpha-1}) + 2 (n-1) + (1 - \alpha) z^{\alpha} \right] \ln z
\right. \notag \\
& \left. \qquad \qquad \qquad \quad
- \left[ \vphantom{\sum} (n-1)^{2} z^{-\alpha} + z - (n-1) z^{\alpha-1} - z^{\alpha} + (n-1) z^{1-\alpha} - (n-1)^{2} z^{-1} \right] \right)
\\
& =
\frac{ 1 }{ ((n-1) + z)^{2} (\ln z)^{2} } 
\notag \\
& \qquad \times
\left( \left[ \vphantom{\sum} (n-1)^{2} (1-\alpha) z^{-\alpha} - (n-1) \alpha (z^{1-\alpha} + z^{\alpha-1}) + 2 (n-1) + (1 - \alpha) z^{\alpha} \right] \ln z
\right. \notag \\
& \left. \qquad \qquad \qquad \qquad \qquad \qquad
- \left[ \vphantom{\sum} (n-1)^{2} (z^{-\alpha} - z^{-1}) - (n-1) (z^{\alpha-1} - z^{1-\alpha}) + z - z^{\alpha} \right] \right) ,
\label{eq:diff_g_z}
\end{align}
where
\begin{itemize}

\item
(a) follows from the fact that
$
\frac{ \mathrm{d} }{ \mathrm{d} x } \left( \frac{ f(x) }{ g(x) } \right) = \frac{ \left[ \frac{ \mathrm{d} f(x) }{ \mathrm{d} x } \right] g(x) - f(x) \left[ \frac{ \mathrm{d} g(x) }{ \mathrm{d} x } \right] }{ (g(x))^{2} }
$
and

\item
(b) follows from the facts that
\begin{align}
\frac{ \partial }{ \partial z } \left( \vphantom{\sum} ((n-1) + z^{\alpha}) (z^{1-\alpha} - 1) \right)
& =
\frac{ \partial }{ \partial z } \left( \vphantom{\sum} (n-1) z^{1-\alpha} - (n-1) + z - z^{\alpha} \right)
\\
& =
\left( \frac{ \partial (n-1) z^{1-\alpha} }{ \partial z } \right) - \left( \frac{ \partial (n-1) }{ \partial z } \right) + \left( \frac{ \partial z }{ \partial z } \right) - \left( \frac{ \partial z^{\alpha} }{ \partial z } \right)
\\
& =
(n-1) \left( \frac{ \partial z^{1-\alpha} }{ \partial z } \right) + \left( \frac{ \partial z }{ \partial z } \right) - \left( \frac{ \partial z^{\alpha} }{ \partial z } \right)
\\
& =
(n-1) (1-\alpha) z^{-\alpha} + 1 - \alpha z^{\alpha-1} ,
\\
\frac{ \partial }{ \partial z } \left( \vphantom{\sum} ((n-1) + z) \ln z \right)
& =
\frac{ \partial }{ \partial z } \left( \vphantom{\sum} (n-1) \ln z + z \ln z \right)
\\
& =
\left( \frac{ \partial (n-1) \ln z }{ \partial z } \right) + \left( \frac{ \partial z \ln z }{ \partial z } \right)
\\
& =
(n-1) \left( \frac{ \partial \ln z }{ \partial z } \right) + \left( \frac{ \partial z \ln z }{ \partial z } \right)
\\
& =
(n-1) z^{-1} + (\ln z + 1) .
\end{align}
\end{itemize}
Substituting $\alpha = 1$ into $\frac{ \partial g(n, z, \alpha) }{ \partial z }$, we readily see that
\begin{align}
&
\left. \frac{ \partial g(n, z, \alpha) }{ \partial z } \right|_{\alpha = 1}
\\
& \quad \overset{\eqref{eq:diff_g_z}}{=}
\frac{ 1 }{ ((n-1) + z)^{2} (\ln z)^{2} } 
\notag \\
& \qquad \times
\left( \left[ \vphantom{\sum} (n-1)^{2} (1-\alpha) z^{-\alpha} - (n-1) \alpha (z^{1-\alpha} + z^{\alpha-1}) + 2 (n-1) + (1 - \alpha) z^{\alpha} \right] \ln z
\right. \notag \\
& \left. \left. \qquad \qquad \qquad \qquad \qquad \qquad \qquad
- \left[ \vphantom{\sum} (n-1)^{2} (z^{-\alpha} - z^{-1}) - (n-1) (z^{\alpha-1} - z^{1-\alpha}) + z - z^{\alpha} \right] \right) \right|_{\alpha = 1}
\\
& \quad =
\frac{ 1 }{ ((n-1) + z)^{2} (\ln z)^{2} } \left( \left[ \vphantom{\sum} - (n-1) (z^{0} + z^{0}) + 2 (n-1) \right] \ln z
\right. \notag \\
& \left. \qquad \qquad \qquad \qquad \qquad \qquad \qquad \qquad \qquad
- \left[ \vphantom{\sum} (n-1)^{2} (z^{-1} - z^{-1}) - (n-1) (z^{0} - z^{0}) + z - z \right] \right)
\\
& \quad =
\frac{ 1 }{ ((n-1) + z)^{2} (\ln z)^{2} } \left( \left[ \vphantom{\sum} - 2 (n-1) + 2 (n-1) \right] \ln z \right)
\\
& \quad =
0 .
\end{align}
On the other hand, substituting $\alpha = \alpha_{1}(n, z) = \frac{ \ln (n-1) }{ \ln z }$ into $\frac{ \partial g(n, z, \alpha) }{ \partial z }$, we have
\begin{align}
&
\left. \frac{ \partial g(n, z, \alpha) }{ \partial z } \right|_{\alpha = \frac{\ln (n-1)}{\ln z}}
\\
& \quad \overset{\eqref{eq:diff_g_z}}{=}
\frac{ 1 }{ ((n-1) + z)^{2} (\ln z)^{2} } 
\notag \\
& \qquad \times
\left( \left[ \vphantom{\sum} (n-1)^{2} (1-\alpha) z^{-\alpha} - (n-1) \alpha (z^{1-\alpha} + z^{\alpha-1}) + 2 (n-1) + (1 - \alpha) z^{\alpha} \right] \ln z
\right. \notag \\
& \left. \left. \qquad \qquad \qquad \qquad \qquad \qquad \quad
- \left[ \vphantom{\sum} (n-1)^{2} (z^{-\alpha} - z^{-1}) - (n-1) (z^{\alpha-1} - z^{1-\alpha}) + z - z^{\alpha} \right] \right) \right|_{\alpha = \frac{\ln (n-1)}{\ln z}}
\\
& \quad =
\frac{ 1 }{ ((n-1) + z)^{2} (\ln z)^{2} } 
\notag \\
& \quad \qquad \times
\left( \left[ \vphantom{\sum} (n-1)^{2} \left( 1 - \frac{\ln (n-1)}{\ln z} \right) z^{-\frac{\ln (n-1)}{\ln z}} - (n-1) \left( \frac{\ln (n-1)}{\ln z} \right) (z^{1-\frac{\ln (n-1)}{\ln z}} + z^{\frac{\ln (n-1)}{\ln z}-1})
\right. \right. \notag \\
& \left. \qquad \qquad \qquad \qquad \qquad \qquad \qquad \qquad \qquad \qquad \qquad \qquad \quad \vphantom{\sum}
+ 2 (n-1) + \left( 1 - \frac{\ln (n-1)}{\ln z} \right) z^{\frac{\ln (n-1)}{\ln z}} \right] \ln z
\notag \\
& \left. \qquad \qquad \qquad \qquad
- \left[ \vphantom{\sum} (n-1)^{2} (z^{-\frac{\ln (n-1)}{\ln z}} - z^{-1}) - (n-1) (z^{\frac{\ln (n-1)}{\ln z}-1} - z^{1-\frac{\ln (n-1)}{\ln z}}) + z - z^{\frac{\ln (n-1)}{\ln z}} \right] \right)
\\
& \quad =
\frac{ 1 }{ ((n-1) + z)^{2} (\ln z)^{2} } 
\notag \\
& \quad \qquad \times
\left( \left[ \vphantom{\sum} (n-1)^{2} \left( 1 - \frac{\ln (n-1)}{\ln z} \right) (n-1)^{-1} - (n-1) \left( \frac{\ln (n-1)}{\ln z} \right) ((n-1)^{-1} z + (n-1) z^{-1})
\right. \right. \notag \\
& \left. \qquad \qquad \qquad \qquad \qquad \qquad \qquad \qquad \qquad \qquad \qquad \qquad \quad \vphantom{\sum}
+ 2 (n-1) + \left( 1 - \frac{\ln (n-1)}{\ln z} \right) (n-1) \right] \ln z
\notag \\
& \left. \qquad \qquad \qquad \quad
- \left[ \vphantom{\sum} (n-1)^{2} ((n-1)^{-1} - z^{-1}) - (n-1) ((n-1) z^{-1} - (n-1)^{-1} z) + z - (n-1) \right] \right)
\\
& \quad =
\frac{ 1 }{ ((n-1) + z)^{2} (\ln z)^{2} } 
\notag \\
& \quad \qquad \times
\left( \left[ \vphantom{\sum} (n-1) \left( 1 - \frac{\ln (n-1)}{\ln z} \right) - \left( \frac{\ln (n-1)}{\ln z} \right) (z + (n-1)^{2} z^{-1})
\right. \right. \notag \\
& \left. \qquad \qquad \qquad \qquad \qquad \qquad \qquad \qquad \qquad \qquad \qquad \qquad \quad \vphantom{\sum}
+ 2 (n-1) + \left( 1 - \frac{\ln (n-1)}{\ln z} \right) (n-1) \right] \ln z
\notag \\
& \left. \qquad \qquad \qquad \qquad \qquad \qquad \qquad \qquad 
- \left[ \vphantom{\sum} (n-1) - (n-1)^{2} z^{-1} - ((n-1)^{2} z^{-1} - z) + z - (n-1) \right] \right)
\\
& \quad =
\frac{ 1 }{ ((n-1) + z)^{2} (\ln z)^{2} } 
\notag \\
& \quad \qquad \times
\left( \left[ \vphantom{\sum} (n-1) (\ln z - \ln (n-1)) - (\ln (n-1)) (z + (n-1)^{2} z^{-1})
\right. \right. \notag \\
& \left. \vphantom{\sum} \qquad \qquad \qquad \qquad \qquad \qquad  \qquad \qquad
+ 2 (n-1) \ln z + (\ln z - \ln (n-1)) (n-1) \right]
\notag \\
& \left. \qquad \qquad \qquad \qquad \qquad \qquad \qquad \qquad \qquad \qquad \qquad \qquad
- \left[ \vphantom{\sum} - (n-1)^{2} z^{-1} - ((n-1)^{2} z^{-1} - z) + z \right] \right)
\\
& \quad =
\frac{ 1 }{ ((n-1) + z)^{2} (\ln z)^{2} } 
\notag \\
& \quad \qquad \times
\left( \left[ \vphantom{\sum} (n-1) \ln z - (n-1) \ln (n-1) - z \ln (n-1)
\right. \right. \notag \\
& \left. \qquad \qquad \qquad \qquad \qquad \vphantom{\sum}
- z^{-1} (n-1)^{2} \ln (n-1) + 2 (n-1) \ln z + (n-1) \ln z - (n-1) \ln (n-1) \right]
\notag \\
& \left. \qquad \qquad \qquad \qquad \qquad \qquad \qquad \qquad \qquad \qquad \qquad
- \left[ \vphantom{\sum} - (n-1)^{2} z^{-1} - (n-1)^{2} z^{-1} + z + z \right] \right)
\\
& \quad =
\frac{ 1 }{ ((n-1) + z)^{2} (\ln z)^{2} } 
\notag \\
& \quad \qquad \times
\left( \vphantom{\sum} (n-1) \ln z - (n-1) \ln (n-1) - z \ln (n-1)
\right. \notag \\
& \qquad \qquad \qquad \qquad
- z^{-1} (n-1)^{2} \ln (n-1) + 2 (n-1) \ln z + (n-1) \ln z - (n-1) \ln (n-1)
\notag \\
& \left. \qquad \qquad \qquad \qquad \qquad \qquad \qquad \qquad \qquad \qquad \qquad \qquad \qquad \qquad \qquad \qquad \quad \vphantom{\sum}
+ 2 (n-1)^{2} z^{-1} - 2 z \right)
\\
& \quad =
\frac{ 1 }{ ((n-1) + z)^{2} (\ln z)^{2} } 
\notag \\
& \quad \qquad \times
\left( \vphantom{\sum} 4 (n-1) \ln z - 2 (n-1) \ln (n-1) - z \ln (n-1)
\right. \notag \\
& \left. \vphantom{\sum} \qquad \qquad \qquad \qquad \qquad \qquad \qquad \qquad \qquad \qquad \qquad
- z^{-1} (n-1)^{2} \ln (n-1) + 2 (n-1)^{2} z^{-1} - 2 z \right)
\\
& \quad =
\frac{ 1 }{ z ((n-1) + z)^{2} (\ln z)^{2} } 
\notag \\
& \quad \qquad \times
\left( \vphantom{\sum} 4 (n-1) z \ln z - 2 z (n-1) \ln (n-1) - z^{2} \ln (n-1) - (n-1)^{2} \ln (n-1) + 2 (n-1)^{2} - 2 z^{2} \right)
\\
& \quad =
\frac{ 1 }{ z ((n-1) + z)^{2} (\ln z)^{2} } \left( \vphantom{\sum} 4 (n-1) z \ln z - ((n-1)^{2} + 2 z (n-1) + z^{2}) \ln (n-1) + 2 (n-1)^{2} - 2 z^{2} \right) \!\!\!
\\
& \quad =
\frac{ 1 }{ z ((n-1) + z)^{2} (\ln z)^{2} } \left( \vphantom{\sum} 4 (n-1) z \ln z - ((n-1) + z)^{2} \ln (n-1) + 2 (n-1)^{2} - 2 z^{2} \right)
\\
& \quad =
\frac{ v(n, z) }{ z ((n-1) + z)^{2} (\ln z)^{2} } ,
\label{eq:v}
\end{align}
where
\begin{align}
v(n, z)
\triangleq
4 (n-1) z \ln z - ((n-1) + z)^{2} \ln (n-1) + 2 (n-1)^{2} - 2 z^{2} .
\end{align}
Thus, it is enough to check the sign of $v(n, z)$ for $z \in (1, +\infty)$.
Substituting $z = 1$ into $v(n, z)$, we see
\begin{align}
v(n, 1)
& =
\left. \left( \vphantom{\sum} 4 (n-1) z \ln z - ((n-1) + z)^{2} \ln (n-1) + 2 (n-1)^{2} - 2 z^{2} \right) \right|_{z = 1}
\\
& =
4 (n-1) \cdot \underbrace{ (1 \ln 1) }_{ = 0 } - ((n-1) + 1)^{2} \ln (n-1) + 2 (n-1)^{2} - 2 \cdot 1^{2}
\\
& =
- n^{2} \ln (n-1) + 2 (n-1)^{2} - 2
\\
& =
- n^{2} \ln (n-1) + (2 n^{2} - 4 n + 2) - 2
\\
& =
- n^{2} \ln (n-1) + 2 n^{2} - 4 n
\\
& =
2 n (n - 2) - n^{2} \ln (n-1)
\\
& =
n \left( \vphantom{\sum} 2 (n - 2) - n \ln (n-1) \right)
\\
& =
n \, w( n ),
\label{eq:v_1}
\end{align}
where
\begin{align}
w( n )
\triangleq
2 (n - 2) - n \ln (n-1) .
\end{align}
Then, we can see that $w( n )$ is strictly decreasing for $n \ge 2$ as follows:
\begin{align}
\frac{ \mathrm{d} w( n ) }{ \mathrm{d} n }
& =
\frac{ \mathrm{d} }{ \mathrm{d} n } \left( \vphantom{\sum} 2 (n - 2) - n \ln (n-1) \right)
\\
& =
\left( \frac{ \mathrm{d} }{ \mathrm{d} n } (2 (n - 2)) \right) - \left( \frac{ \mathrm{d} }{ \mathrm{d} n } (n \ln (n-1)) \right)
\\
& =
2 - \left( \left( \frac{ \mathrm{d} n }{ \mathrm{d} n } \right) \ln (n-1) + n \left( \frac{ \mathrm{d} \ln (n-1) }{ \mathrm{d} n } \right) \right)
\\
& =
2 - \ln (n-1) - \frac{ n }{ n-1 }
\\
& =
\frac{ 2 (n - 1) - n }{ n-1 } - \ln (n-1)
\\
& =
\frac{ n - 2 }{ n-1 } - \ln (n-1)
\\
& \overset{\text{(a)}}{<}
\frac{ n - 2 }{ n-1 } - \left( 1 - \frac{ 1 }{ n-1 } \right)
\\
& =
\frac{ n - 2 }{ n-1 } - \left( \frac{ (n - 1) - 1 }{ n-1 } \right)
\\
& =
\frac{ n - 2 }{ n-1 } - \frac{ n - 2 }{ n-1 }
\\
& =
0 ,
\end{align}
where (a) holds for $n > 2$ from \eqref{eq:ITineq}.
Since $w(2) = 0$ and $w(n)$ is strictly decreasing for $n \ge 2$, we have
\begin{align}
\sgn \! \left( \vphantom{\sum} w(n) \right)
& =
\begin{cases}
0
& \mathrm{if} \ n = 2, \\
-1
& \mathrm{if} \ n \ge 3 ;
\end{cases}
\end{align}
and therefore, we obtain
\begin{align}
\sgn \! \left( \vphantom{\sum} v(n, 1) \right)
& \overset{\eqref{eq:v_1}}{=}
\sgn \! \left( \vphantom{\sum} n \, w(n) \right)
\\
& =
-1
\label{ineq:v_1}
\end{align}
for $n \ge 3$.
Moreover, substituting $z = n-1$ into $v(n, z)$, we readily see that
\begin{align}
v(n, n-1)
& =
\left. \left( \vphantom{\sum} 4 (n-1) z \ln z - ((n-1) + z)^{2} \ln (n-1) + 2 (n-1)^{2} - 2 z^{2} \right) \right|_{z = n-1}
\\
& =
4 (n-1) (n-1) \ln (n-1) - ((n-1) + (n-1))^{2} \ln (n-1) + 2 (n-1)^{2} - 2 (n-1)^{2}
\\
& =
4 (n-1)^{2} \ln (n-1) - 4 (n-1)^{2} \ln (n-1)
\\
& =
0 .
\label{eq:v_n-1}
\end{align}
Next, we calculate the derivatives of $v(n, z)$ with respect to $z$ as follows:
\begin{align}
\frac{ \partial v(n, z) }{ \partial z }
& =
\frac{ \partial }{ \partial z } \left( \vphantom{\sum} 4 (n-1) z \ln z - ((n-1) + z)^{2} \ln (n-1) + 2 (n-1)^{2} - 2 z^{2} \right)
\\
& =
\frac{ \partial }{ \partial z } \left( \vphantom{\sum} 4 (n-1) z \ln z - (n-1)^{2} \ln (n-1) - 2 z (n-1) \ln (n-1) - z^{2} \ln (n-1) + 2 (n-1)^{2} - 2 z^{2} \right)
\\
& =
\frac{ \partial }{ \partial z } \left( \vphantom{\sum} 4 (n-1) z \ln z - 2 z (n-1) \ln (n-1) - z^{2} \ln (n-1) - 2 z^{2} \right)
\\
& =
\left( \frac{ \partial }{ \partial z } 4 (n-1) z \ln z \right) - \left( \frac{ \partial }{ \partial z } 2 z (n-1) \ln (n-1) \right) - \left( \frac{ \partial }{ \partial z } z^{2} \ln (n-1) \right) - \left( \frac{ \partial }{ \partial z } 2 z^{2} \right)
\\
& =
4 (n-1) \left( \frac{ \mathrm{d} }{ \mathrm{d} z } z \ln z \right) - 2 (n-1) \ln (n-1) \left( \frac{ \mathrm{d} }{ \mathrm{d} z } z \right) - \ln (n-1) \left( \frac{ \mathrm{d} }{ \mathrm{d} z } z^{2} \right) - 2 \left( \frac{ \mathrm{d} }{ \mathrm{d} z } z^{2} \right)
\\
& =
4 (n-1) (\ln z + 1) - 2 (n-1) \ln (n-1) - \ln (n-1) (2 z) - 2 (2 z)
\\
& =
4 (n-1) \ln z + 4 (n-1) - 2 (n-1) \ln (n-1) - 2 z \ln (n-1) - 4 z ,
\\
\frac{ \partial^{2} v(n, z) }{ \partial z^{2} }
& =
\frac{ \partial }{ \partial z } \left( \vphantom{\sum} 4 (n-1) \ln z + 4 (n-1) - 2 (n-1) \ln (n-1) - 2 z \ln (n-1) - 4 z \right)
\\
& =
\frac{ \partial }{ \partial z } \left( \vphantom{\sum} 4 (n-1) \ln z - 2 z \ln (n-1) - 4 z \right)
\\
& =
\left( \frac{ \partial }{ \partial z } 4 (n-1) \ln z \right) - \left( \frac{ \partial }{ \partial z } 2 z \ln (n-1) \right) - \left( \frac{ \partial }{ \partial z } 4 z \right)
\\
& =
4 (n-1) \left( \frac{ \mathrm{d} }{ \mathrm{d} z } \ln z \right) - 2 (\ln (n-1)) \left( \frac{ \mathrm{d} }{ \mathrm{d} z } z \right) - 4 \left( \frac{ \mathrm{d} }{ \mathrm{d} z } z \right)
\\
& =
4 (n-1) \left( \frac{1}{z} \right) - 2 \ln (n-1) - 4
\\
& =
\frac{ 4 (n-1) }{ z } - 2 \ln (n-1) - 4 ,
\\
\frac{ \partial^{3} v(n, z) }{ \partial z^{3} }
& =
\frac{ \partial }{ \partial z } \left( \vphantom{\sum} \frac{ 4 (n-1) }{ z } - 2 \ln (n-1) - 4 \right)
\\
& =
\frac{ \partial }{ \partial z } \left( \vphantom{\sum} \frac{ 4 (n-1) }{ z } \right)
\\
& =
4 (n-1) \left( \frac{ \mathrm{d} }{ \mathrm{d} z } \frac{ 1 }{ z } \right)
\\
& =
4 (n-1) \left( - \frac{1}{z^{2}} \right)
\\
& =
- \frac{4 (n-1)}{z^{2}}
\\
& <
0
\qquad (\mathrm{for} \ z \in (0, +\infty)) .
\label{eq:diff3_v}
\end{align}
It follows from \eqref{eq:diff3_v} that $\frac{ \partial^{2} v(n, z) }{ \partial z^{2} }$ is strictly decreasing for $z \in (0, +\infty)$.
Then, we can solve the root of the equation $\frac{ \partial^{2} v(n, z) }{ \partial z^{2} } = 0$ with respect to $z$ as follows:
\begin{align}
&&
\frac{ \partial^{2} v(n, z) }{ \partial z^{2} }
& =
0
\\
& \iff &
\frac{ 4 (n-1) }{ z } - 2 \ln (n-1) - 4
& =
0
\\
& \iff &
\frac{ 4 (n-1) }{ z }
& =
2 \ln (n-1) + 4
\\
& \iff &
\frac{ 2 (n-1) }{ z }
& =
\ln (n-1) + 2
\\
& \iff &
z
& =
\frac{ 2 (n-1) }{ \ln (n-1) + 2 } .
\label{eq:root_v2}
\end{align}
Note that we can see that
\begin{align}
\frac{ 2 (n-1) }{ \ln (n-1) + 2 }
& \overset{\text{(a)}}{<}
\frac{ 2 (n-1) }{ \left(1 - \frac{1}{n-1}\right) + 2 }
\\
& =
\frac{ 2 (n-1) }{ \frac{(n-1) - 1}{n-1} + 2 }
\\
& =
\frac{ 2 (n-1) }{ \frac{(n-2) + 2(n-1)}{n-1} }
\\
& =
\frac{ 2 (n-1)^{2} }{ (n-2) + 2(n-1) }
\\
& =
\frac{ 2 (n-1)^{2} }{ 3(n-1) - 1 }
\\
& =
(n-1) \underbrace{ \frac{ 2 (n-1) }{ 3(n-1) - 1 } }_{ < 1 \ \mathrm{for} \ n \ge 3 }
\\
& <
n-1
\label{ineq:root_v2}
\end{align}
for $n \ge 3$, where (a) holds for $n \ge 3$ from \eqref{eq:ITineq}.
Since
\begin{itemize}
\item
$\frac{ \partial^{2} v(n, z) }{ \partial z^{2} }$ is strictly decreasing for $(0, +\infty)$ (see Eq. \eqref{eq:diff3_v}) and
\item
$\left. \frac{ \partial^{2} v(n, z) }{ \partial z^{2} } \right|_{z = \frac{ 2(n-1) }{ \ln (n-1) + 2 }} = 0$ (see Eq. \eqref{eq:root_v2}),
\end{itemize}
we have
\begin{align}
\sgn \! \left( \frac{ \partial^{2} v(n, z) }{ \partial z^{2} } \right)
=
\begin{cases}
1
& \mathrm{if} \ z \in (0, \frac{ 2(n-1) }{ \ln (n-1) + 2 }) , \\
0
& \mathrm{if} \ z = \frac{ 2(n-1) }{ \ln (n-1) + 2 } , \\
-1
& \mathrm{if} \ z \in (\frac{ 2(n-1) }{ \ln (n-1) + 2 }, +\infty) .
\end{cases}
\label{eq:sign_v2}
\end{align}
It follows from \eqref{eq:sign_v2} that
\begin{itemize}
\item
$\frac{ \partial v(n, z) }{ \partial z }$ is strictly increasing for $z \in (0, \frac{ 2(n-1) }{ \ln (n-1) + 2 }]$ and
\item
$\frac{ \partial v(n, z) }{ \partial z }$ is strictly decreasing for $z \in [\frac{ 2(n-1) }{ \ln (n-1) + 2 }, +\infty)$.
\end{itemize}
We readily see that
\begin{align}
\left. \frac{ \partial v(n, z) }{ \partial z } \right|_{z = n-1}
& =
\left. \left( \vphantom{\sum} 4 (n-1) \ln z + 4 (n-1) - 2 (n-1) \ln (n-1) - 2 z \ln (n-1) - 4 z \right) \right|_{z = n-1}
\\
& =
4 (n-1) \ln (n-1) + 4 (n-1) - 2 (n-1) \ln (n-1) - 2 (n-1) \ln (n-1) - 4 (n-1)
\\
& =
0 .
\label{eq:root_v1_n-1}
\end{align}
We further solve another solution of the equation $\frac{ \partial v(n, z) }{ \partial z } = 0$ with respect to $z$ as follows:
\begin{align}
&&
\frac{ \partial v(n, z) }{ \partial z }
& =
0
\\
& \iff &
4 (n-1) \ln z + 4 (n-1) - 2 (n-1) \ln (n-1) - 2 z \ln (n-1) - 4 z
& =
0
\\
& \iff &
2 (n-1) \ln z + 2 (n-1) - (n-1) \ln (n-1) - z \ln (n-1) - 2 z
& =
0
\\
& \iff &
2 (n-1) \ln z + 2 (n-1) - (n-1) \ln (n-1) - z (\ln (n-1) + 2)
& =
0
\\
& \iff &
2 (n-1) \ln z + 2 (n-1) - (n-1) \ln (n-1)
& =
z (\ln (n-1) + 2)
\\
& \iff &
2 \ln z + 2 - \ln (n-1)
& =
z \left( \frac{ \ln (n-1) + 2 }{ n-1 } \right)
\\
& \iff &
\ln z + 1 - \frac{1}{2} \ln (n-1)
& =
z \left( \frac{ \ln (n-1) + 2 }{ 2(n-1) } \right)
\\
& \iff &
\mathrm{e}^{\ln z + 1 - \frac{1}{2} \ln (n-1)}
& =
\mathrm{e}^{z \left( \frac{ \ln (n-1) + 2 }{ 2(n-1) } \right)}
\\
& \iff &
\mathrm{e}^{\ln z} \cdot \mathrm{e}^{1} \cdot \mathrm{e}^{-\frac{1}{2} \ln (n-1)}
& =
\mathrm{e}^{z \left( \frac{ \ln (n-1) + 2 }{ 2(n-1) } \right)}
\\
& \iff &
z \, \mathrm{e} \, (n-1)^{-\frac{1}{2}}
& =
\mathrm{e}^{z \left( \frac{ \ln (n-1) + 2 }{ 2(n-1) } \right)}
\\
& \iff &
z \, \mathrm{e}^{-z \left( \frac{ \ln (n-1) + 2 }{ 2(n-1) } \right)} 
& =
\mathrm{e}^{-1} \, (n-1)^{\frac{1}{2}}
\\
& \iff &
- z \, \mathrm{e}^{-z \left( \frac{ \ln (n-1) + 2 }{ 2(n-1) } \right)} 
& =
- \mathrm{e}^{-1} \, (n-1)^{\frac{1}{2}}
\\
& \iff &
- z \left( \frac{ \ln (n-1) + 2 }{ 2(n-1) } \right) \mathrm{e}^{-z \left( \frac{ \ln (n-1) + 2 }{ 2(n-1) } \right)} 
& =
- \mathrm{e}^{-1} \, (n-1)^{\frac{1}{2}} \left( \frac{ \ln (n-1) + 2 }{ 2(n-1) } \right)
\\
& \iff &
- z \left( \frac{ \ln (n-1) + 2 }{ 2(n-1) } \right) \mathrm{e}^{-z \left( \frac{ \ln (n-1) + 2 }{ 2(n-1) } \right)} 
& =
- \frac{ \ln (n-1) + 2 }{ 2 \, \mathrm{e} \, \sqrt{n-1} }
\label{eq:both_side_>=-1} \\
& \overset{\text{(a)}}{\iff} &
W_{0} \! \left( - z \left( \frac{ \ln (n-1) + 2 }{ 2(n-1) } \right) \mathrm{e}^{-z \left( \frac{ \ln (n-1) + 2 }{ 2(n-1) } \right)} \right)
& =
W_{0} \! \left( - \frac{ \ln (n-1) + 2 }{ 2 \, \mathrm{e} \, \sqrt{n-1} } \right)
\\
& \overset{\text{(b)}}{\iff} &
- z \left( \frac{ \ln (n-1) + 2 }{ 2(n-1) } \right)
& =
W_{0} \! \left( - \frac{ \ln (n-1) + 2 }{ 2 \, \mathrm{e} \, \sqrt{n-1} } \right)
\\
& \iff &
z
& =
- \frac{ 2 (n-1) W_{0} \! \left( - \frac{ \ln (n-1) + 2 }{ 2 \mathrm{e} \sqrt{n-1} } \right) }{ \ln (n-1) + 2 } ,
\label{eq:root_v1}
\end{align}
where
\begin{itemize}
\item
$W_{0}( \cdot )$ denotes the Lambert $W_{0}$ function, i.e., the inverse function of $f(x) = x \, \mathrm{e}^{x}$ for $x \ge -1$ and
\item
(a) holds for $n \ge 2$ and $z \le \frac{ 2(n-1) }{ \ln (n-1) + 2 }$ since the domain of $W_{0}( \cdot )$ is the interval $[-\frac{1}{\mathrm{e}}, +\infty)$ and the both sides of \eqref{eq:both_side_>=-1} is greater than $- \frac{1}{\mathrm{e}}$ for $n \ge 2$ and $z \ge \frac{ 2(n-1) }{ \ln (n-1) + 2 }$, i.e.,
\begin{align}
(\text{the left-hand side of \eqref{eq:both_side_>=-1}})
& =
- z \left( \frac{ \ln (n-1) + 2 }{ 2(n-1) } \right) \mathrm{e}^{-z \left( \frac{ \ln (n-1) + 2 }{ 2(n-1) } \right)}
\\
& \overset{\text{(c)}}{\ge}
- \left( \frac{ 2(n-1) }{ \ln (n-1) + 2 } \right) \left( \frac{ \ln (n-1) + 2 }{ 2(n-1) } \right) \mathrm{e}^{-\left( \frac{ 2(n-1) }{ \ln (n-1) + 2 } \right) \left( \frac{ \ln (n-1) + 2 }{ 2(n-1) } \right)}
\\
& =
- \, \mathrm{e}^{-1}
\\
& =
- \frac{1}{\mathrm{e}}
\\
(\text{the right-hand side of \eqref{eq:both_side_>=-1}})
& =
- \frac{ \ln (n-1) + 2 }{ 2 \, \mathrm{e} \, \sqrt{n-1} }
\\
& \overset{\text{(d)}}{\ge}
- \frac{ \ln_{(\frac{1}{2})} (n-1) + 2 }{ 2 \, \mathrm{e} \, \sqrt{n-1} }
\\
& =
- \frac{ ( 2 \sqrt{n-1} - 2 ) +2 }{ 2 \, \mathrm{e} \, \sqrt{n-1} }
\\
& =
- \frac{ 2 \sqrt{n-1} }{ 2 \, \mathrm{e} \, \sqrt{n-1} }
\\
& =
- \frac{1}{\mathrm{e}} ,
\end{align}
where
\begin{itemize}
\item
(c) follows from the fact that $f(x) = x \, \mathrm{e}^{x}$ is strictly increasing for $x \ge -1$ and
\item
(d) follows by Lemma \ref{lem:IT_ineq},
\end{itemize}
\item
(b) holds for $z \le \frac{ 2(n-1) }{ \ln (n-1) + 2 }$ since $W_{0}( x \, \mathrm{e}^{x} ) = x$ holds for $x \ge -1$.
\end{itemize}
Since
\begin{itemize}
\item
$W_{0}( x )$ is strictly increasing for $x \ge - \frac{1}{\mathrm{e}}$,
\item
$W_{0}( - \frac{1}{\mathrm{e}} ) = -1$,
\item
$W_{0}( 0 ) = 0$
\item
$- \frac{ \ln (n-1) + 2 }{ 2 \, \mathrm{e} \, \sqrt{n-1} }$ is strictly increasing for $n \ge 2$,
\item
$\left. - \frac{ \ln (n-1) + 2 }{ 2 \, \mathrm{e} \, \sqrt{n-1} } \right|_{n = 2} = - \frac{1}{\mathrm{e}}$, and
\item
$\lim_{n \to +\infty} \left( - \frac{ \ln (n-1) + 2 }{ 2 \, \mathrm{e} \, \sqrt{n-1} } \right) = 0$
\end{itemize}
we see that
\begin{align}
-1 < W_{0} \! \left( - \frac{ \ln (n-1) + 2 }{ 2 \mathrm{e} \sqrt{n-1} } \right) < 0
\end{align}
for $n \ge 3$;
and therefore, we can see that
\begin{align}
0
<
- \frac{ 2 (n-1) W_{0} \! \left( - \frac{ \ln (n-1) + 2 }{ 2 \mathrm{e} \sqrt{n-1} } \right) }{ \ln (n-1) + 2 }
<
\frac{ 2 (n-1) }{ \ln (n-1) + 2 }
\label{ineq:root_v1}
\end{align}
for $n \ge 3$.
Since
\begin{itemize}

\item
$\frac{ \partial v(n, z) }{ \partial z }$ is strictly increasing for $z \in (0, \frac{ 2(n-1) }{ \ln (n-1) + 2 }]$ (see Eq. \eqref{eq:sign_v2}),

\item
$\frac{ \partial v(n, z) }{ \partial z }$ is strictly decreasing for $z \in [\frac{ 2(n-1) }{ \ln (n-1) + 2 }, +\infty)$ (see Eq. \eqref{eq:sign_v2}),

\item
$0 < - \frac{ 2 (n-1) W_{0} \! \left( - \frac{ \ln (n-1) + 2 }{ 2 \mathrm{e} \sqrt{n-1} } \right) }{ \ln (n-1) + 2 } < \frac{ 2 (n-1) }{ \ln (n-1) + 2 } < n-1$ for $n \ge 3$ (see Eqs. \eqref{ineq:root_v2} and \eqref{ineq:root_v1}),

\item
$\left. \frac{ \partial v(n, z) }{ \partial z } \right|_{z = - \frac{ 2 (n-1) W_{0} \! \left( - \frac{ \ln (n-1) + 2 }{ 2 \mathrm{e} \sqrt{n-1} } \right) }{ \ln (n-1) + 2 }} = 0$ (see Eq. \eqref{eq:root_v1}), and

\item
$\left. \frac{ \partial v(n, z) }{ \partial z } \right|_{z = n-1} = 0$ (see Eq. \eqref{eq:root_v1_n-1}),

\end{itemize}
we obtain
\begin{align}
\sgn \! \left( \frac{ \partial v(n, z) }{ \partial z } \right)
=
\begin{cases}
1
& \mathrm{if} \ z \in (- \frac{ 2 (n-1) W_{0} \! \left( - \frac{ \ln (n-1) + 2 }{ 2 \mathrm{e} \sqrt{n-1} } \right) }{ \ln (n-1) + 2 }, n-1) , \\
0
& \mathrm{if} \ z \in \{ - \frac{ 2 (n-1) W_{0} \! \left( - \frac{ \ln (n-1) + 2 }{ 2 \mathrm{e} \sqrt{n-1} } \right) }{ \ln (n-1) + 2 }, n-1 \} , \\
-1
& \mathrm{if} \ z \in (0, - \frac{ 2 (n-1) W_{0} \! \left( - \frac{ \ln (n-1) + 2 }{ 2 \mathrm{e} \sqrt{n-1} } \right) }{ \ln (n-1) + 2 }) \cup (n-1, +\infty) .
\end{cases}
\label{eq:diff1_v}
\end{align}
Since
\begin{itemize}

\item
$v(n, 1) < 0$ for $n \ge 3$ (see Eq. \eqref{eq:v_1}),

\item
$v(n, n-1) = 0$ for $n \ge 3$ (see Eq. \eqref{eq:v_n-1}),

\item
$0 < - \frac{ 2 (n-1) W_{0} \! \left( - \frac{ \ln (n-1) + 2 }{ 2 \mathrm{e} \sqrt{n-1} } \right) }{ \ln (n-1) + 2 } < n-1$ for $n \ge 3$ (see Eqs. \eqref{ineq:root_v2} and \eqref{ineq:root_v1}),

\item
the following monotonicity hold (see Eq. \eqref{eq:diff1_v}):
\begin{itemize}
\item
$v(n, z)$ is strictly decreasing for $z \in (0, - \frac{ 2 (n-1) W_{0} \! \left( - \frac{ \ln (n-1) + 2 }{ 2 \mathrm{e} \sqrt{n-1} } \right) }{ \ln (n-1) + 2 }]$,
\item
$v(n, z)$ is strictly increasing for $z \in [- \frac{ 2 (n-1) W_{0} \! \left( - \frac{ \ln (n-1) + 2 }{ 2 \mathrm{e} \sqrt{n-1} } \right) }{ \ln (n-1) + 2 }, n-1]$, and
\item
$v(n, z)$ is strictly decreasing for $z \in [n-1, +\infty)$,
\end{itemize}

\end{itemize}
we have
\begin{align}
\sgn \! \left( \vphantom{\sum} v(n, z) \right)
& =
\begin{cases}
0
& \mathrm{if} \ z = n-1 , \\
-1
& \mathrm{if} \ z \in [1, n-1) \cup (n-1, +\infty) .
\end{cases}
\label{ineq:v}
\end{align}
Therefore, we obtain
\begin{align}
\sgn \! \left( \left. \frac{ \partial g(n, z, \alpha) }{ \partial z } \right|_{\alpha = \frac{ \ln (n-1) }{ \ln z }} \right)
& \overset{\eqref{eq:v}}{=}
\underbrace{ \sgn \! \left( \left( \frac{ 1 }{ z ((n-1) + z)^{2} (\ln z)^{2} } \right) \right) }_{ = 1 \ \mathrm{for} \ z \in (0, 1) \cup (1, +\infty) } \, \cdot \; \sgn \! \left( \vphantom{\sum} v(n, z) \right)
\\
& \overset{\eqref{ineq:v}}{=}
\begin{cases}
0
& \mathrm{if} \ z = n-1 , \\
-1
& \mathrm{if} \ z \in (1, n-1) \cup (n-1, +\infty) .
\end{cases}
\end{align}
That concludes the proof of Lemma \ref{lem:diff_g_z}.
\end{IEEEproof}

\section{Proof of Lemma \ref{lem:ln(n-1)/2ln(z)}}
\label{app:ln(n-1)/2ln(z)}

\begin{IEEEproof}[Proof of Lemma \ref{lem:ln(n-1)/2ln(z)}]
First note that $\frac{1}{2} \alpha_{1}( n, z ) = \frac{ \ln (n-1) }{ 2 \ln z }$.
In this proof, we show the positivity of $g(n, z, {\textstyle \frac{ \ln (n-1) }{ 2 \ln z }})$ for $n \ge 3$ and $z \in [n-1, (n-1)^{2}]$.
Substituting $z = (n - 1)^{r}$ into $g(n, z, {\textstyle \frac{ \ln (n-1) }{ 2 \ln z }})$, we have
\begin{align}
g(n, z, {\textstyle \frac{ \ln (n-1) }{ 2 \ln z }}) |_{z = (n-1)^{r}}
& =
g(n, (n-1)^{r}, {\textstyle \frac{1}{2 r}})
\\
& \overset{\eqref{eq:g_z}}{=}
\left. \left( (\alpha-1) + \frac{ ((n-1) + z^{\alpha}) (z^{1-\alpha} - 1) }{ ((n-1) + z) \ln z } \right) \right|_{(z, \alpha) = ((n-1)^{r}, \frac{1}{2 r})}
\\
& =
\left( \frac{1}{2 r} - 1 \right) + \frac{ ((n-1) + \left((n-1)^{r}\right)^{\frac{1}{2 r}}) (\left((n-1)^{r}\right)^{1-\frac{1}{2 r}} - 1) }{ ((n-1) + (n-1)^{r}) \ln (n-1)^{r} }
\\
& =
\frac{1 - 2 r}{2 r} + \frac{ ((n-1) + (n-1)^{\frac{1}{2}}) ((n-1)^{\frac{2 r - 1}{2}} - 1) }{ ((n-1) + (n-1)^{r}) \left( r \ln (n-1) \right) }
\\
& =
\frac{1 - 2 r}{2 r} + \frac{ (n-1)^{\frac{2 r - 1}{2} + 1} - (n-1) + (n-1)^{\frac{2 r - 1}{2} + \frac{1}{2}} - (n-1)^{\frac{1}{2}} }{ r ((n-1) + (n-1)^{r}) \ln (n-1) }
\\
& =
\frac{1 - 2 r}{2 r} + \frac{ (n-1)^{r + \frac{1}{2}} - (n-1) + (n-1)^{r} - (n-1)^{\frac{1}{2}} }{ r ((n-1) + (n-1)^{r}) \ln (n-1) }
\\
& =
\frac{1 - 2 r}{2 r} + \frac{ \left[ (n-1)^{r + \frac{1}{2}} + (n-1)^{r} \right] - \left[ (n-1) + (n-1)^{\frac{1}{2}} \right] }{ r ((n-1) + (n-1)^{r}) \ln (n-1) }
\\
& =
\frac{1 - 2 r}{2 r} + \frac{ (n-1)^{r} \left[ (n-1)^{\frac{1}{2}} + 1 \right] - (n-1)^{\frac{1}{2}} \left[ (n-1)^{\frac{1}{2}} + 1 \right] }{ r ((n-1) + (n-1)^{r}) \ln (n-1) }
\\
& =
\frac{1 - 2 r}{2 r} + \frac{ \left( (n-1)^{r} - (n-1)^{\frac{1}{2}} \right) \left( (n-1)^{\frac{1}{2}} + 1 \right) }{ r ((n-1) + (n-1)^{r}) \ln (n-1) }
\\
& =
\frac{1}{2 r} \left( (1 - 2 r) + \frac{ 2 \left( (n-1)^{r} - (n-1)^{\frac{1}{2}} \right) \left( (n-1)^{\frac{1}{2}} + 1 \right) }{ ((n-1) + (n-1)^{r}) \ln (n-1) } \right)
\\
& =
\frac{ (1 - 2 r) ((n-1) + (n-1)^{r}) \ln (n-1) + 2 \left( (n-1)^{r} - (n-1)^{\frac{1}{2}} \right) \left( (n-1)^{\frac{1}{2}} + 1 \right) }{ 2 r ((n-1) + (n-1)^{r}) \ln (n-1) }
\\
& =
\frac{ d(n, r) }{ 2 r ((n-1) + (n-1)^{r}) \ln (n-1) } ,
\label{eq:g_ln(n-1)/2ln(z)_(n-1)^r}
\end{align}
where
\begin{align}
d(n, r)
\triangleq
(1 - 2 r) ((n-1) + (n-1)^{r}) \ln (n-1) + 2 ( (n-1)^{r} - (n-1)^{\frac{1}{2}} ) ( (n-1)^{\frac{1}{2}} + 1 ) .
\label{def:d_n_r}
\end{align}
Since
\begin{align}
\frac{ 1 }{ 2 r ((n-1) + (n-1)^{r}) \ln (n-1) }
>
0
\end{align}
for $n \ge 3$ and $r > 0$, it is enough to check the positivity of $d(n, r)$ for $n \ge 3$ and $1 \le r \le 2$ rather than the right-hand side of \eqref{eq:g_ln(n-1)/2ln(z)_(n-1)^r}.
The derivatives of $d(n, r)$ with respect to $r$ are as follows:
\begin{align}
\frac{ \partial d(n, r) }{ \partial r }
& =
\frac{ \partial }{ \partial r } \left( \vphantom{\sum} (1 - 2 r) ((n-1) + (n-1)^{r}) \ln (n-1) + 2 ( (n-1)^{r} - (n-1)^{\frac{1}{2}} ) ( (n-1)^{\frac{1}{2}} + 1 ) \right)
\\
& =
\frac{ \partial }{ \partial r } \left( \vphantom{\sum} (1 - 2 r) (n-1) \ln (n-1) + (1 - 2 r) (n-1)^{r} \ln (n-1)
\right. \notag \\
& \left. \qquad \qquad \qquad \qquad \qquad \qquad \qquad \qquad
+ \; 2 (n-1)^{r} ((n-1)^{\frac{1}{2}} + 1) - (n-1)^{\frac{1}{2}} ((n-1)^{\frac{1}{2}} + 1) \vphantom{\sum} \right)
\\
& =
\frac{ \partial }{ \partial r } \left( \vphantom{\sum} (1 - 2 r) (n-1) \ln (n-1) + (1 - 2 r) (n-1)^{r} \ln (n-1) + 2 (n-1)^{r} ((n-1)^{\frac{1}{2}} + 1) \right)
\\
& =
\left( \frac{ \mathrm{d} }{ \mathrm{d} r } (1 - 2 r) \right) (n-1) \ln (n-1) + \left( \frac{ \partial }{ \partial r } ((1 - 2 r) (n-1)^{r}) \right) \ln (n-1)
\notag \\
& \qquad \qquad \qquad \qquad \qquad \qquad \qquad \qquad \qquad \qquad \qquad \quad
+ 2 \left( \frac{ \partial }{ \partial r } ((n-1)^{r}) \right) ((n-1)^{\frac{1}{2}} + 1)
\\
& =
( -2 ) (n-1) \ln (n-1) + \left( \left[ \frac{ \mathrm{d} }{ \mathrm{d} r } (1 - 2 r) \right] (n-1)^{r} + (1 - 2 r) \left[ \frac{ \partial }{ \partial r } ((n-1)^{r}) \right]  \right) \ln (n-1)
\notag \\
& \qquad \qquad \qquad \qquad \qquad \qquad \qquad \qquad \qquad \qquad \qquad \quad
+ 2 ( (n-1)^{r} \ln (n-1) ) ((n-1)^{\frac{1}{2}} + 1)
\\
& =
- 2 (n-1) \ln (n-1) + ( [ -2 ] (n-1)^{r} + (1 - 2 r) [ (n-1)^{r} \ln (n-1) ] ) \ln (n-1)
\notag \\
& \qquad \qquad \qquad \qquad \qquad \qquad \qquad \qquad \qquad \quad
+ 2 (n-1)^{r + \frac{1}{2}} \ln (n-1) + 2 (n-1)^{r} \ln (n-1)
\\
& =
- 2 (n-1) \ln (n-1) - 2 (n-1)^{r} \ln (n-1) + (1 - 2 r) (n-1)^{r} (\ln (n-1))^{2}
\notag \\
& \qquad \qquad \qquad \qquad \qquad \qquad \qquad \qquad \qquad \quad
+ 2 (n-1)^{r + \frac{1}{2}} \ln (n-1) + 2 (n-1)^{r} \ln (n-1)
\\
& =
(\ln (n-1)) \left( \vphantom{\sum} - 2 (n-1) + (1 - 2 r) (n-1)^{r} \ln (n-1) + 2 (n-1)^{r + \frac{1}{2}} \right) ,
\label{eq:diff1_d_r} \\
\frac{ \partial^{2} d(n, r) }{ \partial r^{2} }
& =
\frac{ \partial }{ \partial r } \left( (\ln (n-1)) \left( \vphantom{\sum} - 2 (n-1) + (1 - 2 r) (n-1)^{r} \ln (n-1) + 2 (n-1)^{r + \frac{1}{2}} \right) \right)
\\
& =
(\ln (n-1)) \left( \frac{ \partial }{ \partial r } \left( \vphantom{\sum} - 2 (n-1) + (1 - 2 r) (n-1)^{r} \ln (n-1) + 2 (n-1)^{r + \frac{1}{2}} \right) \right)
\\
& =
(\ln (n-1)) \left( \frac{ \partial }{ \partial r } \left( \vphantom{\sum} (1 - 2 r) (n-1)^{r} \ln (n-1) + 2 (n-1)^{r + \frac{1}{2}} \right) \right)
\\
& =
(\ln (n-1)) \left( \left( \frac{ \partial }{ \partial r } ((1 - 2 r) (n-1)^{r}) \right) \ln (n-1) + 2 \left( \frac{ \partial }{ \partial r } ((n-1)^{r + \frac{1}{2}}) \right) \right)
\\
& =
(\ln (n-1)) \left( \left( \left( \frac{ \mathrm{d} }{ \mathrm{d} r } (1 - 2 r) \right) (n-1)^{r} + (1 - 2 r) \left( \frac{ \mathrm{d} }{ \mathrm{d} r } (n-1)^{r} \right) \right) \ln (n-1)
\right. \notag \\
& \left. \qquad \qquad \qquad \qquad \qquad \qquad \qquad \qquad \qquad \qquad \qquad \qquad \qquad
+ 2 \left( \frac{ \partial }{ \partial r } ((n-1)^{r + \frac{1}{2}}) \right) \right)
\\
& =
(\ln (n-1)) \left( \left( \vphantom{\sum} ( -2 ) (n-1)^{r} + (1 - 2 r) (n-1)^{r} \ln (n-1) \right) \ln (n-1)
\right. \notag \\
& \left. \qquad \qquad \qquad \qquad \qquad \qquad \qquad \qquad \qquad \qquad \qquad \qquad \qquad
+ 2 (n-1)^{r + \frac{1}{2}} \ln(n-1) \right)
\\
& =
(\ln (n-1))^{2} \left( \vphantom{\sum} - 2 (n-1)^{r} + (1 - 2 r) (n-1)^{r} \ln (n-1) + 2 (n-1)^{r + \frac{1}{2}} \right)
\\
& =
\underbrace{ (n-1)^{r} (\ln (n-1))^{2} }_{ > 0 } \left( \vphantom{\sum} - 2 + (1 - 2 r) \ln (n-1) + 2 (n-1)^{\frac{1}{2}} \right)
\\
& \overset{\text{(a)}}{=}
\begin{cases}
< 0
& \mathrm{if} \ r > \frac{1}{2} + \frac{ \sqrt{n-1} - 1 }{ \ln (n-1) } \ \mathrm{and} \ n \ge 3 , \\
= 0
& \mathrm{if} \ r = \frac{1}{2} + \frac{ \sqrt{n-1} - 1 }{ \ln (n-1) } \ \mathrm{or} \ n = 2 , \\
> 0
& \mathrm{if} \ r < \frac{1}{2} + \frac{ \sqrt{n-1} - 1 }{ \ln (n-1) } \ \mathrm{and} \ n \ge 3 ,
\end{cases}
\end{align}
where (a) follows from the fact that
\begin{align}
\sgn \! \left( \vphantom{\sum} - 2 + (1 - 2 r) \ln (n-1) + 2 (n-1)^{\frac{1}{2}} \right)
=
\begin{cases}
1
& \mathrm{if} \ r < \frac{1}{2} + \frac{ \sqrt{n-1} - 1 }{ \ln (n-1) } , \\
0
& \mathrm{if} \ r = \frac{1}{2} + \frac{ \sqrt{n-1} - 1 }{ \ln (n-1) } , \\
-1
& \mathrm{if} \ r > \frac{1}{2} + \frac{ \sqrt{n-1} - 1 }{ \ln (n-1) } .
\end{cases}
\label{eq:diff2_d_partial}
\end{align}
We can verify \eqref{eq:diff2_d_partial} as follows:
The derivative of the left-hand side of \eqref{eq:diff2_d_partial} with respect to $r$ is
\begin{align}
\frac{ \partial }{ \partial r } \left( \vphantom{\sum} - 2 + (1 - 2 r) \ln (n-1) + 2 (n-1)^{\frac{1}{2}} \right)
& =
\frac{ \partial }{ \partial r } \left( \vphantom{\sum} - 2 r \ln (n-1) \right)
\\
& =
- 2 \ln (n-1) \left( \frac{ \mathrm{d} }{ \mathrm{d} r } (r) \right)
\\
& =
- 2 \ln (n-1)
\\
& \le
0
\label{eq:diff2_d_partial_diff}
\end{align}
for $n \ge 2$ with equality if and only if $n = 2$.
Further, the left-hand side of \eqref{eq:diff2_d_partial} has can be transformed as follows:
\begin{align}
&&
- 2 + (1 - 2 r) \ln (n-1) + 2 (n-1)^{\frac{1}{2}}
& =
0
\\
& \iff &
(1 - 2 r) \ln (n-1)
& =
2 (1 - (n-1)^{\frac{1}{2}})
\\
& \iff &
1 - 2 r
& =
\frac{ 2 (1 - (n-1)^{\frac{1}{2}}) }{ \ln (n-1) }
\\
& \iff &
r
& =
\frac{1}{2} - \frac{ 1 - \sqrt{n-1} }{ \ln (n-1) }
\\
& \iff &
r
& =
\frac{1}{2} + \frac{ \sqrt{n-1} - 1 }{ \ln (n-1) } ,
\label{eq:diff2_d_partial_root}
\end{align}
Thus, it follows from \eqref{eq:diff2_d_partial_diff} and \eqref{eq:diff2_d_partial_root} that \eqref{eq:diff2_d_partial} holds for $n \ge 3$.
Hence, we have from \eqref{eq:diff2_d_partial} that, if $n \ge 3$, then
\begin{itemize}
\item
$\frac{ \partial d(n, r) }{ \partial r }$ is strictly increasing for $r \in (-\infty, \frac{1}{2} + \frac{ \sqrt{n-1} - 1 }{ \ln (n-1) }]$ and
\item
$\frac{ \partial d(n, r) }{ \partial r }$ is strictly decreasing for $r \in [\frac{1}{2} + \frac{ \sqrt{n-1} - 1 }{ \ln (n-1) }, +\infty)$.
\end{itemize}
Using this monotonicity, we now check the sign of $\frac{ \partial d(n, r) }{ \partial r }$.
Substituting $r = \frac{1}{2}$ into $\frac{ \partial d(n, r) }{ \partial r }$, we see that
\begin{align}
\left. \frac{ \partial d(n, r) }{ \partial r } \right|_{r = \frac{1}{2}}
& \overset{\eqref{eq:diff1_d_r}}{=}
\left. (\ln (n-1)) \left( \vphantom{\sum} - 2 (n-1) + (1 - 2 r) (n-1)^{r} \ln (n-1) + 2 (n-1)^{r + \frac{1}{2}} \right) \right|_{r = \frac{1}{2}}
\\
& =
(\ln (n-1)) \left( \vphantom{\sum} - 2 (n-1) + (1 - 1) (n-1)^{1} \ln (n-1) + 2 (n-1)^{1} \right)
\\
& =
0 .
\label{eq:diff1_d_root1}
\end{align}
Moreover, we can transform the equation $\frac{ \partial d(n, r) }{ \partial r } = 0$ as follows:
\begin{align}
&&
\frac{ \partial d(n, r) }{ \partial r }
& =
0
\\
& \iff &
- 2 (n-1) + (1 - 2 r) (n-1)^{r} \ln (n-1) + 2 (n-1)^{r + \frac{1}{2}}
& =
0
\\
& \iff &
(1 - 2 r) (n-1)^{r} \ln (n-1) + 2 (n-1)^{r + \frac{1}{2}}
& =
2 (n-1)
\\
& \iff &
(n-1)^{r} ((1 - 2 r) \ln (n-1) + 2 (n-1)^{\frac{1}{2}})
& =
2 (n-1)
\\
& \iff &
(1 - 2 r) \ln (n-1) + 2 (n-1)^{\frac{1}{2}}
& =
2 (n-1)^{1-r}
\\
& \iff &
(1 - 2 r) \ln (n-1) + 2 (n-1)^{\frac{1}{2}}
& =
2 \, \mathrm{e}^{\ln(n-1)^{1-r}}
\\
& \iff &
(1 - 2 r) \ln (n-1) + 2 (n-1)^{\frac{1}{2}}
& =
2 \, \mathrm{e}^{(1- r) \ln(n-1)}
\\
& \iff &
- 2 r \ln (n-1) + \ln(n-1) + 2 (n-1)^{\frac{1}{2}}
& =
2 \, \mathrm{e}^{(1- r) \ln(n-1)}
\\
& \iff &
r \ln (n-1) - \frac{1}{2} \ln(n-1) - (n-1)^{\frac{1}{2}}
& =
- \mathrm{e}^{(1- r) \ln(n-1)}
\\
& \iff &
\frac{(2 r - 1) \ln (n-1)}{ 2 } - \sqrt{n-1}
& =
- \mathrm{e}^{(1- r) \ln(n-1)}
\\
& \iff &
\left( \frac{(2 r - 1) \ln (n-1)}{ 2 } - \sqrt{n-1} \right) \mathrm{e}^{\frac{(2 r - 1) \ln (n-1)}{ 2 } - \sqrt{n-1}}
& =
- \mathrm{e}^{(1- r) \ln(n-1)} \, \mathrm{e}^{\frac{(2 r - 1) \ln (n-1)}{ 2 } - \sqrt{n-1}}
\\
& \iff &
\left( \frac{(2 r - 1) \ln (n-1)}{ 2 } - \sqrt{n-1} \right) \mathrm{e}^{\frac{(2 r - 1) \ln (n-1)}{ 2 } - \sqrt{n-1}}
& =
- \mathrm{e}^{\ln \sqrt{n-1} - \sqrt{n-1}}
\label{eq:both_side_>=-1_2} \\
& \overset{\text{(a)}}{\iff} &
W_{0} \! \left( \! \left( \frac{(2 r - 1) \ln (n-1)}{ 2 } - \sqrt{n-1} \right) \mathrm{e}^{\frac{(2 r - 1) \ln (n-1)}{ 2 } - \sqrt{n-1}} \right)
& =
W_{0} \! \left( - \mathrm{e}^{\ln \sqrt{n-1} - \sqrt{n-1}} \right)
\\
& \overset{\text{(b)}}{\iff} &
\frac{(2 r - 1) \ln (n-1)}{ 2 } - \sqrt{n-1}
& =
W_{0} \! \left( - \mathrm{e}^{\ln \sqrt{n-1} - \sqrt{n-1}} \right)
\\
& \iff &
r \ln (n-1)
& =
W_{0} \! \left( - \mathrm{e}^{\ln \sqrt{n-1} - \sqrt{n-1}} \right)
\notag \\
&&& \qquad \qquad
+ \frac{1}{2} \ln (n-1) + \sqrt{n-1}
\\
& \iff &
r
& =
\frac{1}{2} + \frac{ W_{0} \! \left( - \mathrm{e}^{\ln \sqrt{n-1} - \sqrt{n-1}} \right) + \sqrt{n-1} }{ \ln (n-1) },
\label{eq:diff1_d_root2}
\end{align}
where
\begin{itemize}
\item
(a) holds for $n \ge 2$ and $r \ge \frac{1}{2} + \frac{ \sqrt{n-1} - 1 }{ \ln (n-1) }$ since the domain of $W_{0}( \cdot )$ is the interval $[-\frac{1}{\mathrm{e}}, +\infty)$ and the both sides of \eqref{eq:both_side_>=-1_2} is greater than $- \frac{1}{\mathrm{e}}$ for $n \ge 2$ and $r \ge \frac{1}{2} + \frac{ \sqrt{n-1} - 1 }{ \ln (n-1) }$, i.e.,
\begin{align}
(\text{the left-hand side of \eqref{eq:both_side_>=-1_2}})
& =
\left( \frac{(2 r - 1) \ln (n-1)}{ 2 } - \sqrt{n-1} \right)
\notag \\
& \qquad \qquad \qquad \qquad \qquad \times
\exp \left( \frac{(2 r - 1) \ln (n-1)}{ 2 } - \sqrt{n-1} \right)
\\
& =
\left( r \ln (n-1) - \frac{1}{2} \ln (n-1) - \sqrt{n-1} \right)
\notag \\
& \qquad \qquad \qquad \quad \times
\exp \left( r \ln (n-1) - \frac{1}{2} \ln (n-1) - \sqrt{n-1} \right)
\\
& \overset{\text{(d)}}{\ge}
\left( \left( \frac{1}{2} + \frac{ \sqrt{n-1} - 1 }{ \ln (n-1) } \right) \ln (n-1) - \frac{1}{2} \ln (n-1) - \sqrt{n-1} \right)
\notag \\
& \qquad \times
\exp \left( \left( \frac{1}{2} + \frac{ \sqrt{n-1} - 1 }{ \ln (n-1) } \right) \ln (n-1) - \frac{1}{2} \ln (n-1) - \sqrt{n-1} \right)
\\
& =
\left( \frac{1}{2} \ln (n-1) + \sqrt{n-1} - 1 - \frac{1}{2} \ln (n-1) - \sqrt{n-1} \right)
\notag \\
& \qquad \quad \times
\exp \left( \frac{1}{2} \ln (n-1) + \sqrt{n-1} - 1 - \frac{1}{2} \ln (n-1) - \sqrt{n-1} \right)
\\
& =
(-1) \exp (-1)
\\
& =
- \frac{1}{\mathrm{e}} ,
\\
(\text{the right-hand side of \eqref{eq:both_side_>=-1_2}})
& =
- \exp \left( \frac{\ln (n-1)}{2} - \sqrt{n-1} \right)
\\
& \overset{\text{(e)}}{\ge}
- \exp \left( \frac{\ln_{(\frac{1}{2})} (n-1)}{2} - \sqrt{n-1} \right)
\\
& =
- \exp \left( (\sqrt{n-1} - 1) - \sqrt{n-1} \right)
\\
& =
- \frac{1}{\mathrm{e}} ,
\end{align}
where
\begin{itemize}
\item
(d) follows from the facts that $f(x) = x \, \mathrm{e}^{x}$ is strictly increasing for $x \ge -1$ and $r \ge \frac{1}{2} + \frac{ \sqrt{n-1} - 1 }{ \ln (n-1) }$, and
\item
(e) follows from the fact that, for a fixed $x \in (0, 1) \cup (1, +\infty)$, $f_{x}( \alpha ) = \ln_{\alpha} x$ is strictly decreasing for $\alpha \in (-\infty, +\infty)$ (see Lemma \ref{lem:IT_ineq}),
\end{itemize}
\item
(b) holds for $r \ge \frac{1}{2} + \frac{ \sqrt{n-1} - 1 }{ \ln (n-1) }$ since $W_{0}( x \, \mathrm{e}^{x} ) = x$ holds for $x \ge -1$.
\end{itemize}
Note that $\frac{1}{2} + \frac{ \sqrt{n-1} - 1 }{ \ln (n-1) } \le \frac{1}{2} + \frac{ W_{0} \! \left( - \mathrm{e}^{\ln \sqrt{n-1} - \sqrt{n-1}} \right) + \sqrt{n-1} }{ \ln (n-1) }$ for $n \ge 2$ with equality iff $n = 2$.
Since
\begin{itemize}
\item
if $n \ge 3$, then the following monotonicity hold (see Eq. \eqref{eq:diff2_d_partial}):
\begin{itemize}
\item
$\frac{ \partial d(n, r) }{ \partial r }$ is strictly increasing for $r \in (-\infty, \frac{1}{2} + \frac{ \sqrt{n-1} - 1 }{ \ln (n-1) }]$ and
\item
$\frac{ \partial d(n, r) }{ \partial r }$ is strictly decreasing for $r \in [\frac{1}{2} + \frac{ \sqrt{n-1} - 1 }{ \ln (n-1) }, +\infty)$,
\end{itemize}
and
\item
roots of the equation $\frac{ \partial d(n, r) }{ \partial r } = 0$ is at $r \in \{ \frac{1}{2}, \frac{1}{2} + \frac{ W_{0} \! \left( - \mathrm{e}^{\ln \sqrt{n-1} - \sqrt{n-1}} \right) + \sqrt{n-1} }{ \ln (n-1) } \}$ (see Eqs. \eqref{eq:diff1_d_root1} and \eqref{eq:diff1_d_root2}),
\end{itemize}
we obtain
\begin{align}
\sgn \! \left( \frac{ \partial d(n, r) }{ \partial r } \right)
& =
\begin{cases}
1
& \mathrm{if} \ r \in (\frac{1}{2}, \frac{1}{2} + \frac{ W_{0} \! \left( - \mathrm{e}^{\ln \sqrt{n-1} - \sqrt{n-1}} \right) + \sqrt{n-1} }{ \ln (n-1) }) \ \mathrm{and} \ n \ge 3 , \\
0
& \mathrm{if} \ r \in \{ \frac{1}{2}, \frac{1}{2} + \frac{ W_{0} \! \left( - \mathrm{e}^{\ln \sqrt{n-1} - \sqrt{n-1}} \right) + \sqrt{n-1} }{ \ln (n-1) } \} \ \mathrm{or} \ n = 2 , \\
-1
& \mathrm{if} \ r \in (-\infty, \frac{1}{2}) \cup (\frac{1}{2} + \frac{ W_{0} \! \left( - \mathrm{e}^{\ln \sqrt{n-1} - \sqrt{n-1}} \right) + \sqrt{n-1} }{ \ln (n-1) }, +\infty) \ \mathrm{and} \ n \ge 3 ,
\end{cases}
\label{eq:diff1_d_sign}
\end{align}
which implies that, if $n \ge 3$, then
\begin{itemize}
\item
$d(n, r)$ is strictly decreasing for $r \in (-\infty, \frac{1}{2}]$,
\item
$d(n, r)$ is strictly increasing for $r \in [\frac{1}{2}, \frac{1}{2} + \frac{ W_{0} \! \left( - \mathrm{e}^{\ln \sqrt{n-1} - \sqrt{n-1}} \right) + \sqrt{n-1} }{ \ln (n-1) }]$, and
\item
$d(n, r)$ is strictly decreasing for $r \in [\frac{1}{2} + \frac{ W_{0} \! \left( - \mathrm{e}^{\ln \sqrt{n-1} - \sqrt{n-1}} \right) + \sqrt{n-1} }{ \ln (n-1) }, +\infty)$.
\end{itemize}
Using this monotonicity and verifying the positivities of $d(n, 1)$ and $d(n, 2)$ for $n \ge 3$, we prove this lemma.
In order to do so, we first show the positivity of $d(n, 1)$ for $n \ge 3$.
Substituting $r = 1$ into $d(n, r)$, we see
\begin{align}
d(n, 1)
& \overset{\eqref{def:d_n_r}}{=}
\left. \left( \vphantom{\sum} (1 - 2 r) ((n-1) + (n-1)^{r}) \ln (n-1) + 2 ( (n-1)^{r} - (n-1)^{\frac{1}{2}} ) ( (n-1)^{\frac{1}{2}} + 1 ) \right) \right|_{r = 1}
\\
& =
(1 - 2) ((n-1) + (n-1)) \ln (n-1) + 2 ( (n-1) - (n-1)^{\frac{1}{2}} ) ( (n-1)^{\frac{1}{2}} + 1 )
\\
& =
(- 1) (2(n-1)) \ln (n-1) + 2 \sqrt{n-1} ( \sqrt{n-1} - 1 ) ( \sqrt{n-1} + 1 )
\\
& =
- 2 (n-1) \ln (n-1) + 2 \sqrt{n-1} ((n-1) - 1)
\\
& =
- 2 (n-1) \ln (n-1) + 2 \sqrt{n-1} (n-2)
\\
& =
2 \sqrt{n-1} \left( \vphantom{\sum} - \sqrt{n-1} \ln (n-1) + (n-2) \right)
\\
& =
2 \sqrt{n-1} \, d_{1}( n ) ,
\label{eq:d(n,1)}
\end{align}
where
\begin{align}
d_{1}( n )
\triangleq
- \sqrt{n-1} \ln (n-1) + (n-2) .
\end{align}
Since $2 \sqrt{n-1} > 0$ for $n > 1$, it is enough to check the positivity of $d_{1}( n )$ for $n \ge 3$ rather than \eqref{eq:d(n,1)}.
Hence, we analyze $d_{1}( n )$ for $n \ge 3$ as follows:
\begin{align}
d_{1}( 2 )
& =
\left. \left( \vphantom{\sum} - \sqrt{n-1} \ln (n-1) + (n-2) \right) \right|_{n = 2}
\\
& =
- \sqrt{1} \ln (1) + (2-2)
\\
& =
0 ,
\label{eq:d1_n2}
\\
\frac{ \mathrm{d} d_{1}(n) }{ \mathrm{d} n }
& =
\frac{ \mathrm{d} }{ \mathrm{d} n } \left( \vphantom{\sum} - \sqrt{n-1} \ln (n-1) + (n-2) \right)
\\
& =
- \left( \frac{ \mathrm{d} }{ \mathrm{d} n } (\sqrt{n-1} \ln (n-1)) \right) + \left( \frac{ \mathrm{d} }{ \mathrm{d} n } (n-2) \right)
\\
& =
- \left( \frac{ \mathrm{d} }{ \mathrm{d} n } (\sqrt{n-1}) \right) \ln (n-1) - \sqrt{n-1} \left( \frac{ \mathrm{d} }{ \mathrm{d} n } (\ln (n-1)) \right) + 1
\\
& =
- \left( \frac{ 1 }{ 2 \sqrt{n-1} } \right) \ln (n-1) - \sqrt{n-1} \left( \frac{ 1 }{ n-1 } \right) + 1
\\
& =
- \frac{ \ln (n-1) }{ 2 \sqrt{n-1} } - \frac{ 1 }{ \sqrt{n-1} } + 1
\\
& =
\frac{ - \ln (n-1) - 2 + 2 \sqrt{n-1} }{ 2 \sqrt{n-1} }
\\
& =
\frac{ - \ln (n-1) + 2 (\sqrt{n-1} - 1) }{ 2 \sqrt{n-1} }
\\
& =
\frac{ - \ln (n-1) + \ln_{(\frac{1}{2})} (n-1) }{ 2 \sqrt{n-1} }
\\
& \overset{\text{(a)}}{\ge}
\frac{ - \ln (n-1) + \ln (n-1) }{ 2 \sqrt{n-1} }
\\
& =
0 ,
\label{diff_d1}
\end{align}
where (a) holds with equality if and only if $n = 2$ since $\ln_{\alpha} x \ge \ln_{\beta} x$ for $\alpha < \beta$ and $x \in (0, +\infty)$ with equality if and only if $x = 1$.
Since
\begin{itemize}
\item
$d_{1}( 2 ) = 0$ (see Eq. \eqref{eq:d1_n2}) and
\item
$d_{1}( n )$ is strictly increasing for $n \ge 2$ (see Eq. \eqref{diff_d1}),
\end{itemize}
we have $d_{1}( n ) > 0$ for $n \ge 3$; and therefore, we obtain $d(n, 1) > 0$ for $n \ge 3$ from \eqref{eq:d(n,1)}.
Moreover, we second show the positivity of $d(n, 2)$ for $n \ge 3$.
Substituting $r = 2$ into $d(n, r)$ we see
\begin{align}
d(n, 2)
& \overset{\eqref{def:d_n_r}}{=}
\left. \left( \vphantom{\sum} (1 - 2 r) ((n-1) + (n-1)^{r}) \ln (n-1) + 2 ( (n-1)^{r} - (n-1)^{\frac{1}{2}} ) ( (n-1)^{\frac{1}{2}} + 1 ) \right) \right|_{r = 2}
\\
& =
(1 - 4) ((n-1) + (n-1)^{2}) \ln (n-1) + 2 ( (n-1)^{2} - (n-1)^{\frac{1}{2}} ) ( (n-1)^{\frac{1}{2}} + 1 )
\\
& =
- 3 (n-1) (1 + (n-1)) \ln (n-1) + 2 ( (n-1)^{\frac{5}{2}} + (n-1)^{2} - (n-1) - (n-1)^{\frac{1}{2}})
\\
& =
- 3 n (n-1) \ln (n-1) + 2 ( (n-1)^{\frac{1}{2}} ((n-1)^{2} - 1) + (n-1)((n-1) - 1))
\\
& =
- 3 n (n-1) \ln (n-1) + 2 ( (n-1)^{\frac{1}{2}} ((n^{2} - 2n + 1) - 1) + (n-1)(n-2))
\\
& =
- 3 n (n-1) \ln (n-1) + 2 ( (n-1)^{\frac{1}{2}} (n (n - 2)) + (n-1)(n-2))
\\
& =
- 3 n (n-1) \ln (n-1) + 2 (n-2) (n (n-1)^{\frac{1}{2}} + (n-1))
\\
& =
- 3 n (n-1) \ln (n-1) + 2 n (n-2) \sqrt{n-1} + 2 (n-1) (n-2) .
\label{eq:d(n,2)}
\end{align}
Note that
\begin{align}
d(2, 2)
& \overset{\eqref{eq:d(n,2)}}{=}
\left. \left( \vphantom{\sum} - 3 n (n-1) \ln (n-1) + 2 n (n-2) \sqrt{n-1} + 2 (n-1) (n-2) \right) \right|_{n = 2}
\\
& =
- 3 \cdot 2 (2-1) \ln 1 + 2 \cdot 2 (2-2) \sqrt{1} + 2 (2-1) (2-2)
\\
& =
0 .
\label{eq:d(2,2)}
\end{align}
Then, the derivatives of $d(n, 2)$ with respect to $n$ are as follows:
\begin{align}
\frac{ \mathrm{d} d(n, 2) }{ \mathrm{d} n }
& \overset{\eqref{eq:d(n,2)}}{=}
\frac{ \mathrm{d} }{ \mathrm{d} n } \left( \vphantom{\sum} - 3 n (n-1) \ln (n-1) + 2 n (n-2) \sqrt{n-1} + 2 (n-1) (n-2) \right)
\\
& =
- 3 \left( \frac{ \mathrm{d} }{ \mathrm{d} n } (n (n-1) \ln (n-1)) \right) + 2 \left( \frac{ \mathrm{d} }{ \mathrm{d} n } (n (n-2) \sqrt{n-1}) \right) + 2 \left( \frac{ \mathrm{d} }{ \mathrm{d} n } ((n-1) (n-2)) \right)
\\
& =
- 3 \left( \frac{ \mathrm{d} }{ \mathrm{d} n } (n^{2} \ln (n-1) - n \ln (n-1)) \right) + 2 \left( \frac{ \mathrm{d} }{ \mathrm{d} n } (n^{2} \sqrt{n-1} - 2 n \sqrt{n-1}) \right)
\notag \\
& \qquad \qquad \qquad \qquad \qquad \qquad \qquad \qquad \qquad \qquad \qquad \qquad \qquad \qquad
+ 2 \left( \frac{ \mathrm{d} }{ \mathrm{d} n } (n^{2} - 3 n + 2) \right)
\\
& =
- 3 \left( \frac{ \mathrm{d} }{ \mathrm{d} n } (n^{2} \ln (n-1)) \right) + 3 \left( \frac{ \mathrm{d} }{ \mathrm{d} n } (n \ln (n-1)) \right) + 2 \left( \frac{ \mathrm{d} }{ \mathrm{d} n } (n^{2} \sqrt{n-1}) \right)
\notag \\
& \qquad \qquad \qquad \qquad \qquad \qquad \qquad \qquad \qquad \qquad
- 4 \left( \frac{ \mathrm{d} }{ \mathrm{d} n } (n \sqrt{n-1}) \right) + 2 \left( \frac{ \mathrm{d} }{ \mathrm{d} n } (n^{2} - 3 n) \right)
\\
& =
- 3 \left( 2 n \ln (n-1) + \frac{n^{2}}{n-1} \right) + 3 \left( \ln (n-1) + \frac{ n }{ n-1 } \right) + 2 \left(2 n \sqrt{n-1} + \frac{ n^{2} }{ 2 \sqrt{n-1} } \right)
\notag \\
& \qquad \qquad \qquad \qquad \qquad \qquad \qquad \qquad \qquad \qquad \quad
- 4 \left( \sqrt{n-1} + \frac{ n }{ 2 \sqrt{n-1} } \right) + 2 (2 n - 3)
\\
& =
- 6 n \ln (n-1) - \frac{3 n^{2}}{n-1} + 3 \ln (n-1) + \frac{ 3 n }{ n-1 } + 4 n \sqrt{n-1} + \frac{ 2 n^{2} }{ 2 \sqrt{n-1} }
\notag \\
& \qquad \qquad \qquad \qquad \qquad \qquad \qquad \qquad \qquad \qquad \qquad \qquad
- 4 \sqrt{n-1} - \frac{ 4 n }{ 2 \sqrt{n-1} } + 4 n - 6
\\
& =
3 (1 - 2 n) \ln (n-1) - \frac{3 n^{2} - 3 n }{n-1} + 4 (n-1) \sqrt{n-1} + \frac{ n^{2} - 2 n }{ \sqrt{n-1} }  + 4 n - 6
\\
& =
3 (1 - 2 n) \ln (n-1) - \frac{3 n (n-1) }{ (n-1) } + 4 (n-1) \sqrt{n-1} + \frac{ n (n-2) }{ \sqrt{n-1} } + 4 n - 6
\\
& =
3 (1 - 2 n) \ln (n-1) - 3 n + 4 (n-1) \sqrt{n-1} + \frac{ n (n-2) }{ \sqrt{n-1} } + 4 n - 6
\\
& =
3 (1 - 2 n) \ln (n-1) + 4 (n-1) \sqrt{n-1} + \frac{ n (n-2) }{ \sqrt{n-1} } + n - 6
\label{eq:diff1_d(n,2)} \\
\frac{ \mathrm{d}^{2} d(n, 2) }{ \mathrm{d} n^{2} }
& =
\frac{ \mathrm{d} }{ \mathrm{d} n } \left( 3 (1 - 2 n) \ln (n-1) + 4 (n-1) \sqrt{n-1} + \frac{ n (n-2) }{ \sqrt{n-1} } + n - 6 \right)
\\
& =
3 \left( \frac{ \mathrm{d} }{ \mathrm{d} n } ((1 - 2 n) \ln (n-1)) \right) + 4 \left( \frac{ \mathrm{d} }{ \mathrm{d} n } ((n-1) \sqrt{n-1}) \right)
\notag \\
& \qquad \qquad \qquad \qquad \qquad \qquad \qquad \qquad \qquad \qquad
+ \left( \frac{ \mathrm{d} }{ \mathrm{d} n } \left( \frac{ n (n-2) }{ \sqrt{n-1} } \right) \right) + \left( \frac{ \mathrm{d} }{ \mathrm{d} n } (n - 6) \right)
\\
& =
3 \left( \frac{ \mathrm{d} }{ \mathrm{d} n } ((1 - 2 n) \ln (n-1)) \right) + 4 \left( \frac{ \mathrm{d} }{ \mathrm{d} n } ((n-1)^{\frac{3}{2}}) \right) + \left( \frac{ \mathrm{d} }{ \mathrm{d} n } \left( \frac{ n (n-2) }{ \sqrt{n-1} } \right) \right) + 1
\\
& =
3 \left( \frac{ \mathrm{d} }{ \mathrm{d} n } (1 - 2 n) \right) \ln (n-1) + 3 (1 - 2 n) \left( \frac{ \mathrm{d} }{ \mathrm{d} n } (\ln (n-1)) \right) + 4 \left( \frac{ \mathrm{d} }{ \mathrm{d} n } ((n-1)^{\frac{3}{2}}) \right)
\notag \\
& \qquad \qquad \qquad \qquad \qquad \qquad
+ \left( \frac{ \mathrm{d} }{ \mathrm{d} n } (n^{2} - 2n) \right) \frac{1}{\sqrt{n-1}} + n (n-2) \left( \frac{ \mathrm{d} }{ \mathrm{d} n } \left( \frac{1}{\sqrt{n-1}} \right) \right) + 1
\\
& =
3 ( -2 ) \ln (n-1) + 3 (1 - 2 n) \left( \frac{1}{n-1} \right) + 4 \left( \frac{3}{2} (n-1)^{\frac{1}{2}} \right)
\notag \\
& \qquad \qquad \qquad \qquad \qquad \qquad \qquad \qquad
+ (2n - 2) \frac{1}{\sqrt{n-1}} + n (n-2) \left( - \frac{1}{2 (n-1)^{\frac{3}{2}}} \right) + 1
\\
& =
- 6 \ln (n-1) + \frac{3 (1 - 2 n)}{n-1} + 6 \sqrt{n-1} + \frac{2(n-1)}{\sqrt{n-1}} - \frac{n (n-2)}{2 (n-1) \sqrt{n-1} } + 1
\\
& =
- 6 \ln (n-1) + \frac{3 (1 - 2 n)}{n-1} + 6 \sqrt{n-1} + 2 \sqrt{n-1} - \frac{n (n-2)}{2 (n-1) \sqrt{n-1} } + 1
\\
& =
- 6 \ln (n-1) + \frac{3 (1 - 2 n)}{n-1} + 8 \sqrt{n-1} - \frac{n (n-2)}{2 (n-1) \sqrt{n-1} } + 1 ,
\label{eq:diff2_d(n,2)} \\
\frac{ \mathrm{d}^{3} d(n, 2) }{ \mathrm{d} n^{3} }
& =
\frac{ \mathrm{d} }{ \mathrm{d} n } \left( \vphantom{\sum} - 6 \ln (n-1) + \frac{3 (1 - 2 n)}{n-1} + 8 \sqrt{n-1} - \frac{n (n-2)}{2 (n-1) \sqrt{n-1} } + 1 \right)
\\
& =
- 6 \left( \frac{ \mathrm{d} }{ \mathrm{d} n } (\ln (n-1)) \right) + 3 \left( \frac{ \mathrm{d} }{ \mathrm{d} n } \left( \frac{1 - 2 n}{n-1} \right) \right)
\notag \\
& \qquad \qquad \qquad \qquad \qquad \qquad \qquad \qquad
+ 8 \left( \frac{ \mathrm{d} }{ \mathrm{d} n } (\sqrt{n-1}) \right) - \frac{1}{2} \left( \frac{ \mathrm{d} }{ \mathrm{d} n } \left( \frac{n (n-2)}{(n-1) \sqrt{n-1} } \right) \right)
\\
& =
- 6 \left( \frac{1}{n-1} \right) + 3 \left( \frac{ \mathrm{d} }{ \mathrm{d} n } \left( \frac{1 - 2 n}{n-1} \right) \right) + 8 \left( \frac{1}{2\sqrt{n-1}} \right) - \left( \frac{ \mathrm{d} }{ \mathrm{d} n } \left( \frac{n (n-2)}{2 (n-1) \sqrt{n-1} } \right) \right)
\\
& =
- \frac{6}{n-1} + 3 \left( \frac{ \mathrm{d} }{ \mathrm{d} n } \left( \frac{1 - 2 n}{n-1} \right) \right) + \frac{4}{\sqrt{n-1}} - \left( \frac{ \mathrm{d} }{ \mathrm{d} n } \left( \frac{n (n-2)}{2 (n-1) \sqrt{n-1} } \right) \right)
\\
& =
- \frac{6}{n-1} + 3 \left( \frac{ \mathrm{d} }{ \mathrm{d} n } (1 - 2 n) \right) \frac{1}{n-1} + 3 (1 - 2 n) \left( \frac{ \mathrm{d} }{ \mathrm{d} n } \left( \frac{1}{n-1} \right) \right) + \frac{4}{\sqrt{n-1}}
\notag \\
& \qquad \qquad \qquad
- \left( \frac{ \mathrm{d} }{ \mathrm{d} n } (n^{2} - 2n) \right) \frac{1}{2 (n-1) \sqrt{n-1}} - n (n-2) \left( \frac{ \mathrm{d} }{ \mathrm{d} n } \left( \frac{1}{2 (n-1) \sqrt{n-1}} \right) \right)
\\
& =
- \frac{6}{n-1} + 3 ( - 2 ) \frac{1}{n-1} + 3 (1 - 2 n) \left( - \frac{1}{(n-1)^{2}} \right) + \frac{4}{\sqrt{n-1}}
\notag \\
& \quad
- ( 2n - 2 ) \frac{1}{2 (n-1) \sqrt{n-1}} - n (n-2) \left( - \frac{1}{(2 (n-1) \sqrt{n-1})^{2}} \left( \frac{ \mathrm{d} }{ \mathrm{d} n } (2 (n-1) \sqrt{n-1}) \right) \right)
\\
& =
- \frac{6}{n-1} - \frac{6}{n-1} + \frac{3 (2 n - 1)}{(n-1)^{2}} + \frac{4}{\sqrt{n-1}}
\notag \\
& \qquad \qquad
- \frac{1}{\sqrt{n-1}} + \frac{n (n-2)}{4 (n-1)^{3}} \left( \left( \frac{ \mathrm{d} }{ \mathrm{d} n } (2 (n-1)) \right) \sqrt{n-1} + 2 (n-1) \left( \frac{ \mathrm{d} }{ \mathrm{d} n } (\sqrt{n-1}) \right) \right)
\\
& =
- \frac{12}{n-1} + \frac{3 (2 n - 1)}{(n-1)^{2}} + \frac{3}{\sqrt{n-1}} + \frac{n (n-2)}{4 (n-1)^{3}} \left( (2) \sqrt{n-1} + 2 (n-1) \left( \frac{1}{2\sqrt{n-1}} \right) \right)
\\
& =
- \frac{12}{n-1} + \frac{3 (2 n - 1)}{(n-1)^{2}} + \frac{3}{\sqrt{n-1}} + \frac{n (n-2)}{4 (n-1)^{3}} ( 2 \sqrt{n-1} + \sqrt{n-1} )
\\
& =
- \frac{12}{n-1} + \frac{3 (2 n - 1)}{(n-1)^{2}} + \frac{3}{\sqrt{n-1}} + \frac{3 n (n-2)}{4 (n-1)^{2} \sqrt{n-1}}
\\
& =
3 \left( - \frac{4}{n-1} + \frac{2 n - 1}{(n-1)^{2}} + \frac{1}{\sqrt{n-1}} + \frac{n (n-2)}{4 (n-1)^{2} \sqrt{n-1}} \right)
\\
& =
\frac{ 3 }{ 4 } \left( \frac{ - 16 (n-1) \sqrt{n-1} + 4 (2n-1) \sqrt{n-1} + 4 (n-1)^{2} + n (n-2) }{ (n-1)^{2} \sqrt{n-1} } \right)
\\
& =
\frac{ 3 }{ 4 } \left( \frac{ (- 16 (n-1) + 4 (2n-1)) \sqrt{n-1} + (4n^{2} - 8 n + 4) + (n^{2} - 2n) }{ (n-1)^{2} \sqrt{n-1} } \right)
\\
& =
\frac{ 3 }{ 4 } \left( \frac{ (- 16 n + 16 + 8 n - 4) \sqrt{n-1} + (5 n^{2} - 10 n + 4) }{ (n-1)^{2} \sqrt{n-1} } \right)
\\
& =
\frac{ 3 }{ 4 } \left( \frac{ (- 8 n + 12) \sqrt{n-1} + 5 n (n - 2) + 4 }{ (n-1)^{2} \sqrt{n-1} } \right)
\\
& =
\frac{ 3 }{ 4 } \left( \frac{ - 4 (2 n - 3) \sqrt{n-1} + 5 n (n - 2) + 4 }{ (n-1)^{2} \sqrt{n-1} } \right)
\\
& =
\frac{ 3 }{ 4 } \left( \frac{ - 4 (2 n - 3) (\sqrt{n-1} - 1) - 4 (2 n - 3) + 5 n (n - 2) + 4 }{ (n-1)^{2} \sqrt{n-1} } \right)
\\
& =
\frac{ 3 }{ 4 } \left( \frac{ - 2 (2 n - 3) \ln_{(\frac{1}{2})} (n-1) - 4 (2 n - 3) + 5 n (n - 2) + 4 }{ (n-1)^{2} \sqrt{n-1} } \right)
\\
& \overset{\text{(a)}}{\ge}
\frac{ 3 }{ 4 } \left( \frac{ - 2 (2 n - 3) \ln_{0} (n-1) - 4 (2 n - 3) + 5 n (n - 2) + 4 }{ (n-1)^{2} \sqrt{n-1} } \right)
\\
& =
\frac{ 3 }{ 4 } \left( \frac{ - 2 (2 n - 3) ((n-1) - 1) - 4 (2 n - 3) + 5 n (n - 2) + 4 }{ (n-1)^{2} \sqrt{n-1} } \right)
\\
& =
\frac{ 3 }{ 4 } \left( \frac{ - 2 (2 n - 3) (n-2) - 4 (2 n - 3) + 5 n (n - 2) + 4 }{ (n-1)^{2} \sqrt{n-1} } \right)
\\
& =
\frac{ 3 }{ 4 } \left( \frac{ (- 2 (2 n - 3) + 5n) (n-2) - 4 ((2 n - 3) - 1) }{ (n-1)^{2} \sqrt{n-1} } \right)
\\
& =
\frac{ 3 }{ 4 } \left( \frac{ (- 4 n + 6 + 5n) (n-2) - 4 (2 n - 4) }{ (n-1)^{2} \sqrt{n-1} } \right)
\\
& =
\frac{ 3 }{ 4 } \left( \frac{ (n + 6) (n-2) - 8 (n-2) }{ (n-1)^{2} \sqrt{n-1} } \right)
\\
& =
\frac{ 3 }{ 4 } \left( \frac{ ((n+6) - 8) (n-2) }{ (n-1)^{2} \sqrt{n-1} } \right)
\\
& =
\frac{ 3 }{ 4 } \left( \frac{ (n-2)^{2} }{ (n-1)^{2} \sqrt{n-1} } \right)
\\
& =
\begin{cases}
> 0
& \mathrm{if} \ n \ge 3 , \\
= 0
& \mathrm{if} \ n = 2 ,
\end{cases}
\label{eq:diff3_d(n,2)}
\end{align}
where (a) holds with equality if and only if $n = 2$ since $\ln_{\alpha} x \ge \ln_{\beta} x$ for $\alpha < \beta$ and $x \in (0, +\infty)$ with equality if and only if $x = 1$ (see Lemma \ref{lem:IT_ineq}).
Note that
\begin{align}
\left. \frac{ \mathrm{d} d(n, 2) }{ \mathrm{d} n } \right|_{n = 2}
& \overset{\eqref{eq:diff1_d(n,2)}}{=}
\left. \left( \frac{ 3 (1 - 2 n) \sqrt{n-1} \, \ln (n-1) + 5 n (n - 2) + 4 + (n - 6) \sqrt{n-1} }{ \sqrt{n-1} } \right) \right|_{n = 2}
\\
& =
\frac{ 3 (1 - 4) \sqrt{1} \, \ln (1) + 10 (0) + 4 + (-4) \sqrt{1} }{ \sqrt{1} }
\\
& =
4 - 4
\\
& =
0 ,
\label{eq:diff1_d(2,2)} \\
\left. \frac{ \mathrm{d}^{2} d(n, 2) }{ \mathrm{d} n^{2} } \right|_{n = 2}
& \overset{\eqref{eq:diff2_d(n,2)}}{=}
\left. \left( \vphantom{\sum} - 6 \ln (n-1) + \frac{3 (1 - 2 n)}{n-1} + 8 \sqrt{n-1} - \frac{n (n-2)}{2 (n-1) \sqrt{n-1} } + 1 \right) \right|_{n = 2}
\\
& =
\underbrace{ - 6 \ln 1 }_{ = 0 } + \frac{3 (1 - 4)}{1} + 8 \sqrt{1} - \underbrace{ \frac{2 (2-2)}{2 (2-1) \sqrt{1} } }_{ = 0 } + 1
\\
& =
-9 + 8 + 1
\\
& =
0 .
\label{eq:diff2_d(2,2)}
\end{align}
Since
\begin{itemize}
\item
$\frac{ \mathrm{d}^{2} d(n, 2) }{ \mathrm{d} n^{2} }$ is strictly increasing for $n \ge 2$ (see Eq. \eqref{eq:diff3_d(n,2)}) and
\item
$\left. \frac{ \mathrm{d}^{2} d(n, 2) }{ \mathrm{d} n^{2} } \right|_{n = 2} = 0$ (see Eq. \eqref{eq:diff2_d(2,2)}),
\end{itemize}
we get that
\begin{align}
\frac{ \mathrm{d}^{2} d(n, 2) }{ \mathrm{d} n^{2} } \ge 0
\label{eq:diff2_d(n,2)_p}
\end{align}
for $n \ge 2$ with equality if and only if $n = 2$.
Moreover, since
\begin{itemize}
\item
$\frac{ \mathrm{d} d(n, 2) }{ \mathrm{d} n }$ is strictly increasing for $n \ge 2$ (see Eq. \eqref{eq:diff2_d(n,2)_p}) and
\item
$\left. \frac{ \mathrm{d} d(n, 2) }{ \mathrm{d} n } \right|_{n = 2} = 0$ (see Eq. \eqref{eq:diff1_d(2,2)}),
\end{itemize}
we also get that
\begin{align}
\frac{ \mathrm{d} d(n, 2) }{ \mathrm{d} n } \ge 0
\label{eq:diff1_d(n,2)_p}
\end{align}
for $n \ge 2$ with equality if and only if $n = 2$.
Therefore, since
\begin{itemize}
\item
$d(n, 2)$ is strictly increasing for $n \ge 2$ (see Eq. \eqref{eq:diff1_d(n,2)_p}) and
\item
$d(2, 2) = 0$ (see Eq. \eqref{eq:d(2,2)}),
\end{itemize}
we obtain $d(n, 2) > 0$ for $n \ge 3$.

So far, we derived that
\begin{itemize}
\item
if $n \ge 3$, then the following monotonicity hold (see Eq. \eqref{eq:diff1_d_sign}):
\begin{itemize}
\item
$d(n, r)$ is strictly decreasing for $r \in (-\infty, \frac{1}{2}]$,
\item
$d(n, r)$ is strictly increasing for $r \in [\frac{1}{2}, \frac{1}{2} + \frac{ W_{0} \! \left( - \mathrm{e}^{\ln \sqrt{n-1} - \sqrt{n-1}} \right) + \sqrt{n-1} }{ \ln (n-1) }]$, and
\item
$d(n, r)$ is strictly decreasing for $r \in [\frac{1}{2} + \frac{ W_{0} \! \left( - \mathrm{e}^{\ln \sqrt{n-1} - \sqrt{n-1}} \right) + \sqrt{n-1} }{ \ln (n-1) }, +\infty)$,
\end{itemize}
\item
$d(n, 1) > 0$ for $n \ge 3$, and
\item
$d(n, 2) > 0$ for $n \ge 3$.
\end{itemize}
Using the above statements, we now show that, if $n \ge 3$, then $d(n, r) > 0$ for $1 \le r \le 2$.
Note that
\begin{align}
\frac{1}{2} + \frac{ W_{0} \! \left( - \mathrm{e}^{\ln \sqrt{n-1} - \sqrt{n-1}} \right) + \sqrt{n-1} }{ \ln (n-1) }
& \overset{\text{(a)}}{\ge}
\frac{1}{2} + \frac{ -1 + \sqrt{n-1} }{ \ln (n-1) }
\\
& =
\frac{1}{2} \left( 1 + \frac{ 2 (\sqrt{n-1} - 1) }{ \ln (n-1) } \right)
\\
& =
\frac{1}{2} \left( 1 + \frac{ \ln_{(\frac{1}{2})} (n-1) }{ \ln (n-1) } \right)
\\
& \overset{\text{(b)}}{\ge}
\frac{1}{2} \left( 1 + \frac{ \ln (n-1) }{ \ln (n-1) } \right)
\\
& =
1
\end{align}
for $n \ge 2$, where
\begin{itemize}
\item
(a) holds with equality if and only if $n = 2$ since $\left. W_{0} \! \left( - \mathrm{e}^{\ln \sqrt{n-1} - \sqrt{n-1}} \right) \right|_{n = 2} = -1$ and $W_{0} \! \left( - \mathrm{e}^{\ln \sqrt{n-1} - \sqrt{n-1}} \right)$ is strictly increasing for $n \ge 2$, and
\item
(b) holds with equality if and only if $n = 2$ since $\ln_{\alpha} x \ge \ln_{\beta} x$ for $\alpha < \beta$ and $x \in (0, +\infty)$ with equality if and only if $x = 1$ (see Lemma \ref{lem:IT_ineq}).
\end{itemize}
When $\frac{1}{2} + \frac{ W_{0} \! \left( - \mathrm{e}^{\ln \sqrt{n-1} - \sqrt{n-1}} \right) + \sqrt{n-1} }{ \ln (n-1) } \ge 2$, we readily get that $d(n, r) > 0$ for $n \ge 3$ and $1 \le r \le 2$ since
\begin{itemize}
\item
$d(n, r)$ is strictly increasing for $r \in [1, 2]$,
\item
$d(n, 1) > 0$, and $d(n, 2) > 0$.
\end{itemize}
We next prove that $d(n, r) > 0$ for $n \ge 3$ and $1 \le r \le 2$ when $1 < \frac{1}{2} + \frac{ W_{0} \! \left( - \mathrm{e}^{\ln \sqrt{n-1} - \sqrt{n-1}} \right) + \sqrt{n-1} }{ \ln (n-1) } < 2$.
Consider the value $r^{\prime} > 1$ such that $d(n, r^{\prime}) \le 0$.
Since
\begin{itemize}
\item
$d(n, r)$ is strictly increasing for $r \in [1, \frac{1}{2} + \frac{ W_{0} \! \left( - \mathrm{e}^{\ln \sqrt{n-1} - \sqrt{n-1}} \right) + \sqrt{n-1} }{ \ln (n-1) }]$ and
\item
$d(n, 1) > 0$,
\end{itemize}
the value $r^{\prime}$ must be greater than $\frac{1}{2} + \frac{ W_{0} \! \left( - \mathrm{e}^{\ln \sqrt{n-1} - \sqrt{n-1}} \right) + \sqrt{n-1} }{ \ln (n-1) }$.
However, since
\begin{itemize}
\item
$d(n, r)$ is strictly decreasing for $r \in [\frac{1}{2} + \frac{ W_{0} \! \left( - \mathrm{e}^{\ln \sqrt{n-1} - \sqrt{n-1}} \right) + \sqrt{n-1} }{ \ln (n-1) }, 2]$ and
\item
$d(n, 2) > 0$,
\end{itemize}
the value $r^{\prime}$ must be greater than $2$.
Hence, there does not exist $r^{\prime} \in [1, 2]$ such that $d(n, r^{\prime}) \le 0$.
Therefore, we have
\begin{align}
\sgn \! \left( \vphantom{\sum} d(n, r) \right)
=
1
\label{eq:d_positive}
\end{align}
for $n \ge 3$ and $1 \le r \le 2$.
Summarizing the above results, we obrain
\begin{align}
\sgn \! \left( \vphantom{\sum} g(n, z, \textstyle{\frac{\ln(n-1)}{2 \ln z}}) |_{z = (n-1)^{r}} \right)
& \overset{\eqref{eq:g_ln(n-1)/2ln(z)_(n-1)^r}}{=}
\underbrace{ \sgn \! \left( \frac{ 1 }{ 2 r ((n-1) + (n-1)^{r}) \ln (n-1) } \right) }_{=1} \, \cdot \; \sgn \! \left( \vphantom{\sum} d(n, r) \right)
\\
& \overset{\eqref{eq:d_positive}}{=}
1
\end{align}
for $n \ge 3$ and $1 \le r \le 2$.
Therefore, $g(n, z, \textstyle{\frac{\ln(n-1)}{2 \ln z}})$ is always strictly positive for $n \ge 3$ and $z \in [n-1, (n-1)^{2}]$.
That concludes the poof of Lemma \ref{lem:ln(n-1)/2ln(z)}.
\end{IEEEproof}

\end{document}